\documentclass[11 pt, a4paper]{article}
\usepackage[a4paper, hmargin=1.25in,vmargin=1.25in]{geometry}
\usepackage{amsmath,amssymb,amsthm,mathrsfs,amsfonts,mathtools,calculator,dsfont,bm} 
\usepackage[dvipsnames,svgnames, x11names]{xcolor}
\usepackage{natbib,bibentry,bibunits,lmodern}
\usepackage{grffile} 
\usepackage[bottom,hang,flushmargin]{footmisc} 
\usepackage{enumitem} 
\newlist{problems}{enumerate}{1}
\setlist[problems]{label={\arabic*.}, ref={\arabic{part}.\thechapter.\arabic*}}
\usepackage{pdfpages,selectp} 
\usepackage[hyphens]{url}
\usepackage[labelsep=period]{caption} 
\usepackage{subcaption}
\usepackage{fancyvrb} 
\usepackage{environ,trimspaces,xspace}
\usepackage[titletoc]{appendix}
\usepackage{fancyhdr,setspace,titlesec,titling} 
\usepackage{siunitx,booktabs,longtable,threeparttable,multirow,makecell,float,rotating} 
\usepackage{tikz,pgfplots} 
\pgfplotsset{compat=1.11}
\usetikzlibrary{arrows,intersections,plotmarks}
\usetikzlibrary{backgrounds,positioning,fit,petri}
\usetikzlibrary{mindmap,trees,calendar,external,shapes}
\usetikzlibrary{decorations.fractals,decorations.pathmorphing,shadows,shadings,patterns}
\usetikzlibrary{spy,calc} 
\usepgfplotslibrary{groupplots}
\usepackage[bookmarksnumbered,bookmarksopen,psdextra,unicode,hyperindex,breaklinks,hypertexnames=false]{hyperref}
\usepackage[open,openlevel=2,numbered,atend]{bookmark} 
\usepackage{tabularx}
\usepackage{cleveref}
\renewcommand{\thetable}{\Roman{table}}

\definecolor{Ocean}{rgb}{0,0,0.75}
\makeatletter
\renewcommand\paragraph{\@startsection{paragraph}{4}{\z@}%
            {-2.5ex\@plus -1ex \@minus -.25ex}%
            {1.25ex \@plus .25ex}%
            {\normalfont\normalsize\bfseries}}
\makeatother
\hypersetup{colorlinks, linkcolor = {NavyBlue}, citecolor = {NavyBlue}, urlcolor ={NavyBlue}}

\newtheorem{thm}{Theorem}[section]

\newtheorem{lem}{Lemma}[section]
\newtheorem{pro}{Proposition}[section]

\newtheorem{ass}{Assumption}[section]
\theoremstyle{definition}

\newtheorem{rem}{Remark}[section]

\theoremstyle{definition}
\newtheorem{ex}{Example}[section]
\newtheorem*{ex*}{Example}

\newtheoremstyle{exctd}
{\topsep} {\topsep}%
{\upshape}
{}
{\bfseries\scshape}
{.}
{1em}
{\thmname{#1} \thmnumber{ #2}\thmnote{#3} (cont.)}
\theoremstyle{exctd}
\newtheorem*{exctd}{Example}

\newcommand{\amin}{\operatornamewithlimits{arg\,min}}

\newcommand{\transpose}{\text{\scalebox{0.8}{$\intercal$}}}

\newcommand{\convl}{\xrightarrow{L}}

\newcommand{\convp}{\xrightarrow{p}}

\pgfmathdeclarefunction{npdf}{3}{%
\pgfmathparse{1/(#3*sqrt(2*pi))*exp(-((#1-#2)^2)/(2*#3^2))}%
}
\tikzset{
    declare function={
        ncdf(\x,\m,\s)=1/(1 + exp(-0.07056*((\x-\m)/\s)^3 - 1.5976*(\x-\m)/\s));
    }
}


\defaultbibliographystyle{D:/Dropbox/Common/Latex/ecta}
\defaultbibliography{D:/Dropbox/Common/Latex/mybibliography}

\def\choice{AllResultsNo}

\begin{document}
\begin{bibunit}
\pdfbookmark[1]{Title}{title}
\title{A Projection Framework for Testing Shape Restrictions That Form Convex Cones\thanks{We thank three anonymous referees for valuable suggestions that have helped greatly improve this paper. We are indebted to Andres Santos for extensive discussions and constant support. We thank Xiaohong Chen, Denis Chetverikov, Stefan Hoderlein, Tatiana Komarova, Scott Kostyshak, Simon Lee, Sungwon Lee, Ulrich K. M\"{u}ller, Pepe Montiel Olea, Alex Poirier, Brandon Reeves, Kevin Song, Yanglei Song, Xun Tang and numerous seminar participants for their useful comments. We also thank Denis Chetverikov and Simon Lee for sharing their code. Advanced computing resources and consultation provided by Texas A\&M High Performance Research Computing and Marinus Pennings are gratefully acknowledged. A previous version of this paper was circulated under the title ``A General Framework for Inference on Shape Restrictions.''}
}
\author{
Zheng Fang \\ Department of Economics \\ Texas A\&M University\\ zfang@tamu.edu
\and
Juwon Seo\\ Department of Economics\\National University of Singapore\\ ecssj@nus.edu.sg}
\date{First draft: October 16, 2019\\ This draft: May 6, 2021}
\maketitle

\begin{abstract}
This paper develops a uniformly valid and asymptotically nonconservative test based on projection for a class of shape restrictions. The key insight we exploit is that these restrictions form convex cones, a simple and yet elegant structure that has been barely harnessed in the literature. Based on a monotonicity property afforded by such a geometric structure, we construct a bootstrap procedure that, unlike many studies in nonstandard settings, dispenses with estimation of local parameter spaces, and the critical values are obtained in a way as simple as computing the test statistic. Moreover, by appealing to strong approximations, our framework accommodates nonparametric regression models as well as distributional/density-related and structural settings. Since the test entails a tuning parameter (due to the nonstandard nature of the problem), we propose a data-driven choice and prove its validity. Monte Carlo simulations confirm that our test works well.
\end{abstract}


\begin{center}
\end{center}

\newpage

\section{Introduction}

Shape restrictions play a number of fundamental and prominent roles in economics. For example, they often arise as testable implications of economic theory, and may thus serve as plausible restrictions in specifying economic models \citep{Varian1982Demand,Varian1984Production}. In complementary empirical work, they help achieve point identification, tighten identification bounds, improve estimation precision, and develop powerful tests---see \citet{Matzkin1994Handbook} and \citet{ChetverikovSantosAzeem2018Shape} for more detailed discussions.

In this paper, we develop a uniformly valid and asymptotically nonconservative test for a class of shape restrictions. To illustrate, consider the regression model:
\begin{align}\label{Eqn: NP, intro}
Y = \theta_0(Z)+u~,
\end{align}
where $Z\in [0,1]$, $\theta_0:[0,1]\to\mathbf R$, and $E[u|Z]=0$. Suppose that we are interested in testing whether $\theta_0$ is nondecreasing. More formally, if $\theta_0\in \mathbf H\equiv \{\theta:[0,1]\to\mathbf R: \|\theta\|_{\mathbf H}\equiv \{\int_{[0,1]} |\theta(z)|^2\,\mathrm dz\}^{1/2}<\infty\}$ (a mild restriction) and $\Lambda$ is the class of nondecreasing functions in $\mathbf H$, then we may formulate the hypotheses as
\begin{align}\label{Eqn: hypotheses, projection}
\mathrm H_0: \phi(\theta_0)=0 \qquad\qquad \mathrm{vs.}\qquad\qquad \mathrm H_1: \phi(\theta_0)>0~,
\end{align}
where $\phi(\theta)\equiv\min_{\lambda\in\Lambda}\|\theta-\lambda\|_{\mathbf H}$ is the distance from $\theta$ to $\Lambda$. Thus, given an unconstrained (kernel or sieve) estimator $\hat\theta_n$ of $\theta_0$, we may employ a test that rejects $\mathrm H_0$ if $r_n\phi(\hat\theta_n)$ is ``large'' for a suitable $r_n\to\infty$. While conceptually intuitive, construction of critical values turns out to be a delicate and challenging matter. In particular, despite the well-established results on the rate $r_n$ and pointwise asymptotic normality, $r_n\{\hat\theta_n-\theta_0\}$ generically does not converge as a process indexed by $[0,1]$ \citep{ChernozhukovLeeRosen2013Intersection}, rendering the Delta method as in \citet{FangSantos2018HDD} inapplicable.

As a first step, we note that $\Lambda$ being a (closed) convex cone\footnote{By definition, $\Lambda$ is a convex cone if and only if $af+bg\in\Lambda$ whenever $a,b\ge 0$ and $f,g\in\Lambda$.} implies
\begin{align}\label{Eqn: test statistic}
r_n\phi(\hat\theta_n)=\|r_n\{\hat\theta_n-\theta_0\}+r_n\theta_0-\Pi_\Lambda(r_n\{\hat\theta_n-\theta_0\}+r_n\theta_0)\|_{\mathbf H}~,
\end{align}
where $\theta\mapsto\Pi_{\Lambda}(\theta)\equiv\amin_{\lambda\in\Lambda}\|\theta-\lambda\|_{\mathbf H}$ is the projection operator, i.e., $\Pi_{\Lambda}(\theta)$ is the closest (under $\|\cdot\|_{\mathbf H}$) nondecreasing function to $\theta$. Hence, \eqref{Eqn: test statistic} reveals that, in estimating the law of $r_n\phi(\hat\theta_n)$ in order to obtain critical values, it suffices to quantify the variation of $r_n\{\hat\theta_n-\theta_0\}$ and estimate the drift $r_n\theta_0$. Despite the lack of convergence (in $\mathbf H$), $r_n\{\hat\theta_n-\theta_0\}$ as a process may be approximated in law via strong approximations by a number of methods with only mild computation cost, including the simulation method in \citet{ChernozhukovLeeRosen2013Intersection}, the weighted bootstrap in \citet{BelloniChernozhukovChetverikovKato2015New}, and the sieve score bootstrap in \citet{ChenChristensen2018SupNormOptimal}. The treatment of $r_n\theta_0$, on the other hand, consists of the nonstandard step because, as in moment inequality models \citep{AndrewsSoares2010},  $r_n\theta_0$ cannot be consistently estimated in general. In this regard, the convex cone property (but not convexity alone) implies
\begin{align}\label{Eqn: stochastic domiance, intro}
r_n\phi(\hat\theta_n)\le \|r_n\{\hat\theta_n-\theta_0\}+\kappa_n\theta_0-\Pi_\Lambda(r_n\{\hat\theta_n-\theta_0\}+\kappa_n\theta_0)\|_{\mathbf H}~,
\end{align}
whenever $0\le \kappa_n\le r_n$. Hence, a valid critical value may be obtained by bootstrapping the upper bound in \eqref{Eqn: stochastic domiance, intro}, which is possible because $\kappa_n\theta_0$ may be consistently estimated by $\kappa_n\hat\theta_n$ provided $\kappa_n/r_n\to 0$. The very virtue that \eqref{Eqn: stochastic domiance, intro} is an inequality rather than equality may also raise concern for conservativeness of the resulting test. As an extreme case, the choice $\kappa_n=0$ leads to a least favorable test that may be viewed as assuming $\theta_0=0$, suggesting that $\kappa_n$ should not be too small. Along these lines, we develop a data-driven choice of $\kappa_n$ that delivers an asymptotically nonconservative test.


While we started with monotonicity in the univariate model \eqref{Eqn: NP, intro}, the main features of our test are not confined to this special problem. First, the convex cone property is in fact shared by a class of restrictions, e.g., nonnegativity, concavity, Slutsky restriction and supermodularity---see Appendix \ref{App: convex cone}. Regrettably, our framework is not directly applicable absent the property, e.g., quasi-concavity, log-concavity, and $r$-concavity \citep{Kostyshak2017U,KomarovaHidalgo2019Shape}. Second, our framework is applicable to distributional and structural settings \citep{PinkseSchurter2019Auction,Bhattacharya2020Empirical}---see Appendix \ref{App: convex cone}. Third, our framework allows for jointly testing multiple restrictions since intersections of convex cones remain convex cones. This is important because it is common that shape restrictions arise simultaneously. Fourth, while some convex cone restrictions (e.g., monotonicity and concavity) on $\theta_0$ can be characterized as inequalities on derivatives of $\theta_0$, we dispense with derivative estimation because it likely incurs power loss due to a slower convergence rate \citep{Stone1982Global,ChenChristensen2018SupNormOptimal}. Finally, our test does not rely on the least favorable configurations, while being asymptotically nonconservative (thus leading to improved power) and computationally tractable. We stress that, since the norm $\|\cdot\|_{\mathbf H}$ is of $L^2$ nature, our test cannot be inverted to obtain uniform confidence bands. In this sense, our paper complements \citet{HorowitzLee2017Shape}, \citet{FreybergerReeves2017Shape} and \citet{ChenChernozhukovFernandezKostyshakLuo2021Shape}.

The literature on shape restrictions was initiated in the 1950s \citep{BBBB1972}, with much attention since focused on estimation under solely shape restrictions---see, e.g., \citet{HanWangChatterjeeSamworth2019Isotonic} and references therein. There are also post-processing methods that enforce restrictions on unconstrained estimators---see \citet{ChenChernozhukovFernandezKostyshakLuo2021Shape} who studied a number of shape enforcing operators. The projection method in this paper is of post-processing nature. The convex cone structure was recognized in the 1960s as a device to generalize monotone regression, though the focus is on {\it analytic} properties of projections \citep{BBBB1972}. For testing, the structure has barely been exploited beyond identifying the least favorable distributions in parametric settings \citep{Wolak1987Exact,SilvapulleSen2005Constrained}, deriving minimax bounds in univariate Gaussian white noise models \citep{JuditskyNemirovski2002Shape}, and establishing minimaxity for the likelihood ratio test in Gaussian sequence models \citep{WeiWainwrightGuntuboyina2019Geometry}.


Much of the testing literature has been developed by exploiting {\it particular} structures of restrictions, with concavity and especially monotonicity in univariate models being the primary focuses. Despite a sizable literature, existing tests prevalently rely on the least favorable configurations, including \citet{GijbelsHallJonesKoch2000Monotonicity}, \citet{GhosalSenVaart2000Monotonicity}, \citet{HallHeckman2000Monotonicity}, \citet{Durot2003testing} and \citet{Gutknecht2016MonotoneTest} for monotonicity, and \citet{AbrevayaJiang2005Curvature} for concavity---see also
\citet{DumbgenSpokoiny2001Multiscale} and \citet{BaraudHuetLaurent2005Convex} who devised tests for both shapes. \citet{Chetverikov2018Monotonicity} developed, as far as we are aware, the first nonparametric uniformly valid tests designed specifically for monotonicity that avoid the least favorable configurations.


There is also a strand of literature motivated by the {\it common} structure of shape restrictions viewed as inequalities, including \citet{ChernozhukovNeweySantos2019CCMM}, \citet{LeeSongWhang2017Inequal},  \citet{BelloniChernozhukovChetverikovFernandez2019QR} and \citet{Zhu2020Shape}. If the inequalities are based on derivatives, the derivative estimation may then incur power loss. Moreover, all these (uniformly valid) tests, except the least favorable test of \citet{BelloniChernozhukovChetverikovFernandez2019QR}, involve estimation of local parameter spaces. \citet{ChernozhukovNeweySantos2019CCMM} and \citet{Zhu2020Shape}, unlike our setup, allowed for partial identification by working with moment restrictions. By the virtue of their setup, they required estimating the set of minimizers and the use of strong approximations may entail stronger regularity conditions. Similar in spirit to \citet{ChernozhukovNeweySantos2019CCMM} and \citet{Zhu2020Shape}, \citet{KomarovaHidalgo2019Shape} proposed a moment-based test in the univariate model \eqref{Eqn: NP, intro} for shape restrictions that may not form convex cones.


The remainder of the paper is structured as follows. Section \ref{Sec: setup} introduces the setup and some motivating examples. Section \ref{Sec: framework} presents our inferential framework, a data-driven choice of the tuning parameter, and an implementation guide. Section \ref{Sec: simulations} conducts simulation studies, while Section \ref{Sec: conclusion} concludes. Appendix \ref{Sec: main results proof} contains proofs of the main results. Due to space limitation, we relegate the discussions of the convex cone property, an investigation of the special case when $\hat\theta_n$ admits an asymptotic distribution, and presentations of some auxiliary results to the online supplement.

\section{The Setup and Examples}\label{Sec: setup}

Throughout, we denote by $\{X_i\}_{i=1}^n$ the sample with each $X_i$ living in some sample space $\mathcal X$, and by $P$ the joint law of $\{X_i\}_{i=1}^n$ that belongs to some family $\mathbf P$ of distributions on $\mathcal X^n$. The dependence of $P$ and $\mathbf P$ on $n$ is suppressed for notational simplicity. We stress that, under this configuration, $\{X_i\}_{i=1}^n$ need not be i.i.d. In turn, we let $\theta_0$ be the parameter of interest, and, whenever appropriate, make the dependence of $\theta_0$ on $P$ explicit by instead writing $\theta_P$. In order to accommodate Slutsky restriction, we shall work with an abstract Hilbert space (i.e., a complete inner product space) with inner product $\langle\cdot,\cdot\rangle_{\mathbf H}$ and induced norm $\|\cdot\|_{\mathbf H}$. Given the sample $\{X_i\}_{i=1}^n$, our objective is then to test whether $\theta_0$ satisfies the shape in question, i.e., the hypotheses in \eqref{Eqn: hypotheses, projection}. The setup \eqref{Eqn: hypotheses, projection} induces two models: $\mathbf P_0\equiv\{P\in\mathbf P: \phi(\theta_P)=0\}$ and $\mathbf P_1\equiv\mathbf P\backslash\mathbf P_0$.

Before proceeding further, we introduce additional notation. Set $\mathbf R_+\equiv\{x\in\mathbf R: x\ge 0\}$ and $L^2(\mathcal Z)\equiv\{f: \mathcal Z\to\mathbf R: \int_{\mathcal Z} |f(z)|^2\,\mathrm dz<\infty\}$ for $\mathcal Z\subset\mathbf R^{d_z}$. Let $\mathbf M^{m\times k}$ be the space of $m\times k$ matrices, and, for $A\in\mathbf M^{m\times k}$, write its transpose by $A^\transpose$, its trace by $\mathrm{tr}(A)$ if $m=k$, its Moore–Penrose inverse by $A^-$, and its Frobenius norm by $\|A\|\equiv\{\mathrm{tr}(A^\transpose A)\}^{1/2}$. For a sequence $\{h_j\}$ of functions, denote the vector $(h_1,\ldots,h_k)^\transpose$ by $h^k$. For generic families of distributions $\mathbf P_n$, a sequence $\{a_n\}$ of positive scalars, and a sequence $\{\mathbb X_n\}$ of random elements in a normed space $\mathbf D$ with norm $\|\cdot\|_{\mathbf D}$, write $\mathbb X_n=o_p(a_n)$ uniformly in $P\in\mathbf P_n$ if $\lim_{n\to\infty}\sup_{P\in\mathbf P_n}P(\|\mathbb X_n\|_{\mathbf D}>a_n\epsilon)= 0$ for any $\epsilon>0$, and $\mathbb X_n=O_p(a_n)$ uniformly in $P\in\mathbf P_n$ if $\lim_{M\to\infty}\limsup_{n\to\infty}\sup_{P\in\mathbf P_n}P(\|\mathbb X_n\|_{\mathbf D}>Ma_n)= 0$.

\subsection{Examples}\label{Sec: examples}

We now present examples where shape restrictions play important roles. The first example is concerned with nonparametric regression models.

\begin{ex}[Nonparametric Regression]\label{Ex: NP under shape}
Let $X\equiv(Y,Z)\in\mathbf R^{1+d_z}$ satisfy
\begin{align}\label{Eqn: Ex, NP under shape}
Y=\theta_0(Z)+u~,
\end{align}
where $\theta_0: \mathcal Z\subset\mathbf R^{d_z}\to\mathbf R$ with $\mathcal Z$ the support of $Z$, and $E[u|Z]=0$. Here, one may set $\mathbf H=L^2(\mathcal Z)$, let $\Lambda\subset \mathbf H$ consist of (say) monotonic functions, and obtain an unconstrained estimator $\hat\theta_n$ of $\theta_0$ by kernel methods such as local constant/linear/polynomial regression or sieve methods with various basis functions such as splines \citep{ChernozhukovLeeRosen2013Intersection}. The rate $r_n$ equals $(nh_n^{d_z})^{1/2}$ for kernel estimation with bandwidth $h_n$, and $(n/k_n)^{1/2}$ for sieve estimation based on, e.g., B-splines, with $k_n$ (henceforth) the sieve dimension. One may also set up this example as one on the conditional mean $z\mapsto E[Y|Z=z]$. \qed
\end{ex}

Our second example generalizes \eqref{Eqn: Ex, NP under shape} to its instrumental variable (IV) analog.

\begin{ex}[Nonparametric IV Regression]\label{Ex: NPIV under shape}
Let $X\equiv(Y,Z,V)\in\mathbf R^{1+d_z+d_v}$ satisfy \eqref{Eqn: Ex, NP under shape} but with $E[u|V]=0$. Then one may set $\mathbf H$ and $\Lambda$ as in Example \ref{Ex: NP under shape}, and employ the series two-stage least square (2SLS) estimation with, e.g., B-splines, resulting in a rate $r_n$ equal to $(n/k_n)^{1/2} s_n$, where $s_n$ is the smallest singular value of $E[b^{m_n}(V)h^{k_n}(Z)^\transpose]$ with $h^{k_n}$ and $b^{m_n}$ respectively $k_n\times 1$ and $m_n\times 1$ vectors (with $m_n\ge k_n$) of B-splines for $\theta_0$ and the instrument space \citep{NeweyPowell2003NPIV,BlundellChenKristensen2007Engel,ChenChristensen2018SupNormOptimal}. In practice, $s_n$ is unknown but can be replaced with its sample analog. One may conceivably also employ kernel-type estimators \citep{HallHorowitz2005NPIV,DarollesFanFlorensRenault2011NPIV}, though the strong approximation results appear to be lacking.  \qed
\end{ex}

The third example is concerned with nonparametric quantile regression.

\begin{ex}[Nonparametric Quantile Regression]\label{Ex: NPQR under shape}
Let $X\equiv(Y,Z)\in\mathbf R^{1+d_z}$ satisfy
\begin{align}\label{Eqn: Ex, NPQR under shape}
Y=\theta_0(Z,U)~,
\end{align}
where $\theta_0:\mathbf R^{d_z}\times[0,1]\to\mathbf R$, and $U$ is uniformly distributed on $[0,1]$. If $Z$ and $U$ are independent, then $\theta_0$ may be interpreted as the conditional quantile function of $Y$ given $Z$. Here, one may define $\mathbf H$ and $\Lambda$ as previously, and employ the recent sieve estimator of \citet{BelloniChernozhukovChetverikovFernandez2019QR} with $r_n=(n/k_n)^{1/2}$ for, e.g., B-splines.  \qed
\end{ex}

Our final example concerns a restriction in a possibly endogenous regression model.

\begin{ex}[Slutsky Restriction]\label{Ex: Slutsky}
Let $Q\in\mathbf R^{d_q}$ (quantities), $P\in\mathbf R^{d_q}$ (prices), $Y\in\mathbf R$ (income), and $Z\in\mathbf R^{d_z}$ (demographics)  satisfy the following $d_q$ equations:
\begin{align}
Q=g_0(P,Y)+\Gamma_0^\transpose Z + U~,
\end{align}
where $g_0: \mathbf R_+^{d_q+1}\to\mathbf R^{d_q}$ is differentiable, $\Gamma_0\in\mathbf M^{d_z\times d_q}$, and $U\in\mathbf R^{d_q}$ is the error term. The Slutsky matrix of $g_0$ is the mapping $\theta_0: \mathbf R_+^{d_q+1}\to\mathbf M^{d_q\times d_q}$ defined by:
\begin{align}\label{Eqn: Slutsky matrix in ex}
\theta_0(p,y)\equiv \frac{\partial g_0(p,y)}{\partial p^\transpose}+ \frac{\partial g_0(p,y)}{\partial y} g_0(p,y)^\transpose~.
\end{align}
The Slutsky restriction refers to $\theta_0(p,y)$ being negative semidefinite (nsd) at each pair $(p,y)$. For this example, we follow \citet{AguiarSerrano2017Slutsky} by letting $\mathbf H$ be the space of functions $\theta: \mathbf R_+^{d_q+1}\to \mathbf M^{d_q\times d_q}$ with inner product $\langle \cdot,\cdot\rangle_{\mathbf H}$ defined by
\begin{align}\label{Eqn: Slutsky norm in ex}
\langle \theta_1,\theta_2\rangle_{\mathbf H}\equiv \int \mathrm{tr}(\theta_1(p,y)^\transpose\theta_2(p,y))\, \mathrm dp\mathrm dy~,\,\forall\,\theta_1,\theta_2\in\mathbf H~,
\end{align}
and $\Lambda$ the family of mappings $\theta\in\mathbf H$ such that $\theta(p,y)$ is nsd and symmetric for all $(p,y)$. In turn, the nonlinear functional $\theta_0$ may be estimated based on the plug-in principle and a sieve estimator of $g_0$ \citep{DonaldNewey1994Semilinear,BlundellChenKristensen2007Engel}, resulting in a rate $r_n$ as in Example \ref{Ex: NP under shape} (without endogeneity) or \ref{Ex: NPIV under shape} (with endogeneity). \qed
\end{ex}

We refer the reader to Appendix \ref{App: convex cone} that illustrates how our framework also applies to distributional settings where shape restrictions often take the form of various dominance relations and structural models where shape restrictions may serve as testable implications. For ease of reference, we shall call settings where $\hat\theta_n$ admits an asymptotic distribution (in $\mathbf H$) {\it regular} and ones without this property {\it irregular}.


\section{The Inferential Framework}\label{Sec: framework}

\subsection{The General Framework}\label{Sec: general theory}

We commence with our main assumptions in this paper.


\begin{ass}\label{Ass: convex setup}
(i) $\Lambda$ is a known nonempty closed convex set in a Hilbert space $\mathbf H$ with known inner product $\langle \cdot,\cdot\rangle_{\mathbf H}$ and induced norm $\|\cdot\|_{\mathbf H}$; (ii) $\Lambda$ is a cone.
\end{ass}

\begin{ass}\label{Ass: strong approx}
(i) An estimator $\hat\theta_n: \{X_i\}_{i=1}^n\to\mathbf H$ satisfies $\|r_n\{\hat\theta_n-\theta_P\}-\mathbb Z_{n,P}\|_{\mathbf H}=o_p(c_n)$ uniformly in $P\in\mathbf P$ for some $r_n\to\infty$, $\mathbb Z_{n,P}\in\mathbf H$, and $c_n>0$ with $c_n=O(1)$; (ii) $\hat{\mathbb G}_n\in\mathbf H$ is a bootstrap estimator satisfying: $\|\hat{\mathbb G}_{n}-\bar{\mathbb Z}_{n,P}\|_{\mathbf H}=o_p(c_n)$ uniformly in $P\in\mathbf P$, for $\bar{\mathbb Z}_{n,P}$ a copy of $\mathbb Z_{n,P}$ that is independent of $\{X_i\}_{i=1}^n$.
\end{ass}

Assumption \ref{Ass: convex setup} simply abstracts the convex cone feature, where we single out the conic condition for ease of elucidating the roles it plays in this paper. Assumption \ref{Ass: strong approx}(i) requires that $r_n\{\hat\theta_n-\theta_P\}$ be approximated by $\mathbb Z_{n,P}$ (uniformly) at a rate $c_n$. In regular settings, it suffices to have $\sqrt n\{\hat\theta_n-\theta_P\}\convl \mathbb G_P$ uniformly in $P\in\mathbf P$, so that $r_n=\sqrt n$, and $\mathbb Z_{n,P}$ equals $\mathbb G_P$ in law---see Appendix \ref{Sec: special case}. In irregular settings such as Example \ref{Ex: NP under shape}, one may obtain $\hat\theta_n$ by kernel or sieve methods. To illustrate, let $\{Y_i,Z_i\}_{i=1}^n$ be a sample generated by \eqref{Eqn: Ex, NP under shape}, and $h^{k_n}$ be a $k_n\times 1$ vector of B-splines on $\mathcal Z$. Then the sieve estimator of $\theta_P$ is given by $\hat\theta_n=\hat\beta_n^{\transpose}h^{k_n}$ with $\hat\beta_n= [\sum_{i=1}^{n}h^{k_n}(Z_i)h^{k_n}(Z_i)^{\transpose}]^{-}\sum_{i=1}^{n}h^{k_n}(Z_i)Y_i$. Under regularity conditions, one may obtain the linear expansion in $L^2(\mathcal Z)$:
\begin{align}\label{Eqn: implementation, aux1}
r_n \{\hat\theta_n-\theta_P\}=k_n^{-1/2}(h^{k_n})^\transpose (E_P[h^{k_n}(Z)h^{k_n}(Z)^\transpose])^{-1}\frac{1}{\sqrt n}\sum_{i=1}^nh^{k_n}(Z_i)u_i+o_p(\frac{1}{\log n})~,
\end{align}
uniformly in $P\in\mathbf P$. Here, undersmoothing is required in order to deliver \eqref{Eqn: implementation, aux1}. Following \citet{ChernozhukovLeeRosen2013Intersection}, one may then verify Assumption \ref{Ass: strong approx}(i) with $r_n=\sqrt{n/k_n}$, $c_n=1/\log n$, and $\mathbb Z_{n,P}= k_n^{-1/2}(h^{k_n})^\transpose (E_P[h^{k_n}(Z)h^{k_n}(Z)^\transpose])^{-1} G_{n,P}$ for some $G_{n,P}\sim N(0,E_P[u^2h^{k_n}(Z)h^{k_n}(Z)^\transpose])$ . The rate $c_n$ serves to cope with potential degeneracy of the test statistic, an issue inherent in the nonstandard nature of the problem.

Assumption \ref{Ass: strong approx}(ii) demands an analogous approximation for the bootstrap. In regular settings, typically $\hat{\mathbb G}_n=\sqrt n \{\hat\theta_n^*-\hat\theta_n\}$, where $\hat\theta_n^*$ is the same as $\hat\theta_n$ but based on samples drawn from the original data. In Example \ref{Ex: NP under shape}, the expansion \eqref{Eqn: implementation, aux1} suggests
\begin{align}\label{Eqn: implementation, aux2}
\hat{\mathbb G}_n=k_n^{-1/2}(h^{k_n})^\transpose (\frac{1}{n}\sum_{i=1}^nh^{k_n}(Z_i)h^{k_n}(Z_i)^\transpose)^{-1}\frac{1}{\sqrt n}\sum_{i=1}^nW_ih^{k_n}(Z_i)\hat u_i~.
\end{align}
where $\hat u_i\equiv Y_i-\hat\theta_n(Z_i)$ and $\{W_i\}_{i=1}^n$ are (scalar) weights (e.g., standard normals). The verification of Assumption \ref{Ass: strong approx} for Example \ref{Ex: NP under shape} represents a general strategy: establishing an asymptotic linear expansion of $r_n\{\hat\theta_n-\theta_P\}$, and then verifying Assumption \ref{Ass: strong approx} based on the linear term---see \citet{ChernozhukovLeeRosen2013Intersection}, \citet{ChernozhukovNeweySantos2019CCMM}, \citet{ChenChristensen2018SupNormOptimal}, \citet{BelloniChernozhukovChetverikovFernandez2019QR}, and \citet{LiLiao2020Uniform} for more illustrations. We stress that Assumption \ref{Ass: strong approx}(ii) leaves the particular form of $\hat{\mathbb G}_n$ unspecified, and thus accommodates alternative resampling schemes.

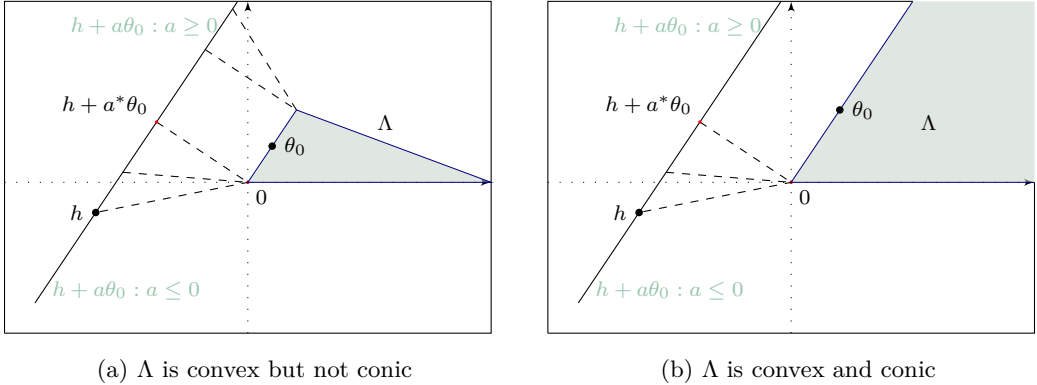
\begin{figure}
\centering \scriptsize
\hspace*{\fill}
\begin{subfigure}[b]{0.45\linewidth}
\begin{tikzpicture}[scale=0.8, >=latex',dot/.style={circle,inner sep=1pt,fill,name=#1}]
\draw (-4,-2.5)--(4,-2.5) -- (4,3) -- (-4,3) -- (-4,-2.5);
\draw[loosely dotted,->] (-4,0)--(4,0);
\draw[loosely dotted,->] (0,-2.5)--(0,3);
\fill[red] (0,0) circle (0.8pt);
\draw (0,0) node[anchor=north west] {$0$};
\fill[fill=Honeydew3,semitransparent] (0,0)--(4/5,6/5)--(4,0)--(0,0);
\draw[line width=0.07mm,NavyBlue] (0,0)--(4/5,6/5)--(4,0)--(0,0);
\draw (2,1) node[anchor=west] {$\Lambda$};
\node[dot=theta,label={east: $\theta_0$}] at (2/5,3/5) {};
\node[dot=h,label={west:$h$}] at (-2.5,-0.5)  {};
\node (zero) at (-5/3,-2.5)  {};
\draw (-3.5,-2) -- (-1/6,3);
\draw (-1.7,2.3) node[anchor=south, semitransparent,SeaGreen4] {$h+a\theta_0:a\ge 0$};
\draw (-2,-1.5) node[anchor=north, semitransparent,SeaGreen4] {$h+a\theta_0:a\le 0$};
\draw[dashed] (h)--(0,0);
\draw[dashed] (-37/18,1/6)--(0,0);
\draw[dashed] (-3/2,1)--(0,0);
\draw[dashed] (-7/10,11/5)--(4/5,6/5);
\draw[dashed] (-1/4,23/8)--(4/5,6/5);
\fill[red] (-3/2,1) circle (0.8pt);
\draw (-3/2,1)  node[anchor=south east] {$h+a^*\theta_0$};
\end{tikzpicture}
\caption{$\Lambda$ is convex but not conic}\label{Fig: convex not cone}
\end{subfigure}
\hspace*{\fill}
\begin{subfigure}[b]{0.45\linewidth}
\begin{tikzpicture}[scale=0.8, >=latex',dot/.style={circle,inner sep=1pt,fill,name=#1}]
\draw (-4,-2.5)--(-4,3)--(2,3);
\draw (-4,-2.5)--(4,-2.5) -- (4,0);
\draw[loosely dotted,->] (-4,0)--(4,0);
\draw[loosely dotted,->] (0,-2.5)--(0,3);
\fill[red] (0,0) circle (0.8pt);
\draw (0,0) node[anchor=north west] {$0$};
\fill[fill=Honeydew3,semitransparent] (0,0)--(2,3)--(4,3)--(4,0)--(0,0);
\draw[line width=0.07mm,NavyBlue] (0,0)--(2,3);
\draw[line width=0.07mm,NavyBlue] (0,0)--(4,0);
\draw (2,1) node[anchor=west] {$\Lambda$};
\node[dot=theta,label={east: $\theta_0$}] at (4/5,6/5) {};
\node[dot=h,label={west:$h$}] at (-2.5,-0.5)  {};
\node (zero) at (-5/3,-2.5)  {};
\draw (-3.5,-2) -- (-1/6,3);
\draw (-1.7,2.3) node[anchor=south, semitransparent,SeaGreen4] {$h+a\theta_0:a\ge 0$};
\draw (-2,-1.5) node[anchor=north, semitransparent,SeaGreen4] {$h+a\theta_0:a\le 0$};
\draw[dashed] (h)--(0,0);
\draw[dashed] (-37/18,1/6)--(0,0);
\draw[dashed] (-3/2,1)--(0,0);
\fill[red] (-3/2,1) circle (0.8pt);
\draw (-3/2,1)  node[anchor=south east] {$h+a^*\theta_0$};
\end{tikzpicture}
\caption{$\Lambda$ is convex and conic}\label{Fig: convex cone}
\end{subfigure}
\hspace*{\fill}
\caption{$\psi_{a,P}(h)\equiv  \|h+a\theta_P-\Pi_\Lambda(h+a\theta_P)\|_{\mathbf H}$ is weakly decreasing in $a\in[0,\infty)$ if $\Lambda$ is convex and conic as in Figure \ref{Fig: convex cone}, but may not be so if it is convex but not conic as in Figure \ref{Fig: convex not cone}.}
\label{Fig: monotonicity}
\end{figure}

The next lemma lays out a number of important building blocks for our development.

\begin{lem}\label{Lem: main}
(i) If Assumption \ref{Ass: convex setup} holds, then any $\hat\theta_n\in\mathbf H$ and $r_n\in\mathbf R_+$ satisfy
\begin{align}
r_n\phi(\hat\theta_n)=\|r_n\{\hat\theta_n-\theta_P\}+r_n\theta_P-\Pi_\Lambda(r_n\{\hat\theta_n-\theta_P\}+r_n\theta_P)\|_{\mathbf H}~. \label{Eqn: test statistic, rep}
\end{align}
(ii) If Assumption \ref{Ass: convex setup} holds, $P\in\mathbf P_0$, and $\kappa_n\in[0,r_n]$, then it follows that
\begin{align}
r_n\phi(\hat\theta_n)\le \|r_n\{\hat\theta_n-\theta_P\}+\kappa_n\theta_P-\Pi_\Lambda(r_n\{\hat\theta_n-\theta_P\}+\kappa_n\theta_P)\|_{\mathbf H}~. \label{Eqn: stochastic domiance}
\end{align}
(iii) If Assumption \ref{Ass: strong approx}(i) holds, $\sup_{P\in\mathbf P} E[\|\mathbb Z_{n,P}\|_{\mathbf H}]<\infty$ uniformly in $n$, and $\kappa_n/r_n=o(c_n)$, then we have that $\kappa_n\hat\theta_n-\kappa_n\theta_P= o_p(c_n)$ uniformly in $P\in\mathbf P$.
\end{lem}

Lemma \ref{Lem: main}(i) highlights the standard, and critically, the nonstandard features of the problem. Specifically, while $r_n\{\hat\theta_n-\theta_P\}$ may be approximated in law by $\hat{\mathbb G}_n$ due to Assumption \ref{Ass: strong approx}, $r_n\theta_P$ cannot be consistently estimated in general, i.e., $r_n\hat\theta_n-r_n\theta_P\neq o_p(1)$. Lemma \ref{Lem: main}(ii) suggests that we may obtain critical values by instead estimating the law of the upper bound in \eqref{Eqn: stochastic domiance}. This is possible because $\kappa_n\theta_P$ can be consistently estimated by $\kappa_n\hat\theta_n$ as shown by Lemma \ref{Lem: main}(iii). We stress that \eqref{Eqn: stochastic domiance} is implied by the convex cone property but not convexity alone---see Figure \ref{Fig: monotonicity}.

To formalize the above discussions, we define maps $\psi_{a,P},\hat\psi_{\kappa_n}:\mathbf H\to\mathbf R$ by
\begin{align}\label{Eqn: psi function}
\psi_{a,P}(h)&\equiv  \|h+a\theta_P-\Pi_\Lambda(h+a\theta_P)\|_{\mathbf H}~,\\
\hat\psi_a(h)&\equiv \|h+a\Pi_\Lambda\hat\theta_n-\Pi_\Lambda(h+a\Pi_\Lambda\hat\theta_n)\|_{\mathbf H}~.
\end{align}
Thus, $\psi_{\kappa_n,P}(r_n\{\hat\theta_n-\theta_P\})$ is precisely the upper bound in \eqref{Eqn: stochastic domiance}, and $\hat\psi_{\kappa_n}(\hat{\mathbb G}_n)$ is its bootstrap analog in which the null hypothesis is enforced through $\Pi_\Lambda\hat\theta_n$ to improve power. Finally, for a significance level $\alpha\in(0,1)$, we define our critical value
\begin{align}\label{Eqn: critical value}
\hat c_{n,1-\alpha}\equiv \inf\{c\in\mathbf R: P(\hat\psi_{\kappa_n}(\hat{\mathbb G}_n)\le c|\{X_i\}_{i=1}^n)\ge 1-\alpha\}~.
\end{align}
As known in the literature \citep{ChernozhukovNeweySantos2019CCMM}, the validity of $\hat c_{n,1-\alpha}$, viewed as a mapping from distributions to the real line, additionally demands a suitable continuity condition, in accord with the continuous mapping theorem. To this end, we let $c_{n,P}(1-\alpha)$ be the $(1-\alpha)$-quantile of $\psi_{\kappa_n,P}(\mathbb Z_{n,P})$ and impose the following:


\begin{ass}\label{Ass: Gaussian}
(i) $\mathbb Z_{n,P}$ is tight and centered Gaussian in $\mathbf H$ for each $n\in\mathbf N$ and $P\in\mathbf P$; (ii) $\sup_{P\in\mathbf P} E[\|\mathbb Z_{n,P}\|_{\mathbf H}]<\infty$ uniformly in $n$; (iii) $c_{n,P}(1-\alpha-\varpi)\ge c_{n,P}(0.5)+\varsigma_n$ for some constant $\varpi,\varsigma_n>0$, each $n$ and $P\in\mathbf P_0$; (iv) $c_n/\varsigma_n^2=O(1)$ as $n\to\infty$.
\end{ass}

Assumption \ref{Ass: Gaussian}(i) formalizes Gaussianity and tightness of each $\mathbb Z_{n,P}$, which is fulfilled in Examples \ref{Ex: NP under shape}--\ref{Ex: Slutsky} as well as most regular settings. Assumption \ref{Ass: Gaussian}(ii) is a mild moment condition that, in our examples, is tantamount to uniform boundedness of eigenvalues of some matrices with growing dimensions. Assumptions \ref{Ass: Gaussian}(iii)(iv) are tied to the natures of densities of Gaussian functionals. Together with Assumptions \ref{Ass: convex setup} and \ref{Ass: Gaussian}(i)-(ii), they in effect amount to the aforementioned continuity condition.

We now state the first main result concerning the size of our test.

\begin{thm}\label{Thm: size control}
Let Assumptions \ref{Ass: convex setup}, \ref{Ass: strong approx}, and \ref{Ass: Gaussian} hold. If $0\le\kappa_n/r_n=o(c_n)$, then
\begin{align}\label{Thm: size control, aux1}
\limsup_{n\to\infty}\sup_{P\in\mathbf P_0} P( r_n \phi(\hat\theta_n)> \hat c_{n,1-\alpha})\le \alpha~,
\end{align}
and, for $\bar{\mathbf P}_0\equiv\{P\in\mathbf P_0: \langle\, \vartheta,\theta_P\rangle_{\mathbf H}=0\,\,\forall\,\vartheta\in\mathbf H\,\,\mathrm{ s.t.}\,\, \sup_{\lambda\in\Lambda}\langle\vartheta,\lambda\rangle_{\mathbf H}\le 0\}$,
\begin{align}\label{Thm: size control, aux2}
\limsup_{n\to\infty}\sup_{P\in\bar{\mathbf P}_0} |P( r_n \phi(\hat\theta_n)> \hat c_{n,1-\alpha})-\alpha|=0~.
\end{align}
\end{thm}

In addition to size control, Theorem \ref{Thm: size control} shows that the limiting rejection rate equals the nominal level, uniformly over $\bar{\mathbf P}_0$. Heuristically, $\bar{\mathbf P}_0$ may be viewed as consisting of the least favorable distributions in $\mathbf P_0$. Indeed, one can show that (i) $\bar{\mathbf P}_0=\{P\in\mathbf P_0: \theta_P=0\}$ if $\Lambda=\mathbf R_+$, and (ii) $\bar{\mathbf P}_0$ contains constant (resp.\ linear) functions if $\Lambda$ is the family of monotone (resp.\ concave) functions in $L^2(\mathcal Z)$. We stress that, the choice $\kappa_n=0$ leads to a least favorable test that amounts to assuming $\theta_P=0$ (or $\theta_P\in\bar{\mathbf P}_0$ by Lemma \ref{Lem: nonconservativeness}). As well documented, least favorable tests, while controlling size uniformly, can be substantially conservative, which has motivated the active development of more powerful tests in nonstandard settings \citep{AndrewsSoares2010,LeeSongWhang2017Inequal}. Intuitively, in view of Lemma \ref{Lem: main}(ii), it is desirable to have $\kappa_n\to\infty$ (to match $r_n\to\infty$), a condition recurrent in nonstandard problems for the sake of nonconservativeness---see \citet{FangSantos2018HDD} and Appendix \ref{Sec: special case}.

We emphasize that our test is in general asymptotically nonsimilar, i.e., there may exist a sequence $\{P_n\}$ of distributions from the null such that
\begin{align}\label{Eqn: similarity}
\liminf_{n\to\infty}P_n(r_n\phi(\hat\theta_n)>\hat c_{n,1-\alpha})<\alpha~.
\end{align}
However, this is, in our view, neither an evidence against nonconservativeness nor a deficiency of our test, analogous to the one-sided $t$-test of $\mathrm H_0: \theta_0\le 0$ whose rejection rates tend to zero at $\theta_0<0$. At a deeper level, \eqref{Eqn: similarity} is in line with the fact that similarity is not a desirable criterion in nonstandard settings, and can lead to tests with very poor power \citep{Lehmann1952Multi,Andrews2012Similar}. Indeed, many powerful tests in these settings are asymptotically nonsimilar---see, e.g., \citet{AndrewsSoares2010}, \citet{LeeSongWhang2017Inequal} and \citet{Chetverikov2018Monotonicity}.

Turning to the power of our test, we define $\mathbf P_{1,n}^\Delta\equiv\{P\in\mathbf P_1: \phi(\theta_P)\ge \Delta/r_n\}$ for $\Delta>0$. The next theorem shows that our test has nontrivial power against $\mathbf P_{1,n}^\Delta$.

\begin{thm}\label{Thm: uniform power}
If Assumptions \ref{Ass: convex setup}, \ref{Ass: strong approx}, and \ref{Ass: Gaussian}(ii) hold and $\kappa_n\ge 0$, then
\begin{align}\label{Thm: uniform power eqn2}
\liminf_{\Delta\to\infty}\liminf_{n\to\infty}\inf_{P\in\mathbf P_{1,n}^\Delta} P( r_n \phi(\hat\theta_n)> \hat c_{n,1-\alpha})=1~.
\end{align}
\end{thm}

If we employ the kernel estimator in Example \ref{Ex: NP under shape}, then Theorem \ref{Thm: uniform power} predicts that our test is powerful against Pitman drifts of order $(nh_n^{d_z})^{-1/2}$. In comparison, the test of \citet{LeeSongWhang2017Inequal} is powerful against drifts of order $(nh_n^{2\nu})^{-1/2}$ or $(nh_n^{d_z/2+2\nu})^{-1/2}$ depending on the drifts, where $\nu=1$ for monotonicity and $\nu=2$ for concavity. Thus, for monotonicity, our convergence rate is faster if $d_z=1$ but may be slower if $d_z>4$; for concavity, our rate is faster if $d_z<4$ but may be slower if $d_z>8$. Theorem \ref{Thm: uniform power} also suggests that $\kappa_n$ affects power through higher order terms, and, we reiterate that asymptotic nonconservativeness requires $\kappa_n\to\infty$ subject to $\kappa_n/r_n=o(c_n)$.

\subsection{Selection of the Tuning Parameter}\label{Sec: tuning parameter}


To motivate, we note that Lemma \ref{Lem: main}(iii) implies: for any $\epsilon>0$,
\begin{align}
P(\frac{\kappa_n}{r_n}\|r_n\{\hat\theta_n-\theta_P\}\|_{\mathbf H}\le c_n\epsilon)=P(\|r_n\{\hat\theta_n-\theta_P\}\|_{\mathbf H}\le \frac{r_nc_n}{\kappa_n}\epsilon)\to 1~.
\end{align}
This suggests that we could choose $\kappa_n$ to be such that $r_nc_n/\kappa_n$ is the $(1-\gamma_n)$-quantile of $\|r_n\{\hat\theta_n-\theta_P\}\|_{\mathbf H}$ with $\gamma_n\downarrow 0$, or the $(1-\gamma_n)$ conditional quantile of $\|\hat{\mathbb G}_n\|_{\mathbf H}$ (given $\{X_i\}_{i=1}^n$) since $\|r_n\{\hat\theta_n-\theta_P\}\|_{\mathbf H}$ is unknown. Formally, we let $\hat\kappa_n\equiv r_nc_n/\hat\tau_{n,1-\gamma_n}$, where
\begin{align}\label{Eqn: tuning parameter, data driven}
\hat\tau_{n,1-\gamma_n}\equiv \inf\{c\in\mathbf R: P(\|\hat{\mathbb G}_n\|_{\mathbf H}\le c|\{X_i\}_{i=1}^n)\ge 1-\gamma_n\}~.
\end{align}
To justify the construction $\hat\kappa_n$, we need to introduce our final assumption.

\begin{ass}\label{Ass: tuning}
$\liminf_{n\to\infty}\inf_{P\in\mathbf P_0} \bar\sigma_{n,P}^2>0$ with $\bar\sigma_{n,P}^2\equiv \sup_{h\in\mathbf H: \|h\|_{\mathbf H}\le 1} E[\langle h,\mathbb Z_{n,P}\rangle_{\mathbf H}^2]$.
\end{ass}

Heuristically, Assumption \ref{Ass: tuning} requires that the coupling variable $\mathbb Z_{n,P}$ for $\hat\theta_n$ be asymptotically non-degenerate. In turn, we can now verify the validity of $\hat\kappa_n$.

\begin{pro}\label{Pro: tuning parameter}
Let Assumptions \ref{Ass: strong approx}, \ref{Ass: Gaussian}(i)-(ii), and \ref{Ass: tuning} hold, and set $\hat\kappa_n\equiv r_nc_n/\hat\tau_{n,1-\gamma_n}$ with $\gamma_n\in(0,1)$ and $\hat\tau_{n,1-\gamma_n}$ as in \eqref{Eqn: tuning parameter, data driven}. If $\gamma_n\to 0$, then $\hat\kappa_n/r_n = o_p(c_n)$ uniformly in $P\in\mathbf P_0$. If $(r_nc_n)^{-2}\log\gamma_n\to 0$, then $\hat\kappa_n\convp\infty$ uniformly in $P\in\mathbf P_0$.
\end{pro}

Analogous rate conditions on $\gamma_n$ have appeared in parametric settings---see \citet{ChenFang2016Rank} and references therein. In nonparametric settings, the practice of obtaining data-driven tuning parameters through quantile estimation has appeared in \citet{ChernozhukovLeeRosen2013Intersection}, \citet{ChernozhukovNeweySantos2019CCMM}, and \citet{FangSantos2018HDD}, though a formal theory seems to be lacking. While the choice of $\gamma_n$ remains technically challenging, the situation somewhat improves because $\gamma_n$ is unit/scale-free, and prior studies such as \citet{FangSantos2018HDD} and \citet{ChenFang2016Rank} have shown that finite sample results are often insensitive to the choice of $\gamma_n$, as also confirmed in our simulations. We recommend $\gamma_n=0.01/\log n$ or $1/n$ for practical implementations.

Finally, the rate $c_n$ (in $\hat\kappa_n$) may be ignored in regular settings, and set to be $1/\log n$ in irregular settings without endogeneity. In Example \ref{Ex: NPIV under shape}, we may let $c_n$ be $(\log k_n)^{-\varsigma}$ for some $\varsigma\in[1/2,1]$ with the sieve 2SLS estimation and $k_n$ basis functions for $\theta_0$ \citep{ChenChristensen2018SupNormOptimal}. We note that $\kappa_n$ affects the critical value $\hat c_{n,1-\alpha}$ and hence the rejection rates monotonically: a smaller (resp.\ larger) $\kappa_n$ leads to less (resp.\ more) rejections---see Lemma \ref{Lem: projection monotonicity}. As a result, if one is uncertain about $k_n$ or $\varsigma$, he/she could simply take $c_n=1/\log n$, thereby only making $\hat c_{n,1-\alpha}$ larger.


\subsection{Implementation and Practical Issues}\label{Sec: implementation}


We next provide a guide for implementing our test. Computation of projections (needed in {\sc Steps} 1 and 2 below) will be discussed in the end.

\noindent\underline{\sc Step 1:} Compute the test statistic $r_n\phi(\hat\theta_n)=r_n\|\hat\theta_n-\Pi_\Lambda(\hat\theta_n)\|_{\mathbf H}$.

This step requires an unconstrained estimator $\hat\theta_n$ of $\theta_P$, which may be obtained by standard estimation procedures---see Examples \ref{Ex: NP under shape}-\ref{Ex: Slutsky} for specific estimators and their rates $r_n$. We stress that our framework imposes no additional structures on $\hat\theta_n$ as far as implementation is concerned. With $\hat\theta_n$ and its projection $\Pi_\Lambda(\hat\theta_n)$ in hand, one may approximate $r_n\phi(\hat\theta_n)$ by, e.g., the trapezoid rule in Examples \ref{Ex: NP under shape}--\ref{Ex: Slutsky} where the $\|\cdot\|_{\mathbf H}$-norm takes the form of an integral. Concretely, in Example \ref{Ex: NP under shape} with $\mathcal Z=[a,b]$ and $a<b$, we approximate $\phi(\hat\theta_n)$ by: for a large $N$ and $\Delta=(b-a)/N$,
\begin{align}\label{Eqn: trapezoid}
\{\frac{1}{2}\frac{b-a}{N} [f^2(z_0)+2f^2(z_1)+\cdots +2f^2(z_{N-1})+f^2(z_N)]\}^{1/2}~,
\end{align}
where $f=\hat\theta_n-\Pi_\Lambda(\hat\theta_n)$ and $z_j=a+j\Delta$.


\noindent\underline{\sc Step 2:} Construct the critical value $\hat c_{n,1-\alpha}$ defined in \eqref{Eqn: critical value}.

The construction requires a bootstrap analog $\hat{\mathbb G}_n$ of $r_n \{\hat\theta_n-\theta_P\}$ and a tuning parameter $\kappa_n$. In regular settings, one often has $\hat{\mathbb G}_n=\sqrt n \{\hat\theta_n^*-\hat\theta_n\}$, where $\hat\theta_n^*$ is the same as $\hat\theta_n$ but based on bootstrap samples. In Examples \ref{Ex: NP under shape}--\ref{Ex: Slutsky}, one may employ various bootstrap schemes in \citet{ChernozhukovLeeRosen2013Intersection}, \citet{ChernozhukovNeweySantos2019CCMM}, \citet{ChenChristensen2018SupNormOptimal}, and \citet{BelloniChernozhukovChetverikovFernandez2019QR}. For $\kappa_n$, we recommend $\hat\kappa_n$ proposed in Section \ref{Sec: tuning parameter} with a suitable $\gamma_n$ (e.g., $\gamma_n=0.01/\log n$ or $1/n$). Summarizing, the critical value $\hat c_{n,1-\alpha}$ may be then obtained as follows:
\begin{enumerate}[label=(\roman*)]
  \item Generate $B$ realizations $\{\hat{\mathbb G}_{n,b}\}_{b=1}^B$ of $\hat{\mathbb G}_n$ (e.g., $B=200$ or larger); e.g., in \eqref{Eqn: implementation, aux2}, generate $\{\{W_{i,b}\}_{i=1}^n\}_{b=1}^B$ that are i.i.d.\ across both $i$ and $b$, and then obtain each $\hat{\mathbb G}_{n,b}$ by evaluating $\hat{\mathbb G}_n$ at $\{W_{i,b}\}_{i=1}^n$, which only involves linear calculations.
  \item Set $\hat\kappa_n=r_nc_n/\hat\tau_{n,1-\gamma_n}$ where $\hat\tau_{n,1-\gamma_n}$ is the $(1-\gamma_n)$-quantile of the $B$ numbers $\|\hat{\mathbb G}_{n,1}\|_{\mathbf H},\ldots, \|\hat{\mathbb G}_{n,B}\|_{\mathbf H}$, and $c_n=1/\log n$ for Example \ref{Ex: NP under shape}. The $\|\cdot\|_{\mathbf H}$-norms (here and below) may be computed by the trapezoid rule as in {\sc Step} 1.
  \item Approximate $\hat c_{n,1-\alpha}$ by the $(1-\alpha)$-quantile of the $B$ numbers
\begin{multline}\label{Eqn: implementation, aux3}
\|\hat{\mathbb G}_{n,1}+\hat\kappa_n\Pi_\Lambda\hat\theta_n-\Pi_\Lambda(\hat{\mathbb G}_{n,1}+\hat\kappa_n\Pi_\Lambda\hat\theta_n)\|_{\mathbf H},\\
\ldots, \|\hat{\mathbb G}_{n,B}+\hat\kappa_n\Pi_\Lambda\hat\theta_n-\Pi_\Lambda(\hat{\mathbb G}_{n,B}+\hat\kappa_n\Pi_\Lambda\hat\theta_n)\|_{\mathbf H}~.
\end{multline}
\end{enumerate}

\noindent\underline{\sc Step 3:} Reject $\mathrm H_0$ if and only if $r_n\phi(\hat\theta_n)>\hat c_{n,1-\alpha}$.

\bigskip


Next, we illustrate the computation of projections. As described in Appendix \ref{App: convex cone}, when closed form expressions do not exist, the projection $\Pi_{\Lambda}(\theta)$ can be computed by solving a linearly constrained quadratic program: for some $A\in\mathbf M^{m\times k}$,
\begin{align}\label{Eqn: LCQP}
\min_{h\in\mathbf R^k}\|h-\vartheta\|\qquad \text{s.t.}\quad Ah\ge 0~,
\end{align}
where $\vartheta\in\mathbf R^k$ is the vector of values $\theta$ takes at the grid points, and $Ah\ge 0$ is the discretized version of the restriction in question. As an extensively studied problem, \eqref{Eqn: LCQP} admits polynomial-time algorithms; e.g., the iteration complexity of the interior point method is $O(\sqrt{m+k}\log(1/\epsilon))$ for an $\epsilon$-accurate solution.


Finally, since \eqref{Eqn: LCQP} is inherently more complicated to compute in higher dimensions, we note a number of strategies for ameliorating the situation. First, the recently developed open-source solver OSQP (\url{https://osqp.org}) is very robust in solving large scale quadratic programming problems. Second, while projection under convexity/concavity is computationally demanding in multivariate settings, the representation result in \citet{Kuosmanen2008Convex}, when coupled with the OSQP solver, can help significantly reduce the computation cost. Last, with regressors more than two or three, one may consider employing semiparametric models, instead of fully nonparametric ones, to alleviate the curse of dimensionality.

\section{Simulation Studies}\label{Sec: simulations}

This section examines the finite sample performance of our test. Due to space limitation, we focus on monotonicity, and defer concavity, monotonicity jointly with concavity, and Slutsky restriction to the online appendix. Throughout, the significance level is  5\%, and, unless otherwise specified, the number of Monte Carlo replications is 3000 while the number of bootstrap repetitions for each replication is 200. All integrals are approximated by the trapezoid rule, and quadratic programs are solved by the OSQP solver. Our test is implemented based on the data-driven choice $\hat\kappa_n$ with $\gamma_n\in\{n^{-1/2}, n^{-3/4}, 1/n, 0.1/\log n, 0.05/\log n, 0.01/\log n, 0.1, 0.05, 0.01\}$.


\subsection{Simulation Designs}

We aim to test whether $\theta_0$ is nondecreasing in Example \ref{Ex: NP under shape} with $d_z\in\{1,2\}$. For $d_z=1$, the regression function $\theta_0:[-1,1]\to\mathbf R$ is specified as: for some $(\mathsf a,\mathsf b,\mathsf c)\in\mathbf R^3$,
\begin{align}\label{Eqn: MC1,aux1}
\theta_0(z)=\mathsf az-\mathsf b\varphi(\mathsf cz)~,
\end{align}
where $\varphi$ is the standard normal pdf. We consider three choices for $(\mathsf a,\mathsf b,\mathsf c)$ under the null hypothesis, namely $(0,0,0)$, $(0.1,0.5,0.5)$ and $(0.5,2,1)$ that are labeled D1, D2 and D3 respectively, and the collection $\{(\mathsf a,\mathsf b,\mathsf c): \mathsf a =0,\mathsf b = 0.2\delta,\mathsf c = 5+0.1\delta,\delta=1,\ldots,10\}$ under the alternative---see Figure \ref{Fig: MC1,theta}. We then draw i.i.d.\ samples $\{Z_i^*,u_i\}_{i=1}^n$ with $n\in\{500,750,1000\}$ from the standard normal distribution in $\mathbf R^2$, and set $Z_i=-1+2\Phi(Z_i^*)\in[-1,1]$ with $\Phi$ the standard normal cdf. For $d_z=2$, the regression function $\theta_0:[0,1]^2\to\mathbf R$ is of the form: for some $(\mathsf a,\mathsf b,\mathsf c)\in\mathbf R^3$,
\begin{align}\label{Eqn: MC2,aux1}
\theta_0(z_1,z_2)=\mathsf a\big(\frac{1}{2}z_1^{\mathsf b}+\frac{1}{2}z_2^{\mathsf b}\big)^{1/\mathsf b}+\mathsf c\log(1+z_1+z_2)~.
\end{align}
We consider three choices of $(\mathsf a,\mathsf b,\mathsf c)$ for the null, namely $(0,0,0)$, $(0.2,1,0)$ and $(0.5,0,0.5)$ that are labeled D1, D2, and D3, respectively, and, for the alternative, set $\mathsf b=0$ and $\mathsf a=\mathsf c=-\Delta\delta $ with $\Delta=0.05$ and $\delta=1,\ldots,10$. Note that the first term on the right-hand side of \eqref{Eqn: MC2,aux1} collapses to $\mathsf a\sqrt{z_1z_2}$ whenever $\mathsf b=0$. In turn, we draw i.i.d.\ samples $\{Z_{1i}^*,Z_{2i}^*,u_i\}_{i=1}^n$ with $n\in\{500,750,1000\}$ from the standard normal distribution in $\mathbf R^3$, and set $Z_i\equiv(Z_{1i},Z_{2i})$ with $Z_{ji}=\Phi(Z_{ji}^*)\in[0,1]$ for all $i$ and $j=1,2$.

\begin{figure}
\centering
\scriptsize
\captionsetup[subfigure]{justification=centering}
\begin{subfigure}[b]{0.49\linewidth}
\resizebox{\textwidth}{!}{
\begin{tikzpicture}[baseline,>=latex']
\begin{axis}[axis equal image,         
             domain=-1:1,        
             samples=100,
             xmin=-1.1, xmax=1.1,
             ymin=-1.1, ymax=0.1,
             enlargelimits=false,
             xlabel=$z$,
             x label style={at={(axis description cs:0.5,-0.02)},anchor=north},
             xtick={-1,-0.5,...,1},
             xticklabels={-1,-0.5,...,1},
             xticklabel style={yshift=0.05cm,anchor=south},
             ytick={-1,-0.5,0},
             yticklabels={-1,-0.5,0},
             every axis plot/.append style={semithick},
            ]
\addplot[NavyBlue]{0};
\addplot[NavyBlue,dashed]{0.1*x-0.5*npdf(0.5*x, 0, 1)};
\addplot[NavyBlue,densely dotted]{0.5*x-2*npdf(x, 0, 1)};
\end{axis}
\end{tikzpicture}
}
\caption{$\mathrm H_0$: D1 (solid), D2 (dashed) and D3 (dotted)}\label{Fig: MC1a}
\end{subfigure}
\hspace*{\fill}
\begin{subfigure}[b]{0.49\linewidth}
\resizebox{\textwidth}{!}{
\begin{tikzpicture}[baseline,>=latex']
\begin{axis}[axis equal image,         
             domain=-1:1,        
             samples=100,
             xmin=-1.1, xmax=1.1,
             ymin=-1.1, ymax=0.1,
             enlargelimits=false,
             xlabel=$z$,
             x label style={at={(axis description cs:0.5,-0.02)},anchor=north},
             xtick={-1,-0.5,...,1},
             xticklabels={-1,-0.5,...,1},
             xticklabel style={yshift=0.05cm,anchor=south},
             ytick={-1,-0.5,0},
             yticklabels={-1,-0.5,0},
             every axis plot/.append style={thin},
             no markers,
            ]
\foreach \b in {0.2,0.4,...,2}{\pgfmathsetmacro{\k}{\b*120}
\edef\temp{\noexpand
\addplot+[solid,color=Blue!\k]{-\b*npdf((5+\b/2)*x, 0, 1)};
}\temp
}
\end{axis}
\end{tikzpicture}
}
\caption{$\mathrm H_1$: $\delta=1,2,\ldots,10$ (from top to bottom)}\label{Fig: MC1b}
\end{subfigure}
\hspace*{\fill}
\caption{The function $\theta_0$ in \eqref{Eqn: MC1,aux1} where, in Figure \ref{Fig: MC1b}, $\mathsf a=0$, $\mathsf b = 0.2\delta$, and $\mathsf c=5+0.1\delta$. Note that the standard deviation of the error is designed to be no smaller than the range of $\theta_0$.}
\label{Fig: MC1,theta}
\end{figure}

To implement our test for \eqref{Eqn: MC1,aux1}, we obtain $\hat\theta_n$ by sieve estimation with cubic B-splines with 3, 5, or 7 interior knots at the equispaced quantiles of $\{Z_i\}_{i=1}^n$, so that the sieve dimension $k_n$ equals $7$, $9$, or $11$, respectively. In multivariate settings, we construct the series functions via tensor product of univariate B-splines. Since the sieve dimension grows quickly as $d_z$ increases (e.g., $k_n=49$ with cubic B-splines, $3$ knots, and $d_z=2$), we employ univariate quadratic as well as cubic B-splines with one or zero knots along each dimension for \eqref{Eqn: MC2,aux1}. Thus, for example, $k_n=9$ with quadratic B-splines and zero knots. In both designs, we compute $\hat{\mathbb G}_{n,b}$ as in \eqref{Eqn: implementation, aux2} by drawing i.i.d.\ weights from the standard normal distribution, and let the coupling rate $c_n=1/\log n$. To alleviate the boundary effects, we evaluate the $L^2$-norms (here and below) over $[-0.9,0.9]$ for \eqref{Eqn: MC1,aux1} and $[0.1,0.9]^2$ for \eqref{Eqn: MC2,aux1}, with step size $0.05$. For ease of reference, we label our test with quadratic B-splines and $j$ knots as FS-Q$j$; similarly, FS-C$j$ is the implementation with cubic B-splines and $j$ knots.

To compare, we implement two alternative nonconservative tests, \citet{LeeSongWhang2017Inequal} and \citet{Chetverikov2018Monotonicity}. The latter also compares with some prior tests in simulations, which show marked power superiority of the author's three tests so that we take them as important benchmarks. For brevity, however, we only present the one-step test, labeled C-OS, and note that the results for the other two tests are very similar to those produced by C-OS, as also observed in \citet{Chetverikov2018Monotonicity}. The details for implementing C-OS in the univariate case are clearly laid out in \citet[p.749]{Chetverikov2018Monotonicity}. For the bivariate case, we compute the test statistic as on p.27 in the arXiv version of \citet{Chetverikov2018Monotonicity}, adopt equispaced empirical quantiles $0.1,0.15,...,0.9$ of each covariate as locations for the weighting function to make the computation feasible, and otherwise follow the implementation in the univariate case.

\citet{LeeSongWhang2017Inequal} considered $L^p$-type statistics. For the sake of comparison, we focus on $p =2$, since different choices of $p$ implicitly aim power at different alternatives. We estimate the first derivatives of $\theta_0$ by local quadratic regression \citep[p.59]{FanGijbels1996Local}, with a kernel $z\mapsto 1.5\max\{1-(2z)^2,0\}$ for \eqref{Eqn: MC1,aux1} and $z\mapsto 0.75\max\{1-z^2,0\}$ for \eqref{Eqn: MC2,aux1}. We choose two bandwidths: a ``large'' one $2s_nn^{-1/(d_z+2q+2)}$ (in the spirit of \citet{LeeSongWhang2017Inequal}) and a ``small'' one $2s_nn^{-1/(d_z+2(q-1)+2)}$ with $q=2$ and $s_n$ the standard derivation of $\{Z_i\}_{i=1}^n$, resulting in two tests labeled LSW-L and LSW-S, respectively. As studentization can be crucial for power (based on unreported simulations), we estimate the standard errors following \citet[p.115]{FanGijbels1996Local}, with the variance of the error estimated by a local polynomial regression of order $q+2$ with bandwidth $2s_nn^{-1/(d_z+2(q+2)+2)}$. Next, we construct critical values based on the empirical bootstrap (as in \citet{LeeSongWhang2017Inequal}) and the tuning parameter $\hat c_n$ in their Section 5.1 with $C_{\mathrm{cs}}=0.4$ (since their results are quite insensitive to other choices of $C_{\mathrm{cs}}$ there). Finally, to ease computation for the bivariate designs, the number of simulation replications is decreased to be 1000 (for LSW only).

{
\setlength{\tabcolsep}{6.5pt}
\renewcommand{\arraystretch}{1.1}
\begin{table}[!ht]
\caption{Empirical Size of Monotonicity Tests for $\theta_0$ in \eqref{Eqn: MC1,aux1} at $\alpha=5\%$} \label{Tab: MC1, size1, main} 
\centering\footnotesize
\begin{threeparttable}
\sisetup{table-number-alignment = center, table-format = 1.3} 
\begin{tabularx}{\linewidth}{@{} cc *{3}{S[round-mode = places,round-precision = 3]}  c *{3}{S[round-mode = places,round-precision = 3]}  c *{3}{S[round-mode = places,round-precision = 3]}@{}} 
\hline
\hline
\multirow{2}{*}{$n$} & \multirow{2}{*}{$\gamma_n$}  & \multicolumn{3}{c}{FS-C3: $k_n=7$} & & \multicolumn{3}{c}{FS-C5: $k_n=9$} & & \multicolumn{3}{c}{FS-C7: $k_n=11$}\\
\cline{3-5} \cline{7-9} \cline{11-13}
 & & {D1} & {D2} & {D3}  & & {D1} & {D2} & {D3} & & {D1} & {D2} & {D3}\\
\hline
\multirow{3}{*}{$500$}   & $1/n$                  & 0.0530 & 0.0157 & 0.0030 & & 0.0577 & 0.0207 & 0.0033 & & 0.0583 & 0.0190 & 0.0027\\
                         & $0.01/\log n$          & 0.0530 & 0.0157 & 0.0030 & & 0.0577 & 0.0207 & 0.0033 & & 0.0583 & 0.0190 & 0.0027\\
                         & $0.01$                 & 0.0530 & 0.0157 & 0.0030 & & 0.0577 & 0.0207 & 0.0033 & & 0.0583 & 0.0193 & 0.0027\\
\cline{2-13}
 \multirow{3}{*}{$750$}  & $1/n$                  & 0.0523 & 0.0097 & 0.0013 & & 0.0560 & 0.0137 & 0.0017 & & 0.0590 & 0.0173 & 0.0030\\
                         & $0.01/\log n$          & 0.0523 & 0.0097 & 0.0013 & & 0.0560 & 0.0137 & 0.0017 & & 0.0590 & 0.0173 & 0.0030\\
                         & $0.01$                 & 0.0523 & 0.0097 & 0.0013 & & 0.0560 & 0.0140 & 0.0020 & & 0.0590 & 0.0173 & 0.0030\\
\cline{2-13}
 \multirow{3}{*}{$1000$} & $1/n$                  & 0.0557 & 0.0107 & 0.0007 & & 0.0560 & 0.0113 & 0.0003 & & 0.0560 & 0.0127 & 0.0007\\
                         & $0.01/\log n$          & 0.0557 & 0.0107 & 0.0007 & & 0.0560 & 0.0113 & 0.0003 & & 0.0560 & 0.0127 & 0.0007\\
                         & $0.01$                 & 0.0560 & 0.0107 & 0.0010 & & 0.0560 & 0.0117 & 0.0003 & & 0.0560 & 0.0130 & 0.0007\\
\hline
 \multicolumn{2}{c}{\multirow{2}{*}{$n$}}& \multicolumn{3}{c}{LSW-S} & & \multicolumn{3}{c}{LSW-L} & & \multicolumn{3}{c}{C-OS}\\
\cline{3-5} \cline{7-9} \cline{11-13}
 & & {D1} & {D2} & {D3}  & & {D1} & {D2} & {D3} & & {D1} & {D2} & {D3}\\
\hline
 \multicolumn{2}{c}{$500$}                        & 0.0600 & 0.0413 & 0.0083 & & 0.0660 & 0.0350 & 0.0037 & & 0.0597  & 0.0413  & 0.0120\\
 \multicolumn{2}{c}{$750$}                        & 0.0567 & 0.0363 & 0.0050 & & 0.0587 & 0.0303 & 0.0057 & & 0.0540  & 0.0337  & 0.0080\\
 \multicolumn{2}{c}{$1000$}                       & 0.0610 & 0.0347 & 0.0047 & & 0.0653 & 0.0347 & 0.0033 & & 0.0493  & 0.0357  & 0.0090\\
\hline
\hline
\end{tabularx}
\begin{tablenotes}[flushleft]
\item {\it Note:} The parameter $\gamma_n$ determines $\hat\kappa_n$ proposed in Section \ref{Sec: tuning parameter} with $c_n=1/\log n$ and $r_n=(n/k_n)^{1/2}$.
\end{tablenotes}
\end{threeparttable}
\end{table}
}

\subsection{Simulation Results}

Tables \ref{Tab: MC1, size1, main} and \ref{Tab: MC2, size} summarize the empirical sizes. Due to space limitation, we only present our tests with $\gamma_n\in\{1/n,0.01/\log n,0.01\}$, and relegate to Tables \ref{Tab: MC1Mon, size1, app}--H.\ref{Tab: MC2Mon, size, app} in Appendix \ref{App: full simulations} the complete set of results. Together, these tables show that our tests are insensitive to the choice of $\gamma_n$. For the univariate design, all tests control size well, though, relatively speaking, the two LSW tests, especially LSW-L, tend to over-reject under D1, while our tests tend to under-reject under D2 and D3. Note, however, that D1 is a least favorable case, D2 is in the ``interior'' (not in the topological sense), and D3 is further into the ``interior.'' Thus, the empirical sizes under D2 and D3 are expected to be smaller than 5\%. For the bivariate design, our tests are over-sized under D1 in small samples, though this feature is also shared by LSW-L and C-OS (to an overall lesser extent). The over-rejection of our tests, in particular FS-C1 (in which case $k_n=25$), is likely because the number of series functions is so ``large'' that the Gaussian approximation is somewhat inaccurate---note that the overall situation improves as $n$ increases. Thus, while undersmoothing requires a ``large'' $k_n$, there lies the tension that it should not be ``too large'' for the sake of distributional approximations.

{
\setlength{\tabcolsep}{3pt}
\renewcommand{\arraystretch}{1.1}
\begin{table}[!ht]
\caption{Empirical Size of Monotonicity Tests for $\theta_0$ in \eqref{Eqn: MC2,aux1} at $\alpha=5\%$} \label{Tab: MC2, size}
\centering\footnotesize
\begin{threeparttable}
\sisetup{table-number-alignment = center, table-format = 1.3} 
\begin{tabularx}{\linewidth}{@{}cc *{3}{S[round-mode = places,round-precision = 3]} c *{3}{S[round-mode = places,round-precision = 3]} c *{3}{S[round-mode = places,round-precision = 3]} c *{3}{S[round-mode = places,round-precision = 3]}@{}} 
\hline
\hline
 \multirow{2}{*}{$n$} & \multirow{2}{*}{$\gamma_n$}& \multicolumn{3}{c}{FS-Q0: $k_n=9$} & & \multicolumn{3}{c}{FS-Q1: $k_n=16$} & & \multicolumn{3}{c}{FS-C0: $k_n=16$} & & \multicolumn{3}{c}{FS-C1: $k_n=25$}\\
\cline{3-5} \cline{7-9} \cline{11-13} \cline{15-17}
&& {D1} & {D2} & {D3}  & & {D1} & {D2} & {D3} & & {D1} & {D2} & {D3} & & {D1} & {D2} & {D3}\\  
\hline
\multirow{3}{*}{$500$} & $1/n$         & 0.0613 & 0.0183 & 0.0000 & & 0.0680 & 0.0300 & 0.0017 & & 0.0670 & 0.0303 & 0.0007 & & 0.0830 & 0.0437 & 0.0040\\
                       & $0.01/\log n$ & 0.0613 & 0.0183 & 0.0000 & & 0.0680 & 0.0300 & 0.0017 & & 0.0670 & 0.0303 & 0.0007 & & 0.0830 & 0.0437 & 0.0040\\
                       & $0.01$        & 0.0627 & 0.0193 & 0.0000 & & 0.0687 & 0.0300 & 0.0020 & & 0.0673 & 0.0307 & 0.0007 & & 0.0837 & 0.0447 & 0.0043\\
\cline{2-17}
 \multirow{3}{*}{$750$}& $1/n$         & 0.0583 & 0.0103 & 0.0000 & & 0.0653 & 0.0247 & 0.0003 & & 0.0640 & 0.0220 & 0.0000 & & 0.0737 & 0.0323 & 0.0003\\
                       & $0.01/\log n$ & 0.0583 & 0.0103 & 0.0000 & & 0.0653 & 0.0247 & 0.0003 & & 0.0640 & 0.0220 & 0.0000 & & 0.0737 & 0.0323 & 0.0003\\
                       & $0.01$        & 0.0597 & 0.0107 & 0.0000 & & 0.0660 & 0.0250 & 0.0003 & & 0.0650 & 0.0223 & 0.0000 & & 0.0740 & 0.0327 & 0.0007\\
\cline{2-17}
\multirow{3}{*}{$1000$}& $1/n$        & 0.0500 & 0.0107 & 0.0000 & & 0.0553 & 0.0210 & 0.0000 & & 0.0527 & 0.0203 & 0.0000 & & 0.0590 & 0.0227 & 0.0003\\
                       & $0.01/\log n$ & 0.0500 & 0.0107 & 0.0000 & & 0.0553 & 0.0210 & 0.0000 & & 0.0527 & 0.0203 & 0.0000 & & 0.0590 & 0.0227 & 0.0003\\
                       & $0.01$        & 0.0500 & 0.0107 & 0.0000 & & 0.0560 & 0.0213 & 0.0000 & & 0.0527 & 0.0203 & 0.0000 & & 0.0590 & 0.0230 & 0.0003\\
\hline
 \multicolumn{2}{c}{\multirow{2}{*}{$n$}}  & \multicolumn{3}{c}{LSW-S} & & \multicolumn{3}{c}{LSW-L} & & \multicolumn{3}{c}{C-OS} & & \multicolumn{3}{c}{ }\\
\cline{3-5} \cline{7-9} \cline{11-13}
 &     & {D1} & {D2} & {D3}  & & {D1} & {D2} & {D3} & & {D1} & {D2} & {D3}\\
\hline
\multicolumn{2}{c}{500}                & 0.0430 & 0.0200 & 0.0000 & & 0.0610  & 0.0260  & 0.0020 & & 0.0683  & 0.0607  & 0.0367\\
\multicolumn{2}{c}{750}                & 0.0540 & 0.0170 & 0.0000 & & 0.0660  & 0.0310  & 0.0000 & & 0.0570  & 0.0453  & 0.0230\\
\multicolumn{2}{c}{1000}               & 0.0440 & 0.0210 & 0.0000 & & 0.0540  & 0.0190  & 0.0000 & & 0.0553  & 0.0423  & 0.0187\\
\hline
\hline
\end{tabularx}
\begin{tablenotes}[flushleft]
\item {\it Note:} The parameter $\gamma_n$ determines $\hat\kappa_n$ proposed in Section \ref{Sec: tuning parameter} with $c_n=1/\log n$ and $r_n=(n/k_n)^{1/2}$.
\end{tablenotes}
\end{threeparttable}
\end{table}
}

\pgfplotstableread{ 
delta alpha MonFiveKn3 MonFiveKn5 MonFiveKn7 MonSevenKn3 MonSevenKn5 MonSevenKn7 MonTenKn3 MonTenKn5 MonTenKn7
0     0.05    0.0530     0.0577     0.0583     0.0523      0.0560      0.0590      0.0557   0.0560     0.0560
1     0.05    0.0697     0.0733     0.0707     0.0773      0.0717      0.0727      0.0857   0.0820     0.0780
2     0.05    0.1203     0.1183     0.1117     0.1563      0.1403      0.1337      0.1890   0.1683     0.1510
3     0.05    0.2143     0.2000     0.1800     0.3090      0.2790      0.2500      0.4087   0.3603     0.3220
4     0.05    0.3633     0.3197     0.2940     0.5107      0.4743      0.4337      0.6580   0.6180     0.5663
5     0.05    0.5357     0.4960     0.4467     0.7113      0.6853      0.6350      0.8597   0.8257     0.7837
6     0.05    0.6977     0.6717     0.6147     0.8787      0.8513      0.8190      0.9573   0.9487     0.9337
7     0.05    0.8323     0.8087     0.7667     0.9560      0.9460      0.9303      0.9890   0.9883     0.9823
8     0.05    0.9213     0.9077     0.8810     0.9873      0.9843      0.9773      0.9983   0.9980     0.9963
9     0.05    0.9640     0.9577     0.9440     0.9973      0.9973      0.9950      1.0000   1.0000     1.0000
10    0.05    0.9860     0.9843     0.9773     0.9993      0.9993      0.9993      1.0000   1.0000     1.0000
}\FirstMon

\pgfplotstableread{
delta alpha   Five1  Seven1   Ten1   Five2   Seven2   Ten2    Fifty2
0     0.05   0.0597  0.0540  0.0493  0.0683  0.0570  0.0553   0.0460
1     0.05   0.0623  0.0600  0.0620  0.0760  0.0600  0.0583   0.0653
2     0.05   0.0767  0.0873  0.1067  0.0797  0.0693  0.0690   0.1023
3     0.05   0.1173  0.1673  0.2260  0.0923  0.0783  0.0860   0.2077
4     0.05   0.1893  0.3010  0.4423  0.0980  0.0983  0.1077   0.4153
5     0.05   0.3210  0.5050  0.6797  0.1123  0.1190  0.1393   0.6890
6     0.05   0.4780  0.7067  0.8657  0.1283  0.1593  0.1877   0.9043
7     0.05   0.6383  0.8530  0.9613  0.1540  0.2120  0.2513   0.9810
8     0.05   0.7813  0.9470  0.9913  0.1810  0.2693  0.3413   0.9993
9     0.05   0.8807  0.9850  0.9987  0.2250  0.3410  0.4530   1.0000
10    0.05   0.9450  0.9957  1.0000  0.2737  0.4260  0.5753   1.0000
}\MainChetverikov

\pgfplotstableread{
delta alpha   FiveOp  FiveUn  SevenOp  SevenUn  TenOp    TenUn
0     0.05    0.0660  0.0600  0.0587   0.0567   0.0653   0.0610
1     0.05    0.0713  0.0607  0.0667   0.0563   0.0790   0.0653
2     0.05    0.0857  0.0660  0.0843   0.0677   0.0973   0.0737
3     0.05    0.1177  0.0787  0.1343   0.0783   0.1467   0.0933
4     0.05    0.1727  0.0993  0.2147   0.1053   0.2473   0.1217
5     0.05    0.2497  0.1293  0.3287   0.1453   0.3817   0.1693
6     0.05    0.3537  0.1717  0.4823   0.2027   0.5753   0.2360
7     0.05    0.4753  0.2280  0.6403   0.2870   0.7690   0.3387
8     0.05    0.6113  0.2973  0.7873   0.4017   0.8967   0.4620
9     0.05    0.7433  0.3777  0.9010   0.5153   0.9647   0.6140
10    0.05    0.8547  0.4803  0.9600   0.6527   0.9897   0.7613
}\MainFirstLSWMon

\pgfplotstableread{ 
delta alpha MonFiveQKn0 MonFiveQKn1 MonSevenQKn0 MonSevenQKn1 MonTenQKn0 MonTenQKn1
0     0.05    0.0613      0.0680       0.0583       0.0653      0.0500     0.0553
1     0.05    0.1110      0.0980       0.1270       0.1097      0.1267     0.0990
2     0.05    0.1987      0.1450       0.2530       0.1773      0.2607     0.1720
3     0.05    0.3173      0.2010       0.4100       0.2730      0.4573     0.2857
4     0.05    0.4597      0.2843       0.5907       0.3850      0.6733     0.4300
5     0.05    0.6147      0.3887       0.7643       0.5207      0.8473     0.5900
6     0.05    0.7457      0.5060       0.8843       0.6493      0.9493     0.7627
7     0.05    0.8573      0.6200       0.9587       0.7757      0.9823     0.8780
8     0.05    0.9313      0.7237       0.9853       0.8720      0.9963     0.9593
9     0.05    0.9790      0.8223       0.9973       0.9387      0.9993     0.9837
10    0.05    0.9937      0.8980       0.9993       0.9760      1.0000     0.9957
}\SecondMonQ

\pgfplotstableread{ 
delta alpha MonFiveCKn0 MonFiveCKn1 MonSevenCKn0 MonSevenCKn1 MonTenCKn0 MonTenCKn1
0     0.05    0.0670      0.0830       0.0640       0.0737      0.0527     0.0590
1     0.05    0.0960      0.1060       0.1090       0.1113      0.1027     0.0970
2     0.05    0.1413      0.1467       0.1827       0.1583      0.1800     0.1527
3     0.05    0.2070      0.1930       0.2737       0.2323      0.2917     0.2347
4     0.05    0.2860      0.2503       0.3917       0.3263      0.4400     0.3477
5     0.05    0.3947      0.3263       0.5243       0.4333      0.6017     0.4857
6     0.05    0.5073      0.4183       0.6517       0.5617      0.7630     0.6297
7     0.05    0.6240      0.5137       0.7750       0.6813      0.8787     0.7787
8     0.05    0.7307      0.6230       0.8763       0.7857      0.9557     0.8813
9     0.05    0.8303      0.7247       0.9413       0.8687      0.9837     0.9510
10    0.05    0.8970      0.8033       0.9737       0.9307      0.9940     0.9800
}\SecondMonC

\pgfplotstableread{
delta alpha   FiveOp  FiveUn  SevenOp  SevenUn  TenOp   TenUn
0     0.05    0.0610  0.0430  0.0660   0.0540   0.0540  0.0440
1     0.05    0.0920  0.0610  0.1000   0.0820   0.0960  0.0780
2     0.05    0.1300  0.0840  0.1670   0.1160   0.1520  0.1300
3     0.05    0.1960  0.1300  0.2430   0.1850   0.2380  0.1830
4     0.05    0.2600  0.1780  0.3450   0.2580   0.3800  0.2620
5     0.05    0.3460  0.2230  0.4750   0.3240   0.5210  0.3500
6     0.05    0.4560  0.2900  0.5810   0.4080   0.6530  0.4460
7     0.05    0.5590  0.3650  0.7200   0.4970   0.7980  0.5640
8     0.05    0.6780  0.4470  0.8230   0.5960   0.8970  0.6620
9     0.05    0.7780  0.5400  0.8900   0.7040   0.9720  0.7960
10    0.05    0.8410  0.6270  0.9410   0.7840   0.9930  0.8680
}\MainSecondLSWMon

\begin{figure}[!h]
\centering\scriptsize
\begin{tikzpicture} 
\begin{groupplot}[group style={group name=myplots,group size=3 by 2,horizontal sep= 0.8cm,vertical sep=1.1cm},
    grid = minor,
    width = 0.375\textwidth,
    xmax=10,xmin=0,
    ymax=1,ymin=0,
    every axis title/.style={below,at={(0.2,0.8)}},
    xlabel=$\delta$,
    x label style={at={(axis description cs:0.95,0.04)},anchor=south},
    xtick={0,2,...,10},
    ytick={0.05,0.5,1},
    tick label style={/pgf/number format/fixed},
    legend style={text=black,cells={align=center},row sep = 3pt,legend columns = -1, draw=none,fill=none},
    cycle list={%
{smooth,tension=0.5,color=SpringGreen4, mark=halfsquare*,every mark/.append style={rotate=270},mark size=1.5pt,line width=0.5pt},
{smooth,tension=0.5,color=DarkOliveGreen4, mark=halfsquare*,every mark/.append style={rotate=90},mark size=1.5pt,line width=0.5pt}, 
{smooth,tension=0.5,color=Honeydew4, mark=10-pointed star,mark size=1.5pt,line width=0.5pt},
{smooth,tension=0.5,color=RoyalBlue1, mark=halfcircle*,every mark/.append style={rotate=90},mark size=1.5pt,line width=0.5pt}, 
{smooth,tension=0.5,color=RoyalBlue2, mark=halfcircle*,every mark/.append style={rotate=180},mark size=1.5pt,line width=0.5pt}, 
{smooth,tension=0.5,color=RoyalBlue3, mark=halfcircle*,every mark/.append style={rotate=270},mark size=1.5pt,line width=0.5pt},
{smooth,tension=0.5,color=RoyalBlue4, mark=halfcircle*,every mark/.append style={rotate=360},mark size=1.5pt,line width=0.5pt},
}
]
\nextgroupplot[legend style = {column sep = 7pt, legend to name = LegendMon1}]
\addplot[smooth,tension=0.5,color=NavyBlue, no markers,line width=0.25pt, densely dotted,forget plot] table[x = delta,y=alpha] from \FirstMon;
\addplot table[x = delta,y=FiveUn] from \MainFirstLSWMon;
\addplot table[x = delta,y=FiveOp] from \MainFirstLSWMon;
\addplot table[x = delta,y=Five1] from \MainChetverikov;
\addplot table[x = delta,y=MonFiveKn3] from \FirstMon;
\addplot table[x = delta,y=MonFiveKn5] from \FirstMon;
\addplot table[x = delta,y=MonFiveKn7] from \FirstMon;
\node[anchor=north] at (axis description cs: 0.25,  0.95) {\fontsize{5}{4}\selectfont $\begin{aligned} d_z&=1\\ n&=500 \end{aligned}$};
\addlegendentry{LSW-S};
\addlegendentry{LSW-L};
\addlegendentry{C-OS};
\addlegendentry{FS-C3};
\addlegendentry{FS-C5};
\addlegendentry{FS-C7};
\nextgroupplot
\node[anchor=north] at (axis description cs: 0.25,  0.95) {\fontsize{5}{4}\selectfont $\begin{aligned} d_z&=1\\ n&=750 \end{aligned}$};
\addplot[smooth,tension=0.5,color=NavyBlue, no markers,line width=0.25pt, densely dotted,forget plot] table[x = delta,y=alpha] from \FirstMon;
\addplot table[x = delta,y=SevenUn] from \MainFirstLSWMon;
\addplot table[x = delta,y=SevenOp] from \MainFirstLSWMon;
\addplot table[x = delta,y=Seven1] from \MainChetverikov;
\addplot table[x = delta,y=MonSevenKn3] from \FirstMon;
\addplot table[x = delta,y=MonSevenKn5] from \FirstMon;
\addplot table[x = delta,y=MonSevenKn7] from \FirstMon;
\nextgroupplot
\node[anchor=north] at (axis description cs: 0.25,  0.95) {\fontsize{5}{4}\selectfont $\begin{aligned} d_z&=1\\ n&=1000 \end{aligned}$};
\addplot[smooth,tension=0.5,color=NavyBlue, no markers,line width=0.25pt, densely dotted,forget plot] table[x = delta,y=alpha] from \FirstMon;
\addplot table[x = delta,y=TenUn] from \MainFirstLSWMon;
\addplot table[x = delta,y=TenOp] from \MainFirstLSWMon;
\addplot table[x = delta,y=Ten1] from \MainChetverikov;
\addplot table[x = delta,y=MonTenKn3] from \FirstMon;
\addplot table[x = delta,y=MonTenKn5] from \FirstMon;
\addplot table[x = delta,y=MonTenKn7] from \FirstMon;
\nextgroupplot[legend style = {column sep = 3.5pt, legend to name = LegendMon2}]
\node[anchor=north] at (axis description cs: 0.25,  0.95) {\fontsize{5}{4}\selectfont $\begin{aligned} d_z&=2\\ n&=500 \end{aligned}$};
\addplot[smooth,tension=0.5,color=NavyBlue, no markers,line width=0.25pt, densely dotted,forget plot] table[x = delta,y=alpha] from \FirstMon;
\addplot table[x = delta,y=FiveUn] from \MainSecondLSWMon;
\addplot table[x = delta,y=FiveOp] from \MainSecondLSWMon;
\addplot table[x = delta,y=Five2] from \MainChetverikov;
\addplot table[x = delta,y=MonFiveQKn0] from \SecondMonQ;
\addplot table[x = delta,y=MonFiveQKn1] from \SecondMonQ;
\addplot table[x = delta,y=MonFiveCKn0] from \SecondMonC;
\addplot table[x = delta,y=MonFiveCKn1] from \SecondMonC;
\addlegendentry{LSW-S};
\addlegendentry{LSW-L};
\addlegendentry{C-OS};
\addlegendentry{FS-Q0};
\addlegendentry{FS-Q1};
\addlegendentry{FS-C0};
\addlegendentry{FS-C1};
\nextgroupplot
\node[anchor=north] at (axis description cs: 0.25,  0.95) {\fontsize{5}{4}\selectfont $\begin{aligned} d_z&=2\\ n&=750 \end{aligned}$};
\addplot[smooth,tension=0.5,color=NavyBlue, no markers,line width=0.25pt, densely dotted,forget plot] table[x = delta,y=alpha] from \FirstMon;
\addplot table[x = delta,y=SevenUn] from \MainSecondLSWMon;
\addplot table[x = delta,y=SevenOp] from \MainSecondLSWMon;
\addplot table[x = delta,y=Seven2] from \MainChetverikov;
\addplot table[x = delta,y=MonSevenQKn0] from \SecondMonQ;
\addplot table[x = delta,y=MonSevenQKn1] from \SecondMonQ;
\addplot table[x = delta,y=MonSevenCKn0] from \SecondMonC;
\addplot table[x = delta,y=MonSevenCKn1] from \SecondMonC;
\nextgroupplot
\node[anchor=north] at (axis description cs: 0.25,  0.95) {\fontsize{5}{4}\selectfont $\begin{aligned} d_z&=2\\ n&=1000 \end{aligned}$};
\addplot[smooth,tension=0.5,color=NavyBlue, no markers,line width=0.25pt, densely dotted,forget plot] table[x = delta,y=alpha] from \FirstMon;
\addplot table[x = delta,y=TenUn] from \MainSecondLSWMon;
\addplot table[x = delta,y=TenOp] from \MainSecondLSWMon;
\addplot table[x = delta,y=Ten2] from \MainChetverikov;
\addplot table[x = delta,y=MonTenQKn0] from \SecondMonQ;
\addplot table[x = delta,y=MonTenQKn1] from \SecondMonQ;
\addplot table[x = delta,y=MonTenCKn0] from \SecondMonC;
\addplot table[x = delta,y=MonTenCKn1] from \SecondMonC;
\end{groupplot}
\node at ($(myplots c2r1) + (0,-2.25cm)$) {\ref{LegendMon1}};
\node at ($(myplots c2r2) + (0,-2.25cm)$) {\ref{LegendMon2}};
\end{tikzpicture}
\caption{Empirical power of our test (with $\gamma_n=0.01/\log n$), LSW-L, LSW-S, and C-OS for the designs \eqref{Eqn: MC1,aux1} and \eqref{Eqn: MC2,aux1}, where corresponding to $\delta=0$ are the empirical sizes under D1. Note that FS-Q1 and FS-C1 nearly overlap each other in the second row.} \label{Fig: MC}
\end{figure}
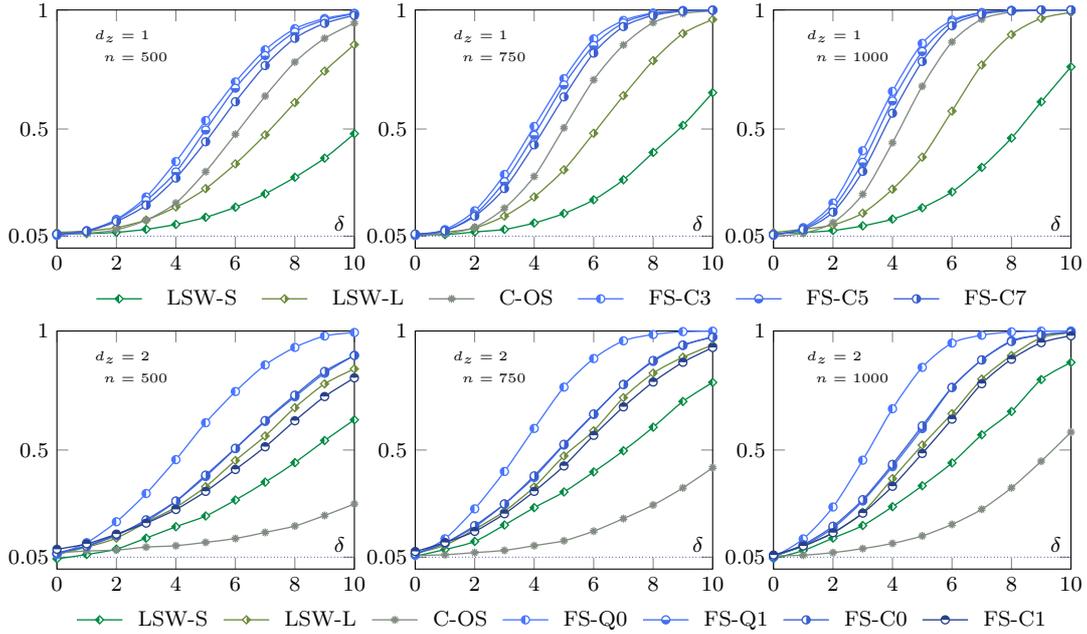

Figure \ref{Fig: MC} depicts the power curves, where we only show our tests with $\gamma_n=0.01/\log n$ for brevity and the fact that other choices of $\gamma_n$ lead to very similar curves---see Figure \ref{Fig: MC1, FullMon} in Appendix \ref{App: full simulations}. For $d_z=1$, our tests are moderately more powerful than C-OS, across sample sizes and the number of interior knots, and they are all considerably more powerful than LSW-L and in particular LSW-S. For $d_z=2$, our tests remain competitive in terms of power, though LSW-L is more powerful than FS-C1 (the least powerful among our tests). Interestingly, C-OS is the least powerful of all tests, which may be explained by the fact that only part of the discordance between regressors and the outcome is being picked up through the indicator function in the test function $b(s)$---see p.27 in the arXiv version of \citet{Chetverikov2018Monotonicity}. We note that the power of our tests is overall decreasing in $k_n$, which is consistent with Theorem \ref{Thm: uniform power} since $r_n=\sqrt{n/k_n}$.

{
\setlength{\tabcolsep}{5.25pt}
\renewcommand{\arraystretch}{1.1}
\begin{table}[!ht]
\caption{Run-times (in Seconds) of Monotonicity Tests} \label{Tab: MC, runtime}
\centering\footnotesize
\sisetup{table-number-alignment = center, table-format = 1.2} 
\begin{tabularx}{\linewidth}{@{}cc *{5}{S[round-mode = places,round-precision = 2]} c  *{5}{S[round-mode = places,round-precision = 2]} c@{}} 
\hline
\hline
&\multirow{2}{*}{$n$}& \multicolumn{5}{c}{Design \eqref{Eqn: MC1,aux1}} & & \multicolumn{5}{c}{Design \eqref{Eqn: MC2,aux1}} & \\
\cline{3-7} \cline{9-13}
&& {FS-C3} & {FS-C7} & {LSW-L} & {LSW-S} & {C-OS} & & {FS-Q0} & {FS-C1} & {LSW-L} & {LSW-S} & {C-OS} &\\  
\hline
&$500$        & 0.1236 & 0.1461 & 22.2861  & 22.2216  & 0.0822 & & 0.3984 & 0.4488 & 38.8922  & 39.2504  & 7.1872  &\\
&$750$        & 0.1280 & 0.1310 & 56.8640  & 56.6187  & 0.1566 & & 0.4126 & 0.5008 & 128.8489 & 128.1390 & 17.3866 &\\
&$1000$       & 0.1390 & 0.1558 & 103.5839 & 102.1905 & 0.3057 & & 0.3893 & 0.4221 & 219.2428 & 222.7889 & 29.4911 &\\
\hline
\hline
\end{tabularx}
\end{table}
}

Finally, we compare the run-times of a single replication based on the design D1 in both univariate and bivariate cases. For brevity, we only report our tests with the smallest and the largest $k_n$, based on $\gamma_n=0.01/\log n$. All numbers in Table \ref{Tab: MC, runtime} are obtained by running MATLAB R2019b in a Windows 10 PC with 16 GB RAM and an Intel\textsuperscript{\textregistered} Core\textsuperscript{\tiny TM} i7-7700 processor having 4 cores and 3.60 GHz base speed. Overall, Table \ref{Tab: MC, runtime} shows that our tests are relatively simple to implement, in both univariate and bivariate settings, and their computation cost increases only modestly as we move from $d_z=1$ to $d_z=2$. We stress that the computational complexity of quadratic programs involved in our tests depends on the fineness of discretization, not the sample size.

\section{Conclusion}\label{Sec: conclusion}

In this paper, we have developed a uniformly valid and asymptotically nonconservative test for a class of shape restrictions, which is applicable to nonparametric regression models as well as parametric, distributional, and structural settings. The key insight we exploit is that these restrictions form convex cones in Hilbert spaces, a structure that enables us to employ a projection-based test whose properties may be analyzed in an elegant, transparent, and unifying way. In particular, while the problem is inherently nonstandard, we are able to develop a bootstrap procedure that may be implemented in a way as simple as computing the test statistic.


\titleformat{\section}{\Large\center}{{\sc Appendix} \thesection}{1em}{}
\setcounter{section}{0}
\renewcommand \thesection{\Alph{section}}
\numberwithin{equation}{section}
\numberwithin{ass}{section}
\numberwithin{figure}{section}
\numberwithin{table}{section}

\section{Proofs of Main Results}\label{Sec: main results proof}

\noindent{\sc Proof of Lemma \ref{Lem: main}:} Part (i) is immediate because $h\mapsto\Pi_\Lambda(h)$ is positively homogeneous by Assumption \ref{Ass: convex setup} and Theorem 5.6-(7) in \citet{Deutsch2012Best}, while part (ii) follows by Lemma \ref{Lem: projection monotonicity}, part (i), $\theta_P\in\Lambda$, and $\kappa_n\in[0,r_n]$. For part (iii), note that $\kappa_n\hat\theta_n-\kappa_n\theta_P=r_n \{\hat\theta_n-\theta_P\}\cdot (\kappa_n/r_n)$. By Assumption \ref{Ass: strong approx}(i), $\sup_{P\in\mathbf P}E[\|\mathbb Z_{n,P}\|_{\mathbf H}]<\infty$ uniformly in $n$ and Markov's inequality, we have: uniformly in $P\in\mathbf P$,
\begin{align}\label{Eqn: main lemma, aux2}
\| r_n \{\hat\theta_n-\theta_P\}\|_{\mathbf H}\le \| r_n \{\hat\theta_n-\theta_P\}-\mathbb Z_{n,P}\|_{\mathbf H}+\|\mathbb Z_{n,P}\|_{\mathbf H}=O_p(1)~.
\end{align}
Part (iii) now follows from combining \eqref{Eqn: main lemma, aux2} and  $\kappa_n/r_n=o(c_n)$. \qed

\noindent{\sc Proof of Theorem \ref{Thm: size control}:} We structure our proof in four steps.

\noindent\underline{Step 1:} Build up a strong approximation for $ r_n\phi(\hat\theta_n)$ that is valid uniformly in $P\in\mathbf P$.

We make use of the map $\psi_{a,P}$ defined in \eqref{Eqn: psi function}, and let $\mathbb G_{n,P}\equiv r_n\{\hat\theta_n-\theta_P\}$. By Lemma \ref{Lem: main} and the definition of $\psi_{a,P}$, we may rewrite the test statistic:
\begin{align}\label{Eqn: size control, aux1}
r_n\phi(\hat\theta_n)=\|\mathbb G_{n,P}+r_n\theta_P-\Pi_\Lambda(\mathbb G_{n,P}+r_n\theta_P)\|_{\mathbf H}=\psi_{r_n,P}(\mathbb G_{n,P})~.
\end{align}
By Theorem 3.16 in \citet{AliprantisandBorder2006} and Assumption \ref{Ass: strong approx}(i), we have:
\begin{align}\label{Eqn: size control, aux2}
|\psi_{r_n,P}(\mathbb G_{n,P})-\psi_{r_n,P}(\mathbb Z_{n,P})|\le \|\mathbb G_{n,P}- \mathbb Z_{n,P}\|_{\mathbf H}=o_p(c_n)~,
\end{align}
uniformly in $P\in\mathbf P$. Combining results \eqref{Eqn: size control, aux1} and \eqref{Eqn: size control, aux2}, we thus obtain the strong approximation for our test statistic: uniformly in $P\in\mathbf P$,
\begin{align}\label{Eqn: size control, aux3}
r_n\phi(\hat\theta_n)=\psi_{r_n,P}(\mathbb Z_{n,P})+o_p(c_n)~.
\end{align}

\noindent\underline{Step 2:} Build up a strong approximation of $ \hat\psi_{\kappa_n}(\hat{\mathbb G}_n)$ that is valid uniformly in $P\in\mathbf P_0$.

First, Theorem 3.16 in \citet{AliprantisandBorder2006} implies: for each $P\in\mathbf P_0$,
\begin{align}\label{Eqn: size control, aux5}
|\hat\psi_{\kappa_n}(\hat{\mathbb G}_n)- \psi_{\kappa_n,P}(\hat{\mathbb G}_n)|
\le \kappa_n\|\Pi_\Lambda\hat\theta_n-\theta_P\|_{\mathbf H}\le  \| \kappa_n \{\hat\theta_n-\theta_P\}\|_{\mathbf H} = o_p(c_n)~,
\end{align}
where the second inequality is due to $\theta_P=\Pi_\Lambda(\theta_P)$ for all $P\in\mathbf P_0$, Assumption \ref{Ass: convex setup}(i), Lemma 6.54-d in \citet{AliprantisandBorder2006}, and the final step is due to Lemma \ref{Lem: main}(iii). Again by Theorem 3.16 in \citet{AliprantisandBorder2006}, we have
\begin{align}\label{Eqn: size control, aux8}
| \psi_{\kappa_n,P}(\hat{\mathbb G}_n)-\psi_{\kappa_n,P}(\bar{\mathbb Z}_{n,P})|\le \|\hat{\mathbb G}_n- \bar{\mathbb Z}_{n,P}\|_{\mathbf H}=o_p(c_n)~,
\end{align}
uniformly in $P\in\mathbf P$, where the equality is due to Assumption \ref{Ass: strong approx}(ii). We thus obtain by \eqref{Eqn: size control, aux5} and \eqref{Eqn: size control, aux8} and the triangle inequality that: uniformly in $P\in\mathbf P_0$,
\begin{align}\label{Eqn: size control, aux9}
\hat\psi_{\kappa_n}(\hat{\mathbb G}_n) = \psi_{\kappa_n,P}(\bar{\mathbb Z}_{n,P}) +o_p(c_n)~.
\end{align}

\noindent\underline{Step 3:} Control the estimation error of $\hat c_{n,1-\alpha}$. By results \eqref{Eqn: size control, aux3} and \eqref{Eqn: size control, aux9}, we may select a sequence of positive scalars $\epsilon_n=o(c_n)$ (sufficiently slow) such that, as $n\to\infty$,
\begin{gather}
\sup_{P\in\mathbf P_0} P(|r_n\phi(\hat\theta_n)-\psi_{r_n,P}(\mathbb Z_{n,P})|>\epsilon_n)=o(1)~, \label{Eqn: size control, aux11a}\\
\sup_{P\in\mathbf P_0} P(|\hat\psi_{\kappa_n}(\hat{\mathbb G}_n) - \psi_{\kappa_n,P}(\bar{\mathbb Z}_{n,P})|>\epsilon_n)=o(1)~.\label{Eqn: size control, aux11b}
\end{gather}
By Markov's inequality, Fubini's theorem, and \eqref{Eqn: size control, aux11b}, we have: for each $\eta>0$,
\begin{multline}\label{Eqn: size control, aux12}
\sup_{P\in\mathbf P_0} P(P(|\hat\psi_{\kappa_n}(\hat{\mathbb G}_n) - \psi_{\kappa_n,P}(\bar{\mathbb Z}_{n,P})|>\epsilon_n|\{X_i\}_{i=1}^n)>\eta)\\
\le \sup_{P\in\mathbf P_0}\frac{1}{\eta} P(|\hat\psi_{\kappa_n}(\hat{\mathbb G}_n) - \psi_{\kappa_n,P}(\bar{\mathbb Z}_{n,P})|>\epsilon_n)=o(1)~.
\end{multline}
Thus, we may select a sequence of positive scalars $\eta_n\downarrow 0$ such that
\begin{align}\label{Eqn: size control, aux13}
P(|\hat\psi_{\kappa_n}(\hat{\mathbb G}_n) - \psi_{\kappa_n,P}(\bar{\mathbb Z}_{n,P})|>\epsilon_n|\{X_i\}_{i=1}^n)=o_p(\eta_n)~,
\end{align}
uniformly in $P\in\mathbf P_0$. Since $\bar{\mathbb Z}_{n,P}$ is independent of $\{X_i\}_{i=1}^n$ by Assumption \ref{Ass: strong approx}(ii), the conditional cdf of $\psi_{\kappa_n,P}(\bar{\mathbb Z}_{n,P})$ given $\{X_i\}_{i=1}^n$ is precisely its unconditional analog. Thus, we may conclude by Lemma 11 in \citet{ChernozhukovLeeRosen2013Intersection} and result \eqref{Eqn: size control, aux13} that
\begin{align}\label{Eqn: size control, aux14}
\liminf_{n\to\infty}\inf_{P\in\mathbf P_0} P(\hat c_{n,1-\alpha}+\epsilon_n\ge c_{n,P}(1-\alpha-\eta_n))=1~.
\end{align}

\noindent\underline{Step 4:} Conclude with the help of a partial anti-concentration inequality.

To begin with, note that by results \eqref{Eqn: size control, aux11a} and \eqref{Eqn: size control, aux14}, we have
\begin{align}\label{Eqn: size control, aux15}
\limsup_{n\to\infty}\sup_{P\in\mathbf P_0}& P(r_n\phi(\hat\theta_n)>\hat c_{n,1-\alpha})\notag\\
&\le \limsup_{n\to\infty} \sup_{P\in\mathbf P_0}P(r_n\phi(\hat\theta_n)>\hat c_{n,1-\alpha},|r_n\phi(\hat\theta_n)-\psi_{r_n,P}(\mathbb Z_{n,P})|\le \epsilon_n)\notag\\
&\le \limsup_{n\to\infty}\sup_{P\in\mathbf P_0}P(\psi_{r_n,P}(\mathbb Z_{n,P})>c_{n,P}(1-\alpha-\eta_n)-2\epsilon_n)\notag\\
&\le \limsup_{n\to\infty}\sup_{P\in\mathbf P_0}P(\psi_{\kappa_n,P}(\mathbb Z_{n,P})>c_{n,P}(1-\alpha-\eta_n)-2\epsilon_n)~,
\end{align}
where the final step follows by Lemma \ref{Lem: projection monotonicity} and $0\le\kappa_n\le r_n$ for all large $n$ (due to $\kappa_n/r_n=o(c_n)$ and $c_n=O(1)$). In turn, we note that, since $\eta_n\downarrow 0$ and $\epsilon_n=o(c_n)$, it follows by Assumption \ref{Ass: Gaussian}(iii), Proposition \ref{Pro: anti concentration}, and result \eqref{Eqn: size control, aux15} that
\begin{multline}\label{Eqn: size control, aux18}
\limsup_{n\to\infty}\sup_{P\in\mathbf P_0} P(r_n\phi(\hat\theta_n)>\hat c_{n,1-\alpha})\le \limsup_{n\to\infty}\sup_{P\in\mathbf P_0}P(\psi_{\kappa_n,P}(\mathbb Z_{n,P})>c_{n,P}(1-\alpha-\eta_n))\\
\le\limsup_{n\to\infty}\sup_{P\in\mathbf P_0}\{\alpha+\eta_n\}=\alpha~,
\end{multline}
as desired for the first claim. For the second claim, it thus suffices to show
\begin{align}\label{Eqn: size control, aux19}
\liminf_{n\to\infty}\inf_{P\in\bar{\mathbf P}_0} P(r_n\phi(\hat\theta_n)>\hat c_{n,1-\alpha})\ge \alpha~.
\end{align}
For this, we note that, by simple manipulations,
\begin{align}\label{Eqn: size control, aux20}
\liminf_{n\to\infty}\inf_{P\in\bar{\mathbf P}_0}& P(r_n\phi(\hat\theta_n)>\hat c_{n,1-\alpha})\notag\\
&\ge \liminf_{n\to\infty}\inf_{P\in\bar{\mathbf P}_0}P(r_n\phi(\hat\theta_n)>\hat c_{n,1-\alpha},|r_n\phi(\hat\theta_n)-\psi_{r_n,P}(\mathbb Z_{n,P})|\le \epsilon_n)\notag\\
& = \liminf_{n\to\infty}\inf_{P\in\bar{\mathbf P}_0}P(\psi_{r_n,P}(\mathbb Z_{n,P})-\epsilon_n>\hat c_{n,1-\alpha})~,
\end{align}
where the last step is due to result \eqref{Eqn: size control, aux11a} and $\bar{\mathbf P}_0\subset \mathbf P_0$. Moreover, another application of Lemma 11 in \citet{ChernozhukovLeeRosen2013Intersection} to \eqref{Eqn: size control, aux13} yields
\begin{align}\label{Eqn: size control, aux21}
\liminf_{n\to\infty}\inf_{P\in\mathbf P_0} P(\hat c_{n,1-\alpha}\le c_{n,P}(1-\alpha+\eta_n)+\epsilon_n)=1~.
\end{align}
By the definition of $\bar{\mathbf P}_0$ and Lemma \ref{Lem: nonconservativeness}, we also note that, for all $n$ and $P\in\bar{\mathbf P}_0$,
\begin{align}\label{Eqn: size control, aux22}
\psi_{r_n,P}(\mathbb Z_{n,P}) = \psi_{\kappa_n,P}(\mathbb Z_{n,P})=\|\mathbb Z_{n,P}-\Pi_\Lambda(\mathbb Z_{n,P})\|_{\mathbf H}~.
\end{align}
Combining results \eqref{Eqn: size control, aux20}, \eqref{Eqn: size control, aux21}, and \eqref{Eqn: size control, aux22} with Proposition \ref{Pro: anti concentration} yields
\begin{align}\label{Eqn: size control, aux23}
\liminf_{n\to\infty}\inf_{P\in\bar{\mathbf P}_0}& P(r_n\phi(\hat\theta_n)>\hat c_{n,1-\alpha})\notag\\
&\ge \liminf_{n\to\infty}\inf_{P\in\bar{\mathbf P}_0}P(\psi_{r_n,P}(\mathbb Z_{n,P})-\epsilon_n>c_{n,P}(1-\alpha+\eta_n)+\epsilon_n)\notag\\
& = \liminf_{n\to\infty}\inf_{P\in\bar{\mathbf P}_0}P(\psi_{\kappa_n,P}(\mathbb Z_{n,P})>c_{n,P}(1-\alpha+\eta_n)+2\epsilon_n)\notag\\
& = \liminf_{n\to\infty}\inf_{P\in\bar{\mathbf P}_0}P(\psi_{\kappa_n,P}(\mathbb Z_{n,P})>c_{n,P}(1-\alpha+\eta_n)-2\epsilon_n)~.
\end{align}
Since $c_{n,P}(1-\alpha+\eta_n)-\epsilon_n< c_{n,P}(1-\alpha+\eta_n)$, we thus obtain by result \eqref{Eqn: size control, aux23}, the definition of quantiles, and $\eta_n=o(1)$ that
\begin{align}\label{Eqn: size control, aux24}
\liminf_{n\to\infty}\inf_{P\in\bar{\mathbf P}_0} P(r_n\phi(\hat\theta_n)>\hat c_{n,1-\alpha})\ge \liminf_{n\to\infty}\{\alpha-\eta_n\}=\alpha~,
\end{align}
which, together with the first claim, establishes the second claim of the theorem. \qed

\noindent{\sc Proof of Theorem \ref{Thm: uniform power}:} First, by Assumption \ref{Ass: convex setup} and Lemma \ref{Lem: projection monotonicity}, we have
\begin{align}\label{Eqn: uniform power, aux1}
\|\hat{\mathbb G}_n+\kappa_n\Pi_\Lambda\hat\theta_n &-\Pi_\Lambda(\hat{\mathbb G}_n+\kappa_n\Pi_\Lambda\hat\theta_n)\|_{\mathbf H} \le  \|\hat{\mathbb G}_n-\Pi_\Lambda\hat{\mathbb G}_n\|_{\mathbf H} \notag\\
&\le \|\hat{\mathbb G}_n\|_{\mathbf H}\le \|\hat{\mathbb G}_n-\bar{\mathbb Z}_{n,P}\|_{\mathbf H}+\|\bar{\mathbb Z}_{n,P}\|_{\mathbf H}~,
\end{align}
where the second inequality follows from Assumption \ref{Ass: convex setup} and Theorem 5.6(5) in \citet{Deutsch2012Best}, and the third inequality is due to the triangle inequality. By Assumptions \ref{Ass: strong approx} and \ref{Ass: Gaussian}(ii), we in turn have from \eqref{Eqn: uniform power, aux1} that, uniformly in $P\in\mathbf P$,
\begin{align}\label{Eqn: uniform power, aux2}
\|\hat{\mathbb G}_n+\kappa_n\Pi_\Lambda\hat\theta_n-\Pi_\Lambda(\hat{\mathbb G}_n+\kappa_n\Pi_\Lambda\hat\theta_n)\|_{\mathbf H}
=o_p(c_n)+O_p(1)=O_p(1)~.
\end{align}
By the definition of $\hat c_{n,1-\alpha}$, we note that, for $M>0$ and uniformly in $P\in\mathbf P$,
\begin{align}\label{Eqn: uniform power, aux3}
P(\hat c_{n,1-\alpha}>M)&\le P(P(\|\hat{\mathbb G}_n+\kappa_n\Pi_\Lambda\hat\theta_n-\Pi_\Lambda(\hat{\mathbb G}_n+\kappa_n\Pi_\Lambda\hat\theta_n)\|_{\mathbf H}>M|\{X_i\}_{i=1}^n)>\alpha)\notag\\
&\le\frac{1}{\alpha} P(\|\hat{\mathbb G}_n+\kappa_n\Pi_\Lambda\hat\theta_n-\Pi_\Lambda(\hat{\mathbb G}_n+\kappa_n\Pi_\Lambda\hat\theta_n)\|_{\mathbf H}>M) ~,
\end{align}
where the second inequality holds by Markov's inequality and Fubini's theorem. It follows from results \eqref{Eqn: uniform power, aux2} and \eqref{Eqn: uniform power, aux3} that $\hat c_{n,1-\alpha}=O_p(1)$ uniformly in $P\in\mathbf P$.

Next, we bound $r_n\phi(\hat\theta_n)$ from below. By Theorem 3.16 in \citet{AliprantisandBorder2006} and the triangle inequality, we have: uniformly in $P\in\mathbf P$,
\begin{multline}\label{Eqn: uniform power, aux5}
|r_n\phi(\hat\theta_n)-r_n\phi(\theta_P)|\le \|r_n\{\hat\theta_n-\theta_P\}\|_{\mathbf H}\\
\le \|r_n\{\hat\theta_n-\theta_P\}-\mathbb Z_{n,P}\|_{\mathbf H}+ \|\mathbb Z_{n,P}\|_{\mathbf H}\le o_p(c_n)+O_p(1)=O_p(1)~,
\end{multline}
where the third inequality follows by Assumptions \ref{Ass: strong approx}(i) and \ref{Ass: Gaussian}(ii), and the last step is due to $c_n=O(1)$. It follows from result \eqref{Eqn: uniform power, aux5} and the definition of $\mathbf P_{1,n}^\Delta$ that
\begin{align}\label{Eqn: uniform power, aux6}
r_n\phi(\hat\theta_n)=r_n\phi(\theta_P)+r_n\phi(\hat\theta_n)-r_n\phi(\theta_P)\ge \Delta+O_p(1)~,
\end{align}
uniformly in $P\in\mathbf P_{1,n}^\Delta$. The theorem thus follows from combining result \eqref{Eqn: uniform power, aux6} and the order $\hat c_{n,1-\alpha}=O_p(1)$ uniformly in $P\in\mathbf P$ that we have established .\qed

\titleformat{\section}{\normalfont\Large\bfseries}{\thesection}{1em}{}

\addcontentsline{toc}{section}{References}
\putbib
\end{bibunit}

\clearpage\newpage

\begin{bibunit}

\begin{appendices}
\titleformat{\section}{\Large\center}{{\sc Appendix} \thesection}{1em}{}
\setcounter{page}{1}
\setcounter{section}{1}
\renewcommand{\thetable}{\thesection.\Roman{table}}
\emptythanks
\phantomsection
\pdfbookmark[1]{Appendix Title}{title1}
\title{\vspace{-2cm}Supplement to ``A Projection Framework for Testing Shape Restrictions That Form Convex Cones''}
\author{
Zheng Fang \\ zfang@tamu.edu \\ Department of Economics \\ Texas A\&M University
\and
Juwon Seo \\ ecssj@nus.edu.sg \\ Department of Economics\\National University of Singapore}
\date{ }
\maketitle

\vspace{-0.3in}

This supplement is organized as follows. Appendix \ref{App: convex cone} discusses particular shape restrictions with the convex cone property, Appendix \ref{Sec: special case} specializes our test to the regular case where $r_n\{\hat\theta_n-\theta_0\}$ converges, Appendix \ref{App: auxiliary results} collects additional proofs and auxiliary results, and Appendix \ref{App: more simulations} presents additional simulation studies and an empirical application. Appendix \ref{Sec: appendix, ex} verifies the main assumptions for our examples, Appendix \ref{App: proofs for the special case} provides proofs for Appendix \ref{Sec: special case}, while Appendix \ref{App: full simulations} contains simulation results omitted from the main text and Appendix \ref{App: more simulations}, all of which are relegated to the arXiv version of this paper (https://arxiv.org/abs/1910.07689) due to space limitation. For ease of reference, we centralize some notation in the table below.

{\renewcommand{\arraystretch}{1}
\begin{table}[h]
\begin{center}
\begin{tabularx}{\textwidth}{cX}
\hline\hline
$ a \lesssim b$                 & For some constant $M$ that is universal in the proof, $a\leq Mb$.\\
$a^{(j)}$                       & The $j$th coordinate of a vector $a\in\mathbf R^d$.\\
$a^{(-j)}$                      & The vector in $\mathbf R^{d-1}$ obtained by deleting the $j$th entry of $a\in\mathbf R^d$.\\
$a\wedge b$                     & For $a,b\in\mathbf R^d$, $a\wedge b\equiv (\min\{a^{(1)}, b^{(1)}\},\ldots,\min\{a^{(d)}, b^{(d)}\})$.\\
$a\vee b$                       & For $a,b\in\mathbf R^d$, $a\vee b\equiv (\max\{a^{(1)}, b^{(1)}\},\ldots,\max\{a^{(d)}, b^{(d)}\})$.\\
$ a\Lambda$                     & For a set $\Lambda$ in a vector space and $a\in\mathbf R$, $a\Lambda\equiv\{a\lambda: \lambda\in\Lambda\}$.\\
$\Lambda+\theta$                & For a set $\Lambda$ and an element $\theta$ in a vector space, $\Lambda+\theta\equiv\{\lambda+\theta: \lambda\in\Lambda\}$.\\
$\overline{\Lambda}$            & For a set $\Lambda$ in a topological space, $\overline\Lambda$ is the closure of $\Lambda$.\\
$\Phi$ & The standard normal cdf.\\
$\|f\|_\infty$                  & For a function $f:  T\to\mathbf M^{m\times k}$, $\|f\|_\infty\equiv\sup_{t\in T}\sqrt{\mathrm{tr}(f(t)^\transpose f(t))}$.\\
$\ell^\infty(T)$       & For a nonempty set $T$, $\ell^\infty(T) \equiv \{f:T\to\mathbf R:    \|f\|_\infty < \infty\}$.\\
\hline\hline
\end{tabularx}
\end{center}
\end{table}
\vspace{-0.1in}}

\section{Shape Restrictions as Convex Cones}\label{App: convex cone}

In this section, we discuss the convex cone property for some shape restrictions and provide details in formulating the linearly constrained quadratic program \eqref{Eqn: LCQP}, along with additional references omitted from the main text. For ease of exposition, we shall work with $\mathbf H=L^2([0,1]^d)$ except in Example \ref{Ex: Slutsky aux}. In turn, we let $\{z_j\}_{j=1}^k$ be a collection of grid points over $[0,1]^d$, based on which we approximate the $\|\cdot\|_{\mathbf H}$-norms via numerical integration; e.g., if $d=2$, then we may take $\{(s/N,t/N): s=0,\ldots,N, t=0,\ldots,N\}$ with some suitably chosen $N$. Finally, let $\vartheta\equiv [\theta(z_1),\ldots,\theta(z_k)]^\transpose$ and define $D_k\in\mathbf M^{(k-1)\times k}$ as the matrix such that $D_k\vartheta=[\theta(z_2)-\theta(z_1),\ldots,\theta(z_k)-\theta(z_{k-1})]^\transpose$.

\begin{ex}[Monotonicity]\label{Ex: monotonicity}
Let $\Lambda$ be the class of nondecreasing functions in $\mathbf H$. The convex cone property of $\Lambda$ is well known---see, for example, Theorem 7.1 in \citet{BBBB1972}. To compute the projections onto $\Lambda$, let $\theta\in\mathbf H$. If $d=1$, then $\Pi_\Lambda\theta$ may be approximated over $\{z_j\}_{j=1}^k$ by $h^*$ that solves
\begin{align}\label{Eqn: quadratic1}
\min_{h\in\mathbf R^k}\|h-\vartheta\|\qquad \text{s.t.}\quad D_kh\ge 0~.
\end{align}
If $d=2$, then $\Pi_\Lambda\theta$ is approximated by solving the same problem in \eqref{Eqn: quadratic1} but subject to $Ah\ge 0$, where $A=[A_1^\transpose, A_2^\transpose]^\transpose$ such that $A_1h\ge 0$ enforces the monotonicity with respect to the first coordinate and $A_2h\ge 0$ enforces the second. Computations in higher dimensions are analogous though more complicated.


There is a large literature on estimation by imposing solely shape restrictions, mostly based on the maximum likelihood and least squares principles---see, e.g., \citet{HanWangChatterjeeSamworth2019Isotonic} and references therein. Alternatively, monotonicity may be enforced by applying certain operators, such as projection \citep{MammenMarronTurlachWand2001Projection} and monotone rearrangement \citep{ChernozhukovFernandezGalichon2009Improving}, to unconstrained estimators. To retain smoothness, smoothed monotone estimators have also been developed---see, e.g., \citet{MammenMarronTurlachWand2001Projection} and \citet{HallHuang2001Monotonicity}. Finally, as discussed in the Introduction, an overwhelming majority of existing tests, with the notable exception of \citet{Chetverikov2018Monotonicity}, are based on least favorite configurations and limited to univariate settings. \qed
\end{ex}

\begin{ex}[Concavity/convexity]\label{Ex: convexity}
Let $\Lambda$ be the family of concave functions in $\mathbf H$, and $\theta\in\mathbf H$ be given. Proposition 3 in \citet{LimGlynn2012Convex} implies that $\Lambda$ is a closed convex cone. If $d=1$ and $\{z_j\}$ are equidistanced, then the projection $\Pi_\Lambda\theta$ may be approximated over $\{z_j\}_{j=1}^k$ by $h^*$ that solves
\begin{align}\label{Eqn: quadratic contraint convex}
\min_{h\in\mathbf R^k}\|h-\vartheta\|\qquad \text{s.t.}\quad D_{k-1}D_kh\le 0~.
\end{align}
Unfortunately, \eqref{Eqn: quadratic contraint convex} is not readily generalizable to multivariate settings. As formalized in \citet{Kuosmanen2008Convex}, the projection $\Pi_\Lambda\theta$ may be approximated by the map $z\mapsto \min_{j=1}^k\{a_j^*+z^\transpose b_j^*\}$, where $\{a_j^*,b_j^*\}_{j=1}^k$ solve the problem
\begin{equation}
\begin{aligned}\label{Eqn: quadratic contraint convex2}
\min_{ a_i\in\mathbf R,b_i\in\mathbf R^d}  \quad \{\sum_{i=1}^k[\theta(z_i)&-a_i-b_i^\transpose z_i]^2\}^{1/2}\\
 \text{s.t.} & \quad a_i+b_i^\transpose z_i\le a_j+b_j^\transpose z_i \text{ for }i,j=1,1,\ldots,k~.
\end{aligned}
\end{equation}
Note that the number of effective constraints in \eqref{Eqn: quadratic contraint convex2} is $k(k-1)$. An attractive feature of the formulation in \eqref{Eqn: quadratic contraint convex2} is that the joint test of monotonicity and concavity amounts to the same problem but subject to the {\it additional} constraints $b_j\ge 0$ for all $j$.

As with monotonicity, there are three general estimation strategies: estimation under solely convexity/concavity \citep{HanWellner2016Multivariate}, smoothing \citep{HallHuang2001Monotonicity,MammenMarronTurlachWand2001Projection}, and post-processing \citep{ChenChernozhukovFernandezKostyshakLuo2021Shape}. The studies on testing are less extensive than monotonicity, and share the features that most of them are conservative and/or limited to univariate settings---see the Introduction for references. \citet{ChenKato2019Jackknife} developed a bootstrap version of \citet{AbrevayaJiang2005Curvature}, which, despite its nonconservativeness, is computationally intensive to implement. \citet{SongChenKato2020Stratified} proposed an ``incomplete'' version of \citet{ChenKato2019Jackknife}, which, as documented in their simulations, ``is consistently on the conservative side.'' \qed
\end{ex}

\begin{ex}[Slutsky Restriction]\label{Ex: Slutsky aux}
For simplicity, let us consider the setup of Example \ref{Ex: Slutsky}, and note that $\Lambda$ being a convex cone is well known in linear algebra (see also \citet[p.195]{AguiarSerrano2017Slutsky}). The projection $\Pi_\Lambda\theta$ of $\theta\in\mathbf H$ onto $\Lambda$ admits a closed form expression. Specifically, for $\theta_{\sigma}\equiv (\theta+\theta^\transpose)/2$ the symmetric part of $\theta$, let $\theta_\sigma(t)= \mathbb U(t)\mathbb S(t)\mathbb U(t)^\transpose$ where $\mathbb S(t)\equiv\mathrm{diag}(\lambda_1(t),\ldots,\lambda_{d_q}(t))$ and $\mathbb U$ satisfies $\mathbb U(t)\mathbb U(t)^\transpose =I_{d_q}$ for all $t\equiv(p,y)$. Here, $\mathrm{diag}(a_1,\ldots,a_{d_q})\in\mathbf M^{d_q\times d_q}$ is the diagonal matrix whose diagonal entries are $a_1,\ldots, a_{d_q}$. In turn, letting $\mathbb S_-(t)\equiv\mathrm{diag}(\lambda_{1,-}(t),\ldots,\lambda_{d_q,-}(t))$ with $\lambda_{j,-}\equiv\min\{\lambda_j,0\}$ for all $j=1,\ldots,d_q$, we have:  for all $t\equiv(p,y)$,
\begin{align}
(\Pi_\Lambda\theta)(t) = \mathbb U(t)\mathbb S_-(t)\mathbb U(t)^\transpose~.
\end{align}

\citet{Hoderlein2011Rational} and \citet{DetteHoderleinNeumeyer2016Slutsky} developed tests for fixed $(p,y)$. As theory predicts the restriction for all $(p,y)$, one may employ these tests by discretizing the data. However, discretization entails an extra tuning parameter whose choice may be a delicate matter. Moreover, \citet{DetteHoderleinNeumeyer2016Slutsky}'s test, as the authors noted, is in general conservative, while validity of \citet{Hoderlein2011Rational}'s test has not been formally proven---see \citet{ChenFang2016Rank} for the challenges involved in a related but different problem.\qed
\end{ex}

\begin{ex}[Supermodularity]
Let $d\ge 2$ and $\Lambda\subset\mathbf H$ be the set of supermodular functions, i.e., $f\in\Lambda$ if and only if, for any $y,z\in[0,1]^d$,
\begin{align}
f(y)+f(z)\le f(y\vee z)+f(y\wedge z)~.
\end{align}
By Lemma 2.6.1 in \citet{Topkis1998Supermodularity}, $\Lambda$ is a closed convex cone. Consider $d=2$ first, and pick $\theta\in\mathbf H$. For simplicity, let $\vartheta$ be the vectorization of the matrix $\Theta^\transpose$ such that the $(i,j)$th entry of $\Theta$ is $\theta(i/n,j/n)$, for $i,j=0,\ldots,N$. Then, following \citet{Beresteanu2007Shape}, computing $\Pi_\Lambda(\theta)$ amounts to solving: for $k\equiv N+1 $,
\begin{align}\label{Eqn: quadratic, supermodularity}
\min_{h\in\mathbf R^{k^2}}\|h-\vartheta\|\qquad \text{s.t.}\quad (D_k\otimes D_k)\vartheta\ge 0~,
\end{align}
where the number of constraints is $N^2$. If $d\ge 3$, then the equivalence of supermodularity and pairwise supermodularity \citep{Topkis1998Supermodularity} implies that each pair of covariates must satisfy the constraint in \eqref{Eqn: quadratic, supermodularity}. Despite its importance in economics, econometric studies are rather limited.  \citet{Chetverikov2018Monotonicity}'s test on monotonicity, as the author noted, may be adapted to handle supermodularity. Interestingly, separability of a function $\theta_0$ in its arguments is equivalent to $\theta_0$ being supermodular and submodular \citep[Theorem 2.6.4]{Topkis1998Supermodularity}, and thus also shares the convex cone property. \qed
\end{ex}

\begin{ex}[Nonnegativity]\label{Shape: Nonnegativity}
Let $\Lambda\subset \mathbf H$ be the family of nonnegative functions, and $\theta\in\mathbf H$. As well known (see, e.g., \citet[p.65]{Deutsch2012Best}), $\Lambda$ is a convex cone and the projection of $\theta$ onto $\Lambda$ is given by: for any $t\in [0,1]^d$,
\begin{align}
(\Pi_\Lambda\theta)(t)=\max\{\theta(t),0\}~.
\end{align}
There are numerous studies on nonnegativity, such as (conditional) moment inequalities characterizing choice probabilities or payoffs \citep{Pakes_Porter_Ho_Ishii2015}, (conditional) stochastic dominance for ordering uncertain prospects \citep{LintonSongWhang2010Improved}, Lorenz dominance for measuring inequality \citep{SunBeare2019Lorenz}, and inequalities constraining equilibrium bid distributions or winning probabilities in auction models \citep{GuerrePerrigneVuong2009NPID}. \qed
\end{ex}

\begin{ex}[Joint Restrictions]
Shape restrictions often arise simultaneously in economics---see, e.g., \citet{AitSahaliaDuarte2003Option}. Existing tests, however, mostly focus on particular restrictions, and a multiple testing based on these tests is generally conservative. In contrast, our framework allows for jointly testing restrictions as intersections of convex cones remain convex cones. For example, letting $\Lambda$ consist of monotonic and supermodular functions leads to a joint test of monotonicity and supermodularity, for which the constraints in the quadratic program are obtained by vertically stacking the individual $A$ matrices in \eqref{Eqn: LCQP}.
\qed
\end{ex}

We conclude by making a few remarks. First, just as the $t$-test is inconsistent in testing $\mathrm H_0: \theta_0<0$ vs.\ $\mathrm H_1: \theta_0\ge 0$ for a mean parameter $\theta_0$, a level $\alpha$ test for a ``strict'' restriction such as strict concavity is generally inconsistent. Assumption \ref{Ass: convex setup}(i) ensures that ``equality'' is included under $\mathrm H_0$. We note that closedness of $\Lambda$ (in $\mathbf H$) may require identifying shape restrictions through equivalent classes; e.g., for monotonicity in $L^2([0,1])$, we have $\theta\in\Lambda$ if $\theta=\vartheta$ almost everywhere for some monotonic function $\vartheta\in L^2([0,1])$. Second, the convex cone property depends on a proper choice of the parameter; e.g., the range restriction $\Lambda_0\equiv\{f\in L^2([0,1]): f(x)\le 1\,\forall\,x\in[0,1]\}$ is not a convex cone, but we may consider $\theta_0\equiv 1-f_0$ if $f_0$ is the original parameter, and define $\Lambda\equiv\{g\in L^2([0,1]): g(x)\ge 0\,\forall\,x\in[0,1]\}$. It may be necessary to choose a parameter that involves some derivative(s); e.g., in Example \ref{Ex: Slutsky}, Assumption \ref{Ass: convex setup} holds for the Slutsky matrix $\theta_0$ (which involves derivatives of $g_0$) but not for $g_0$ itself. Third, while Section \ref{Sec: examples} is centered around regression models as a result of their popularity and the space limitation, our framework is also applicable to other settings, such as those concerning densities/distributions, including monotonicity of densities \citep{Fang2019KW}, likelihood ratio ordering \citep{BeareandMoon2015} and stochastic monotonicity \citep{Lee_Linton_Whang2009monotonicity}. Note that, in the presence of covariates (as controls), some of these results are not directly applicable. Alternatively, one may apply our test in structural models where shape restrictions arise as testable implications---see, e.g., \citet{PinkseSchurter2019Auction}. Finally, in implementing our test, one may be prompted to ignore some features of $\theta_0$ that coexist with the shape restriction but invalidate Assumption \ref{Ass: convex setup} when incorporated. For example, if $\theta_0\in L^2([0,1])$ and $0\le \theta_0(x)\le 1$ with $\theta_0(0)=0$ and $\theta_0(1)=1$, then testing monotonicity on $\theta_0$ without the equality constraints may result in power loss---note that Theorem 1.6 in \citet{BBBB1972} implies that projection preserves the range.


\section{The Special Case}\label{Sec: special case}

The aim of this section is twofold. First, we show that, when $\hat\theta_n$ admits an asymptotic distribution, Assumptions \ref{Ass: strong approx} and \ref{Ass: Gaussian} can be simplified to conditions that may be more familiar to practitioners. Second, we expound the point that, even in this special case, our test improves upon existing inferential methods along several dimensions.

We need additional notation and concepts. Specifically, define
\begin{align}
\mathrm{BL}_1(\mathbf H)=\{f: \mathbf H\to\mathbf R: |f(x)|\le 1,|f(x)-f(y)|\le\|x-y\|_{\mathbf H} \text{ for all }x,y\in\mathbf H\}~,
\end{align}
and denote the tangent cone $T_{\theta_P}$ of $\Lambda$ at $\theta_P\in\Lambda\subset\mathbf H$ by $T_{\theta_P}\equiv \overline{ \bigcup_{\alpha> 0}\alpha\{\Lambda- \theta_P\}}$. In turn, define a map $\phi_{\theta_P}': \mathbf H\to\mathbf R$ by $\phi_{\theta_P}'(h)\equiv\|h-\Pi_{T_{\theta_P}}h\|_{\mathbf H}$, which is in fact the so-called Hadamard directional derivative of $\phi$. Since only the functional form of $\phi_{\theta_P}'$ is relevant to us here, we refer the reader to \citet{FangSantos2018HDD} for detailed discussions of this concept.

We next impose an analog of Assumption \ref{Ass: strong approx} as follows.

\begin{ass}\label{Ass: strong approx, regular}
(i) $\sup_{f\in\mathrm{BL}_1(\mathbf H)} |E_P[f( r_n\{\hat\theta_n-\theta_P\})]-E[f(\mathbb G_P)]|=o(1)$ uniformly in $P\in\mathbf P$ for an estimator $\hat\theta_n: \{X_i\}_{i=1}^n\to\mathbf H$; (ii) $\hat{\mathbb G}_n\in\mathbf H$ is a bootstrap estimator satisfying $\sup_{f\in\mathrm{BL}_1(\mathbf H)} |E[f(\hat{\mathbb G}_n)|\{X_i\}_{i=1}^n]-E[f(\mathbb G_P)]|=o_p(1)$ uniformly in $P\in\mathbf P$.
\end{ass}

Assumption \ref{Ass: strong approx, regular} simply requires uniform convergence in distribution and uniform validity of bootstrap, which may be verified by appealing to existing results \citep{GineZinn1991GaussianCharac,Sheehy_Wellner1992uniform}. Assumption \ref{Ass: strong approx, regular} in fact automatically implies a weak version of Assumption \ref{Ass: strong approx} obtained by replacing the independence condition in Assumption \ref{Ass: strong approx}(ii) with an asymptotical independence condition characterized as: uniformly in $P\in\mathbf P$,
\begin{align}\label{Eqn: asymptotic independence}
\sup_{f\in\mathrm{BL}_1(\mathbf H)} |E[f(\bar{\mathbb Z}_{n,P})|\{X_i\}_{i=1}^n]-E[f(\bar{\mathbb Z}_{n,P})]|=o_p(1)~.
\end{align}

\begin{pro}\label{Pro: strong approximation, special}
Let $\mathbf H$ be a separable Hilbert space. If Assumption \ref{Ass: strong approx, regular} holds, then (i) the above weak version of Assumption \ref{Ass: strong approx} follows, with $c_n=1$ and $\mathbb Z_{n,P}$ copies of $\mathbb G_P$,\footnote{We are indebted to Andres Santos for suggesting this result and sketching the proof.} and (ii) $\psi_{\kappa_n,P}(\mathbb Z_{n,P})\convl \phi_{\theta_P}'(\mathbb G_P)$ for all $P\in\mathbf P_0$, provided $\kappa_n\to\infty$.
\end{pro}

Since our results in Section \ref{Sec: general theory} remain valid under the weak version of Assumption \ref{Ass: strong approx} by Lemma \ref{Lem: asymptotic independence}, Proposition \ref{Pro: strong approximation, special}(i) implies that our test is applicable to this special case subject to Assumptions \ref{Ass: convex setup}, \ref{Ass: strong approx, regular}, and \ref{Ass: Gaussian}. Proposition \ref{Pro: strong approximation, special}(ii) further implies that, if $\kappa_n\to\infty$, then the coupling variables $\{\psi_{\kappa_n,P}(\mathbb Z_{n,P})\}$ admit a limit in distribution. Therefore, one may replace Assumption \ref{Ass: Gaussian}(iii) with $c_P(1-\alpha-\varpi)\ge c_P(0.5)+\varsigma$ for some $\varsigma>0$ and $c_P(\tau)$ the $\tau$-quantile of $\phi_{\theta_P}'(\mathbb G_P)$, which is effectively the same as requiring that $\phi_{\theta_P}'(\mathbb G_P)$ be continuous and strictly increasing at $c_P(1-\alpha)$ as imposed in \citet{FangSantos2018HDD}. In turn, Assumption \ref{Ass: Gaussian}(iv) then reduces to $c_n=O(1)$ and so the coupling order $o_p(c_n)$ becomes $o_p(1)$.


We next compare our test to some existing ones. Employing a generalized Delta method, \citet{FangSantos2018HDD} obtained that, under Assumptions \ref{Ass: convex setup}(i) and \ref{Ass: strong approx, regular}(i),
\begin{align}\label{Eqn: projection limit}
 r_n \phi(\hat\theta_n)\convl  \phi_{\theta_P}'(\mathbb G_P)\equiv\|\mathbb G_P-\Pi_{T_{\theta_P}}\mathbb G_P\|_{\mathbf H}~,
\end{align}
for each $P\in\mathbf P_0$. Exploiting the insight that the limit in \eqref{Eqn: projection limit} is the composition of $\phi_{\theta_P}'$ and $\mathbb G_P$, \citet{FangSantos2018HDD} then showed that a general consistent bootstrap of the limit in \eqref{Eqn: projection limit} may be obtained by constructing $\hat\phi_n'(\hat{\mathbb G}_n)$, a composition of a suitably consistent estimator $\hat\phi_n'$ of $\phi_{\theta_P}'$ with a consistent bootstrap $\hat{\mathbb G}_n$ for $\mathbb G_P$.


\begin{figure}[t]
\centering
\begin{minipage}[b]{0.3\linewidth}
\begin{tikzpicture}[>=latex',scale=1.1]
\draw (0,-1)--(-2.5,-1)--(-2.5,3)--(1.5,3)--(1.5,0); 
\shadedraw[left color=Honeydew3,right color=Honeydew3] (-2,1)rectangle (1,2); 
\draw (0.5,1.5) node {\tiny $\Lambda$};
\fill[red] (-2,2) circle (0.8pt);
\draw (-2.1,2.1) node {\tiny $\theta_1$}; 
\draw[->,NavyBlue] (-2,2)--(-1.3,2);
\draw[->,NavyBlue] (-2,2)--+(270:0.7);
\draw[->,NavyBlue] (-2,2)--+(315:0.7);
\fill[very nearly transparent] (0,0) rectangle (1.5,-1); 
\draw (0.75,-0.5) node {\tiny $T_{\theta_1}$};
\draw (0,0)--(1.5,0);
\draw (0,0)--(0,-1);
\draw[->,LightSlateGrey] (0,0)--(0.7,0);
\draw[->,LightSlateGrey] (0,0)--+(270:0.7);
\draw[->,LightSlateGrey] (0,0)--+(315:0.7);
\draw[->,densely dashed] (-1,0.8)--+(315:0.7);
\draw[loosely dotted] (-2.5,0)--(0,0);
\draw (1.5,0) node[anchor=west] {\tiny $0$};
\draw[loosely dotted] (0,0)--(0,3);
\draw (0,3) node[anchor=south] {\tiny $0$};
\end{tikzpicture}
\end{minipage}
\hfill
\begin{minipage}[b]{0.3\linewidth}
\begin{tikzpicture}[>=latex',scale=1.1]
\draw (-2.5,0)--(-2.5,3)--(1.5,3)--(1.5,0); 
\shadedraw[left color=Honeydew3,right color=Honeydew3] (-2,1)rectangle (1,2); 
\draw (0.5,1.5) node {\tiny $\Lambda$};
\fill[red] (-1,2) circle (0.8pt);
\draw (-1,2) node[anchor=south] {\tiny $\theta_2$}; 
\draw[->,NavyBlue] (-1,2)--(-0.3,2);
\draw[->,NavyBlue] (-1,2)--+(180:0.7);
\draw[->,NavyBlue] (-1,2)--+(225:0.7);
\draw[->,NavyBlue] (-1,2)--+(270:0.7);
\draw[->,NavyBlue] (-1,2)--+(315:0.7);
\fill[very nearly transparent] (-2.5,-1) rectangle (1.5,0);
\draw (-1,-0.5) node {\tiny $T_{\theta_2}$};
\draw[->,LightSlateGrey] (0,0)--(0.7,0);
\draw[->,LightSlateGrey] (0,0)--+(180:0.7);
\draw[->,LightSlateGrey] (0,0)--+(225:0.7);
\draw[->,LightSlateGrey] (0,0)--+(270:0.7);
\draw[->,LightSlateGrey] (0,0)--+(315:0.7);
\draw[->,densely dashed] (-0.8,0.8)--+(315:0.7);
\draw (-2.5,0)--(1.5,0);
\draw (1.5,0) node[anchor=west] {\tiny $0$};
\draw[loosely dotted] (0,-1)--(0,3);
\draw (0,3) node[anchor=south] {\tiny $0$};
\end{tikzpicture}
\end{minipage}
\hfill
\begin{minipage}[b]{0.3\linewidth}
\begin{tikzpicture}[>=latex',scale=1.1]
\shadedraw[left color=Honeydew3,right color=Honeydew3,semitransparent] (-2,1)rectangle (1,2); 
\draw (0.5,1.5) node {\tiny $\Lambda$};
\fill[red] (-1,1.5) circle (0.8pt);
\draw (-1,1.55) node[anchor=west] {\tiny $\theta_3$}; 
\draw[->,NavyBlue] (-1,1.5)--(-0.3,1.5);
\draw[->,NavyBlue] (-1,1.5)--+(45:0.7);
\draw[->,NavyBlue] (-1,1.5)--+(90:0.7);
\draw[->,NavyBlue] (-1,1.5)--+(135:0.7);
\draw[->,NavyBlue] (-1,1.5)--+(180:0.7);
\draw[->,NavyBlue] (-1,1.5)--+(225:0.7);
\draw[->,NavyBlue] (-1,1.5)--+(270:0.7);
\draw[->,NavyBlue] (-1,1.5)--+(315:0.7);
\fill[very nearly transparent] (-2.5,-1) rectangle (1.5,3);
\draw (-1,0.5) node {\tiny $T_{\theta_3}$};
\draw[->,LightSlateGrey] (0,0)--(0.7,0);
\draw[->,LightSlateGrey] (0,0)--+(45:0.7);
\draw[->,LightSlateGrey] (0,0)--+(90:0.7);
\draw[->,LightSlateGrey] (0,0)--+(135:0.7);
\draw[->,LightSlateGrey] (0,0)--+(180:0.7);
\draw[->,LightSlateGrey] (0,0)--+(225:0.7);
\draw[->,LightSlateGrey] (0,0)--+(270:0.7);
\draw[->,LightSlateGrey] (0,0)--+(315:0.7);
\draw[->,densely dashed] (-0.5,0.85)--+(315:0.5);
\draw[loosely dotted] (-2.5,0)--(1.5,0);
\draw (1.5,0) node[anchor=west] {\tiny $0$};
\draw[loosely dotted] (0,-1)--(0,3);
\draw (0,3) node[anchor=south] {\tiny $0$};
\end{tikzpicture}
\end{minipage}
\caption{The tangent cone $T_\theta$ depends on $\theta$ discontinuously. As $\theta$ moves from the corner at $\theta_1$ but still stays on the boundary at $\theta_2$, $T_\theta$ changes from the fourth orthant $T_{\theta_1}$  to the half plane $T_{\theta_2}$. In turn, as $\theta$ moves into the interior at $\theta_3$ from $\theta_2$, $T_{\theta}$ becomes the entire plane $T_{\theta_3}$.}
\label{Fig: tangent cone}
\end{figure}
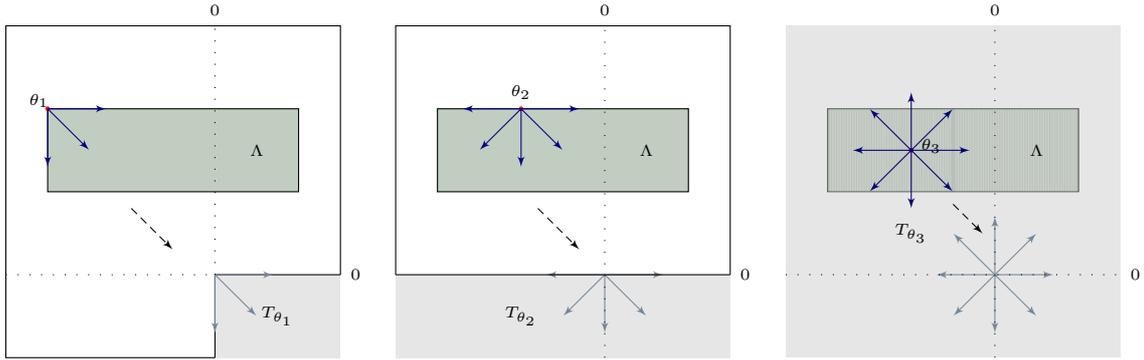

While the bootstrap $\hat{\mathbb G}_n$ is often straightforward to construct as in Section \ref{Sec: general theory}, obtaining a suitable estimator $\hat\phi_n'$ turns out to be nontrivial. The challenge involved may be understood in view of the discontinuity of the cone-valued map $\theta\mapsto T_{\theta}$, as illustrated in Figure \ref{Fig: tangent cone}. In this regard, \citet{FangSantos2018HDD} proposed the following concrete estimator: for any $h\in\mathbf H$ and some $\kappa_n\uparrow\infty$,
\begin{align}\label{Eqn: analytic estimator}
\hat\phi_n'(h)=\sup_{\theta\in\Lambda: r_n\|\theta-\Pi_\Lambda \hat\theta_n\|_{\mathbf H}\le \kappa_n} \|h-\Pi_{T_\theta} h\|_{\mathbf H}~.
\end{align}
Evaluating the supremum in \eqref{Eqn: analytic estimator}, however, may be computationally costly as it entails estimation of a local parameter space, i.e., $T_{\theta_P}$. Alternatively, one may employ a numerical estimator following \citet{Hong_Li2018numerical}, but there are no data-driven procedures to date for selecting the step size (needed to carry out the numerical differentiation). This raises substantive concerns because the resulting bootstrap may be sensitive to the choice of the step size, as documented in \citet{MastenPoireir2017Breakdown} and \citet{ChenFang2016Rank}. One may also appeal to the $m$ out of $n$ bootstrap or subsampling, but the choice of the sub-sample size may be difficult, among other issues---see Remark 3.1 in \citet{ChenFang2016Rank}.

While our development is undertaken outside the scope of the Delta method, there is an intriguing connection to the general theory of \citet{FangSantos2018HDD}, as we now flesh out. To this end, recall our bootstrap statistic $\hat\psi_{\kappa_n}(\hat{\mathbb G}_n)$.

\begin{pro}\label{Pro: derivative estimator}
Let Assumptions \ref{Ass: convex setup} and \ref{Ass: strong approx, regular}(i) hold. If $\kappa_n\to\infty$ and $\kappa_n/ r_n\to 0$, then it follows that $\hat\psi_{\kappa_n}(h)\convp \phi_{\theta_P}'(h)$ for any $h\in\mathbf H$ and $P\in\mathbf P_0$.
\end{pro}

Since $h\mapsto\hat\psi_{\kappa_n}(h)$ is Lipschitz continuous, Proposition \ref{Pro: derivative estimator} implies that $\hat\psi_{\kappa_n}$ is consistent in estimating $\phi_{\theta_P}'$ in the sense of \citet{FangSantos2018HDD}---see their Remark 3.4. Therefore, when $r_n\{\hat\theta_n-\theta_P\}$ converges in distribution, our test is effectively the test of \citet{FangSantos2018HDD} (with respect to their general theory), but based on a derivative estimator that is new and simpler relative to \eqref{Eqn: analytic estimator}. We stress that the computational advantage hinges on the convex cone property but not convexity alone. In accord with previous discussions, Proposition \ref{Pro: derivative estimator} also shows that, by letting $\kappa_n\to\infty$ (in addition to $\kappa_n/r_n\to 0$), our test is not conservative in the sense that it is pointwise (in $P$) asymptotically exact as in \citet{FangSantos2018HDD}.

\section{More Proofs and Auxiliary Results}\label{App: auxiliary results}

\noindent{\sc Proof of Proposition \ref{Pro: tuning parameter}:} Let $\hat F_n$ be the conditional cdf of $\|\hat{\mathbb G}_n\|_{\mathbf H}$ given $\{X_i\}_{i=1}^n$, and let $F_{n,P}$ be the cdf of $\|\mathbb Z_{n,P}\|_{\mathbf H}$. Note that $F_{n,P}$ is also the cdf of $\|\bar{\mathbb Z}_{n,P}\|_{\mathbf H}$ since $\bar{\mathbb Z}_{n,P}$ is a copy of $\mathbb Z_{n,P}$ by Assumption \ref{Ass: strong approx}(ii). As a first step, we show that $\hat F_n$ and $F_{n,P}$ are suitably close in probability. By Assumptions \ref{Ass: strong approx}(ii), we obtain
\begin{align}\label{Eqn: tuning parameter, aux1}
P(\|\hat{\mathbb G}_n-\bar{\mathbb Z}_{n,P}\|_{\mathbf H}> \delta_n)=o(1)~,
\end{align}
for some $\delta_n=o(c_n)$, uniformly in $P\in\mathbf P$. Fix $\eta>0$. By Markov's inequality, Fubini's theorem, and result \eqref{Eqn: tuning parameter, aux1}, we may in turn have that, uniformly in $P\in\mathbf P$,
\begin{multline}\label{Eqn: tuning parameter, aux2}
P(P(|\|\hat{\mathbb G}_n\|_{\mathbf H}-\|\bar{\mathbb Z}_{n,P}\|_{\mathbf H}|> \delta_n|\{X_i\}_{i=1}^n)>\eta)\\
\le\frac{1}{\eta} P(|\|\hat{\mathbb G}_n\|_{\mathbf H}-\|\bar{\mathbb Z}_{n,P}\|_{\mathbf H}|> \delta_n)\le\frac{1}{\eta} P(\|\hat{\mathbb G}_n-\bar{\mathbb Z}_{n,P}\|_{\mathbf H}> \delta_n)=o(1)~.
\end{multline}
Since $\eta>0$ is arbitrary, we may therefore conclude from \eqref{Eqn: tuning parameter, aux2} that
\begin{align}\label{Eqn: tuning parameter, aux3}
P(|\|\hat{\mathbb G}_n\|_{\mathbf H}-\|\bar{\mathbb Z}_{n,P}\|_{\mathbf H}|> \delta_n|\{X_i\}_{i=1}^n)=o_p(1)~.
\end{align}
By simple manipulations, we then have: for all $t\in\mathbf R$,
\begin{align}\label{Eqn: tuning parameter, aux4}
\hat F_n(t)- F_{n,P}(t)&=P(\|\hat{\mathbb G}_n\|_{\mathbf H}\le t|\{X_i\}_{i=1}^n)-P(\|\bar{\mathbb Z}_{n,P}\|_{\mathbf H}\le t)\notag\\
&\le P(\|\bar{\mathbb Z}_{n,P}\|_{\mathbf H}\le t+ \delta_n|\{X_i\}_{i=1}^n)-P(\|\bar{\mathbb Z}_{n,P}\|_{\mathbf H}\le t) \notag\\
&\qquad \qquad \qquad \qquad  + P(|\|\hat{\mathbb G}_n\|_{\mathbf H}-\|\bar{\mathbb Z}_{n,P}\|_{\mathbf H}|> \delta_n|\{X_i\}_{i=1}^n)\notag\\
&\le P(|\|\bar{\mathbb Z}_{n,P}\|_{\mathbf H}-t|\le  \delta_n ) + o_p(1)~,
\end{align}
uniformly in $P\in\mathbf P$, where the second inequality follows by $\bar{\mathbb Z}_{n,P}$ being independent of $\{X_i\}_{i=1}^n$ (so that $P(\|\bar{\mathbb Z}_{n,P}\|_{\mathbf H}\le t)=P(\|\bar{\mathbb Z}_{n,P}\|_{\mathbf H}\le t|\{X_i\}_{i=1}^n)$) and result \eqref{Eqn: tuning parameter, aux3}. By analogous arguments, we also have: for all $t\in\mathbf R$,
\begin{align}\label{Eqn: tuning parameter, aux5}
 F_{n,P}(t)-\hat F_n(t) \le P(|\|\bar{\mathbb Z}_{n,P}\|_{\mathbf H}-t|\le  \delta_n ) + o_p(1)~,
\end{align}
uniformly in $P\in\mathbf P$. Combining results \eqref{Eqn: tuning parameter, aux4} and \eqref{Eqn: tuning parameter, aux5}, we arrive at:
\begin{align}\label{Eqn: tuning parameter, aux6}
|\hat F_n(t)- F_{n,P}(t)| \le P(|\|\bar{\mathbb Z}_{n,P}\|_{\mathbf H}-t|\le  \delta_n ) + o_p(1)~,
\end{align}
for all $t\in\mathbf R$, uniformly in $P\in\mathbf P$, where the $o_p(1)$ term does not involve $t$.

Let $m_{n,P}$ be the median of $F_{n,P}$. By Assumptions \ref{Ass: Gaussian}(i) and \ref{Ass: tuning}, we may apply Lemma \ref{Lem: density bound, norm} to conclude that, for any $t>m_{n,P}+\delta_n$,
\begin{multline}\label{Eqn: tuning parameter, aux7}
P(|\|\bar{\mathbb Z}_{n,P}\|_{\mathbf H}-t|\le  \delta_n)=\int_{t- \delta_n}^{t+ \delta_n} F_{n,P}'(r)\,\mathrm dr\\
\le \int_{t- \delta_n}^{t+ \delta_n} \frac{2r-m_{n,P}}{(r-m_{n,P})^2}\,\mathrm dr\le 2 \delta_n\frac{2(t-\delta_n)-m_{n,P}}{(t- \delta_n-m_{n,P})^2}~,
\end{multline}
where the second inequality (in the second line) follows by $r\mapsto (2r-m_{n,P})/(r-m_{n,P})^2$ being decreasing on $(m_{n,P},\infty)$. Since $\bar{\mathbb Z}_{n,P}$ is a copy of $\mathbb Z_{n,P}$, by \citet{Kwapien1994Median} and Assumption \ref{Ass: Gaussian}(ii), we then have: for some constant $\zeta>0$,
\begin{align}\label{Eqn: tuning parameter, aux8}
\sup_{P\in\mathbf P} m_{n,P}\le \sup_{P\in\mathbf P}E_P[\|\bar{\mathbb Z}_{n,P}\|_{\mathbf H}]\le \zeta<\infty~.
\end{align}
Since $\delta_n=o(c_n)=o(1)$ due to $c_n=O(1)$, we obtain from results \eqref{Eqn: tuning parameter, aux7} and \eqref{Eqn: tuning parameter, aux8} that, for all $n$ large so that $\delta_n\le 1$ and for all $t\ge \zeta+2$,
\begin{multline}\label{Eqn: tuning parameter, aux9}
\sup_{P\in\mathbf P_0}P(|\|\bar{\mathbb Z}_{n,P}\|_{\mathbf H}-t|\le  \delta_n)\\
\le 2\delta_n\{\frac{2}{t-\delta_n-m_{n,P}}+\frac{m_{n,P}}{(t-\delta_n-m_{n,P})^2}\}\le 2 \delta_n(2+\zeta)~.
\end{multline}
Exploiting $\delta_n=o(1)$ again, we may combine \eqref{Eqn: tuning parameter, aux6} and \eqref{Eqn: tuning parameter, aux9} to conclude
\begin{align}\label{Eqn: tuning parameter, aux10}
|\hat F_n(t)- F_{n,P}(t)| = o_p(1)~,
\end{align}
uniformly in $P\in\mathbf P_0$ and $t\in[\zeta+2,\infty)$.

Next, we aim to prove the first claim of the proposition. Let $M>\zeta+2$ be any large constant. By Lemma 6.10 in \citet{AliprantisandBorder2006}, we have
\begin{align}\label{Eqn: tuning parameter, aux11}
\|\mathbb Z_{n,P}\|_{\mathbf H}=\sup_{h\in\mathbf H_1}\langle h,\mathbb Z_{n,P}\rangle_{\mathbf H}~,
\end{align}
where $\mathbf H_1\equiv\{h\in\mathbf H:\|h\|_{\mathbf H}\le 1\}$. In turn, it follows from result \eqref{Eqn: tuning parameter, aux11} that
\begin{align}\label{Eqn: tuning parameter, aux12}
F_{n,P}(M)=P(\sup_{h'\in\mathbf H_1}\langle h',\mathbb Z_{n,P}\rangle_{\mathbf H}\le M) \le P(\langle h,\mathbb Z_{n,P}\rangle_{\mathbf H}\le M) = \Phi(\frac{M}{\sigma_{n,P}(h)})~,
\end{align}
for all $h\in\mathbf H_1$, where $\sigma_{n,P}^2(h)\equiv E[\langle h,\mathbb Z_{n,P}\rangle_{\mathbf H}^2]$. By the definition of $\bar\sigma_{n,P}^2$, we may then select a sequence $\{h_j\}$ in $\mathbf H_1$ such that $\sigma_{n,P}^2(h_j)\to\bar \sigma_{n,P}^2$ as $j\to\infty$. By continuity of $\sigma\mapsto \Phi(M/\sigma)$, we thus obtain from \eqref{Eqn: tuning parameter, aux12} that
\begin{align}\label{Eqn: tuning parameter, aux13}
 F_{n,P}(M)\le \Phi(\frac{M}{\bar\sigma_{n,P}})~,
\end{align}
for any $P\in\mathbf P_0$ and $n$. By Assumption \ref{Ass: tuning}, we may select some constant $\underline\sigma>0$ such that $\inf_{P\in\mathbf P_0}\bar\sigma_{n,P}>\underline\sigma$ for large $n$. By result \eqref{Eqn: tuning parameter, aux13}, we then must have
\begin{align}\label{Eqn: tuning parameter, aux14}
 F_{n,P}(M)\le \Phi(\frac{M}{\underline\sigma})<1~,
\end{align}
for any $P\in\mathbf P_0$ and $n$. Now, by the definition of $\hat\tau_{n,1-\gamma_n}$, we note that
\begin{align}\label{Eqn: tuning parameter, aux14a}
P(\hat\tau_{n,1-\gamma_n}\le M)\le P(\hat F_n(M)\ge 1-\gamma_n) = P(o_p(1) + F_{n,P}(M)\ge 1-\gamma_n)~,
\end{align}
uniformly in $P\in\mathbf P_0$, where the second equality follows by result \eqref{Eqn: tuning parameter, aux10} since $M\ge \zeta+2$ by choice. Combining results \eqref{Eqn: tuning parameter, aux14} and \eqref{Eqn: tuning parameter, aux14a}, we therefore conclude that
\begin{align}\label{Eqn: tuning parameter, aux15}
\limsup_{n\to\infty}\sup_{P\in\mathbf P_0}P(\hat\tau_{n,1-\gamma_n}\le M) = 0~,
\end{align}
whenever $\gamma_n\to 0$. Since $M$ is arbitrary, $\hat\tau_{n,1-\gamma_n}\convp\infty$ uniformly in $P\in\mathbf P_0$ and so the first claim of the proposition follows.

For the second claim, define $\psi: \mathbf H\to\ell^\infty(\mathbf H_1)$ by: for each $h\in\mathbf H$ and $t\in\mathbf H_1$,
\begin{align}
\psi(h)(t)\equiv\langle t,h\rangle_{\mathbf H}~.
\end{align}
By Corollary 6.55 (the Riesz representation theorem) and Lemma 6.10 in \citet{AliprantisandBorder2006}, $\sup_{t\in\mathbf H_1} |\psi(\bar{\mathbb Z}_{n,P})(t)|=\|\bar{\mathbb Z}_{n,P}\|_{\mathbf H}$. Clearly, $\psi$ is linear and continuous. In turn, by Assumption \ref{Ass: Gaussian}(i), $\psi(\bar{\mathbb Z}_{n,P})$ is tight and centered Gaussian in $\ell^\infty(\mathbf H_1)$ by Lemma 2.2.2 in \citet{Bogavcev1998gaussian}. By Example 1.5.10 in \citet{Vaart1996} and Proposition 2.1.12 in \citet{Gine_Nickl2014mathematical}, $\{\psi(\bar{\mathbb Z}_{n,P})(t): t\in \mathbf H_1\}$ is separable as a process; it also has finite median by \eqref{Eqn: tuning parameter, aux8}. By Proposition A.2.4 in \citet{Vaart1996} and \eqref{Eqn: tuning parameter, aux8}, we have: for some absolute constant $C>0$,
\begin{align}\label{Eqn: tuning parameter, aux16}
E[\|\bar{\mathbb Z}_{n,P}\|_{\mathbf H}^2]\le C (E[\|\bar{\mathbb Z}_{n,P}\|_{\mathbf H}])^2 \le C\zeta^2~.
\end{align}
By Proposition A.2.1 in \citet{Vaart1996} and result \eqref{Eqn: tuning parameter, aux16}, we may thus conclude that, for all $x>0$, all $n$ and all $P\in\mathbf P_0$,
\begin{align}\label{Eqn: tuning parameter, aux17}
P(\|\bar{\mathbb Z}_{n,P}\|_{\mathbf H}> x)\le 2\exp\{-\frac{x^2}{8E[\|\bar{\mathbb Z}_{n,P}\|_{\mathbf H}^2]}\}\le 2\exp\{-\frac{x^2}{8C\zeta^2}\} ~.
\end{align}

By the definition of $\hat\tau_{n,1-\gamma_n}$ and the triangle inequality, we have
\begin{multline}\label{Eqn: tuning parameter, aux18}
\gamma_n<P(\|\hat{\mathbb G}_n\|_{\mathbf H}>\hat\tau_{n,1-\gamma_n}-\delta_n|\{X_i\}_{i=1}^n)\\
 \le P(\|\bar{\mathbb Z}_{n,P}\|_{\mathbf H}>\hat\tau_{n,1-\gamma_n}-\delta_n-e_{n,P}|\{X_i\}_{i=1}^n) ~,
\end{multline}
where $e_{n,P}\equiv \|\hat{\mathbb G}_n-\bar{\mathbb Z}_{n,P}\|_{\mathbf H}=o_p(c_n)$ uniformly in $P\in\mathbf P_0$ (by Assumption \ref{Ass: strong approx}(ii)). By result \eqref{Eqn: tuning parameter, aux15} and $c_n=O(1)$ by Assumption \ref{Ass: strong approx}(i), we note that
\begin{align}\label{Eqn: tuning parameter, aux19}
\liminf_{n\to\infty}\inf_{P\in\mathbf P_0}P(\hat\tau_{n,1-\gamma_n}-\delta_n-e_{n,P}>0)= 1~.
\end{align}
Since $\bar{\mathbb Z}_{n,P}$ is independent of $\{X_i\}_{i=1}^n$, we may conclude from results \eqref{Eqn: tuning parameter, aux17}, \eqref{Eqn: tuning parameter, aux18} and \eqref{Eqn: tuning parameter, aux19} that, with probability approaching 1 and uniformly in $P\in\mathbf P_0$,
\begin{align}\label{Eqn: tuning parameter, aux20}
\gamma_n\le 2\exp\{-\frac{(\hat\tau_{n,1-\gamma_n}-\delta_n-e_{n,P})^2}{8C\zeta^2}\}~.
\end{align}
Taking natural logarithms on both sides of \eqref{Eqn: tuning parameter, aux20} plus simple algebra yield:
\begin{align}\label{Eqn: tuning parameter, aux21}
\frac{1}{8C\zeta^2} (\frac{\hat\tau_{n,1-\gamma_n}}{ r_nc_n}-\frac{\delta_n}{ r_nc_n}-\frac{e_{n,P}}{r_nc_n})^2  \le  -\frac{\log\gamma_n}{r_n^2c_n^2}+\frac{\log 2}{r_n^2c_n^2}~.
\end{align}
Suppose $(r_nc_n)^{-2}\log\gamma_n\to 0$. Then we must have $r_nc_n\to\infty$ since $\gamma_n\to 0$ (and so $\log\gamma_n\to-\infty$). Since also $\delta_n=o(c_n)$ and $e_{n,P}=o_p(c_n)$ uniformly in $P\in\mathbf P_0$, we obtain from \eqref{Eqn: tuning parameter, aux21} that $\hat\tau_{n,1-\gamma_n}/ (r_nc_n) \convp 0$ and hence $\kappa_n\equiv r_nc_n/\hat\tau_{n,1-\gamma_n} \convp\infty$ uniformly in $P\in\mathbf P_0$. This completes the proof of the second claim of the proposition. \qed

\begin{lem}\label{Lem: projection monotonicity}
Let Assumption \ref{Ass: convex setup} hold and $\theta_0\in\Lambda$. Define $\psi_{a}(h)\equiv  \|h+a\theta_0-\Pi_\Lambda(h+a\theta_0)\|_{\mathbf H}$ for $h\in\mathbf H$ and $a\ge 0$. Then
$a\mapsto\psi_{a}(h)$ is weakly decreasing on $[0,\infty)$.
\end{lem}
\noindent{\sc Proof:} The lemma immediately follows if we can show that
\begin{align}\label{Eqn: projection monotonicity, aux1}
\psi_{a}(h)=\min_{|a'|\le a}\|h+a'\theta_0-\Pi_\Lambda(h+a'\theta_0)\|_{\mathbf H}~.
\end{align}
Let $\Lambda_1^\circ\equiv\{h^*\in\mathbf H: \langle h^*,\lambda\rangle_{\mathbf H}\le 0\text{ for all }\lambda\in\Lambda, \|h^*\|_{\mathbf H}\le 1\}$. By Assumption \ref{Ass: convex setup} and \citet[p.125-7]{Deutsch2012Best}, we then have: for all $h\in\mathbf H$,
\begin{multline}\label{Eqn: projection monotonicity, aux2}
\min_{|a'|\le a}\|h+a'\theta_0-\Pi_\Lambda(h+a'\theta_0)\|_{\mathbf H}=\min_{|a'|\le a}\max_{h^*\in\Lambda_1^\circ} \langle h^*,h+a'\theta_0\rangle_{\mathbf H}\\
=\min_{|a'|\le a}\max_{h^*\in\Lambda_1^\circ} \{\langle h^*,h\rangle_{\mathbf H}+a'\langle h^*,\theta_0\rangle_{\mathbf H}\}~.
\end{multline}
In turn, by Theorems 49.A and 49.B in \citet{Zeidler1990III}, we obtain
\begin{align}\label{Eqn: projection monotonicity, aux3}
\min_{|a'|\le a}\max_{h^*\in\Lambda_1^\circ} \{\langle h^*,h\rangle_{\mathbf H}+a'\langle h^*,\theta_0\rangle_{\mathbf H}\}=\max_{h^*\in\Lambda_1^\circ} \min_{|a'|\le a}\{\langle h^*,h\rangle_{\mathbf H}+a'\langle h^*,\theta_0\rangle_{\mathbf H}\}~.
\end{align}
Since $\langle h^*,\theta_0\rangle_{\mathbf H}\le 0$ for all $h^*\in\Lambda_1^\circ$, it follows from result \eqref{Eqn: projection monotonicity, aux2} that
\begin{multline}\label{Eqn: projection monotonicity, aux4}
\max_{h^*\in\Lambda_1^\circ} \min_{|a'|\le a}\{\langle h^*,h\rangle_{\mathbf H}+a'\langle h^*,\theta_0\rangle_{\mathbf H}\}=\max_{h^*\in\Lambda_1^\circ} \{\langle h^*,h\rangle_{\mathbf H}+a\langle h^*,\theta_0\rangle_{\mathbf H}\}\\
=\max_{h^*\in\Lambda_1^\circ} \{\langle h^*,h+a\theta_0\rangle_{\mathbf H}\}=\|(h+a\theta_0)-\Pi_\Lambda (h+a\theta_0)\|_{\mathbf H}~,
\end{multline}
where the last step is by \citet[p.125-7]{Deutsch2012Best}. The equality in \eqref{Eqn: projection monotonicity, aux1} then follows from combining \eqref{Eqn: projection monotonicity, aux2}, \eqref{Eqn: projection monotonicity, aux3}, and \eqref{Eqn: projection monotonicity, aux4}. \qed

\begin{lem}\label{Lem: nonconservativeness}
Let Assumption \ref{Ass: convex setup} hold and $\bar{\mathbf P}_0$ be as in Theorem \ref{Thm: size control}. Then it follows that, for any $h\in\mathbf H$, $a\in\mathbf R_+$ and $P\in\bar{\mathbf P}_0$,
\begin{align}
\Pi_\Lambda(h+a\theta_P)=\Pi_\Lambda(h)+a\theta_P~.
\end{align}
\end{lem}
\noindent{\sc Proof:} Let $\Lambda^\circ\equiv\{\vartheta\in\mathbf H: \sup_{\lambda\in\Lambda}\langle\vartheta,\lambda\rangle_{\mathbf H}\le 0\}$. Fix any $h\in\mathbf H$, $a\in\mathbf R_+$, and $P\in\bar{\mathbf P}_0$. By Assumption \ref{Ass: convex setup}, $\Pi_\Lambda(h)+a\theta_P\in\Lambda$. First, note that, for any $\lambda\in\Lambda$,
\begin{align}\label{Eqn: nonconservativeness, aux1}
\langle h+a\theta_P-\{\Pi_\Lambda(h)+a\theta_P\},\lambda\rangle_{\mathbf H} =  \langle h-\Pi_\Lambda(h),\lambda \rangle_{\mathbf H}\le 0~,
\end{align}
where the inequality follows by Assumption \ref{Ass: convex setup} and Theorem 4.7 in \citet{Deutsch2012Best}. Next, for $\lambda_0\equiv \Pi_\Lambda(h)+a\theta_P\in\Lambda$, we have
\begin{multline}\label{Eqn: nonconservativeness, aux2}
\langle h+a\theta_P-\{\Pi_\Lambda(h)+a\theta_P\},\lambda_0\rangle_{\mathbf H} \\
 = \langle h-\Pi_\Lambda(h), \Pi_\Lambda(h)\rangle_{\mathbf H} + a\langle h-\Pi_\Lambda(h), \theta_P\rangle_{\mathbf H}  = 0~,
\end{multline}
where the second equality is due to $\langle h-\Pi_\Lambda(h), \Pi_\Lambda(h)\rangle_{\mathbf H}=0$ by Assumption \ref{Ass: convex setup} and Theorem 4.7 in \citet{Deutsch2012Best}, $h-\Pi_\Lambda(h)\in\Lambda^\circ$ by Assumption \ref{Ass: convex setup} and Theorem 5.6 in \citet{Deutsch2012Best}, and the definition of $\bar{\mathbf P}_0$. The conclusion of the lemma then follows from applying Theorem 4.7 in \citet{Deutsch2012Best} to \eqref{Eqn: nonconservativeness, aux1} and \eqref{Eqn: nonconservativeness, aux2}. \qed

\begin{pro}\label{Pro: anti concentration}
Let Assumptions \ref{Ass: convex setup} and \ref{Ass: Gaussian} hold, and $\psi_{a,P}$ be defined as in \eqref{Eqn: psi function}. Then for any sequence $\{\epsilon_n\}$ of positive scalars satisfying $\epsilon_n=o(c_n)$,
\begin{align}\label{Eqn: anti concentration, aux}
\limsup_{n\to\infty}\sup_{P\in\mathbf P_0}\sup_{x\in [c_{n,P}(0.5)+\varsigma_n,\infty)} P(|\psi_{\kappa_n,P}(\mathbb Z_{n,P})-x|\le \epsilon_n)=0~.
\end{align}
\end{pro}
\noindent{\sc Proof:} Let $\{\epsilon_n\}$ be an arbitrary sequence of positive scalars satisfying $\epsilon_n=o(c_n)$ as $n\to\infty$. Fix $n\in\mathbf N$ and $P\in\mathbf P_0$ for the moment. Let $\Lambda_1^\circ\equiv\{t\in\mathbf H: \langle t,\lambda\rangle_{\mathbf H}\le 0\text{ for all }\lambda\in\Lambda, \|t\|_{\mathbf H}\le 1\}$. By Assumption \ref{Ass: convex setup} and \citet[p.125-7]{Deutsch2012Best}, we may then write: for $e_t(n,P)\equiv\kappa_n\langle t,\theta_P\rangle_{\mathbf H}$,
\begin{align}\label{Eqn: anti concentration, aux1}
\psi_{\kappa_n,P}(\mathbb Z_{n,P})=\max_{t\in\Lambda_1^\circ}\{\langle t,\mathbb Z_{n,P}\rangle_{\mathbf H}+e_t(n,P)\}~.
\end{align}
Since $0\in\Lambda_1^\circ$ and $\langle t,\mathbb Z_{n,P}\rangle_{\mathbf H}+e_t(n,P)=0$ at $t=0$, the maximum in \eqref{Eqn: anti concentration, aux1} must be attained at $t\in\Lambda_1^\circ$ such that $\langle t,\mathbb Z_{n,P}\rangle_{\mathbf H}+e_t(n,P)\ge 0$. Moreover, $\langle t,\mathbb Z_{n,P}\rangle_{\mathbf H}\le \|\mathbb Z_{n,P}\|_{\mathbf H}$ for all $t\in\Lambda_1^\circ$ by the Cauchy--Schwarz inequality. Therefore, whenever $\| \mathbb Z_{n,P}\|_{\mathbf H}\le M$ with $M>0$, the maximum in \eqref{Eqn: anti concentration, aux1} must be attained at some $t\in\Lambda_1^\circ$ with $e_t(n,P)\ge -M$. It follows that, whenever $\| \mathbb Z_{n,P}\|_{\mathbf H}\le M$,
\begin{align}\label{Eqn: anti concentration, aux2}
\psi_{\kappa_n,P}(\mathbb Z_{n,P}) = \max_{t\in \Lambda_{1,M}^\circ(n,P)}\{\langle t,\mathbb Z_{n,P}\rangle_{\mathbf H}+e_t(n,P)\}~.
\end{align}
where $\Lambda_{1,M}^\circ(n,P)\equiv\{t\in\Lambda_1^\circ: e_t(n,P)\ge -M\}$. Hence, for any $x\in\mathbf R$,
\begin{align}\label{Eqn: anti concentration, aux3}
P(|\psi_{\kappa_n,P}&(\mathbb Z_{n,P})-x|\le \epsilon_n)\notag\\
&\le P(|\max_{t\in \Lambda_{1,M}^\circ(n,P)}\{\langle t,\mathbb Z_{n,P}\rangle_{\mathbf H}+e_t(n,P)\}-x|\le \epsilon_n )+P(\|\mathbb Z_{n,P}\|_{\mathbf H}>M)\notag\\
&\le P(|\max_{t\in \Lambda_{1,M}^\circ(n,P)}\{\langle t,\mathbb Z_{n,P}\rangle_{\mathbf H}+e_t(n,P)\}-x|\le \epsilon_n )+\frac{\zeta}{M}~,
\end{align}
for some constant $\zeta>0$ satisfying $\sup_{P\in\mathbf P}E[\|\mathbb Z_{n,P}\|_{\mathbf H}]<\zeta$, where the existence of $\zeta$ is guaranteed by Markov's inequality and Assumption \ref{Ass: Gaussian}(ii).

We next aim to control the first term on the right-hand side of \eqref{Eqn: anti concentration, aux3} by bounding the density of $\max_{t\in \Lambda_{1,M}^\circ(n,P)}\{\langle t,\mathbb Z_{n,P}\rangle_{\mathbf H}+e_t(n,P)\}$. To this end, let $F_{n,P,M}$ be the cdf of $\max_{t\in \Lambda_{1,M}^\circ(n,P)}\{\langle t,\mathbb Z_{n,P}\rangle_{\mathbf H}+e_t(n,P)\}$. We proceed with some useful facts. First, by Assumption \ref{Ass: Gaussian}(i), Lemma 1.3.2 in \citet{Vaart1996}, and the corollary to Theorem I.3.1 in \citet{Vakhania_Tarieladze_Chobanyan1987probability}, $\mathbb Z_{n,P}$ is a centered Radon Gaussian variable in $\mathbf H$. Second, for $\underline r_M(n,P)\equiv\inf\{r\in\mathbf R: F_{n,P,M}(r)>0\}$, Theorem 11.1 in \citet{Davydov1998local} in turn implies that $F_{n,P,M}$ is absolutely continuous on $(\underline{r}_M(n,P),\infty)$ so that it admits a density on $(\underline{r}_M(n,P),\infty)$ which we denote by $f_{n,P,M}$. Third, by Proposition 11.2 in \citet{Davydov1998local}, we may assume without loss of generality that $\Lambda_{1,M}^\circ(n,P)$ is countable. Fourth, since $e_t(n,P)\le 0$ for any $t\in\Lambda_1^\circ$ and $P\in\mathbf P_0$, we have $e_{t,M}(n,P)\equiv e_t(n,P)+M\le M$, which, together with the Cauchy--Schwarz inequality and $\mathbb Z_{n,P}\in\mathbf H$ (by Assumption \ref{Ass: strong approx}(i)), implies
\begin{align}\label{Eqn: anti concentration, aux4}
\max_{t\in \Lambda_{1,M}^\circ(n,P)}\{\langle t,\mathbb Z_{n,P}\rangle_{\mathbf H}+e_{t,M}(n,P)\}\le \|\mathbb Z_{n,P}\|_{\mathbf H}+M<\infty~,
\end{align}
almost surely. Fifth, for $\bar\sigma_{n,P,M}^2\equiv\sup_{t\in \Lambda_{1,M}^\circ(n,P)} E[\langle t,\mathbb Z_{n,P}\rangle_{\mathbf H}^2]$, we shall show towards the end of the proof that, for all large $M>0$,
\begin{align}\label{Eqn: anti concentration, aux5}
\bar\sigma_{n,P,M}^2>0~.
\end{align}
In what follows, we fix any such large $M$. Sixth, for any $r>\underline{r}_M(n,P)$, we note
\begin{multline}\label{Eqn: anti concentration, aux6}
F_{n,P,M}(r)\equiv P(\max_{t\in \Lambda_{1,M}^\circ(n,P)}\{\langle t,\mathbb Z_{n,P}\rangle_{\mathbf H}+e_t(n,P)\le r) \\
= P(\max_{t\in \Lambda_{1,M}^\circ(n,P)}\{\langle t,\mathbb Z_{n,P}\rangle_{\mathbf H}+e_{t,M}(n,P)\}\le r+M)~.
\end{multline}
Seventh, for $m_{n,P,M}$ the median of $F_{n,P,M}$, we have by the quantile equivariance that the median of $\max_{t\in \Lambda_{1,M}^\circ(n,P)}\{\langle t,\mathbb Z_{n,P}\rangle_{\mathbf H}+e_{t,M}(n,P)\}$ is $m_{n,P,M}+M$. Note that $m_{n,P,M}\ge \underline{r}_M(n,P)\ge 0$ because $\max_{t\in \Lambda_{1,M}^\circ(n,P)}\{\langle t,\mathbb Z_{n,P}\rangle_{\mathbf H}+e_t(n,P)\}\ge 0$.

With the above preparations, we may apply Theorem 2.2.2 in \citet{Yurinsky1995Gaussian} with $b=m_{n,P,M}+M$ and $u=r+M$ to conclude:
\begin{align}\label{Eqn: anti concentration, aux7}
f_{n,P,M}(r) = F_{n,P,M}'(r) \le \frac{2(r+M)-(m_{n,P,M}+M)}{[(r+M)-(m_{n,P,M}+M)]^2}=\frac{2r-m_{n,P,M}+M}{(r-m_{n,P,M})^2}
\end{align}
for all $r>m_{n,P,M}$. By the choice of $\epsilon_n$ and $c_n=O(1)$, we note that
\begin{align}\label{Eqn: anti concentration, aux7a}
\epsilon_n=o(c_n)=o(\sqrt{c_n/\varsigma_n^2}\sqrt{c_n}\varsigma_n) =o (\varsigma_n)~,
\end{align}
as $n\to\infty$. Therefore, we have $\epsilon_n\le \varsigma_n/2$ for all $n$ sufficiently large, so that
\begin{align}\label{Eqn: anti concentration, aux8}
x-\epsilon_n-m_{n,P,M}\ge m_{n,P,M}+\varsigma_n - \epsilon_n-m_{n,P,M}\ge\frac{\varsigma_n}{2}
\end{align}
whenever $x\ge m_{n,P,M}+\varsigma_n$. Since $r\mapsto(2r-m_{n,P,M}+M)/(r-m_{n,P,M})^2$ is decreasing on $(m_{n,P,M},\infty)$, we may thus conclude by the fundamental theorem of calculus and results \eqref{Eqn: anti concentration, aux6} and \eqref{Eqn: anti concentration, aux7} that, for all $x\ge m_{n,P,M}+\varsigma_n$ and $n$ large,
\begin{multline}\label{Eqn: anti concentration, aux9}
P(|\max_{t\in \Lambda_{1,M}^\circ(n,P)}\{\langle t,\mathbb Z_{n,P}\rangle_{\mathbf H}+e_t(n,P)\}-x|\le \epsilon_n )\\
 = \int_{x-\epsilon_n}^{x+\epsilon_n}f_{n,P,M}(r)\,\mathrm dr
\le 2\epsilon_n \frac{2(m_{n,P,M}+\varsigma_n/2)-m_{n,P,M}+M}{(\varsigma_n/2)^2}~.
\end{multline}

Since $\Lambda_{1,M}^\circ(n,P)\subset \Lambda_1^\circ$, we obtain in view of \eqref{Eqn: anti concentration, aux1} and Lemma \ref{Lem: projection monotonicity} that
\begin{align}\label{Eqn: anti concentration, aux10}
\max_{t\in \Lambda_{1,M}^\circ(n,P)}\{\langle t,\mathbb Z_{n,P}\rangle_{\mathbf H}+e_t(n,P)\}\le \psi_{k_n,P}(\mathbb Z_{n,P})
\le \psi_{0,P}(\mathbb Z_{n,P})=\|\mathbb Z_{n,P}\|_{\mathbf H}~.
\end{align}
By result \eqref{Eqn: anti concentration, aux10}, \citet{Kwapien1994Median} and  Assumption \ref{Ass: Gaussian}(ii), we note
\begin{align}\label{Eqn: anti concentration, aux11}
m_{n,P,M}\le m_{n,P}\equiv c_{n,P}(0.5)\le E[\|\mathbb Z_{n,P}\|_{\mathbf H}] \le \zeta~,
\end{align}
where we remind the reader our choice of $\zeta$ from \eqref{Eqn: anti concentration, aux3}. Combining results \eqref{Eqn: anti concentration, aux3}, \eqref{Eqn: anti concentration, aux9} and \eqref{Eqn: anti concentration, aux10}, we thus obtain that
\begin{align}\label{Eqn: anti concentration, aux12}
\sup_{P\in\mathbf P_0}\sup_{x\in[c_{n,P}(0.5)+\varsigma_n)}P(|\psi_{\kappa_n,P}(\mathbb Z_{n,P})-x|\le \epsilon_n)\lesssim \epsilon_n\frac{\zeta+\varsigma_n+M}{\varsigma_n^2}+\frac{\zeta}{M}~.
\end{align}
Since $\epsilon_n=o(c_n)$, we may select a sequence $a_n\downarrow 0$ (sufficiently slow) such that $\epsilon_n=o(a_nc_n)$. In turn, by setting $M\equiv M_n=a_n^{-1}$ which diverges to infinity, we may then conclude by Assumption \ref{Ass: Gaussian}(iv) and results \eqref{Eqn: anti concentration, aux7a} and \eqref{Eqn: anti concentration, aux12} that
\begin{align}\label{Eqn: anti concentration, aux12a}
\sup_{P\in\mathbf P_0}\sup_{x\in[c_{n,P}(0.5)+\varsigma_n)}P(|\psi_{\kappa_n,P}(\mathbb Z_{n,P})-x|\le \epsilon_n)\to 0~.
\end{align}

It remains to prove \eqref{Eqn: anti concentration, aux5}. For this, we fix $n$ and $P\in\mathbf P_0$ in what follows. Let $\bar\sigma_{n,P}^2\equiv\sup_{t\in\Lambda_1^\circ} E[\langle t,\mathbb Z_{n,P}\rangle_{\mathbf H}^2]$. Then we must have $\bar\sigma_{n,P}^2>0$. Indeed, suppose by way of contradiction that $\bar\sigma_{n,P}^2=0$. This implies $\langle t,\mathbb Z_{n,P}\rangle_{\mathbf H}=0$ almost surely for all $t\in\Lambda_1^\circ$. By result \eqref{Eqn: anti concentration, aux1} and Proposition 11.2 in \citet{Davydov1998local}, we have $\psi_{\kappa_n} (\mathbb Z_{n,P}) = 0$ almost surely. Then all quantiles of $\psi_{\kappa_n} (\mathbb Z_{n,P})$ are equal to zero, contradicting Assumption \ref{Ass: Gaussian}(iii). Next, fix $\eta>0$. Then we may select some $t_{n,P}\in\Lambda_1^\circ$ such that
\begin{align}\label{Eqn: anti concentration, aux13}
\bar\sigma_{n,P}^2\le E[\langle t_{n,P},\mathbb Z_{n,P}\rangle_{\mathbf H}^2]+\eta~.
\end{align}
Moreover, by choosing $M\ge \kappa_n\|\theta_P\|_{\mathbf H}$, we may employ the Cauchy--Schwarz inequality and $\|t_{n,P}\|_{\mathbf H}\le 1$ (due to $t_{n,P}\in\Lambda_1^\circ$) to obtain that
\begin{align}\label{Eqn: anti concentration, aux14}
|e_{t_{n,P}}(n,P)|\equiv |\kappa_n\langle t_{n,P},\theta_P\rangle_{\mathbf H}|\le\kappa_n\|\theta_P\|_{\mathbf H}\le M~.
\end{align}
In turn, it follows from result \eqref{Eqn: anti concentration, aux14} that $t_{n,P}\in \Lambda_{1,M}^\circ(n,P)$ so that
\begin{align}\label{Eqn: anti concentration, aux15}
E[\langle t_{n,P},\mathbb Z_{n,P}\rangle_{\mathbf H}^2]\le \sup_{t\in \Lambda_{1,M}^\circ(n,P)}E[\langle t,\mathbb Z_{n,P}\rangle_{\mathbf H}^2] = \bar\sigma_{n,P,M}^2~.
\end{align}
Combining results \eqref{Eqn: anti concentration, aux13} and \eqref{Eqn: anti concentration, aux15}, we may then conclude that
\begin{align}\label{Eqn: anti concentration, aux16}
\bar\sigma_{n,P}^2\le \bar\sigma_{n,P,M}^2 + \eta \le \bar\sigma_{n,P}^2+\eta
\end{align}
whenever $M\ge \kappa_n\|\theta_P\|_{\mathbf H}$. Since $\eta$ is arbitrary, result \eqref{Eqn: anti concentration, aux16} implies that $\bar\sigma_{n,P,M}^2\to \bar\sigma_{n,P}^2$ as $M\to\infty$. This, together with $\bar\sigma_{n,P}^2>0$, implies \eqref{Eqn: anti concentration, aux5}. \qed

\begin{lem}\label{Lem: density bound, norm}
Let $\mathbf D$ be a Banach space with norm $\|\cdot\|_{\mathbf D}$ and $\mathbf D_1^*\equiv\{x^*\in\mathbf D^*: \sup_{\|x\|_{\mathbf D}\le 1}|x^*(x)|\le 1\}$, the unit ball in the topological dual $\mathbf D^*$ of $\mathbf D$. If $\mathbb G\in\mathbf D$ is a tight centered Gaussian variable such that $\sup_{x^*\in\mathbf D_1^*}E[x^*(\mathbb G)^2]>0$, then the cdf $F$ of $\|\mathbb G\|_{\mathbf D}$ is absolutely continuous on $(0,\infty)$, and, for any $r> m_F$ with $m_F$ the median of $F$,
\begin{align}\label{Eqn: density bound, norm, aux}
F'(r) \le  \frac{2r-m_F}{(r-m_F)^2} ~.
\end{align}
\end{lem}
\noindent{\sc Proof:} Since $\mathbb G$ is tight and $\mathbf D$ is Banach, Lemma 1.3.2 in \citet{Vaart1996} and the corollary to Theorem I.3.1 in \citet{Vakhania_Tarieladze_Chobanyan1987probability} imply that $\mathbb G$ is Radon. Hence, since $\mathbb G$ is centered Gaussian, we know by the remark following Proposition 7.4 in \citet{Davydov1998local} that the support $\mathbf D_0$  of $\mathbb G$ is a closed separable subspace of $\mathbf D$ and hence a separable Banach space under $\|\cdot\|_{\mathbf D}$. Therefore, by Proposition 1.12.17 in \citet{BogachevSmolyanov2017TVS}, it follows that, for all $x\in\mathbf D_0$,
\begin{align}\label{Eqn: density bound, norm, aux1}
\|x\|_{\mathbf D}=\sup_{n=1}^\infty x_n^*(x)~,
\end{align}
where $\{x_n^*\}_{n=1}^\infty$ live in $\mathbf D_{0,1}^*$, the unit ball of the topological dual space $\mathbf D_0^*$ of $\mathbf D_0$. By the Hahn--Banach extension theorem (see, for example, Theorem 5.53 in \citet{AliprantisandBorder2006}), each $x_n^*$ admits an extension that belongs to $\mathbf D_1^*$, which we continue to denote by $x_n^*$ with some abuse of notation. In other words, \eqref{Eqn: density bound, norm, aux2} holds with $\{x_n^*\}_{n=1}^\infty$ living in $\mathbf D_1^*$. Since $P(\mathbb G\in\mathbf D_0)=1$, we then obtain that, almost surely,
\begin{align}\label{Eqn: density bound, norm, aux2}
\|\mathbb G\|_{\mathbf D}=\sup_{n=1}^\infty x_n^*(\mathbb G)~.
\end{align}
For each $n$, we have $E[x_n^*(\mathbb G)]=0$ due to $\mathbb G$ being centered. Moreover, the supremum in \eqref{Eqn: density bound, norm, aux2} is finite almost surely. Since $\sup_{x^*\in\mathbf D_1^*}E[x^*(\mathbb G)^2]>0$ by assumption, Theorem 2.2.1 in \citet{Yurinsky1995Gaussian} implies that $F$ is absolutely continuous on $(r_0,\infty)$ with $r_0\equiv \inf\{r\in\mathbf R: F(r)>0\}$. Since the support $\mathbf D_0$ of $\mathbb G$ as a subspace includes $0$ (in $\mathbf D$), we have by Problem 11.3 in \citet{Davydov1998local} that $r_0=0$. This proves the first claim. The second claim follows immediately by applying Theorem 2.2.2-(a) in \citet{Yurinsky1995Gaussian} with $b=m_F$ and noting that $t\equiv \Phi^{-1}(F(m_F))\ge \Phi^{-1}(0.5) = 0$. \qed

\section{More Simulation Studies and Empirical Application}\label{App: more simulations}

\subsection{More Simulation Studies}\label{App: more simulations1}

This section conducts more simulation studies for three restrictions: concavity/convexity, monotonicity jointly with convexity, and Slutsky restriction. For the first two, we shall compare to the test by \citet{LeeSongWhang2017Inequal} which is asymptotically nonconservative and meanwhile computationally manageable---see the discussions of other existing tests in Example \ref{Ex: convexity}. For the Slutsky restriction, one may also adopt the nonconservative test by \citet{ChernozhukovNeweySantos2019CCMM}. However, its implementation requires nonlinearly constrained optimization(in addition to optimization over the estimated set of minimizers) in each bootstrap repetition, and the computation cost grows quickly with the relevant dimension \citep[p.617]{Zhu2020Shape}. By restricting to linear (in $g_0$ in the context of Example \ref{Ex: Slutsky}) constraints, \citet{Zhu2020Shape} developed a computationally simpler inferential framework, which unfortunately excludes the Slutsky restriction. For these reasons, we shall only implement our test for the Slutsky restriction. We stress, however, that \citet{ChernozhukovNeweySantos2019CCMM} accommodated partial identification while we cannot.

The first set of simulations makes use of exactly the same univariate design \eqref{Eqn: MC1,aux1} in Section \ref{Sec: simulations}, and we aim to test whether $\theta_0$ is convex, and whether $\theta_0$ is nondecreasing {\it and} convex. The implementation of our tests remains unchanged other than adjusting linear constraints in quadratic programs accordingly. Following \citet[p.59]{FanGijbels1996Local}, the LSW tests are implemented similarly as before but now based on local polynomial regression of order $q=3$ for both restrictions (so that the bandwidths are evaluated at $q=3$). Note in particular that, for the joint test of monotonicity and convexity, we estimate the first and second derivatives of $\theta_0$ in a single local polynomial regression of order $3$, instead of two separate regressions, for ease of computation. Thus, in assessing that ``additional restrictions help improve power,'' one should compare the resulting power curves to those for convexity, rather than those for monotonicity in Section \ref{Sec: simulations} which are associated with a different convergence rate $r_n$ (through its dependence on $q$).

The second set of simulations are based on the same design for \eqref{Eqn: MC2,aux1} except
\begin{align}\label{Eqn: MC2 supp,aux1}
\theta_0(z_1,z_2)=\mathsf a\big(\frac{1}{2}z_1^{\mathsf b}+\frac{1}{2}z_2^{\mathsf b}\big)^{1/\mathsf b}+\mathsf c\log(1+5(z_1+z_2))~,
\end{align}
where we adopt the same set of choices for $(\mathsf a,\mathsf b,\mathsf c)$ but with $\Delta=0.05$ replaced by $\Delta=0.2$, so that the power of the implemented tests is close to 1 as $\delta$ increases from $1$ to $10$. We then aim to test concavity of $\theta_0$. To ease computation, the $L^2$-integrals for our test are evaluated over $[0.1,0.9]^2$ but now with marginal step size $0.1$. The LSW tests are based on the Hessian matrix $z\mapsto\Theta_0(z)$ of $\theta_0$ so that, in the notation of LSW, $J=1$ and $v_{\tau,1}(z)=a_{\tau}^\transpose\Theta_0(z)a_\tau$ with $a_\tau\equiv[\cos(\tau),\sin(\tau)]^\transpose$. To reduce computation cost, we approximate the resulting triple integrals over $z\in[0.1,0.9]^2$ with marginal step size $0.1$ and over $\tau$ based on 500 draws from the uniform distribution on $[0,2\pi]$. As with the LSW tests for \eqref{Eqn: MC2,aux1}, the number of Monte Carlo simulation replications for the LSW tests in the bivariate design \eqref{Eqn: MC2 supp,aux1} is decreased to be 1000.

{
\begin{table}[!ht]
\setlength{\tabcolsep}{5.25pt}
\renewcommand{\arraystretch}{1.1}
\caption{Empirical Size of Shape Tests for $\theta_0$ in \eqref{Eqn: MC1,aux1} at $\alpha=5\%$} \label{Tab: MC1, size1, supp} 
\centering\footnotesize
\begin{threeparttable}
\sisetup{table-number-alignment = center, table-format = 1.3} 
\begin{tabularx}{\linewidth}{@{}c @{\hspace{3pt}} cc *{3}{S[round-mode = places,round-precision = 3]} @{\hspace{4.25pt}} c *{3}{S[round-mode = places,round-precision = 3]} @{\hspace{4.25pt}} c *{3}{S[round-mode = places,round-precision = 3]}@{}} 
\hline
\hline
\multirow{2}{*}{Shape} & \multirow{2}{*}{$n$} & \multirow{2}{*}{$\gamma_n$}  & \multicolumn{3}{c}{FS-C3: $k_n=7$} & & \multicolumn{3}{c}{FS-C5: $k_n=9$} & & \multicolumn{3}{c}{FS-C7: $k_n=11$}\\
\cline{4-6} \cline{8-10} \cline{12-14}
& & & {D1} & {D2} & {D3}  & & {D1} & {D2} & {D3} & & {D1} & {D2} & {D3}\\
\hline \multirow{9}{*}{Con}&
\multirow{3}{*}{$500$}    & $1/n$                  & 0.0480 & 0.0423 & 0.0093 & & 0.0563 & 0.0473 & 0.0157 & & 0.0533 & 0.0453 & 0.0173\\
&                         & $0.01/\log n$          & 0.0480 & 0.0423 & 0.0093 & & 0.0563 & 0.0473 & 0.0157 & & 0.0533 & 0.0453 & 0.0173\\
&                         & $0.01$                 & 0.0487 & 0.0423 & 0.0093 & & 0.0563 & 0.0477 & 0.0157 & & 0.0533 & 0.0453 & 0.0173\\
\cline{3-14}
& \multirow{3}{*}{$750$}  & $1/n$                  & 0.0577 & 0.0460 & 0.0070 & & 0.0623 & 0.0547 & 0.0107 & & 0.0593 & 0.0553 & 0.0190\\
&                         & $0.01/\log n$          & 0.0577 & 0.0460 & 0.0070 & & 0.0623 & 0.0547 & 0.0107 & & 0.0593 & 0.0553 & 0.0190\\
&                         & $0.01$                 & 0.0577 & 0.0460 & 0.0077 & & 0.0623 & 0.0547 & 0.0110 & & 0.0593 & 0.0557 & 0.0203\\
\cline{3-14}
& \multirow{3}{*}{$1000$} & $1/n$                  & 0.0523 & 0.0440 & 0.0053 & & 0.0550 & 0.0467 & 0.0097 & & 0.0540 & 0.0443 & 0.0127\\
&                         & $0.01/\log n$          & 0.0523 & 0.0440 & 0.0053 & & 0.0550 & 0.0467 & 0.0097 & & 0.0540 & 0.0443 & 0.0127\\
&                         & $0.01$                 & 0.0523 & 0.0443 & 0.0053 & & 0.0550 & 0.0470 & 0.0097 & & 0.0540 & 0.0450 & 0.0127\\
\cline{2-14} \multirow{9}{*}{Mon-Con}&
\multirow{3}{*}{$500$}    & $1/n$                  & 0.0500 & 0.0263 & 0.0070 & & 0.0540 & 0.0323 & 0.0110 & & 0.0543 & 0.0317 & 0.0133\\
&                         & $0.01/\log n$          & 0.0500 & 0.0263 & 0.0070 & & 0.0540 & 0.0323 & 0.0110 & & 0.0543 & 0.0317 & 0.0133\\
&                         & $0.01$                 & 0.0503 & 0.0263 & 0.0073 & & 0.0543 & 0.0327 & 0.0113 & & 0.0543 & 0.0320 & 0.0140\\
\cline{3-14}
& \multirow{3}{*}{$750$}  & $1/n$                  & 0.0563 & 0.0257 & 0.0053 & & 0.0590 & 0.0340 & 0.0080 & & 0.0573 & 0.0343 & 0.0173\\
&                         & $0.01/\log n$          & 0.0563 & 0.0257 & 0.0053 & & 0.0590 & 0.0340 & 0.0080 & & 0.0573 & 0.0343 & 0.0173\\
&                         & $0.01$                 & 0.0563 & 0.0257 & 0.0053 & & 0.0590 & 0.0347 & 0.0080 & & 0.0573 & 0.0343 & 0.0180\\
\cline{3-14}
& \multirow{3}{*}{$1000$} & $1/n$                  & 0.0550 & 0.0220 & 0.0037 & & 0.0553 & 0.0287 & 0.0057 & & 0.0533 & 0.0300 & 0.0103\\
&                         & $0.01/\log n$          & 0.0550 & 0.0220 & 0.0037 & & 0.0553 & 0.0287 & 0.0057 & & 0.0533 & 0.0300 & 0.0103\\
&                         & $0.01$                 & 0.0550 & 0.0227 & 0.0037 & & 0.0557 & 0.0293 & 0.0060 & & 0.0533 & 0.0300 & 0.0103\\
\hline
\multirow{2}{*}{Shape} & \multicolumn{2}{c}{\multirow{2}{*}{Tests}}& \multicolumn{3}{c}{$n=500$} & & \multicolumn{3}{c}{$n=750$} & & \multicolumn{3}{c}{$n=1000$}\\
\cline{4-6} \cline{8-10} \cline{12-14}
& & & {D1} & {D2} & {D3}  & & {D1} & {D2} & {D3} & & {D1} & {D2} & {D3}\\
\hline \multirow{2}{*}{Con}&
  \multicolumn{2}{c}{LSW-S}                        & 0.0593 & 0.0583 & 0.0480 & & 0.0633 & 0.0577 & 0.0487 & & 0.0567 & 0.0547 & 0.0457\\
& \multicolumn{2}{c}{LSW-L}                        & 0.0633 & 0.0657 & 0.0497 & & 0.0643 & 0.0640 & 0.0467 & & 0.0580 & 0.0583 & 0.0460\\
\cline{2-14}
\multirow{2}{*}{Mon-Con}&
  \multicolumn{2}{c}{LSW-S}                        & 0.0653 & 0.0570 & 0.0297 & & 0.0653 & 0.0517 & 0.0320 & & 0.0597 & 0.0483 & 0.0263\\
& \multicolumn{2}{c}{LSW-L}                        & 0.0680 & 0.0570 & 0.0303 & & 0.0687 & 0.0530 & 0.0310 & & 0.0647 & 0.0543 & 0.0263\\
\hline
\hline
\end{tabularx}
\begin{tablenotes}[flushleft]
\item {\it Note:} ``Con'' refers to ``Convexity,'' and ``Mon-Con'' refers to ``Monotonicity and Convexity.'' The parameter $\gamma_n$ determines $\hat\kappa_n$ proposed in Section \ref{Sec: tuning parameter} with $c_n=1/\log n$ and $r_n=(n/k_n)^{1/2}$.
\end{tablenotes}
\end{threeparttable}
\end{table}
}

{
\setlength{\tabcolsep}{4pt}
\renewcommand{\arraystretch}{1.1}
\begin{table}[!ht]
\caption{Empirical Size of Concavity Tests for $\theta_0$ in \eqref{Eqn: MC2 supp,aux1} at $\alpha=5\%$} \label{Tab: MC2, size, supp}
\centering\footnotesize
\begin{threeparttable}
\sisetup{table-number-alignment = center, table-format = 1.3} 
\begin{tabularx}{\linewidth}{@{}c@{\hspace{2.5pt}}c@{\hspace{2.5pt}} *{3}{S[round-mode = places,round-precision = 3]} @{\hspace{1pt}}c *{3}{S[round-mode = places,round-precision = 3]} @{\hspace{1pt}}c *{3}{S[round-mode = places,round-precision = 3]} @{\hspace{1pt}}c *{3}{S[round-mode = places,round-precision = 3]}@{}} 
\hline
\hline
\multirow{2}{*}{$n$} & \multirow{2}{*}{$\gamma_n$}& \multicolumn{3}{c}{FS-Q0: $k_n=9$} & & \multicolumn{3}{c}{FS-Q1: $k_n=16$} & & \multicolumn{3}{c}{FS-C0: $k_n=16$} & & \multicolumn{3}{c}{FS-C1: $k_n=25$}\\
\cline{3-5} \cline{7-9} \cline{11-13} \cline{15-17}
&& {D1} & {D2} & {D3}  & & {D1} & {D2} & {D3} & & {D1} & {D2} & {D3} & & {D1} & {D2} & {D3}\\  
\hline
\multirow{3}{*}{$500$} & $1/n$        & 0.0620 & 0.0607 & 0.0147 & & 0.0687 & 0.0667 & 0.0287 & & 0.0697 & 0.0667 & 0.0283 & & 0.0830 & 0.0810 & 0.0443\\
                      & $0.01/\log n$ & 0.0620 & 0.0607 & 0.0147 & & 0.0687 & 0.0667 & 0.0287 & & 0.0697 & 0.0667 & 0.0283 & & 0.0830 & 0.0810 & 0.0443\\
                      & $0.01$        & 0.0630 & 0.0610 & 0.0147 & & 0.0690 & 0.0673 & 0.0293 & & 0.0707 & 0.0680 & 0.0293 & & 0.0843 & 0.0823 & 0.0447\\
\cline{2-17}
\multirow{3}{*}{$750$} & $1/n$        & 0.0643 & 0.0627 & 0.0107 & & 0.0730 & 0.0733 & 0.0267 & & 0.0720 & 0.0733 & 0.0237 & & 0.0687 & 0.0707 & 0.0347\\
                      & $0.01/\log n$ & 0.0643 & 0.0627 & 0.0107 & & 0.0730 & 0.0733 & 0.0267 & & 0.0720 & 0.0733 & 0.0237 & & 0.0687 & 0.0707 & 0.0347\\
                      & $0.01$        & 0.0650 & 0.0630 & 0.0107 & & 0.0737 & 0.0737 & 0.0280 & & 0.0740 & 0.0743 & 0.0243 & & 0.0693 & 0.0713 & 0.0357\\
\cline{2-17}
\multirow{3}{*}{$1000$}& $1/n$        & 0.0567 & 0.0590 & 0.0037 & & 0.0667 & 0.0663 & 0.0177 & & 0.0693 & 0.0670 & 0.0137 & & 0.0660 & 0.0647 & 0.0267\\
                      & $0.01/\log n$ & 0.0567 & 0.0590 & 0.0037 & & 0.0667 & 0.0663 & 0.0177 & & 0.0693 & 0.0670 & 0.0137 & & 0.0660 & 0.0647 & 0.0267\\
                      & $0.01$        & 0.0573 & 0.0593 & 0.0037 & & 0.0673 & 0.0670 & 0.0180 & & 0.0703 & 0.0680 & 0.0140 & & 0.0667 & 0.0650 & 0.0273\\
\hline
 \multicolumn{2}{c}{\multirow{2}{*}{$n$}}  & \multicolumn{3}{c}{LSW-S} & & \multicolumn{3}{c}{LSW-L} & &   & &  \\
\cline{3-5} \cline{7-9}
 &     & {D1} & {D2} & {D3}  & & {D1} & {D2} & {D3} & &   &   &  \\
\hline
\multicolumn{2}{c}{500}               & 0.0460 & 0.0490 & 0.0430 & & 0.0590  & 0.0550  & 0.0490 & &   &   & \\
\multicolumn{2}{c}{750}               & 0.0680 & 0.0560 & 0.0490 & & 0.0710  & 0.0740  & 0.0480 & &   &   & \\
\multicolumn{2}{c}{1000}              & 0.0530 & 0.0530 & 0.0430 & & 0.0620  & 0.0510  & 0.0370 & &   &   & \\
\hline
\hline
\end{tabularx}
\begin{tablenotes}[flushleft]
\item {\it Note:} The parameter $\gamma_n$ determines $\hat\kappa_n$ proposed in Section \ref{Sec: tuning parameter} with $c_n=1/\log n$ and $r_n=(n/k_n)^{1/2}$.
\end{tablenotes}
\end{threeparttable}
\end{table}
}

Tables \ref{Tab: MC1, size1, supp}--\ref{Tab: MC2, size, supp} summarize the empirical sizes with $\gamma_n\in\{1/n,0.01/\log n,0.01\}$---see also Tables \ref{Tab: MC1Con, size1, app}--\ref{Tab: MC1MonCon, size1, app} in Appendix \ref{App: full simulations}. Once again, our tests are insensitive to the choice of $\gamma_n$. In the univariate case, our tests control sizes well across shapes, sample sizes, and the number of knots, while LSW's tests for monotonicity jointly with convexity are slightly over-sized. In the bivariate case, our tests, especially FS-C1 (in which case the sieve dimension is 25), tend to over-reject, though to an overall lesser extent as $n$ increases. The size distortions in small samples may be explained by the fact that the Gaussian approximation is inaccurate due to a ``large'' number of regressors being used in the sieve estimation. On the other hand, LSW-L and in particular LSW-S exhibit overall less size distortions compared to our tests except FS-Q0.

In turn, Figures \ref{Fig: MC1, supp}--\ref{Fig: MC2, supp} depict the power curves, where we only show our tests with $\gamma_n=0.01/\log n$ due to space limitation and the fact that other choices of $\gamma_n$ enjoy very similar curves---see also Figures \ref{Fig: MC1, FullCon}--\ref{Fig: MC1, FullMonCon} and \ref{Fig: MC2, FullCon} in Appendix \ref{App: full simulations}. Overall, our tests appear to be significantly more powerful than the LSW tests across shapes, sample sizes and the number of interior knots, in both univariate and bivariate designs.  The power of the LSW tests in the bivariate case is less than 25\% across sample sizes. The substantial power gaps are in line with the fact that the LSW tests entail estimation of the second derivatives of $\theta_0$, which admit slower rates of convergence. We note, however, that our test of convexity in the design \eqref{Eqn: MC1,aux1} has power slightly below $5\%$ when $\delta=1$. This is a setting where $\theta_0$ is visually close to being convex. By further imposing monotonicity, the power discrepancies at $\delta=1$ then vanish---see the second row in Figure \ref{Fig: MC1, supp}.

\pgfplotstableread{ 
delta alpha ConFiveKn3 ConFiveKn5 ConFiveKn7 ConSevenKn3 ConSevenKn5 ConSevenKn7 ConTenKn3 ConTenKn5 ConTenKn7
0     0.05    0.0480     0.0563     0.0533      0.0577      0.0623      0.0593     0.0523    0.0550    0.0540
1     0.05    0.0390     0.0467     0.0453      0.0433      0.0480      0.0500     0.0433    0.0417    0.0447
2     0.05    0.0523     0.0573     0.0550      0.0557      0.0593      0.0633     0.0720    0.0647    0.0683
3     0.05    0.0783     0.0787     0.0733      0.0997      0.1010      0.0940     0.1363    0.1217    0.1097
4     0.05    0.1153     0.1147     0.1017      0.1713      0.1633      0.1423     0.2360    0.2027    0.1803
5     0.05    0.1853     0.1663     0.1560      0.2813      0.2503      0.2290     0.3737    0.3250    0.2817
6     0.05    0.2690     0.2387     0.2143      0.4157      0.3593      0.3300     0.5540    0.4863    0.4260
7     0.05    0.3803     0.3320     0.3003      0.5710      0.5057      0.4610     0.7150    0.6523    0.5860
8     0.05    0.4897     0.4470     0.3997      0.7090      0.6490      0.5960     0.8427    0.7957    0.7397
9     0.05    0.6080     0.5597     0.5040      0.8177      0.7753      0.7227     0.9183    0.8950    0.8607
10    0.05    0.7200     0.6713     0.6240      0.9017      0.8673      0.8320     0.9653    0.9540    0.9347
}\FirstCon

\pgfplotstableread{ 
delta alpha MonConFiveKn3 MonConFiveKn5 MonConFiveKn7 MonConSevenKn3 MonConSevenKn5 MonConSevenKn7 MonConTenKn3 MonConTenKn5 MonConTenKn7
0     0.05     0.0500        0.0540        0.0543        0.0563         0.0590          0.0573        0.0550       0.0553      0.0533
1     0.05     0.0617        0.0653        0.0647        0.0710         0.0687          0.0710        0.0757       0.0690      0.0707
2     0.05     0.1127        0.1097        0.1070        0.1383         0.1290          0.1247        0.1730       0.1520      0.1390
3     0.05     0.1983        0.1783        0.1687        0.2897         0.2590          0.2320        0.3850       0.3380      0.3040
4     0.05     0.3410        0.2957        0.2807        0.4920         0.4527          0.4033        0.6463       0.5990      0.5483
5     0.05     0.5147        0.4740        0.4310        0.7000         0.6633          0.6230        0.8547       0.8207      0.7807
6     0.05     0.6840        0.6577        0.5987        0.8713         0.8433          0.8117        0.9580       0.9463      0.9280
7     0.05     0.8237        0.7967        0.7550        0.9523         0.9390          0.9280        0.9880       0.9870      0.9793
8     0.05     0.9140        0.9037        0.8760        0.9860         0.9840          0.9757        0.9980       0.9983      0.9967
9     0.05     0.9637        0.9560        0.9420        0.9973         0.9980          0.9957        0.9997       1.0000      0.9997
10    0.05     0.9877        0.9840        0.9757        0.9987         0.9993          0.9990        1.0000       1.0000      1.0000
}\FirstMonCon

\pgfplotstableread{
delta alpha   FiveOp  FiveUn  SevenOp  SevenUn   TenOp   TenUn
0     0.05    0.0633  0.0593   0.0643   0.0633   0.0580  0.0567
1     0.05    0.0643  0.0557   0.0690   0.0630   0.0593  0.0607
2     0.05    0.0717  0.0593   0.0670   0.0647   0.0660  0.0603
3     0.05    0.0757  0.0637   0.0827   0.0687   0.0740  0.0640
4     0.05    0.0847  0.0713   0.0920   0.0783   0.0940  0.0700
5     0.05    0.1000  0.0773   0.1110   0.0860   0.1170  0.0810
6     0.05    0.1160  0.0813   0.1443   0.0947   0.1480  0.0927
7     0.05    0.1390  0.0967   0.1777   0.1080   0.1940  0.1040
8     0.05    0.1707  0.1083   0.2163   0.1290   0.2513  0.1287
9     0.05    0.2137  0.1233   0.2747   0.1487   0.3220  0.1447
10    0.05    0.2707  0.1453   0.3497   0.1867   0.4100  0.1860
}\MainFirstLSWCon

\pgfplotstableread{
delta alpha   FiveOp  FiveUn  SevenOp   SevenUn  TenOp    TenUn
0     0.05    0.0680  0.0653   0.0687   0.0653   0.0647   0.0597
1     0.05    0.0710  0.0663   0.0710   0.0633   0.0700   0.0633
2     0.05    0.0747  0.0663   0.0723   0.0677   0.0750   0.0697
3     0.05    0.0883  0.0737   0.0880   0.0750   0.0913   0.0750
4     0.05    0.0997  0.0777   0.1117   0.0867   0.1133   0.0903
5     0.05    0.1240  0.0910   0.1437   0.0987   0.1550   0.1043
6     0.05    0.1563  0.1093   0.1920   0.1190   0.2157   0.1253
7     0.05    0.2043  0.1333   0.2650   0.1493   0.3000   0.1577
8     0.05    0.2650  0.1520   0.3467   0.1883   0.4100   0.2083
9     0.05    0.3337  0.1887   0.4543   0.2410   0.5390   0.2683
10    0.05    0.4297  0.2340   0.5940   0.3037   0.6987   0.3603
}\MainFirstLSWMonCon

\begin{figure}[!h]
\centering\scriptsize
\begin{tikzpicture}[every text node part/.style={align=center}] 
\begin{groupplot}[group style={group name=my plots,group size=3 by 2,horizontal sep= 0.8cm,vertical sep=0.5cm},
    grid = minor,
    width = 0.375\textwidth,
    xmax=10,xmin=0,
    ymax=1,ymin=0,
    every axis title/.style={below,at={(0.2,0.8)}},
    xlabel=$\delta$,
    x label style={at={(axis description cs:0.95,0.04)},anchor=south},
    xtick={0,2,...,10},
    ytick={0.05,0.5,1},
    tick label style={/pgf/number format/fixed},
    legend style={text=black,cells={align=center},row sep = 3pt,legend columns = -1, draw=none,fill=none},
    cycle list={%
{smooth,tension=0.5,color=RoyalBlue1, mark=halfcircle*,every mark/.append style={rotate=90},mark size=1.5pt,line width=0.5pt}, 
{smooth,tension=0.5,color=RoyalBlue3, mark=halfcircle*,every mark/.append style={rotate=180},mark size=1.5pt,line width=0.5pt},
{smooth,tension=0.5,color=RoyalBlue4, mark=halfcircle*,every mark/.append style={rotate=270},mark size=1.5pt,line width=0.5pt},
{smooth,tension=0.5,color=DarkOliveGreen4, mark=halfsquare*,every mark/.append style={rotate=90},mark size=1.5pt,line width=0.5pt}, 
{smooth,tension=0.5,color=SpringGreen4, mark=halfsquare*,every mark/.append style={rotate=270},mark size=1.5pt,line width=0.5pt},
}
]
\nextgroupplot
\node[anchor=north,font=\fontsize{5}{4}\selectfont] at (axis description cs: 0.25,  0.95) {Convexity \\ $n=500$};
\addplot[smooth,tension=0.5,color=NavyBlue, no markers,line width=0.25pt, densely dotted,forget plot] table[x = delta,y=alpha] from \FirstCon;
\addplot table[x = delta,y=ConFiveKn3] from \FirstCon;
\addplot table[x = delta,y=ConFiveKn5] from \FirstCon;
\addplot table[x = delta,y=ConFiveKn7] from \FirstCon;
\addplot table[x = delta,y=FiveOp] from \MainFirstLSWCon;
\addplot table[x = delta,y=FiveUn] from \MainFirstLSWCon;
\nextgroupplot
\node[anchor=north,font=\fontsize{5}{4}\selectfont] at (axis description cs: 0.25,  0.95) {Convexity \\ $n=750$};
\addplot[smooth,tension=0.5,color=NavyBlue, no markers,line width=0.25pt, densely dotted,forget plot] table[x = delta,y=alpha] from \FirstCon;
\addplot table[x = delta,y=ConSevenKn3] from \FirstCon;
\addplot table[x = delta,y=ConSevenKn5] from \FirstCon;
\addplot table[x = delta,y=ConSevenKn7] from \FirstCon;
\addplot table[x = delta,y=SevenOp] from \MainFirstLSWCon;
\addplot table[x = delta,y=SevenUn] from \MainFirstLSWCon;
\nextgroupplot
\node[anchor=north,font=\fontsize{5}{4}\selectfont] at (axis description cs: 0.25,  0.95) {Convexity \\ $n=1000$};
\addplot[smooth,tension=0.5,color=NavyBlue, no markers,line width=0.25pt, densely dotted,forget plot] table[x = delta,y=alpha] from \FirstCon;
\addplot table[x = delta,y=ConTenKn3] from \FirstCon;
\addplot table[x = delta,y=ConTenKn5] from \FirstCon;
\addplot table[x = delta,y=ConTenKn7] from \FirstCon;
\addplot table[x = delta,y=TenOp] from \MainFirstLSWCon;
\addplot table[x = delta,y=TenUn] from \MainFirstLSWCon;
\nextgroupplot[legend style = {column sep = 3.5pt, legend to name = LegendCon1}]
\node[anchor=north,font=\fontsize{5}{4}\selectfont] at (axis description cs: 0.25,  0.95) {Monotonicity\\ and Convexity \\ $n=500$};
\addplot[smooth,tension=0.5,color=NavyBlue, no markers,line width=0.25pt, densely dotted,forget plot] table[x = delta,y=alpha] from \FirstMonCon;
\addplot table[x = delta,y=MonConFiveKn3] from \FirstMonCon;
\addplot table[x = delta,y=MonConFiveKn5] from \FirstMonCon;
\addplot table[x = delta,y=MonConFiveKn7] from \FirstMonCon;
\addplot table[x = delta,y=FiveOp] from \MainFirstLSWMonCon;
\addplot table[x = delta,y=FiveUn] from \MainFirstLSWMonCon;
\addlegendentry{FS-C3};
\addlegendentry{FS-C5};
\addlegendentry{FS-C7};
\addlegendentry{LSW-L};
\addlegendentry{LSW-S};
\nextgroupplot
\node[anchor=north,font=\fontsize{5}{4}\selectfont] at (axis description cs: 0.25,  0.95) {Monotonicity\\ and Convexity \\ $n=750$};
\addplot[smooth,tension=0.5,color=NavyBlue, no markers,line width=0.25pt, densely dotted,forget plot] table[x = delta,y=alpha] from \FirstMonCon;
\addplot table[x = delta,y=MonConSevenKn3] from \FirstMonCon;
\addplot table[x = delta,y=MonConSevenKn5] from \FirstMonCon;
\addplot table[x = delta,y=MonConSevenKn7] from \FirstMonCon;
\addplot table[x = delta,y=SevenOp] from \MainFirstLSWMonCon;
\addplot table[x = delta,y=SevenUn] from \MainFirstLSWMonCon;
\nextgroupplot
\node[anchor=north,font=\fontsize{5}{4}\selectfont] at (axis description cs: 0.25,  0.95) {Monotonicity\\ and Convexity \\ $n=1000$};
\addplot[smooth,tension=0.5,color=NavyBlue, no markers,line width=0.25pt, densely dotted,forget plot] table[x = delta,y=alpha] from \FirstMonCon;
\addplot table[x = delta,y=MonConTenKn3] from \FirstMonCon;
\addplot table[x = delta,y=MonConTenKn5] from \FirstMonCon;
\addplot table[x = delta,y=MonConTenKn7] from \FirstMonCon;
\addplot table[x = delta,y=TenOp] from \MainFirstLSWMonCon;
\addplot table[x = delta,y=TenUn] from \MainFirstLSWMonCon;
\end{groupplot}
\node at ($(myplots c2r2) + (0,-1.75cm)$) {\ref{LegendCon1}};
\end{tikzpicture}
\caption{Empirical power of shape tests for \eqref{Eqn: MC1,aux1} where our tests are implemented with $\gamma_n=0.01/\log n$ and corresponding to $\delta=0$ are the empirical sizes under D1.} \label{Fig: MC1, supp}
\end{figure}

\pgfplotstableread{ 
delta alpha ConFiveQKn0 ConFiveQKn1 ConSevenQKn0 ConSevenQKn1 ConTenQKn0 ConTenQKn1
0     0.05    0.0620      0.0687      0.0643        0.0730      0.0567     0.0667
1     0.05    0.1033      0.0923      0.1227        0.1173      0.1277     0.1087
2     0.05    0.1657      0.1417      0.2033        0.1770      0.2480     0.1920
3     0.05    0.2553      0.2027      0.3347        0.2730      0.4007     0.3263
4     0.05    0.3563      0.2913      0.4767        0.3800      0.5810     0.4613
5     0.05    0.4817      0.3827      0.6270        0.5103      0.7543     0.6243
6     0.05    0.6153      0.4873      0.7770        0.6463      0.8857     0.7700
7     0.05    0.7413      0.6017      0.8850        0.7700      0.9507     0.8757
8     0.05    0.8383      0.7123      0.9467        0.8750      0.9817     0.9523
9     0.05    0.9080      0.8170      0.9787        0.9343      0.9953     0.9807
10    0.05    0.9527      0.8817      0.9943        0.9737      0.9990     0.9920
}\SecondConQ

\pgfplotstableread{ 
delta alpha ConFiveCKn0 ConFiveCKn1 ConSevenCKn0 ConSevenCKn1 ConTenCKn0 ConTenCKn1
0     0.05    0.0697      0.0830       0.0720       0.0687      0.0693     0.0660
1     0.05    0.0977      0.1000       0.1217       0.0980      0.1210     0.0930
2     0.05    0.1500      0.1300       0.1877       0.1270      0.2137     0.1350
3     0.05    0.2240      0.1640       0.2943       0.1887      0.3433     0.1987
4     0.05    0.3150      0.2033       0.4077       0.2497      0.4990     0.2717
5     0.05    0.4113      0.2540       0.5450       0.3250      0.6673     0.3613
6     0.05    0.5280      0.3127       0.6857       0.4137      0.8033     0.4897
7     0.05    0.6480      0.3850       0.8063       0.5130      0.9033     0.6137
8     0.05    0.7537      0.4593       0.8940       0.6113      0.9633     0.7317
9     0.05    0.8480      0.5397       0.9503       0.7093      0.9843     0.8287
10    0.05    0.9067      0.6223       0.9840       0.7990      0.9953     0.9050
}\SecondConC

\pgfplotstableread{
delta alpha  FiveOp  FiveUn  SevenOp  SevenUn   TenOp   TenUn
0     0.05   0.0590  0.0460  0.0710   0.0680    0.0620  0.0530
1     0.05   0.0660  0.0550  0.0760   0.0670    0.0610  0.0570
2     0.05   0.0780  0.0670  0.0880   0.0640    0.0700  0.0600
3     0.05   0.0840  0.0710  0.0960   0.0760    0.0800  0.0710
4     0.05   0.0990  0.0740  0.1110   0.0800    0.1030  0.0770
5     0.05   0.0950  0.0750  0.1180   0.0930    0.1160  0.0850
6     0.05   0.1050  0.0710  0.1300   0.0920    0.1320  0.0960
7     0.05   0.1150  0.0860  0.1490   0.1040    0.1540  0.1010
8     0.05   0.1320  0.0870  0.1720   0.1240    0.1740  0.1040
9     0.05   0.1400  0.0980  0.2070   0.1330    0.1920  0.1200
10    0.05   0.1600  0.1040  0.2290   0.1400    0.2370  0.1460
}\MainSecondLSWCon

\begin{figure}[!h]
\centering\scriptsize
\begin{tikzpicture}[every text node part/.style={align=center}] 
\begin{groupplot}[group style={group name=myplots,group size=3 by 1,horizontal sep= 0.8cm,vertical sep=0.5cm},
    grid = minor,
    width = 0.375\textwidth,
    xmax=10,xmin=0,
    ymax=1,ymin=0,
    every axis title/.style={below,at={(0.2,0.8)}},
    xlabel=$\delta$,
    x label style={at={(axis description cs:0.95,0.04)},anchor=south},
    xtick={0,2,...,10},
    ytick={0.05,0.5,1},
    tick label style={/pgf/number format/fixed},
    legend style={text=black,cells={align=center},row sep = 3pt,legend columns = -1, draw=none,fill=none},
    cycle list={%
{smooth,tension=0.5,color=RoyalBlue1, mark=halfcircle*,every mark/.append style={rotate=90},mark size=1.5pt,line width=0.5pt}, 
{smooth,tension=0.5,color=RoyalBlue2, mark=halfcircle*,every mark/.append style={rotate=180},mark size=1.5pt,line width=0.5pt}, 
{smooth,tension=0.5,color=RoyalBlue3, mark=halfcircle*,every mark/.append style={rotate=270},mark size=1.5pt,line width=0.5pt},
{smooth,tension=0.5,color=RoyalBlue4, mark=halfcircle*,every mark/.append style={rotate=360},mark size=1.5pt,line width=0.5pt},
{smooth,tension=0.5,color=DarkOliveGreen4, mark=halfsquare*,every mark/.append style={rotate=90},mark size=1.5pt,line width=0.5pt}, 
{smooth,tension=0.5,color=SpringGreen4, mark=halfsquare*,every mark/.append style={rotate=270},mark size=1.5pt,line width=0.5pt},
}
]
\nextgroupplot[legend style = {column sep = 3.5pt, legend to name = LegendCon2}]
\node[anchor=north,font=\fontsize{5}{4}\selectfont] at (axis description cs: 0.25,  0.95) {Concavity \\ $n=500$};
\addplot[smooth,tension=0.5,color=NavyBlue, no markers,line width=0.25pt, densely dotted,forget plot] table[x = delta,y=alpha] from \SecondConQ;
\addplot table[x = delta,y=ConFiveQKn0] from \SecondConQ;
\addplot table[x = delta,y=ConFiveQKn1] from \SecondConQ;
\addplot table[x = delta,y=ConFiveCKn0] from \SecondConC;
\addplot table[x = delta,y=ConFiveCKn1] from \SecondConC;
\addplot table[x = delta,y=FiveOp] from \MainSecondLSWCon;
\addplot table[x = delta,y=FiveUn] from \MainSecondLSWCon;
\addlegendentry{FS-Q0};
\addlegendentry{FS-Q1};
\addlegendentry{FS-C0};
\addlegendentry{FS-C1};
\addlegendentry{LSW-L};
\addlegendentry{LSW-S};
\nextgroupplot
\node[anchor=north,font=\fontsize{5}{4}\selectfont] at (axis description cs: 0.25,  0.95) {Concavity \\ $n=750$};
\addplot[smooth,tension=0.5,color=NavyBlue, no markers,line width=0.25pt, densely dotted,forget plot] table[x = delta,y=alpha] from \SecondConQ;
\addplot table[x = delta,y=ConSevenQKn0] from \SecondConQ;
\addplot table[x = delta,y=ConSevenQKn1] from \SecondConQ;
\addplot table[x = delta,y=ConSevenCKn0] from \SecondConC;
\addplot table[x = delta,y=ConSevenCKn1] from \SecondConC;
\addplot table[x = delta,y=SevenOp] from \MainSecondLSWCon;
\addplot table[x = delta,y=SevenUn] from \MainSecondLSWCon;
\nextgroupplot
\node[anchor=north,font=\fontsize{5}{4}\selectfont] at (axis description cs: 0.25,  0.95) {Concavity \\ $n=1000$};
\addplot[smooth,tension=0.5,color=NavyBlue, no markers,line width=0.25pt, densely dotted,forget plot] table[x = delta,y=alpha] from \SecondConQ;
\addplot table[x = delta,y=ConTenQKn0] from \SecondConQ;
\addplot table[x = delta,y=ConTenQKn1] from \SecondConQ;
\addplot table[x = delta,y=ConTenCKn0] from \SecondConC;
\addplot table[x = delta,y=ConTenCKn1] from \SecondConC;
\addplot table[x = delta,y=TenOp] from \MainSecondLSWCon;
\addplot table[x = delta,y=TenUn] from \MainSecondLSWCon;
\end{groupplot}
\node at ($(myplots c2r1) + (0,-2.4cm)$) {\ref{LegendCon2}};
\end{tikzpicture}
\caption{Empirical power of concavity tests for \eqref{Eqn: MC2 supp,aux1} where our tests are implemented with $\gamma_n=0.01/\log n$ and corresponding to $\delta=0$ are the empirical sizes under D1.} \label{Fig: MC2, supp}
\end{figure}

{
\setlength{\tabcolsep}{3.2pt}
\renewcommand{\arraystretch}{1.2}
\begin{table}[!ht]
\caption{Empirical Size of Testing Slutsky Restriction on $g_0$ in \eqref{Eqn: MC3, aux1} at $\alpha=5\%$}\label{Tab: MC3, size}
\centering\footnotesize
\begin{threeparttable}
\sisetup{table-number-alignment = center, table-format = 1.3} 
\begin{tabularx}{\linewidth}{@{}cc *{3}{S[round-mode = places,round-precision = 3]} c *{3}{S[round-mode = places,round-precision = 3]} c *{3}{S[round-mode = places,round-precision = 3]} c *{3}{S[round-mode = places,round-precision = 3]}@{}} 
\hline
\hline
\multirow{2}{*}{$n$} & \multirow{2}{*}{$\gamma_n$} & \multicolumn{3}{c}{FS-Q0: $k_n=27$} & & \multicolumn{3}{c}{FS-Q1: $k_n=64$} & & \multicolumn{3}{c}{FS-C0: $k_n=64$} & & \multicolumn{3}{c}{FS-C1: $k_n=125$}\\
\cline{3-5} \cline{7-9} \cline{11-13} \cline{15-17}
& & {D1} & {D2} & {D3}  & & {D1} & {D2} & {D3} & & {D1} & {D2} & {D3} & & {D1} & {D2} & {D3}\\  
\hline
\multirow{3}{*}{$1000$}& $1/n$         & 0.0717 & 0.0373 & 0.0220 & & 0.0920 & 0.0560 & 0.0373 & & 0.0980 & 0.0580 & 0.0393 & & 0.1533 & 0.1013 & 0.0723\\
                       & $0.01/\log n$ & 0.0717 & 0.0373 & 0.0220 & & 0.0920 & 0.0560 & 0.0373 & & 0.0980 & 0.0580 & 0.0393 & & 0.1533 & 0.1013 & 0.0723\\
                       & $0.01$        & 0.0727 & 0.0373 & 0.0223 & & 0.0920 & 0.0567 & 0.0373 & & 0.0983 & 0.0583 & 0.0393 & & 0.1550 & 0.1013 & 0.0723\\
\hline
\multirow{3}{*}{$3000$}& $1/n$         & 0.0540 & 0.0193 & 0.0087 & & 0.0647 & 0.0260 & 0.0137 & & 0.0650 & 0.0237 & 0.0123 & & 0.0830 & 0.0350 & 0.0163\\
                       & $0.01/\log n$ & 0.0540 & 0.0193 & 0.0087 & & 0.0647 & 0.0260 & 0.0137 & & 0.0650 & 0.0237 & 0.0123 & & 0.0830 & 0.0350 & 0.0163\\
                       & $0.01$        & 0.0543 & 0.0193 & 0.0087 & & 0.0650 & 0.0260 & 0.0137 & & 0.0653 & 0.0240 & 0.0123 & & 0.0840 & 0.0357 & 0.0167\\
\hline
\multirow{3}{*}{$5000$}& $1/n$         & 0.0560 & 0.0200 & 0.0083 & & 0.0670 & 0.0223 & 0.0093 & & 0.0660 & 0.0210 & 0.0083 & & 0.0667 & 0.0233 & 0.0090\\
                       & $0.01/\log n$ & 0.0560 & 0.0200 & 0.0083 & & 0.0670 & 0.0223 & 0.0093 & & 0.0660 & 0.0210 & 0.0083 & & 0.0667 & 0.0233 & 0.0090\\
                       & $0.01$        & 0.0560 & 0.0203 & 0.0083 & & 0.0673 & 0.0223 & 0.0093 & & 0.0663 & 0.0213 & 0.0083 & & 0.0670 & 0.0233 & 0.0090\\
\hline
\hline
\end{tabularx}
\begin{tablenotes}[flushleft]
\item {\it Note:} The parameter $\gamma_n$ determines $\hat\kappa_n$ proposed in Section \ref{Sec: tuning parameter} with $c_n=1/\log n$ and $r_n=(n/k_n)^{1/2}$.
\end{tablenotes}
\end{threeparttable}
\end{table}
}

\pgfplotstableread{
delta alpha  TenQ0   TenQ1   TenC0   TenC1   ThirtyQ0  ThirtyQ1 ThirtyC0 ThirtyC1  FiftyQ0 FiftyQ1 FiftyC0 FiftyC1
0     0.05   0.0717  0.0920  0.0980  0.1533  0.0540    0.0647   0.0650   0.0830    0.0560  0.0670  0.0660  0.0667
1     0.05   0.0760  0.1037  0.1090  0.1753  0.0710    0.0700   0.0667   0.0857    0.0753  0.0783  0.0750  0.0747
2     0.05   0.0883  0.1083  0.1137  0.1743  0.1127    0.0843   0.0830   0.0913    0.1667  0.0967  0.1010  0.0930
3     0.05   0.1140  0.1097  0.1230  0.1843  0.2200    0.1097   0.1123   0.1087    0.3737  0.1443  0.1573  0.1210
4     0.05   0.1533  0.1270  0.1377  0.1960  0.4237    0.1627   0.1740   0.1300    0.7247  0.2420  0.2677  0.1567
5     0.05   0.2237  0.1477  0.1580  0.2047  0.6880    0.2313   0.2597   0.1647    0.9453  0.3970  0.4590  0.2257
6     0.05   0.3260  0.1700  0.1840  0.2137  0.8960    0.3433   0.3963   0.2087    0.9947  0.6357  0.7080  0.3257
7     0.05   0.4527  0.2063  0.2253  0.2400  0.9803    0.5047   0.5730   0.2770    1.0000  0.8513  0.8990  0.4677
8     0.05   0.6080  0.2560  0.2803  0.2620  0.9987    0.6880   0.7553   0.3607    1.0000  0.9640  0.9833  0.6307
9     0.05   0.7407  0.3057  0.3387  0.2903  1.0000    0.8447   0.8953   0.4723    1.0000  0.9943  0.9987  0.8037
10    0.05   0.8530  0.3817  0.4150  0.3240  1.0000    0.9483   0.9680   0.5973    1.0000  0.9993  0.9993  0.9170
}\Slutsky

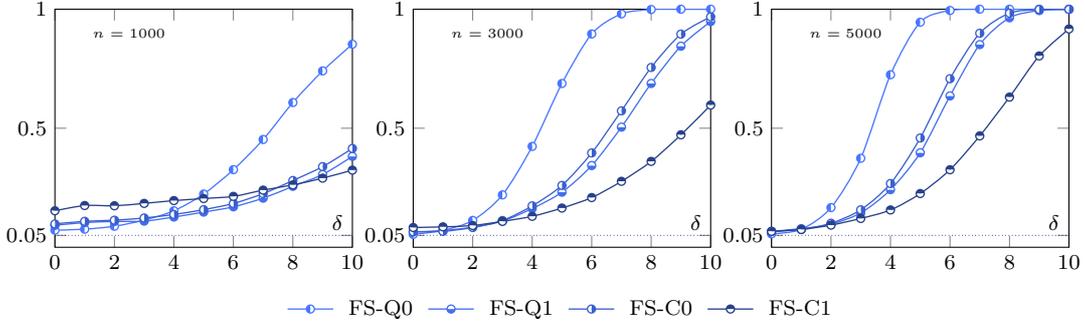
\begin{figure}[!h]
\centering\scriptsize
\begin{tikzpicture}[every text node part/.style={align=center}] 
\begin{groupplot}[group style={group name=myplots,group size=3 by 5,horizontal sep= 0.8cm,vertical sep=0.5cm},
    grid = minor,
    width = 0.375\textwidth,
    xmax=10,xmin=0,
    ymax=1,ymin=0,
    every axis title/.style={below,at={(0.2,0.8)}},
    xlabel=$\delta$,
    x label style={at={(axis description cs:0.95,0.04)},anchor=south},
    xtick={0,2,...,10},
    ytick={0.05,0.5,1},
    tick label style={/pgf/number format/fixed},
    legend style={text=black,cells={align=center},row sep = 3pt,legend columns = -1, draw=none,fill=none},
    cycle list={%
{smooth,tension=0.5,color=RoyalBlue1, mark=halfcircle*,every mark/.append style={rotate=90},mark size=1.5pt,line width=0.5pt}, 
{smooth,tension=0.5,color=RoyalBlue2, mark=halfcircle*,every mark/.append style={rotate=180},mark size=1.5pt,line width=0.5pt}, 
{smooth,tension=0.5,color=RoyalBlue3, mark=halfcircle*,every mark/.append style={rotate=270},mark size=1.5pt,line width=0.5pt},
{smooth,tension=0.5,color=RoyalBlue4, mark=halfcircle*,every mark/.append style={rotate=360},mark size=1.5pt,line width=0.5pt},
}
]
\nextgroupplot[legend style = {column sep = 3.5pt, legend to name = LegendSlutsky}]
\node[anchor=north,font=\fontsize{5}{4}\selectfont] at (axis description cs: 0.25,  0.95) {$n=1000$};
\addplot[smooth,tension=0.5,color=NavyBlue, no markers,line width=0.25pt, densely dotted,forget plot] table[x = delta,y=alpha] from \Slutsky;
\addplot table[x = delta,y=TenQ0] from \Slutsky;
\addplot table[x = delta,y=TenQ1] from \Slutsky;
\addplot table[x = delta,y=TenC0] from \Slutsky;
\addplot table[x = delta,y=TenC1] from \Slutsky;
\addlegendentry{FS-Q0};
\addlegendentry{FS-Q1};
\addlegendentry{FS-C0};
\addlegendentry{FS-C1};
\nextgroupplot
\node[anchor=north,font=\fontsize{5}{4}\selectfont] at (axis description cs: 0.25,  0.95) {$n=3000$};
\addplot[smooth,tension=0.5,color=NavyBlue, no markers,line width=0.25pt, densely dotted,forget plot]table[x = delta,y=alpha] from \Slutsky;
\addplot table[x = delta,y=ThirtyQ0] from \Slutsky;
\addplot table[x = delta,y=ThirtyQ1] from \Slutsky;
\addplot table[x = delta,y=ThirtyC0] from \Slutsky;
\addplot table[x = delta,y=ThirtyC1] from \Slutsky;
\nextgroupplot
\node[anchor=north,font=\fontsize{5}{4}\selectfont] at (axis description cs: 0.25,  0.95) {$n=5000$};
\addplot[smooth,tension=0.5,color=NavyBlue, no markers,line width=0.25pt, densely dotted,forget plot] table[x = delta,y=alpha] from \Slutsky;
\addplot table[x = delta,y=FiftyQ0] from \Slutsky;
\addplot table[x = delta,y=FiftyQ1] from \Slutsky;
\addplot table[x = delta,y=FiftyC0] from \Slutsky;
\addplot table[x = delta,y=FiftyC1] from \Slutsky;
\end{groupplot}
\node at ($(myplots c2r1) + (0,-2.4cm)$) {\ref{LegendSlutsky}};
\end{tikzpicture}
\caption{Empirical power of testing Slutsky restriction on $g_0$ in \eqref{Eqn: MC3, aux2} with $\gamma_n=0.01/\log n$, where corresponding to $\delta=0$ are the empirical sizes under D1.} \label{Fig: MC3}
\end{figure}

Our final set of Monte Carlo simulations concerns Slutsky restriction based on Example \ref{Ex: Slutsky} with $d_q=2$. Concretely, we draw i.i.d.\ samples $\{P_{1i}^*,P_{2i}^*,Y_i^*,Z_i^*,U_{1i},U_{2i}\}_{i=1}^n$ from the standard normal distribution in $\mathbf R^6$ and set $P_i=[P_{1i},P_{2i}]^\transpose$ with $P_{ji}=1+\Phi(P_{ji}^*)$, $Y_i=\Phi(Y_i^*)$, $Z_i=\Phi(Z_i^*)$, and $U_i=[U_{1i},U_{2i}]^\transpose$ for all $i$ and $j=1,2$. In turn, we let $\Gamma_0 = [1,1]^\transpose$ and consider three specifications for $g_0\equiv[g_{10},g_{20}]^\transpose$ under the null:
\begin{align}\label{Eqn: MC3, aux1}
g_{j0}(p_1,p_2,y)=\mathsf ap_j^{\frac{1}{\mathsf b-1}}\frac{y}{p_1^{\mathsf b/(\mathsf b-1)}+p_2^{\mathsf b/(\mathsf b-1)}}+\mathsf c~,\quad j=1,2~,
\end{align}
with $(\mathsf a,\mathsf b,\mathsf c)=(0,0.5,0.5),(0.5,0,0)$, and $(1,0.5,0)$, labeled D1, D2, and D3, respectively. Note that D1 is a least favorable case, while D2 and D3 may be respectively rationalized by a Cobb--Douglas and a CES (constant elasticity of substitution) utility function. For specifications under the alternative, we choose
\begin{align}\label{Eqn: MC3, aux2}
[g_{10}(p_1,p_2,y),\,g_{20}(p_1,p_2,y)]=[\exp\{(p_1-1.5)0.1\delta\},\,\exp\{-(p_2-1.5)0.1\delta\}]~,
\end{align}
where $\delta=1,\ldots,10$. The resulting Slutsky matrix $\theta_0(p_1,p_2,y)$ at each $(p_1,p_2,y)$ (as defined in \eqref{Eqn: Slutsky matrix in ex}) has one of its eigenvalues positive and the other negative.

To implement our test, we construct a vector $h^{k_n}$ of series functions via tensor product of univariate B-splines, obtain $\hat g_n$ by regressing $\{Q_i\}_{i=1}^n$ on $\{h^{k_n}(P_i,Y_i),Z_i\}_{i=1}^n$, and then derive $\hat\theta_n$ by differentiating $\hat g_n$. The whole procedure can be streamlined by the commands \texttt{spmak}, \texttt{fnval}, and \texttt{fnder} provided by the Curve Fitting Toolbox in MATLAB. A practical issue of grave concern is, however, that estimation of $\theta_0$ now involves trivariate nonparametric functions, resulting in potentially too large a sieve dimension $k_n$ (e.g., $k_n=125$ for FS-C1). For this reason, we employ the same set of B-splines as in the bivariate design, but experiment with $n\in \{1000, 3000,5000\}$. In turn, we evaluate the integrals (see \eqref{Eqn: Slutsky norm in ex}) over $[1.1,1.9]^2\times[0.1,0.9]$ with marginal step size $0.05$. Finally, we construct the critical values based on the sieve score bootstrap with i.i.d.\ standard normals as weights---see Appendix \ref{Sec: Slutsky, appendix} (note that our designs are configured without endogeneity for simplicity).

Table \ref{Tab: MC3, size} and Figure \ref{Fig: MC3} report partial results of our simulations---see Table \ref{Tab: MC3, size, app} and Figure \ref{Fig: MC3, FullSlutsky} in Appendix \ref{App: full simulations} for the full set of results. Not surprisingly, our tests exhibit marked size distortions when the sieve dimension is ``large'' relative to the sample size, but otherwise control size reasonably well. As emphasized previously, Gaussian approximation may be inaccurate if $k_n$ is ``too large.''  On the other hand, the power performance is influenced by $k_n$ through two channels: accuracy of the Gaussian approximation and the rate $r_n=\sqrt{n/k_n}$. This may explain the relative low power of our tests when $n=1000$, though all power curves improve as $n$ increases.

To conclude, we report the run-times of a single replication based on designs D1 in the computing environment of Section \ref{Sec: simulations}. As before, we only report our tests with the smallest and the largest $k_n$, based on $\gamma_n=0.01/\log n$. Overall, Table \ref{Tab: MC, runtime2} supports our previous claim on the relative computational simplicity of our tests---when comparing run-times across shapes and the dimensions of covariates, keep in mind that the fineness of discretization varies. When working with real data sets, one may increase the number of grid points and the number of bootstrap repetitions, as the computational cost is no more than one Monte Carlo simulation replication.

{
\setlength{\tabcolsep}{1.5pt}
\renewcommand{\arraystretch}{1.1}
\begin{table}[!ht]
\caption{Run-times (in Seconds) of Shape Tests} \label{Tab: MC, runtime2}
\centering\scriptsize
\begin{threeparttable}
\sisetup{table-format = 3.2,table-number-alignment = center} 
\begin{tabularx}{\linewidth}{@{}cc *{2}{S[round-mode = places,round-precision = 2]} c *{2}{S[round-mode = places,round-precision = 2]} c@{\hspace{1.25pt}} *{2}{S[round-mode = places,round-precision = 2]} c *{2}{S[round-mode = places,round-precision = 2]} c@{\hspace{1.25pt}} *{2}{S[round-mode = places,round-precision = 2]} c *{2}{S[round-mode = places,round-precision = 2]} c@{\hspace{1.25pt}} *{2}{S[round-mode = places,round-precision = 2]} c@{}} 
\hline
\hline
&\multirow{3}{*}{$n$}& \multicolumn{5}{c}{Convexity: \eqref{Eqn: MC1,aux1}} & & \multicolumn{5}{c}{Mon-Con: \eqref{Eqn: MC1,aux1}} & & \multicolumn{5}{c}{Concavity: \eqref{Eqn: MC2 supp,aux1}} & & \multicolumn{2}{c}{Slutsky: \eqref{Eqn: MC3, aux1}} & \\
\cline{3-7} \cline{9-13} \cline{15-19} \cline{21-22}
&& \multicolumn{2}{c}{FS} & & \multicolumn{2}{c}{LSW} & & \multicolumn{2}{c}{FS} & & \multicolumn{2}{c}{LSW} & & \multicolumn{2}{c}{FS} & & \multicolumn{2}{c}{LSW} & & \multicolumn{2}{c}{FS} & \\
\cline{3-4} \cline{6-7} \cline{9-10} \cline{12-13} \cline{15-16} \cline{18-19} \cline{21-22}
&& {C3} & {C7} && {L} & {S} & & {Q0} & {C1} && {L} & {S} & & {Q0} & {C1} && {L} & {S} & & {Q0} & {C1} & \\  
\cline{2-19} \cline{21-22}
&$500$        & 0.2386 & 0.2424 && 23.0194  & 22.3220  & & 0.2444 & 0.2676 && 23.6102  & 23.2546   && 16.0498 & 17.0906 && 13.1564 & 13.2857 & & 9.5914 & 16.8517& \\
&$750$        & 0.2525 & 0.2625 && 56.4153  & 57.4263  & & 0.2597 & 0.2851 && 57.5583  & 57.5890   && 14.9603 & 16.5335 && 38.7568 & 38.7796 & & 9.8024 & 17.2299& \\
&$1000$       & 0.2513 & 0.2574 && 102.2136 & 101.6382 & & 0.2599 & 0.2569 && 101.4130 & 102.7763  && 16.1228 & 17.1188 && 68.0818 & 69.1066 & & 10.4518 & 19.0047& \\
\hline
\hline
\end{tabularx}
\begin{tablenotes}[flushleft]
\item  {\it Note:} The sample sizes for the Slutsky restriction from top to bottom should be 1000, 3000 and 5000.
\end{tablenotes}
\end{threeparttable}
\end{table}
}

\subsection{Empirical Application}

To further illustrate the implementation of our test, we revisit the problem of option pricing functions under shape restrictions in financial economics. As forcefully argued in the literature, parametric models are barely grounded in financial theory and may be inadequate in capturing key aspects of the relationship under consideration. This has spurred a line of research on nonparametric estimation of option pricing functions under shape restrictions \citep{AitSahaliaDuarte2003Option,BirkePilz2009NoArbitrage}. In particular, completeness of the market and absence of arbitrage opportunities imply two prominent restrictions: monotonicity and convexity of the call/put option price with respect to the strike price of the option, at a specific  valuation date and for the same time-to-expiration. Below we complement the literature by testing the validity of these restrictions.


We approach the problem in the setup of Example \ref{Ex: NP under shape} following the aforementioned studies, where $Y$ denotes the option price and $Z$ the corresponding strike price. We aim to test three shape restrictions on $\theta_0$, i.e., monotonicity, convexity, and monotonicity jointly with convexity. While $\theta_0$ should be convex for both call and put options, $\theta_0$ should be nonincreasing for the former and nondecreasing for the latter. We make use of the data set analyzed in \citet{BeareSchmidt2016Pricing}, which consists of prices for European call and put options written on the S\&P 500 index---see Section 4 in \citet{BeareSchmidt2016Pricing} for detailed descriptions of the data set. We select two dates for our test problems: October 22, 2008 which has the maximal number of call options ($n=93$), and October 19, 2011 which has the maximal number of put options ($n=143$). Such small sample sizes, while not uncommon in practice, may raise concerns on the performance of our test. In unreported simulations based on the univariate designs in Section \ref{Sec: simulations}, we found that, with the sample size equal to $n=100$ and $\gamma_n\in\{0.01/\log n,1/n\}$, series estimation based on quadratic B-splines with two interior knots (labeled Q2) and cubic B-splines with one knot (labeled C1) delivers null rejection rates no larger than $0.068$ (at $5\%$ nominal level) and reasonable power (over $0.5$ at $\delta=10$). Thus, our implementation below will be based on these choices of splines and knots.

The remaining details of the implementation are the same as those in Sections \ref{Sec: simulations} and \ref{App: more simulations1} (for the univariate designs) beyond the following changes. First, the strike prices are converted via the affine transformation $z\mapsto 2(z-a)/(b-a)-1$, with $a$ and $b$ respectively the minimal and maximal strike prices in the data. As a result, the converted strike prices fall within the range $[-1,1]$ (to be consistent with Section \ref{Sec: simulations}) without changing the shape restrictions under consideration. Second, the number of bootstrap repetitions is increased to $1000$, while the step size for numerical integration is decreased to $0.01$. These changes echo our previous remark that, in applications, ``one may increase the number of grid points and the number of bootstrap repetitions, as the computational cost is no more than one Monte Carlo simulation replication.''

Table \ref{Tab: application} reports the $p$-values of our test (with $\gamma_n=0.01/\log n$). We fail to reject the three null hypotheses for both call and put options, at all conventional significance levels. In some cases, there are sizable discrepancies in the $p$-values across Q2 and C1 (for the same shape). This may be explained by the small sample issue, which is also in line with our simulation results for the Slutsky restrictions (those with high ratios of $k_n/n$). Overall, though, our findings point to strong evidences of the presence of the three shape restrictions (in the present rather restrictive setting).

{
\setlength{\tabcolsep}{4.05pt}
\renewcommand{\arraystretch}{1.1}
\begin{table}[!ht]
\caption{Testing Shape Restrictions of Option Pricing Functions: $p$-Values} \label{Tab: application}
\centering\scriptsize
\sisetup{table-format = 3.2,table-number-alignment = center} 
\begin{tabularx}{\linewidth}{@{}c *{2}{S[round-mode = places,round-precision = 2]} c *{2}{S[round-mode = places,round-precision = 2]} c@{\hspace{1.25pt}} *{2}{S[round-mode = places,round-precision = 2]} c *{2}{S[round-mode = places,round-precision = 2]} c@{\hspace{1.25pt}} *{2}{S[round-mode = places,round-precision = 2]} c *{2}{S[round-mode = places,round-precision = 2]}  c@{}} 
\hline
\hline
& \multicolumn{8}{c}{Call options} & & \multicolumn{8}{c}{Put options}  & \\
\cline{2-9} \cline{11-18}
& \multicolumn{2}{c}{{Monotonicity}} & & \multicolumn{2}{c}{Convexity} & & \multicolumn{2}{c}{Mon-Con} & & \multicolumn{2}{c}{Monotonicity} & & \multicolumn{2}{c}{Convexity} & & \multicolumn{2}{c}{Mon-Con}  & \\
\cline{2-3} \cline{5-6} \cline{8-9} \cline{11-12} \cline{14-15} \cline{17-18}
& {Q2} & {C1} && {Q2} & {C1} & & {Q2} & {C1} && {Q2} & {C1} & & {Q2} & {C1} && {Q2} & {C1} & \\
\cline{1-19}
& 0.702 & 0.22 && 0.552  & 0.139  & & 0.606 & 0.30 &&  0.573  & 0.756    && 0.364  & 0.891  && 0.722  &  0.862 & \\
\hline
\hline
\end{tabularx}
\end{table}
}

\end{appendices}

\titleformat{\section}{\normalfont\Large\bfseries}{\thesection}{1em}{}

\addcontentsline{toc}{section}{References}
\putbib
\end{bibunit}


\clearpage\newpage

\begin{bibunit}

\begin{appendices}
\titleformat{\section}{\Large\center}{{\sc Appendix} \thesection}{1em}{}
\setcounter{page}{1}
\setcounter{section}{5}
\renewcommand{\thetable}{\thesection.\Roman{table}}
\emptythanks

\phantomsection
\pdfbookmark[1]{Appendix Title}{title2}
\title{Second Supplement to ``A Projection Framework for Testing Shape Restrictions That Form Convex Cones''}
\author{
Zheng Fang \\ Department of Economics \\ Texas A\&M University\\ zfang@tamu.edu
\and
Juwon Seo\\ Department of Economics\\National University of Singapore\\ ecssj@nus.edu.sg}
\date{ }
\maketitle

\vspace{-0.3in}

This supplement collects results omitted from the main text and the online supplement due to space limitation. In particular, Appendix \ref{Sec: appendix, ex} verifies the main assumptions for our examples, Appendix \ref{App: proofs for the special case} provides proofs for Appendix \ref{Sec: special case}, and Appendix \ref{App: full simulations} collects the complete set of simulation results for Section \ref{Sec: simulations} and Appendix \ref{App: more simulations}.

{\renewcommand{\arraystretch}{1}
\begin{table}[h]
\begin{center}
\begin{tabularx}{\textwidth}{cX}
\hline\hline
$ a \lesssim b$                 & For some constant $M$ that is universal in the proof, $a\leq Mb$.\\
$a^{(j)}$                       & The $j$-th coordinate of a vector $a\in\mathbf R^d$.\\
$a^{(-j)}$                      & The vector in $\mathbf R^{d-1}$ obtained by deleting the $j$-th entry of $a\in\mathbf R^d$.\\
$a\wedge b$, $a\vee b$          & For $a,b\in\mathbf R$, $a\wedge b=\min\{a,b\}$ and $a\vee b=\max\{a,b\}$.\\
$ a\Lambda$                     & For a set $\Lambda$ in a vector space and $a\in\mathbf R$, $a\Lambda\equiv\{a\lambda: \lambda\in\Lambda\}$.\\
$\Lambda+\theta$                & For a set $\Lambda$ and an element $\theta$ in a vector space, $\Lambda+\theta\equiv\{\lambda+\theta: \lambda\in\Lambda\}$.\\
$\Lambda+\Gamma$                & For sets $\Lambda$ and $\Gamma$ in a vector space, $\Lambda+\Gamma\equiv\{\lambda+\gamma: \lambda\in\Lambda,\gamma\in\Gamma\}$.\\
$\overline{\Lambda}$            & For a set $\Lambda$ in a topological space, $\overline\Lambda$ is the closure of $\Lambda$.\\
$A^-$                           & The Moore-Penrose inverse of a matrix $A$.\\
$A_l^-$                         & For a matrix $A\in\mathbf M^{m\times k}$, $A_l^-\equiv (A^\transpose A)^- A^\transpose$. \\
$\sigma_{\min}(A)$              & The smallest singular value of a matrix $A$. \\
$\sigma_{\max}(A)$              & The largest singular value of a matrix $A$. \\
$\lambda_{\min}(A)$              & The smallest eigenvalue of a matrix $A$ for a symmetric matrix $A$. \\
$\lambda_{\max}(A)$              & The largest eigenvalue of a matrix $A$ for a symmetric matrix $A$. \\
$\|A\|$                         & For a matrix $A$, $\|A\|\equiv\sqrt{\mathrm{tr}(A^\transpose A)}$.\\
$\|A\|_{o}$                       & For a matrix $A$, $\|A\|_{o}\equiv \sup\{\|Ax\|/\|x\|: x\neq 0\}\equiv\sigma_{\max}(A)$.\\
$A\le B$                        & For matrices $A$ and $B$, $A\le B$ iff $B-A$ is positive semidefinite.\\
$\|X \|_{P,r}$                & The $L_r$ norm of a random variable $X$ under measure $P$.\\
$\overset{d}{=}$              & Equality in distribution. \\
$\|X\|_{P,\infty}$         & The essential supremum norm of $X$ under measure $P$.\\
$\|f\|_\infty$                  & For a function $f:  T\to\mathbf M^{m\times k}$, $\|f\|_\infty\equiv\sup_{t\in T}\sqrt{\mathrm{tr}(f(t)^\transpose f(t))}$.\\
$\ell^\infty(T)$       & For a nonempty set $T$, $\ell^\infty(T) \equiv \{f:T\to\mathbf R:    \|f\|_\infty < \infty\}$.\\
\hline\hline
\end{tabularx}
\end{center}
\end{table}
}

\section{Discussions of Examples \ref{Ex: NP under shape}-\ref{Ex: Slutsky}} \label{Sec: appendix, ex}

In this section, we verify our main assumptions for Examples \ref{Ex: NP under shape}-\ref{Ex: Slutsky}. Throughout, we assume that  the underlying probability space is sufficiently rich for the sake of strong approximations. This is, however, not a restriction as one may always augment the original space by taking a suitable product space \citep{ChernozhukovLeeRosen2013Intersection}.

\subsection{Nonparametric Instrumental Variable Regression (NPIV)}\label{Sec: npiv, app}

Since Example \ref{Ex: NP under shape} is a special case of Example \ref{Ex: NPIV under shape}, we focus on the latter for simplicity---see also \citet{ChernozhukovLeeRosen2013Intersection} and \citet{BelloniChernozhukovChetverikovKato2015New} for strong approximation results for the former. We proceed by introducing some notation. Let $\{h_k\}_{k=1}^\infty$ and $\{b_m\}_{m=1}^\infty$ be basis functions for approximating $\theta_0$ and the instrument space respectively. Define $\underline Y_n\equiv (Y_1,\ldots,Y_n)^\transpose$, $h^{k_n}\equiv (h_1,\ldots,h_{k_n})^\transpose$, and $b^{m_n}\equiv(b_1,\ldots,b_{m_n})^\transpose$, and set
\begin{align}
H_n\equiv [h^{k_n}(Z_1),\ldots,h^{k_n}(Z_n)]^\transpose~,\, B_n\equiv [b^{m_n}(V_1),\ldots,b^{m_n}(V_n)]^\transpose~.
\end{align}
Here, we require $m_n\ge k_n$, just as in classical 2SLS estimation. Further define
\begin{multline}
\Phi_{n,P}\equiv E[h^{k_n}(Z)h^{k_n}(Z)^\transpose]~,\,\Psi_{n,P}\equiv E[b^{m_n}(V)b^{m_n}(V)^\transpose]~,\\
\Pi_{n,P}\equiv E[b^{m_n}(V)h^{k_n}(Z)^\transpose]~,
\end{multline}
and their respective sample analogs:
\begin{align}
\hat\Phi_n\equiv \frac{H_n^\transpose H_n}{n},\,\hat\Psi_n\equiv \frac{B_n^\transpose B_n}{n}~,\,\hat\Pi_n\equiv \frac{B_n^\transpose H_n}{n}~.
\end{align}
In turn, we estimate $\theta_0$ by $\hat\theta_n\equiv\hat\beta_n^\transpose h^{k_n}$, where $\hat\beta_n$ is the series 2SLS estimator, i.e.,
\begin{align}\label{Eqn: series 2SLS}
\hat\beta_n\equiv [\hat\Pi_n^\transpose\hat\Psi_n^- \hat\Pi_n]^- \hat\Pi_n^\transpose\hat\Psi_n^- \frac{1}{n}B_n^\transpose\underline Y_n~.
\end{align}
Thus, if $Z=V$ and $h^{k_n}=b^{m_n}$, then $\hat\beta_n$ reduces to the series LS estimator.

Next, let $\mathcal H_{k_n}$ and $\mathcal B_{m_n}$ be the subspaces spanned by $h_1,\ldots,h_{k_n}$ and $b_1,\ldots,b_{m_n}$ respectively. Then we denote by $\mathrm{Proj}_k$ the projection operator in $L^2(Z)$ onto $\mathcal H_{k_n}$, and by $\mathrm{Proj}_m$ the projection operator in $L^2(V)$ onto $\mathcal B_{m_n}$. Thus, if $\theta\in L^2(Z)$, then
\begin{align}
\mathrm{Proj}_k\theta=\gamma_{n,P}^\transpose h^{k_n}~,\,\gamma_{n,P}\equiv \Phi_{n,P}^{-1} E_P[h^{k_n}(Z)\theta(Z)]~.
\end{align}
Let $\mathrm{Proj}_{m,k}: L^2(Z)\to L^2(Z)$ be the 2SLS projection operator define by
\begin{align}
\mathrm{Proj}_{m,k}\theta=\beta_{n,P}^\transpose h^{k_n}~,
\end{align}
where $\beta_{n,P}\equiv (\Pi_{n,P}^\transpose \Psi_{n,P}^{-1}\Pi_{n,P})^{-1}  \Pi_{n,P}^\transpose \Psi_{n,P}^{-1} E_P[b^{m_n}(V)\theta(Z)]$. Moreover, for $a_{n,P}(z)\equiv \theta_P(z)-h^{k_n}(z)^\transpose\beta_{n,P}$ which is the 2SLS projection residual, let
\begin{align}
A_{n,P}\equiv [a_{n,P}(Z_1),\ldots,a_{n,P}(Z_n)]^\transpose~.
\end{align}
Finally, we define the conditional expectation operator $\Upsilon_P: L^2(Z)\to L^2(V)$ by
\begin{align}
\Upsilon_P\theta\equiv E_P[\theta(Z)|V]~.
\end{align}

With these notation in hand, we now impose the following sufficient conditions, where the Hilbert space $\mathbf H$ is the space of squared integrable functions on the support of $\mathcal Z$ with respect to the Lebesgue measure.

\begin{ass}\label{Ass: series 2SLS estimation}
(i) (a) $\{Y_i,Z_i,V_i\}_{i=1}^n$ are i.i.d., generated by \eqref{Eqn: Ex, NP under shape} and governed by $P\in\mathbf P$; (b) The operator $\Upsilon_P: L^2(Z)\to L^2(V)$ is injective for each $P\in\mathbf P$; (c) $\|\cdot\|_{\mathbf H}\lesssim \|\cdot\|_{L^2(Z)}$; (d) The support $\mathcal Z$ of $Z$ is bounded uniformly in $P\in\mathbf P$.


\noindent (ii) $\{h_k\}_{j=1}^\infty$ are functions on $\mathcal Z$ satisfying (a) $\sup_{n\in\mathbf N}\sup_{P\in\mathbf P}\lambda_{\max}(\Phi_{n,P})<\infty$; (b) $\sup_{P\in\mathbf P}\|h^{k_n}\|_{P,\infty}\le\xi_n$ where $\{\xi_{n}\}$ is bounded from below; (c) $\|\theta_0-\gamma_{n,P}^{\transpose}h^{k_n}\|_{P,\infty}=O(\delta_n)$ for some $\delta_n=o(1)$, uniformly in $P\in\mathbf P$; (d) $\|\Upsilon_P(\theta_0-\gamma_{n,P}^\transpose h^{k_n})\|_{L^2(V)}\lesssim s_n \|\theta_0-\gamma_{n,P}^\transpose h^{k_n}\|_{L^2(Z)}$ for some $s_n>0$, all $n$ and $P\in\mathbf P$.


\noindent (iii) $\{b_m\}_{m=1}^\infty$ are functions on $\mathcal V$ satisfying (a) the eigenvalues of $\Psi_{n,P}$ are bounded above and away from zero uniformly in $n$ and $P\in\mathbf P$; (b) $\sup_{P\in\mathbf P}\|b^{m_n}\|_{P,\infty}\le\xi_n$; (c) $\inf_{P\in\mathbf P}\sigma_{\min}(\Pi_{n,P})\gtrsim s_n>0$ for each $n$; (d) $\|\mathrm{Proj}_{m,k}(\theta_0-\gamma_{n,P}^\transpose h^{k_n})\|_\infty\lesssim \|\theta_0-\gamma_{n,P}^\transpose h^{k_n}\|_\infty$ uniformly in $P\in\mathbf P$.

\noindent (iv) (a) $\sup_{P\in\mathbf P}\|E_P[|u|^3|V]\|_{P,\infty} <\infty$; (b) $E_P[u^2|V]>\underline\sigma^2$ almost surely for some absolute constant $\underline\sigma^2>0$ and for all $P\in\mathbf P$.

\noindent (v) (a) $2\le k_n\le m_n\le c_0k_n$ for some $c_0\ge 1$; (b) $\varpi_n=o(1)$ with $\varpi_n$ defined as
\begin{align}\label{Eqn: coupling rate, npiv}
\varpi_n\equiv \frac{\sqrt n s_n}{\xi_n}\delta_n + (\frac{\xi_n m_n^2}{\sqrt n})^{1/3} + s_n^{-1}\sqrt{\frac{\xi_n^2 m_n \log m_n}{n}} + \delta_n\sqrt{(\xi_n^2\log m_n)\vee m_n}~;
\end{align}
(c) $\xi_n^3\{(\log m_n)/n\}^{1/2}=o(1)$.

\noindent (vi) (a) $\{W_i\}_{i=1}^\infty$ is an i.i.d. sequence of random variables; (b) $\{W_i\}_{i=1}^n$ are independent of $\{X_i\}_{i=1}^n$ for all $n$; (c) $E[W_1]=0$, $\mathrm{Var}(W_1)=1$ and $E[|W_1|^{3}]<\infty$.

\end{ass}

Assumption \ref{Ass: series 2SLS estimation} is essentially due to \citet{ChenChristensen2018SupNormOptimal} who study general functionals of $\theta_0$ but are concerned with pointwise in $P$ results. Assumption \ref{Ass: series 2SLS estimation}(i) is standard. In particular, Assumption \ref{Ass: series 2SLS estimation}(i)-(b) is the so-called $L^2(Z)$-completeness condition that is necessary for point identification of $\theta_0$. While $L^p$-completeness cannot be nontrivially tested \citep{Canay_Santos_Shaikh2013testability}, \citet{Andrews2017Completeness} show that, as a restriction lying between completeness and bounded completeness, $L^2$-completeness is generic in the sense that the set of distributions for which $L^2$-completeness fails is ``shy'' within a certain set of distributions, a notion generalizing the concept of Lebesgue null sets to infinite dimensional settings. Given Assumption \ref{Ass: series 2SLS estimation}(i)-(d), Assumption \ref{Ass: series 2SLS estimation}(i)-(c) is satisfied if the density of $Z$ is bounded away from zero on the support, uniformly in $P\in\mathbf P$. Assumptions \ref{Ass: series 2SLS estimation}(ii) and (iii) are mostly standard in series estimation. In particular, Assumption \ref{Ass: series 2SLS estimation}(ii)-(c) may be verified by results from approximation theory in conjunction with Proposition 3.1 in \citet{DevoreLotentz1993constructive}---see \citet{BelloniChernozhukovChetverikovKato2015New} and \citet{ChenChristensen2018SupNormOptimal}. The possibility that $s_n$ may approach zero reflects the fundamental issue that the NPIV model may be ill-posed \citep{NeweyPowell2003NPIV,CarrascoFlorensRenault2007LinearInverse}. Given Assumptions \ref{Ass: series 2SLS estimation}(i)-(b), (ii)-(a) and (iii)-(a)(d), $\sigma_{\min}^{-1}(\Pi_{n,P})$ is equivalent to, up to constants, the sieve measure of ill-posedness \citep{BlundellChenKristensen2007Engel}---see Lemma A.1 in \citet{ChenChristensen2018SupNormOptimal} and Corollary 11.6.5 in \citet{Bernstein2018Matrix}. Assumptions \ref{Ass: series 2SLS estimation}(ii)-(d) and (iii)-(d) are mild---see \citet[p.56]{ChenChristensen2018SupNormOptimal} for more discussions. Assumption \ref{Ass: series 2SLS estimation}(iv) imposes mild moment restrictions. Assumption \ref{Ass: series 2SLS estimation}(iv)-(b) may be dispensed with at the cost of potentially slowing down the coupling rate. Assumption \ref{Ass: series 2SLS estimation}(v) regulates the tuning parameters, approximation errors of the basis functions, and the degree of ill-posedness. Finally, Assumption \ref{Ass: series 2SLS estimation}(vi) supplies the multiplier-type bootstrap weights.


As a first step, we derive the uniform (in $P\in\mathbf P$) Bahadur representation for $\hat\beta_n$.

\begin{lem}\label{Lem: Bahadur, series 2SLS beta}
If Assumptions \ref{Ass: series 2SLS estimation}(i)-(a)(b), (ii), (iii)-(a)(b)(c), (iv)-(a) and (v)-(a)(b) hold, then it follows that, uniformly in $P\in\mathbf P$,
\begin{multline}
\hat\beta_n-\beta_{n,P}=(\Pi_{n,P}^\transpose \Psi_{n,P}^{-1}\Pi_{n,P})^{-1}  \Pi_{n,P}^\transpose \Psi_{n,P}^{-1}\frac{1}{n}\sum_{i=1}^{n}b^{m_n}(V_i)u_i \\ + O_p(s_n^{-2}\sqrt{\frac{\xi_{n}^2m_n\log (m_n)}{n^2}} +s_n^{-1}\delta_n\sqrt{\frac{(\xi_{n}^2\log m_n)\vee m_n}{n}})~.
\end{multline}
\end{lem}
\noindent{\sc Proof:} Throughout, Assumptions \ref{Ass: series 2SLS estimation}(i)-(b) and (v)-(a) are silently imposed. Then, by Assumptions \ref{Ass: series 2SLS estimation}(i)-(a) and (iii)-(a)(b), we may invoke Lemma 6.2 in \citet{BelloniChernozhukovChetverikovKato2015New} and Markov's inequality to conclude that, uniformly in $P\in\mathbf P$,
\begin{align}\label{Eqn: Bahadur, series 2SLS beta, aux1}
\|\hat\Psi_n-\Psi_{n,P}\|_o=O_p(\sqrt{\frac{\xi_{n}^2\log m_n}{n}})~,
\end{align}
where we also exploited $\xi_n^2\log(m_n)/n=o(1)$ by Assumption \ref{Ass: series 2SLS estimation}(v)-(b). By result \eqref{Eqn: Bahadur, series 2SLS beta, aux1} and Assumption \ref{Ass: series 2SLS estimation}(iii)-(a), it follows from Lemma \ref{Lem: inverse matrix} that
\begin{align}\label{Eqn: Bahadur, series 2SLS beta, aux2}
\|\hat\Psi_n^--\Psi_{n,P}^{-1}\|_o=O_p(\sqrt{\frac{\xi_{n}^2\log m_n}{n}})~,
\end{align}
uniformly in $P\in\mathbf P$. Assumption \ref{Ass: series 2SLS estimation}(iii)-(a) and Proposition 3.2 in \citet{HemmenAndo1980Ideal} (see also Problem X.5.5 in \citet{Bhatia1997Matrix}) together imply from result \eqref{Eqn: Bahadur, series 2SLS beta, aux2} that, uniformly in $P\in\mathbf P$,
\begin{align}\label{Eqn: Bahadur, series 2SLS beta, aux3}
\|\hat\Psi_n^{-1/2}-\Psi_{n,P}^{-1/2}\|_o
\le \frac{1}{\{\lambda_{\min}(\Psi_{n,P}^{-1})\}^{1/2}}\|\hat\Psi_n^--\Psi_{n,P}^{-1}\|_o\lesssim O_p(\sqrt{\frac{\xi_{n}^2\log m_n}{n}})~.
\end{align}
Moreover, by Assumptions \ref{Ass: series 2SLS estimation}(i)-(a), (ii)-(a)(b), (iii)-(a)(b) and (v)-(b), we may invoke Corollary E.1 in \citet{Kato2013QuasiB} to conclude that, uniformly in $P\in\mathbf P$,
\begin{align}\label{Eqn: Bahadur, series 2SLS beta, aux4}
\|\hat\Pi_n-\Pi_{n,P}\|_o=O_p(\sqrt{\frac{\xi_{n}^2\log (m_n)}{n}})~.
\end{align}
By Assumptions \ref{Ass: series 2SLS estimation}(ii)-(a) and (iii)-(a), we note also that Lemma \ref{Lem: covariance bound} implies
\begin{align}\label{Eqn: Bahadur, series 2SLS beta, aux5}
\sup_{n\in\mathbf N}\sup_{P\in\mathbf P}\|\Pi_{n,P}\|_o< \infty~.
\end{align}

Given results \eqref{Eqn: Bahadur, series 2SLS beta, aux3}, \eqref{Eqn: Bahadur, series 2SLS beta, aux4} and \eqref{Eqn: Bahadur, series 2SLS beta, aux5}, together with Assumption \ref{Ass: series 2SLS estimation}(iii)-(a), we may then conclude by Lemma \ref{Lem: matrix product} that, uniformly in $P\in\mathbf P$,
\begin{align}\label{Eqn: Bahadur, series 2SLS beta, aux6}
\|\hat\Psi_n^{-1/2}\hat\Pi_n-\Psi_{n,P}^{-1/2}\Pi_{n,P}\|_o=O_p(\sqrt{\frac{\xi_{n}^2\log (m_n)}{n}})~.
\end{align}
By Assumptions \ref{Ass: series 2SLS estimation}(iii)-(a)(c) and Corollary 11.6.5 in \citet{Bernstein2018Matrix}, we note
\begin{align}\label{Eqn: Bahadur, series 2SLS beta, aux7}
\sigma_{\min}(\Psi_{n,P}^{-1/2}\Pi_{n,P})\ge \sigma_{\max}(\Psi_{n,P})^{-1/2}\sigma_{\min}(\Pi_{n,P})\gtrsim s_n~,
\end{align}
uniformly in $P\in\mathbf P$. Define the event $\mathcal A_{n,P}$ as
\begin{multline}\label{Eqn: Bahadur, series 2SLS beta, aux8}
\mathcal A_{n,P}\equiv\{\|\hat\Psi_n^{-1/2}\hat\Pi_n-\Psi_{n,P}^{-1/2}\Pi_{n,P}\|_o\le\frac{1}{2} \sigma_{\min}(\Psi_{n,P}^{-1/2}\Pi_{n,P})~,\\
\hat\Psi_n^{-1/2}\hat\Pi_n\text{ has full column rank}\}~.
\end{multline}
Results \eqref{Eqn: Bahadur, series 2SLS beta, aux6} and \eqref{Eqn: Bahadur, series 2SLS beta, aux7}, Lemma \ref{Lem: inverse matrix}, and Assumption \ref{Ass: series 2SLS estimation}(v)-(b) then imply
\begin{align}\label{Eqn: Bahadur, series 2SLS beta, aux9}
\limsup_{n\to\infty}\sup_{P\in\mathbf P}P(\mathcal A_{n,P}^c)=0~.
\end{align}
By results \eqref{Eqn: Bahadur, series 2SLS beta, aux6}, \eqref{Eqn: Bahadur, series 2SLS beta, aux7} and \eqref{Eqn: Bahadur, series 2SLS beta, aux9}, we in turn have by Lemma F.4 in \citet{ChenChristensen2018SupNormOptimal} that, uniformly in $P\in\mathbf P$,
\begin{align}\label{Eqn: Bahadur, series 2SLS beta, aux10}
\|(\hat\Psi_n^{-1/2}\hat\Pi_n)_l^--(\Psi_{n,P}^{-1/2}\Pi_{n,P})_l^-\|_o=O_p(s_n^{-2}\sqrt{\frac{\xi_{n}^2\log (m_n)}{n}})~.
\end{align}
Moreover, result \eqref{Eqn: Bahadur, series 2SLS beta, aux7} and Fact 8.3.33 in \citet{Bernstein2018Matrix} imply:
\begin{align}\label{Eqn: Bahadur, series 2SLS beta, aux11}
\|(\Psi_{n,P}^{-1/2}\Pi_{n,P})_l^-\|_o\le \sigma_{\min}(\Psi_{n,P}^{-1/2}\Pi_{n,P})^{-1}=O(s_n^{-1})~,
\end{align}
uniformly in $P\in\mathbf P$. In turn, given results \eqref{Eqn: Bahadur, series 2SLS beta, aux3}, \eqref{Eqn: Bahadur, series 2SLS beta, aux10} and \eqref{Eqn: Bahadur, series 2SLS beta, aux11}, we may obtain by Lemma \ref{Lem: matrix product} that, uniformly in $P\in\mathbf P$,
\begin{align}\label{Eqn: Bahadur, series 2SLS beta, aux12}
\|(\hat\Psi_n^{-1/2}\hat\Pi_n)_l^-\hat\Psi_n^{-1/2}-(\Psi_{n,P}^{-1/2}\Pi_{n,P})_l^-\Psi_{n,P}^{-1/2}\|_o=O_p(s_n^{-2}\sqrt{\frac{\xi_{n}^2\log (m_n)}{n}})~,
\end{align}
where we also exploited $\|\Psi_{n,P}^{-1/2}\|_o\le\sigma_{\min}(\Psi_{n,P})^{-1/2}<\infty$ uniformly in $n$ and $P\in\mathbf P$ (by Assumption \ref{Ass: series 2SLS estimation}(iii)-(a)) and boundedness of $\{s_n\}$ (by result \eqref{Eqn: Bahadur, series 2SLS beta, aux5}).

Now, by Jensen's inequality and Assumption \ref{Ass: series 2SLS estimation}(i)-(a), we have
\begin{multline} \label{Eqn: Bahadur, series 2SLS beta, aux15}
E_P[\|\frac{B_n^\transpose U_n}{n}\|]\le \{\frac{1}{n}E_P[b^{m_n}(V)^\transpose b^{m_n}(V) u^2 ]\}^{1/2}\\
 \lesssim \frac{1}{\sqrt n }\{E_P[\|b^{m_n}(V)\|^2]\}^{1/2}
\lesssim  \sqrt{\frac{m_n}{n}}~,
\end{multline}
where the second inequality follows by Assumption \ref{Ass: series 2SLS estimation}(iv)-(a), and the third inequality by Assumption (iii)-(a) and Lemma \ref{Lem: basis norm}. In turn, we thus obtain from results \eqref{Eqn: Bahadur, series 2SLS beta, aux12} and \eqref{Eqn: Bahadur, series 2SLS beta, aux15} that, uniformly in $P\in\mathbf P$,
\begin{align}\label{Eqn: Bahadur, series 2SLS beta, aux16}
\|[(\hat\Psi_n^{-1/2}\hat\Pi_n)_l^-\hat\Psi_n^{-1/2}&-(\Psi_{n,P}^{-1/2}\Pi_{n,P})_l^-\Psi_{n,P}^{-1/2}]\frac{B_n^\transpose U_n}{n}\|\notag\\
&\le\|(\hat\Psi_n^{-1/2}\hat\Pi_n)_l^-\hat\Psi_n^{-1/2}-(\Psi_{n,P}^{-1/2}\Pi_{n,P})_l^-\Psi_{n,P}^{-1/2}\|_o\|\frac{B_n^\transpose U_n}{n}\| \notag\\
 &=  O_p(s_n^{-2}\sqrt{\frac{\xi_{n}^2m_n\log (m_n)}{n^2}})~.
\end{align}
Since $\hat\Pi_n^\transpose\hat\Psi_n^-\hat\Pi_n=(\hat\Psi_n^{-1/2}\hat\Pi_n)^\transpose \hat\Psi_n^{-1/2}\hat\Pi_n$, we know by result \eqref{Eqn: Bahadur, series 2SLS beta, aux9} that
\begin{align}\label{Eqn: Bahadur, series 2SLS beta, aux17}
\liminf_{n\to\infty}\inf_{P\in\mathbf P}P(\hat\Pi_n^\transpose\hat\Psi_n^-\hat\Pi_n\text{ is invertible})=1~.
\end{align}
Let $\mathcal E_n\equiv \{\hat\Pi_n^\transpose\hat\Psi_n^-\hat\Pi_n\text{ is invertible}\}$. Since $(\hat\Pi_n^\transpose\hat\Psi_n^-\hat\Pi_n)^- \hat\Pi_n^\transpose\hat\Psi_n^-\hat\Pi_n=I_{k_n}$ under the event $\mathcal E_n$, we may thus write by simple algebra that, under $\mathcal E_n$,
\begin{align}\label{Eqn: Bahadur, series 2SLS beta, aux18}
\hat\beta_n-\beta_{n,P}=[\hat\Pi_n^\transpose\hat\Psi_n^- \hat\Pi_n]^- \hat\Pi_n^\transpose\hat\Psi_n^- B_n^\transpose U_n/n+[\hat\Pi_n^\transpose\hat\Psi_n^- \hat\Pi_n]^- \hat\Pi_n^\transpose\hat\Psi_n^- B_n^\transpose A_{n,P}/n~.
\end{align}
The lemma then follows from combining Lemma \ref{Lem: npiv, bias}, \eqref{Eqn: Bahadur, series 2SLS beta, aux16}, \eqref{Eqn: Bahadur, series 2SLS beta, aux17}, and \eqref{Eqn: Bahadur, series 2SLS beta, aux18}. \qed

\begin{pro}[Strong Approximation for Series 2SLS Estimators]\label{Pro: strong approx, series npiv}
Assumptions \ref{Ass: series 2SLS estimation}(i), (ii), (iii), (iv)-(a) and (v)-(a)(b) together imply Assumption \ref{Ass: strong approx}(i) with $r_n=\sqrt n s_n/\xi_n$, $\hat\theta_n=\hat\beta_n^\transpose h^{k_n}$, $c_n=\varpi_n\ell_n$ for $\varpi_n$ as in \eqref{Eqn: coupling rate, npiv} and any sequence $\{\ell_n\}$ of positive scalars that tend to infinity (slowly), and
\begin{align}
\mathbb Z_{n,P} =\frac{s_n}{\xi_n}(h^{k_n})^\transpose (\Pi_{n,P}^\transpose \Psi_{n,P}^{-1}\Pi_{n,P})^{-1}\Pi_{n,P}^\transpose \Psi_{n,P}^{-1} G_{n,P}~,
\end{align}
where $G_{n,P}\sim N(0,\Sigma_{n,P})$ for $\Sigma_{n,P}\equiv E_P[b^{m_n}(V)b^{m_n}(V)^\transpose u^2]$.
\end{pro}
\begin{rem}\label{Rem: rate estimation}
As clear from the proof, one may replace $r_n=\sqrt n s_n/\xi_n$ with $r_n=\sqrt n s_{n,P}/\xi_n$ \ref{Pro: bootstrap, series npiv} for $s_{n,P}\equiv\sigma_{\min}(\Pi_{n,P})$. Since $s_{n,P}$ is unknown, in practice one may in turn replace it by $\hat s_n\equiv\sigma_{\min}(\hat\Pi_n)$. The resulting difference is asymptotically negligible by Weyl's inequality (see, for example, Fact 11.16.40 in \citet{Bernstein2018Matrix}), result \eqref{Eqn: Bahadur, series 2SLS beta, aux4}, Assumption \ref{Ass: series 2SLS estimation}(v)-(b) and $\sup_{P\in\mathbf P}E[\|\mathbb Z_{n,P}\|]$ being bounded uniformly in $n$. \qed
\end{rem}

\noindent{\sc Proof of Proposition \ref{Pro: strong approx, series npiv}:} Let $\Delta_{n,P}\equiv\sum_{i=1}^{n} E_P[\| b^{m_n}(V_i) u_i/\sqrt n\|^3]$. By Assumptions \ref{Ass: series 2SLS estimation}(i)-(a), (iii)-(b) and (iv)-(a) and the law of iterated expectations, we have:
\begin{align}\label{Eqn: strong approx, series npiv, aux1}
\Delta_{n,P}= E_P[\frac{\|b^{m_n}(V_1)u\|^3}{\sqrt n}]\lesssim E_P[\frac{\xi_n\|b^{m_n}(V_1)\|^2}{\sqrt n}]\lesssim \frac{\xi_nm_n}{\sqrt n}~,
\end{align}
where the final step follows by Assumption \ref{Ass: series 2SLS estimation}(iii)-(a) and Lemma \ref{Lem: basis norm}. By Assumption \ref{Ass: series 2SLS estimation}(i)-(a), we may apply Yurinskii's coupling \citep[Theorem 10.10]{Pollard2002User} to conclude that, for any $\epsilon>0$, there is some $G_{n,P}\sim N(0,\Sigma_{n,P})$ satisfying
\begin{align}\label{Eqn: strong approx, series npiv, aux2}
P(\|\frac{1}{\sqrt n}\sum_{i=1}^{n}b^{m_n}(V_i) u_i- G_{n,P}\|>3 \epsilon)\lesssim \eta_{n,P}(1+\frac{|\log(1/\eta_{n,P})|}{m_n}) ~,
\end{align}
where $\eta_{n,P}\equiv\Delta_{n,P}m_n/\epsilon^3$. By result \eqref{Eqn: strong approx, series npiv, aux1}, Assumption \ref{Ass: series 2SLS estimation}(v)-(b) and $x\mapsto x|\log(1/x)|$ being increasing on $(0,x_0)$ for some small $x_0>0$, it follows from \eqref{Eqn: strong approx, series npiv, aux2} that
\begin{align}\label{Eqn: strong approx, series npiv, aux2a}
P(\|\frac{1}{\sqrt n}\sum_{i=1}^{n}b^{m_n}(V_i) u_i- G_{n,P}\|>3 \epsilon)\lesssim \eta_n(1+\frac{|\log(1/\eta_n)|}{m_n}) ~,
\end{align}
where $\eta_n\equiv \xi_nm_n^2n^{-1/2}/\epsilon^3$. Setting $\epsilon\equiv M(\xi_nm_n^2/\sqrt n)^{1/3}$ with $M>0$ in \eqref{Eqn: strong approx, series npiv, aux2a} yields
\begin{align}\label{Eqn: strong approx, series npiv, aux2b}
P(\|\frac{1}{\sqrt n}\sum_{i=1}^{n}b^{m_n}(V_i) u_i- G_{n,P}\|>3 M(\xi_nm_n^2/\sqrt n)^{1/3})\lesssim \frac{1}{M^3}(1+\frac{|3\log M|}{m_n}) ~,
\end{align}
In turn, we may conclude by \eqref{Eqn: strong approx, series npiv, aux2b} that, uniformly in $P\in\mathbf P$,
\begin{align}\label{Eqn: strong approx, series npiv, aux3}
\|\frac{1}{\sqrt n}\sum_{i=1}^{n}b^{m_n}(V_i) u_i-G_{n,P}\|=O_p((\frac{\xi_nm_n^2}{\sqrt n})^{1/3})~.
\end{align}

Next, by result \eqref{Eqn: npiv, residual, aux8} and simple algebra, we have that, uniformly in $P\in\mathbf P$,
\begin{multline}\label{Eqn: strong approx, series npiv, aux4}
\|r_n\{\hat\theta_n-\theta_P\} -r_n (h^{k_n})^\transpose  \{\hat\beta_n-\beta_{n,P}\}\|_{\mathbf H}=\|r_n \{\theta_P-\beta_{n,P}^\transpose h^{k_n}\}\|_{\mathbf H}\\
\lesssim \|r_n \{\theta_P-\beta_{n,P}^\transpose h^{k_n}\}\|_\infty \lesssim \|r_n \{\theta_P-\gamma_{n,P}^\transpose h^{k_n}\}\|_\infty \lesssim  O(r_n \delta_n)~.
\end{multline}
By Assumption \ref{Ass: series 2SLS estimation}(iii)-(a) and result \eqref{Eqn: Bahadur, series 2SLS beta, aux11}, we note that
\begin{multline}\label{Eqn: bootstrap, series npiv, aux5}
\|(\Pi_{n,P}^\transpose \Psi_{n,P}^{-1}\Pi_{n,P})^{-1}\Pi_{n,P}^\transpose \Psi_{n,P}^{-1}\|_o\\ =
\|(\Psi_{n,P}^{-1/2}\Pi_{n,P})_l^-\Psi_{n,P}^{-1/2}\|_o\le \|(\Psi_{n,P}^{-1/2}\Pi_{n,P})_l^-\|\|\Psi_{n,P}^{-1/2}\|\lesssim s_n^{-1}~.
\end{multline}
In turn, by the Cauchy--Schwarz inequality, Assumption \ref{Ass: series 2SLS estimation}(ii)-(b), Lemma \ref{Lem: Bahadur, series 2SLS beta}, and results \eqref{Eqn: strong approx, series npiv, aux3} and \eqref{Eqn: bootstrap, series npiv, aux5}, we may obtain that
\begin{align}\label{Eqn: strong approx, series npiv, aux5}
\|r_n & (h^{k_n})^\transpose  \{\hat\beta_n-\beta_{n,P}\}  - \frac{s_n}{\xi_n} (h^{k_n})^\transpose (\Pi_{n,P}^\transpose \Psi_{n,P}^{-1}\Pi_{n,P})^{-1}\Pi_{n,P}^\transpose \Psi_{n,P}^{-1}G_{n,P}\|_{\mathbf H}\notag\\
&\lesssim \xi_n\{\frac{s_n}{\xi_n}s_n^{-1}O_p((\frac{\xi_nm_n^2}{\sqrt n})^{1/3})+  r_n O_p(s_n^{-2}\sqrt{\frac{\xi_{n}^2m_n\log (m_n)}{n^2}}  + s_n^{-1}\delta_n\sqrt{\frac{(\xi_{n}^2\log m_n)\vee m_n}{n}})\}\notag\\
&=O_p((\frac{\xi_nm_n^2}{\sqrt n})^{1/3} + s_n^{-1}\sqrt{\frac{\xi_{n}^2m_n\log (m_n)}{n}} + \delta_n\sqrt{(\xi_{n}^2\log m_n)\vee m_n})~,
\end{align}
uniformly in $P\in\mathbf P$. The conclusion of the proposition then follows from results \eqref{Eqn: strong approx, series npiv, aux4} and \eqref{Eqn: strong approx, series npiv, aux5}, together with the triangle inequality and Assumption \ref{Ass: series 2SLS estimation}(v)-(b). \qed

The Bahadur representation in Lemma \ref{Lem: Bahadur, series 2SLS beta} suggests a natural bootstrap, namely, the sieve score bootstrap that is proposed in \citet{ChenPouzo2015SieveWald} and \citet{ChenChristensen2018SupNormOptimal}. Concretely, for $\hat u_i\equiv Y_i-\hat\theta_n(Z_i)$, let
\begin{align}\label{Eqn: bootstrap npiv}
\hat{\mathbb G}_n\equiv\frac{s_n}{\xi_n}(h^{k_n})^\transpose [\hat\Pi_n^\transpose\hat\Psi_n^-\hat\Pi_n]^-\hat\Pi_n^\transpose\hat\Psi_n^{-} \frac{1}{\sqrt n}\sum_{i=1}^{n} W_i b^{m_n}(V_i)\hat u_i~,
\end{align}
where $\{W_i\}$ are bootstrap weights satisfying Assumption \ref{Ass: series 2SLS estimation}(vi).

\begin{pro}[Sieve Score Bootstrap for Series 2SLS Estimators]\label{Pro: bootstrap, series npiv}
Assumption \ref{Ass: series 2SLS estimation} implies Assumption \ref{Ass: strong approx}(ii) with $\hat{\mathbb G}_n$ given by \eqref{Eqn: bootstrap npiv} with
\begin{align}
c_n= \{(\frac{\xi_nm_n^2}{\sqrt n})^{1/3}+
  s_n^{-1}\xi_n\sqrt{\frac{m_n}{n}}+ \delta_n+ (\xi_n^3\sqrt{\frac{\log m_n}{n}})^{1/2}\}\ell_n'~,
\end{align}
where $\{\ell_n'\}$ is any sequence of positive scalars that tends to infinity (slowly).
\end{pro}
\begin{rem}
For Assumption \ref{Ass: strong approx} overall, one should take the maximum of the coupling rates in Propositions \ref{Pro: strong approx, series npiv} and \ref{Pro: bootstrap, series npiv}. The two propositions are stated in terms of two (potentially) different rates because they may be of independent interest. \qed
\end{rem}
\noindent{\sc Proof:} We proceed in three steps.

\noindent\underline{Step 1:} Derive a Gaussian approximation of $\hat{\mathbb G}_n$ conditional on the data.

Let $\hat\Delta_n\equiv\sum_{i=1}^{n} E[\|W_i b^{m_n}(V_i) \hat u_i/\sqrt n\|^3|\{X_i\}_{i=1}^n]$. By Assumption \ref{Ass: series 2SLS estimation}(vi) and the triangle inequality, we note that
\begin{align}\label{Eqn: bootstrap, series npiv, aux1}
\hat\Delta_n & \lesssim n^{-3/2} \sum_{i=1}^{n} \|b^{m_n}(V_i) \hat u_i\|^3\notag\\
&\lesssim n^{-3/2} \sum_{i=1}^{n} \|b^{m_n}(V_i) (\hat u_i-u_i)\|^3+n^{-3/2} \sum_{i=1}^{n} \|b^{m_n}(V_i) u_i\|^3\notag\\
&\le n^{-3/2} \max_{i=1}^n |\hat u_i-u_i|^3\sum_{i=1}^{n} \|b^{m_n}(V_i)\|^3+n^{-3/2} \sum_{i=1}^{n} \|b^{m_n}(V_i)\|^3 | u_i|^3  ~.
\end{align}
By Assumptions \ref{Ass: series 2SLS estimation}(i)-(a), (ii)-(c), (iii)-(a)(b), (iv)-(a) and (v)-(b), and Lemmas \ref{Lem: basis norm} and \ref{Lem: npiv, residual}, we may in turn have from \eqref{Eqn: bootstrap, series npiv, aux1} that, uniformly in $P\in\mathbf P$,
\begin{multline}\label{Eqn: bootstrap, series npiv, aux2}
\hat\Delta_n \lesssim n^{-1/2}(O_p(s_n^{-1}\xi_n\sqrt{\frac{m_n}{n}}+\delta_n))^3 \xi_n O_p(m_n)+ n^{-1/2}\xi_nO_p(m_n)\\
=O_p(\frac{\xi_nm_n}{\sqrt n})~.
\end{multline}
Letting $\hat\Sigma_n\equiv\sum_{i=1}^{n}b^{m_n}(V_i)b^{m_n}(V_i)^\transpose \hat u_i^2/n$, we may then apply Theorem 10.8 in \citet{Pollard2002User} to conclude that, for each $\epsilon>0$, there exists a random vector $\hat G_{m_n}\sim N(0,\hat\Sigma_n)$ conditional on the data that satisfies
\begin{align}\label{Eqn: bootstrap, series npiv, aux3}
P(\|\frac{1}{\sqrt n}\sum_{i=1}^{n}W_i b^{m_n}(V_i) \hat u_i- \hat G_{m_n}\|>3  \epsilon|\{X_i\}_{i=1}^n)
\le C_0  \hat\eta_n(1+\frac{|\log(1/\hat\eta_n)|}{m_n}) ~,
\end{align}
where $\hat\eta_n\equiv \hat\Delta_n m_n\epsilon^{-3}$ and $C_0>0$ is some universal constant. Setting $\epsilon=M(\xi_nm_n^2/\sqrt n)^{1/3}$ in \eqref{Eqn: bootstrap, series npiv, aux3} and given \eqref{Eqn: bootstrap, series npiv, aux2}, we may apply Fubini's theorem and Markov's inequality to \eqref{Eqn: bootstrap, series npiv, aux3} to conclude that, unconditionally and uniformly in $P\in\mathbf P$,
\begin{align}\label{Eqn: bootstrap, series npiv, aux4}
\|\frac{1}{\sqrt n}\sum_{i=1}^{n}W_i b^{m_n}(V_i) \hat u_i- \hat G_{m_n}\|=O_p((\frac{\xi_nm_n^2}{\sqrt n})^{1/3})~.
\end{align}

Next, by results \eqref{Eqn: Bahadur, series 2SLS beta, aux12} and \eqref{Eqn: bootstrap, series npiv, aux5} and the triangle inequality, we have
\begin{align}\label{Eqn: bootstrap, series npiv, aux6}
\|(\hat\Psi_n^{-1/2}\hat\Pi_n)_l^-\hat\Psi_n^{-1/2}\|_o\le O_p(s_n^{-2}\sqrt{\frac{\xi_{n}^2\log (m_n)}{n}})+ O(s_n^{-1})=O_p(s_n^{-1})~,
\end{align}
uniformly in $P\in\mathbf P$, where the last step exploited Assumption \ref{Ass: series 2SLS estimation}(v)-(b). Define
\begin{align}
\hat{\mathbb Z}_n\equiv\frac{s_n}{\xi_n}(h^{k_n})^\transpose [\hat\Pi_n^\transpose\hat\Psi_n^-\hat\Pi_n]^-\hat\Pi_n^\transpose\hat\Psi_n^{-} \hat G_{m_n}~.
\end{align}
By results \eqref{Eqn: bootstrap, series npiv, aux4} and \eqref{Eqn: bootstrap, series npiv, aux6}, Assumption \ref{Ass: series 2SLS estimation}(ii)-(b) and the Cauchy--Schwarz inequality, we then obtain that, uniformly in $P\in\mathbf P$,
\begin{multline}\label{Eqn: bootstrap, series npiv, aux7}
\|\hat{\mathbb G}_n-\hat{\mathbb Z}_n\|_{\mathbf H}=\|\frac{s_n}{\xi_n}(h^{k_n})^\transpose (\hat\Psi_n^{-1/2}\hat\Pi_n)_l^-\hat\Psi_n^{-1/2}(\frac{1}{\sqrt n}\sum_{i=1}^{n}W_i b^{m_n}(V_i) \hat u_i-\hat G_{m_n})\|_{\mathbf H}\\
\le \frac{s_n}{\xi_n}\xi_n O_p(s_n^{-1}) O_p((\frac{\xi_nm_n^2}{\sqrt n})^{1/3}) =O_p((\frac{\xi_nm_n^2}{\sqrt n})^{1/3})~.
\end{multline}

\noindent\underline{Step 2:} Control the estimation error of $\hat\Sigma_n\equiv\sum_{i=1}^{n}b^{m_n}(V_i)b^{m_n}(V_i)^\transpose \hat u_i^2/n$.

First, define the ``infeasible'' variance estimator
\begin{align}
\tilde\Sigma_n\equiv \frac{1}{n}\sum_{i=1}^{n}b^{m_n}(V_i)b^{m_n}(V_i)^\transpose u_i^2~.
\end{align}
Then by simple algebra and the triangle inequality, we may obtain:
\begin{multline}\label{Eqn: bootstrap, series npiv, aux8}
\|\hat\Sigma_n-\tilde \Sigma_n\|_o\le \|\frac{1}{n}\sum_{i=1}^{n}b^{m_n}(V_i)b^{m_n}(V_i)^\transpose (\hat u_i-u_i)^2\|_o\\
+2 \|\frac{1}{n}\sum_{i=1}^{n}b^{m_n}(V_i)b^{m_n}(V_i)^\transpose (\hat u_i-u_i)u_i\|_o~.
\end{multline}
By result \eqref{Eqn: Bahadur, series 2SLS beta, aux1} and the triangle inequality, we have
\begin{align}\label{Eqn: bootstrap, series npiv, aux9}
\|\frac{1}{n}\sum_{i=1}^{n}b^{m_n}(V_i)b^{m_n}(V_i)^\transpose\|_o \le O_p(\sqrt{\frac{\xi_n^2\log m_n}{n}})+ \|\Psi_{n,P}\|_o=O_p(1)~,
\end{align}
uniformly in $P\in\mathbf P$, where the last step follows by Assumption \ref{Ass: series 2SLS estimation}(iii)-(a) and (v)-(b). It follows from Lemmas \ref{Lem: matrix operator norm} and \ref{Lem: npiv, residual} and result \eqref{Eqn: bootstrap, series npiv, aux9} that, uniformly in $P\in\mathbf P$,
\begin{multline}\label{Eqn: bootstrap, series npiv, aux10}
\|\frac{1}{n}\sum_{i=1}^{n}b^{m_n}(V_i)b^{m_n}(V_i)^\transpose (\hat u_i-u_i)^2\|_o\\ \le \max_{i=1}^n |\hat u_i-u_i|^2\|\frac{1}{n}\sum_{i=1}^{n}b^{m_n}(V_i)b^{m_n}(V_i)^\transpose\|_o
=O_p(s_n^{-2}\xi_n^2\frac{m_n}{n}+\delta_n^2)~.
\end{multline}
Next, by Assumption \ref{Ass: series 2SLS estimation}(iv)-(a), $\sup_{P\in\mathbf P}E_P[|u|^3]<\infty$. Therefore, by the triangle inequality, Lemma \ref{Lem: npiv, variance} and Assumption \ref{Ass: series 2SLS estimation}(v)-(b), we have
\begin{multline}\label{Eqn: bootstrap, series npiv, aux11}
\|\frac{1}{n}\sum_{i=1}^{n}b^{m_n}(V_i)b^{m_n}(V_i)^\transpose|u_i|\|_o\le o_p(1) + E_P[|u_1|\|b^{m_n}(V_1)b^{m_n}(V_1)^\transpose\|_o]\\
=o_p(1) + E_P\big[E[|u_1||V_1]\|b^{m_n}(V_1)b^{m_n}(V_1)^\transpose\|_o\big]=O_p(1)~,
\end{multline}
where the last step is due to Assumption \ref{Ass: series 2SLS estimation}(iii)-(a) and (iv)-(a). In turn, it follows from \eqref{Eqn: bootstrap, series npiv, aux11} and Lemmas \ref{Lem: matrix operator norm} and \ref{Lem: npiv, residual} that, uniformly in $P\in\mathbf P$,
\begin{multline}\label{Eqn: bootstrap, series npiv, aux12}
\|\frac{1}{n}\sum_{i=1}^{n}b^{m_n}(V_i)b^{m_n}(V_i)^\transpose (\hat u_i-u_i)u_i\|_o\\
\le \max_{i=1}^n|\hat u_i-u_i| \|\frac{1}{n}\sum_{i=1}^{n}b^{m_n}(V_i)b^{m_n}(V_i)^\transpose |u_i|\|_o=O_p(s_n^{-1}\xi_n\sqrt{\frac{m_n}{n}}+\delta_n)~.
\end{multline}
Results \eqref{Eqn: bootstrap, series npiv, aux10} and \eqref{Eqn: bootstrap, series npiv, aux12} then yield from \eqref{Eqn: bootstrap, series npiv, aux8} that, uniformly in $P\in\mathbf P$,
\begin{align}\label{Eqn: bootstrap, series npiv, aux13}
\|\hat\Sigma_n-\tilde \Sigma_n\|_o=O_p(s_n^{-1}\xi_n\sqrt{\frac{m_n}{n}}+\delta_n)~,
\end{align}
where we also exploited $\delta_n=o(1)$ by Assumption \ref{Ass: series 2SLS estimation}(ii)-(c) and $s_n^{-1}\xi_n\sqrt{m_n/n}=o(1)$ by Assumption \ref{Ass: series 2SLS estimation}(v)-(b). By Lemma \ref{Lem: npiv, variance} (with $\varsigma=1$) and result \eqref{Eqn: bootstrap, series npiv, aux13}, together with the triangle inequality, we may therefore conclude that, uniformly in $P\in\mathbf P$,
\begin{multline}\label{Eqn: bootstrap, series npiv, aux14}
\|\hat\Sigma_n- \Sigma_{n,P}\|_o\le \|\hat\Sigma_n- \tilde\Sigma_n\|_o+\|\tilde\Sigma_n- \Sigma_{n,P}\|_o\\
=O_p\big(s_n^{-1}\xi_n\sqrt{\frac{m_n}{n}}+\delta_n+(\xi_n^3\sqrt{\frac{\log m_n}{n}})^{1/2}\big).
\end{multline}
By Assumptions \ref{Ass: series 2SLS estimation}(iii)-(a) and (iv)-(b) and the law of iterated expectations, $\lambda_{\min}(\Sigma_{n,P})$ is bounded away from zero uniformly in $n$ and $P\in\mathbf P$. Therefore, we obtain by result \eqref{Eqn: bootstrap, series npiv, aux14} and Proposition 3.2 in \citet{HemmenAndo1980Ideal} (see also Problem X.5.5 in \citet{Bhatia1997Matrix}) that, uniformly in $P\in\mathbf P$,
\begin{multline}\label{Eqn: bootstrap, series npiv, aux15}
\|\hat\Sigma_n^{1/2}- \Sigma_{n,P}^{1/2}\|_o\le \frac{1}{\{\lambda_{\min}(\Sigma_{n,P})\}^{1/2}}\|\hat\Sigma_n- \Sigma_{n,P}\|_o\\ \lesssim O_p\big(s_n^{-1}\xi_n\sqrt{\frac{m_n}{n}}+\delta_n+(\xi_n^3\sqrt{\frac{\log m_n}{n}})^{1/2}\big)~.
\end{multline}

\noindent\underline{Step 3:} Remove the dependence of $\hat{\mathbb Z}_n$ on $\{X_i\}_{i=1}^n$.

By Lemma \ref{Lem: Gaussian coupling}, we may write $\hat G_{m_n}=\hat\Sigma_n^{1/2}\bar G_{m_n}$ where $\bar G_{m_n}\sim N(0,I_{m_n})$ is independent of the data $\{X_i\}_{i=1}^n$. In turn, let
\begin{align}
\bar{\mathbb Z}_{n,P}=\frac{s_n}{\xi_n}(h^{k_n})^\transpose(\Psi_{n,P}^{-1/2}\Pi_{n,P})_l^-\Psi_{n,P}^{-1/2}\Sigma_{n,P}^{1/2}\bar G_{m_n}~.
\end{align}
Clearly, $\bar{\mathbb Z}_{n,P}$ is centered Gaussian in $\mathbf H$ and independent of $\{X_i\}_{i=1}^n$. Further define
\begin{align}
\hat\Omega_n\equiv (\hat\Psi_n^{-1/2}\hat\Pi_n)_l^-\hat\Psi_n^{-1/2}\hat\Sigma_n^{1/2}~,\,\Omega_{n,P}\equiv (\Psi_{n,P}^{-1/2}\Pi_{n,P})_l^-\Psi_{n,P}^{-1/2}\Sigma_{n,P}^{1/2}~.
\end{align}
Note that $\sup_{P\in\mathbf P}\|\Sigma_{n,P}^{1/2}\|_o=\sup_{P\in\mathbf P}\|\Sigma_{n,P}\|_o^{1/2}$ which are bounded uniformly in $n$ by Assumptions \ref{Ass: series 2SLS estimation}(iii)-(a) and (iv)-(a). Therefore, by results \eqref{Eqn: Bahadur, series 2SLS beta, aux12}, \eqref{Eqn: bootstrap, series npiv, aux5} and \eqref{Eqn: bootstrap, series npiv, aux15}, we may apply Lemma \ref{Lem: matrix product} to conclude that
\begin{multline}\label{Eqn: bootstrap, series npiv, aux16}
\|\hat\Omega_n-\Omega_{n,P}\|_o=O_p(s_n^{-2}\sqrt{\frac{\xi_{n}^2\log (m_n)}{n}})\\
 + O_p(s_n^{-2}\xi_n\sqrt{\frac{m_n}{n}}+s_n^{-1}\delta_n+s_n^{-1}(\xi_n^3\sqrt{\frac{\log m_n}{n}})^{1/2})~,
\end{multline}
uniformly in $P\in\mathbf P$. By Assumption \ref{Ass: series 2SLS estimation}(i)-(c) and Jensen's inequality, we have
\begin{align}\label{Eqn: bootstrap, series npiv, aux17}
E[\|\hat{\mathbb Z}_n&-\bar{\mathbb Z}_{n,P}\|_{\mathbf H}|\{X_i\}_{i=1}^n]\lesssim \{E[\int \big(\frac{s_n}{\xi_n}h^{k_n}(z)^\transpose (\hat\Omega_n-\Omega_{n,P})\bar G_{m_n}\big)^2\,\mathrm{d}z|\{X_i\}_{i=1}^n]\}^{1/2}\notag\\
&=\{\int \frac{s_n^2}{\xi_n^2}h^{k_n}(z)^\transpose (\hat\Omega_n-\Omega_{n,P})E[\bar G_{m_n}\bar G_{m_n}^\transpose ] (\hat\Omega_n-\Omega_{n,P})^\transpose h^{k_n}(z)
 \,\mathrm{d}z\}^{1/2}\notag\\
&=\{\int \frac{s_n^2}{\xi_n^2}h^{k_n}(z)^\transpose (\hat\Omega_n-\Omega_{n,P})(\hat\Omega_n-\Omega_{n,P})^\transpose h^{k_n}(z)
 \,\mathrm{d}z\}^{1/2}~,
\end{align}
where the first equality follows by Fubini's theorem and the independence between $\bar G_{m_n}$ and $\{X_i\}_{i=1}^n$, and the second equality is due to $\bar G_{m_n}\sim N(0,I_{m_n})$. By Assumption \ref{Ass: series 2SLS estimation}(i)-(d) and (ii)-(b), we obtain from \eqref{Eqn: bootstrap, series npiv, aux16} and \eqref{Eqn: bootstrap, series npiv, aux17} that
\begin{multline}\label{Eqn: bootstrap, series npiv, aux18}
E[\|\hat{\mathbb Z}_n-\bar{\mathbb Z}_{n,P}\|_{\mathbf H}|\{X_i\}_{i=1}^n]\lesssim \{\int \frac{s_n^2}{\xi_n^2}\|h^{k_n}(z)\|^2 \|\hat\Omega_n-\Omega_{n,P}\|_o^2\,\mathrm{d}z\}^{1/2}\le s_n \|\hat\Omega_n-\Omega_{n,P}\|_o\\
=O_p(s_n^{-1}\sqrt{\frac{\xi_{n}^2\log (m_n)}{n}}+s_n^{-1}\xi_n\sqrt{\frac{m_n}{n}}+\delta_n+(\xi_n^3\sqrt{\frac{\log m_n}{n}})^{1/2})~,
\end{multline}
uniformly in $P\in\mathbf P$. In turn, by Fubini's theorem, Markov's inequality and $\log x\le x$ for $x>0$, we obtain from \eqref{Eqn: bootstrap, series npiv, aux18} that, unconditionally and uniformly in $P\in\mathbf P$,
\begin{align}\label{Eqn: bootstrap, series npiv, aux19}
\|\hat{\mathbb Z}_n-\bar{\mathbb Z}_{n,P}\|_{\mathbf H}
=O_p(s_n^{-1}\xi_n\sqrt{\frac{m_n}{n}}+\delta_n+(\xi_n^3\sqrt{\frac{\log m_n}{n}})^{1/2})~.
\end{align}
The conclusion of the proposition then follows from results \eqref{Eqn: bootstrap, series npiv, aux7}, \eqref{Eqn: bootstrap, series npiv, aux19}, the triangle inequality and Assumption \ref{Ass: series 2SLS estimation}(v)-(b)(c). \qed

\begin{lem}\label{Lem: npiv, bias}
If Assumptions \ref{Ass: series 2SLS estimation}(i)-(a)(b), (ii), (iii)-(a)(b)(c) and (v)-(a)(b) hold, then it follows that, uniformly in $P\in\mathbf P$,
\begin{align}
[\hat\Pi_n^\transpose\hat\Psi_n^-\hat\Pi_n]^-\hat\Pi_n^\transpose\hat\Psi_n^{-} \frac{B_n^\transpose A_{n,P}}{n}=O_p(s_n^{-1}\delta_n\sqrt{\frac{(\xi_{n}^2\log m_n)\vee m_n}{n}})~.
\end{align}
\end{lem}
\noindent{\sc Proof:} We adapt the proof of Lemma A.3 in \citet{ChenChristensen2018SupNormOptimal}, mostly by making their arguments uniform in $P\in\mathbf P$. Let $\mathcal E_n\equiv \{\hat\Pi_n^\transpose\hat\Psi_n^-\hat\Pi_n\text{ is invertible}\}$ as in the proof of Lemma \ref{Lem: Bahadur, series 2SLS beta}. Since $(\hat\Pi_n^\transpose\hat\Psi_n^-\hat\Pi_n)^- \hat\Pi_n^\transpose\hat\Psi_n^-\hat\Pi_n=I_{k_n}$ under the event $\mathcal E_n$, we may thus write by simple algebra that, under $\mathcal E_n$,
\begin{multline}\label{Eqn: npiv, bias, aux2}
[\hat\Pi_n^\transpose\hat\Psi_n^-\hat\Pi_n]^-\hat\Pi_n^\transpose\hat\Psi_n^{-} \frac{B_n^\transpose A_{n,P}}{n}=\{ (\hat\Psi_n^{-1/2}\hat\Pi_n)_l^- \hat\Psi_n^{-1/2}\Psi_{n,P}^{1/2}-(\Psi_{n,P}^{-1/2}\Pi_{n,P})_l^-\}\Psi_{n,P}^{-1/2}\frac{B_n^\transpose D_{n,P}}{n}\\
+ (\Psi_{n,P}^{-1/2}\Pi_{n,P})_l^- \big\{\Psi_{n,P}^{-1/2}\big(\frac{B_n^\transpose D_{n,P}}{n}-E_P[b^{m_n}(V)d_{n,P}(Z)]\big)\big\}~,
\end{multline}
where $D_{n,P}\equiv(d_{n,P}(Z_1),\ldots,d_{n,P}(Z_n))^\transpose$ with $d_{n,P}(z)\equiv \theta_0(z)-h^{k_n}(z)^\transpose \gamma_{n,P}$. Result \eqref{Eqn: Bahadur, series 2SLS beta, aux17} allows us to focus on the event $\mathcal E_n$.

Before dealing with the right hand side of \eqref{Eqn: npiv, bias, aux2}, we need some preparations. First, as in the proof of Lemma F.9 in \citet{ChenChristensen2018SupNormOptimal}, we have
\begin{multline}\label{Eqn: npiv, bias, aux4}
E_P[\|\Psi_{n,P}^{-1/2}\big(\frac{B_n^\transpose D_{n,P}}{n}-E_P[b^{m_n}(V)d_{n,P}(Z)]\big)\|^2]\\
 \le \frac{1}{n} E_P[\|\Psi_{n,P}^{-1/2}b^{m_n}(V)\|^2d_{n,P}^2(Z)] \lesssim \frac{1}{n}m_n\delta_n^2~,
\end{multline}
where the first inequality follows by Jensen's inequality, Assumption \ref{Ass: series 2SLS estimation}(i)-(a) and the fact $E[\|X-E[X]\|^2]\le E[\|X\|^2]$ for any random vector $X\in\mathbf R^d$, and the second inequality by Lemma \ref{Lem: basis norm} and Assumptions \ref{Ass: series 2SLS estimation}(ii)-(c) and (iii)-(a). Second, by result \eqref{Eqn: npiv, bias, aux4} and the triangle inequality, we in turn have that, uniformly in $P\in\mathbf P$,
\begin{align}\label{Eqn: npiv, bias, aux5}
\|\Psi_{n,P}^{-1/2}\frac{B_n^\transpose D_{n,P}}{n}\|\le \|\Psi_{n,P}^{-1/2}E_P[b^{m_n}(V)d_{n,P}(Z)]\|+ O_p(\delta_n\sqrt{\frac{m_n}{n}})~.
\end{align}
Recalling the definitions of $\Upsilon_P$ and $\mathrm{Proj}_m$, we note that
\begin{multline}\label{Eqn: npiv, bias, aux6}
\|\Psi_{n,P}^{-1/2}E_P[b^{m_n}(V)d_{n,P}(Z)]\|= \|\Psi_{n,P}^{-1/2}E_P[b^{m_n}(V)\Upsilon_P(d_{n,P})(V)]\|\\
=\|\mathrm{Proj}_m \big(\Upsilon_P d_{n,P}\big)\|_{L_2(V)}\le \|\Upsilon_P d_{n,P}\|_{L^2(V)}\le s_nO(\delta_n)~,
\end{multline}
where the first equality follows by the law of iterated expectations, the second equality by direct calculations, the first inequality by the projection theorem, and the second inequality by Assumption \ref{Ass: series 2SLS estimation}(ii)-(c)(d). By results \eqref{Eqn: npiv, bias, aux5} and \eqref{Eqn: npiv, bias, aux6} and Assumption \ref{Ass: series 2SLS estimation}(v)-(b), we then obtain that, uniformly in $P\in\mathbf P$,
\begin{align}\label{Eqn: npiv, bias, aux7}
\|\Psi_{n,P}^{-1/2}\frac{B_n^\transpose D_{n,P}}{n}\|=O_p(\delta_n\sqrt{\frac{m_n}{n}})+O_p(s_n\delta_n)=O_p(s_n\delta_n)~.
\end{align}
Third, by Assumption \ref{Ass: series 2SLS estimation}(iii)-(a) and result \eqref{Eqn: Bahadur, series 2SLS beta, aux12}, we have: uniformly in $P\in\mathbf P$,
\begin{align}\label{Eqn: npiv, bias, aux8}
\|(\hat\Psi_n^{-1/2}\hat\Pi_n)_l^- \hat\Psi_n^{-1/2} & \Psi_{n,P}^{1/2}-(\Psi_{n,P}^{-1/2}\Pi_{n,P})_l^-\|_o \notag\\
&\le \|(\hat\Psi_n^{-1/2}\hat\Pi_n)_l^- \hat\Psi_n^{-1/2}-(\Psi_{n,P}^{-1/2}\Pi_{n,P})_l^-\Psi_{n,P}^{-1/2}\|_o\|\Psi_{n,P}^{1/2}\|_o\notag\\
&=O_p(s_n^{-2}\sqrt{\frac{\xi_{n}^2\log (m_n)}{n}})~.
\end{align}

Given the above preparations, we now obtain by results \eqref{Eqn: npiv, bias, aux7} and \eqref{Eqn: npiv, bias, aux8} that
\begin{multline}\label{Eqn: npiv, bias, aux9}
\|\{ (\hat\Psi_n^{-1/2}\hat\Pi_n)_l^- \hat\Psi_n^{-1/2}\Psi_{n,P}^{1/2}-(\Psi_{n,P}^{-1/2}\Pi_{n,P})_l^-\}\Psi_{n,P}^{-1/2}\frac{B_n^\transpose D_{n,P}}{n}\|_o\\
\le O_p(s_n^{-2}\sqrt{\frac{\xi_{n}^2\log (m_n)}{n}}) O_p(s_n\delta_n)=O_p(s_n^{-1}\delta_n\sqrt{\frac{\xi_{n}^2\log (m_n)}{n}})~,
\end{multline}
uniformly in $P\in\mathbf P$. Results \eqref{Eqn: Bahadur, series 2SLS beta, aux11} and \eqref{Eqn: npiv, bias, aux4} imply: uniformly in $P\in\mathbf P$,
\begin{multline}\label{Eqn: npiv, bias, aux10}
\|(\Psi_{n,P}^{-1/2}\Pi_{n,P})_l^- \big\{\Psi_{n,P}^{-1/2}\big(\frac{B_n^\transpose D_{n,P}}{n}-E_P[b^{m_n}(V)d_{n,P}(Z)]\big)\big\}\|_o\\
\le O(s_n^{-1}) O_p(\delta_n\sqrt{\frac{m_n}{n}}) =O_p(s_n^{-1}\delta_n\sqrt{\frac{m_n}{n}}) ~.
\end{multline}
The lemma then follows from combining \eqref{Eqn: npiv, bias, aux2}, \eqref{Eqn: npiv, bias, aux9} and \eqref{Eqn: npiv, bias, aux10}. \qed

\begin{lem}\label{Lem: npiv, residual}
If Assumptions \ref{Ass: series 2SLS estimation}(i)-(a)(b), (ii), (iii)-(a)(b)(c)(d), (iv)-(a) and (v)-(a)(b) hold, then it follows that, uniformly in $P\in\mathbf P$,
\begin{align}
\max_{i=1}^n|\hat u_i-u_i|=O_p(s_n^{-1}\xi_n\sqrt{\frac{m_n}{n}}+\delta_n)~.
\end{align}
\end{lem}
\noindent{\sc Proof:} By definition, we may write
\begin{align}\label{Eqn: npiv, residual, aux1}
\hat u_i-u_i=\theta_P(Z_i)-\hat\theta_n(Z_i)=h^{k_n}(Z_i)^\transpose (\beta_{n,P}-\hat\beta_n)+a_{n,P}(Z_i)~.
\end{align}
By simple algebra, we note that
\begin{align}\label{Eqn: npiv, residual, aux2}
(\Pi_{n,P}^\transpose \Psi_{n,P}^{-1}\Pi_{n,P})^{-1}  \Pi_{n,P}^\transpose \Psi_{n,P}^{-1}\frac{1}{n}\sum_{i=1}^{n}b^{m_n}(V_i)u_i
=(\Psi_{n,P}^{-1/2}\Pi_{n,P})_l^-\Psi_{n,P}^{-1/2}\frac{B_n^\transpose U_n}{n}~.
\end{align}
By Assumption \ref{Ass: series 2SLS estimation}(iii)-(a) and results \eqref{Eqn: Bahadur, series 2SLS beta, aux11} and \eqref{Eqn: Bahadur, series 2SLS beta, aux15}, we obtain
\begin{align}\label{Eqn: npiv, residual, aux3}
\|(\Psi_{n,P}^{-1/2}\Pi_{n,P})_l^-\Psi_{n,P}^{-1/2}\frac{B_n^\transpose U_n}{n}\|\le O(s_n^{-1}) O(1) O_p(\sqrt{\frac{m_n}{n}})=O_p(s_n^{-1}\sqrt{\frac{m_n}{n}})~,
\end{align}
uniformly in $P\in\mathbf P$. It follows from Lemma \ref{Lem: Bahadur, series 2SLS beta}, result \eqref{Eqn: npiv, residual, aux3}, and the triangle inequality that, uniformly in $P\in\mathbf P$,
\begin{align}\label{Eqn: npiv, residual, aux4}
\|\beta_{n,P}-\hat\beta_n\|\le O_p(s_n^{-1}\sqrt{\frac{m_n}{n}})+O_p(s_n^{-2}\sqrt{\frac{\xi_{n}^2m_n\log (m_n)}{n^2}} )=O_p(s_n^{-1}\sqrt{\frac{m_n}{n}})~.
\end{align}
where the last step is due to Assumption \ref{Ass: series 2SLS estimation}(v)-(b). By the Cauchy--Schwarz inequality, Assumption \ref{Ass: series 2SLS estimation}(ii)-(b) and result \eqref{Eqn: npiv, residual, aux4}, we thus have: uniformly in $P\in\mathbf P$,
\begin{align}\label{Eqn: npiv, residual, aux5}
\max_{i=1}^n|h^{k_n}(Z_i)^\transpose (\beta_{n,P}-\hat\beta_n)|\le \xi_n O_p(s_n^{-1}\sqrt{\frac{m_n}{n}})~.
\end{align}

On the other hand, by definition we note that
\begin{align}\label{Eqn: npiv, residual, aux6}
a_{n,P}\equiv\theta_P-\mathrm{Proj}_{m,k}\theta_P=\theta_P-\mathrm{Proj}_k\theta_P-(\mathrm{Proj}_{m,k}\theta_P-\mathrm{Proj}_k\theta_P)~.
\end{align}
Exploiting $\mathrm{Proj}_{m,k}\theta=\theta$ for any $\theta\in\mathcal H_{k_n}$, we have
\begin{align}\label{Eqn: npiv, residual, aux7}
\mathrm{Proj}_{m,k}\theta_P-\mathrm{Proj}_k\theta_P=\mathrm{Proj}_{m,k}(\theta_P-\mathrm{Proj}_k\theta_P)~.
\end{align}
By Assumptions \ref{Ass: series 2SLS estimation}(ii)-(c) and (iii)-(d) and the triangle inequality, we may then obtain from results \eqref{Eqn: npiv, residual, aux6} and \eqref{Eqn: npiv, residual, aux7} that, uniformly in $P\in\mathbf P$,
\begin{align}\label{Eqn: npiv, residual, aux8}
\|a_{n,P}\|_\infty\lesssim \|\theta_P-\mathrm{Proj}_k\theta_P\|_\infty=O(\delta_n)~.
\end{align}
The conclusion of the lemma then follows from combining results \eqref{Eqn: npiv, residual, aux1}, \eqref{Eqn: npiv, residual, aux5} and \eqref{Eqn: npiv, residual, aux8}, together with the triangle inequality. \qed

\begin{lem}\label{Lem: npiv, variance}
Let Assumptions \ref{Ass: series 2SLS estimation}(i)-(a) and (iii)-(a)(b) hold. If there is some constant $\varsigma>0$ such that $\sup_{P\in\mathbf P}E_P[|u|^{2+\varsigma}]<\infty$, then, uniformly in $P\in\mathbf P$,
\begin{gather*}
\|\frac{1}{n}\sum_{i=1}^{n} |u_i| b^{m_n}(V_i)b^{m_n}(V_i)^\transpose- E_P[|u|b^{m_n}(V)b^{m_n}(V)^\transpose]\|_o=O_p(\big(\xi_n^{1+1/\varsigma}\sqrt{\frac{\log m_n}{n}}\big)^{\frac{2\varsigma}{2\varsigma+1}})~,\\
\|\frac{1}{n}\sum_{i=1}^{n} u_i^2 b^{m_n}(V_i)b^{m_n}(V_i)^\transpose- E_P[u^2b^{m_n}(V)b^{m_n}(V)^\transpose]\|_o=O_p(\big(\xi_n^{1+2/\varsigma}\sqrt{\frac{\log m_n}{n}}\big)^{\frac{\varsigma}{\varsigma+1}})~.
\end{gather*}
\end{lem}
\noindent{\sc Proof:} 
We closely follow the proof of Lemma 3.1 in \citet{ChenChristensen2015Optimal} but make their arguments uniform in $P\in\mathbf P$. Let $\{M_n\}$ be a sequence of positive scalars to be chosen, and, for $i=1,\ldots,n$, set
\begin{gather*}
\Xi_{1,i}\equiv |u_i| b^{m_n}(V_i)b^{m_n}(V_i)^\transpose 1\{\|u_i b^{m_n}(V_i)b^{m_n}(V_i)^\transpose\|_o\le M_n^2\}~,\\
\Xi_{2,i}\equiv |u_i| b^{m_n}(V_i)b^{m_n}(V_i)^\transpose 1\{\|u_i b^{m_n}(V_i)b^{m_n}(V_i)^\transpose\|_o> M_n^2\}~.
\end{gather*}
Then simple manipulations reveal that
\begin{multline}\label{Eqn: npiv, variance, aux1}
\frac{1}{n}\sum_{i=1}^{n} |u_i| b^{m_n}(V_i)b^{m_n}(V_i)^\transpose-E_P[|u| b^{m_n}(V)b^{m_n}(V)^\transpose]\\
=\frac{1}{n}\sum_{i=1}^{n}(\Xi_{1,i}-E_P[\Xi_{1,i}])+\frac{1}{n}\sum_{i=1}^{n}(\Xi_{2,i}-E_P[\Xi_{2,i}])~.
\end{multline}
By construction, we have $\|\Xi_{1,i}\|_o\le M_n^2$ and hence, by the triangle inequality and Jensen's inequality \citep[p.40]{Tropp2015Introduction},
\begin{align}\label{Eqn: npiv, variance, aux2}
\|\Xi_{1,i}-E_P[\Xi_{1,i}]\|_o\le \|\Xi_{1,i}\|_o+E_P[\|\Xi_{1,i}\|_o]\le 2M_n^2~,
\end{align}
for all $i=1,\ldots,n$. Moreover, we have: for all $i=1,\ldots,n$,
\begin{multline}\label{Eqn: npiv, variance, aux3}
E_P[(\Xi_{1,i}-E_P[\Xi_{1,i}])^2]\le E_P[\Xi_{1,i}\Xi_{1,i}^\transpose]\\
=E_P[u_i^2 \|b^{m_n}(V_i)\|^2 b^{m_n}(V_i)b^{m_n}(V_i)^\transpose 1\{\|u_i b^{m_n}(V_i)b^{m_n}(V_i)^\transpose\|_o\le M_n^2\}]~.
\end{multline}
For generic vectors $a\in\mathbf R^d$ and $b\in\mathbf R^p$, we note the simple fact that
\begin{align}\label{Eqn: npiv, variance, aux4}
\|ab^\transpose\|_o=\sup_{x\in\mathbf R^p: \|x\|=1}\|ab^\transpose x\|=\|a\|\sup_{x\in\mathbf R^p: \|x\|=1}|b^\transpose x|=\|a\|\|b\|~.
\end{align}
In view of \eqref{Eqn: npiv, variance, aux4}, we may thus obtain from \eqref{Eqn: npiv, variance, aux3} that
\begin{align}\label{Eqn: npiv, variance, aux5}
E_P[(\Xi_{1,i}-E_P[\Xi_{1,i}])^2]&\le M_n^2 E_P[|u_i| b^{m_n}(V_i)b^{m_n}(V_i)^\transpose 1\{\|u_i b^{m_n}(V_i)b^{m_n}(V_i)^\transpose\|_o\le M_n^2\}]\notag\\
&\le M_n^2 E_P\big[E_P[|u_i||V_i] b^{m_n}(V_i)b^{m_n}(V_i)^\transpose \big]\notag\\
&\lesssim M_n^2 E_P\big[ b^{m_n}(V_i)b^{m_n}(V_i)^\transpose \big]\lesssim M_n^2 I_{m_n}~,
\end{align}
where the last line follows by $\sup_{P\in\mathbf P}E_P[|u|^{2+\varsigma}]<\infty$ with $\varsigma>0$ and Assumption \ref{Ass: series 2SLS estimation}(iii)-(a). Since eigenvalues and singular values of any positive semidefinite matrix coincide, we obtain by result \eqref{Eqn: npiv, variance, aux4} and Corollary III.2.3 in \citet{Bhatia1997Matrix} that
\begin{align}\label{Eqn: npiv, variance, aux6}
\|E_P[(\Xi_{1,i}-E_P[\Xi_{1,i}])^2]\|_o\lesssim M_n^2~.
\end{align}
Given  \eqref{Eqn: npiv, variance, aux2} and \eqref{Eqn: npiv, variance, aux6} and Assumption \ref{Ass: series 2SLS estimation}(i)-(a), we may invoke Theorem 6.6.1 in \citet{Tropp2015Introduction} and Markov's inequality to conclude that: uniformly in $P\in\mathbf P$,
\begin{align}\label{Eqn: npiv, variance, aux7}
E_P[\|\frac{1}{n}\sum_{i=1}^{n}(\Xi_{1,i}-E_P[\Xi_{1,i}])\|_o]\lesssim M_n\sqrt{\frac{\log m_n}{n}}~.
\end{align}

For the second term on the right side of \eqref{Eqn: npiv, variance, aux1}, we note that
\begin{align}\label{Eqn: npiv, variance, aux8}
\|\Xi_{2,i}\|_o\le |u_i|\xi_n^2 1\{|u_i| \xi_n^2>M_n^2\}~.
\end{align}
By Assumption \ref{Ass: series 2SLS estimation}(i)-(a), result \eqref{Eqn: npiv, variance, aux8} and Jensen's inequality, we thus obtain
\begin{multline}\label{Eqn: npiv, variance, aux9}
E_P[\|\frac{1}{n}\sum_{i=1}^{n}(\Xi_{2,i}-E_P[\Xi_{2,i}])\|_o]\le 2 E_P[|u|\xi_n^2 1\{|u| \xi_n^2>M_n^2\}]\\
\le 2 \frac{\xi_n^2}{(M_n^2/\xi_n^2)^\varsigma} E_P[|u|^{1+\varsigma}]\lesssim \frac{\xi_n^{2+2\varsigma}}{M_n^{2\varsigma}}~,
\end{multline}
uniformly in $P\in\mathbf P$, where the last step follows by $\sup_{P\in\mathbf P}E_P[|u|^{2+\varsigma}]<\infty$. Now, we choose $M_n$ to be such that the upper bounds in \eqref{Eqn: npiv, variance, aux7} and \eqref{Eqn: npiv, variance, aux9} are equal, i.e.,
\begin{align}\label{Eqn: npiv, variance, aux10}
M_n=\xi_n^{\frac{2+2\varsigma}{2\varsigma+1}}(\frac{n}{\log m_n})^{\frac{1}{2(2\varsigma+1)}}~.
\end{align}
Combining results \eqref{Eqn: npiv, variance, aux1}, \eqref{Eqn: npiv, variance, aux7}, \eqref{Eqn: npiv, variance, aux9} and \eqref{Eqn: npiv, variance, aux10} then yields
\begin{multline}\label{Eqn: npiv, variance, aux11}
E_P[\|\frac{1}{n}\sum_{i=1}^{n} |u_i| b^{m_n}(V_i)b^{m_n}(V_i)^\transpose-E_P[|u| b^{m_n}(V)b^{m_n}(V)^\transpose]\|_o]\\
\lesssim \big(\xi_n^{1+1/\varsigma}\sqrt{\frac{\log m_n}{n}}\big)^{\frac{2\varsigma}{2\varsigma+1}}~.
\end{multline}
The first claim then follows from \eqref{Eqn: npiv, variance, aux11} and Jensen's inequality. The proof of the second claim is analogous and thus omitted. \qed

\subsection{Nonparametric Quantile Regression}\label{App: NPQR}

We now construct strong approximations for Example \ref{Ex: NPQR under shape} following the recent work by \citet{BelloniChernozhukovChetverikovFernandez2019QR}. Our parameter of interest is $\theta_0: \mathcal T\to\mathbf R$ where $\mathcal T\equiv  \mathcal Z\times \mathcal U$ with $\mathcal Z$ (a subset of) the support of $Z$ and $\mathcal U\subset (0,1)$ a closed interval. Thus, the results presented here allow us to conduct inference on shape restrictions with respect to the quantile index, to the covariates or to both jointly. We note that \citet{ChernozhukovLeeRosen2013Intersection} also obtain strong approximations uniform in covariates but for a fixed quantile (see their Example 4), which may be of interest if the shape restriction in question is imposed with respect to the covariates.

Before proceeding further, we introduce some notation. Let $\{h_k\}_{k=1}^\infty$ be a sequence of basis functions on $\mathcal Z$, $h^{k_n}\equiv (h_1,\ldots,h_{k_n})^\transpose$, $\hat\Phi_n\equiv\sum_{i=1}^{n}h^{k_n}(Z_i)h^{k_n}(Z_i)^\transpose/n$ and $\Phi_{n,P}\equiv E_P[h^{k_n}(Z)h^{k_n}(Z)^\transpose]$. Moreover, we denote by $u\mapsto\beta_{n,P}(u)$ the series coefficient process, which is characterized as the solution to the approximation problem:
\begin{align}
\min_{\beta\in\mathbf R^{k_n}} E_P[\rho_u(Y-h^{k_n}(Z)^\transpose\beta)-\rho_u(Y-\theta_P(Z,u))]~,
\end{align}
where $\rho_u(y)\equiv (u-1\{y\le 0\})y$. Accordingly, we let $u\mapsto\hat\beta_n(u)$ be the series estimator defined as the solution to the problem
\begin{align}
\min_{\beta\in\mathbf R^{k_n}} \frac{1}{n}\sum_{i=1}^{n}\rho_u(Y_i-h^{k_n}(Z_i)^\transpose\beta)~.
\end{align}
Let $a_{n,P}(z,u)\equiv\theta_P(z,u)-h^{k_n}(z)^\transpose\beta_{n,P}(u)$ be the series approximation error, and let $f_{Y|Z}(\cdot, z)$ be the conditional density of $Y$ given $Z=z$, where the dependence on $P$ is suppressed. Denote by $\mathcal Y_z$ be the support of $f_{Y|Z}(\cdot, z)$ (given $Z=z$), and by $D_yf_{Y|Z}$ the derivative of the function $y\mapsto f_{Y|Z}(y,z)$. Finally, define the following (Jacobian) matrix that plays crucial roles in quantile regression:
\begin{align}
J_{n,P}(u)\equiv E_P[f_{Y|Z}(\theta_P(Z,u),Z)h^{k_n}(Z)h^{k_n}(Z)^\transpose]~.
\end{align}

Given the above notation, we impose that following assumption that is taken from \citet{BelloniChernozhukovChetverikovFernandez2019QR} (with only minor modifications).

\begin{ass} \label{Ass: series npqr}
(i) (a) $\{Y_i,Z_i, U_i\}_{i=1}^n$ are i.i.d., generated according to \eqref{Eqn: Ex, NPQR under shape} and governed by $P\in\mathbf P$; (b) The dimension $d_z$ of $Z$ is fixed (and does not involve $n$ and $P\in\mathbf P$); (c) The support of $Z$ is bounded in $\mathbf R^{d_z}$ uniformly in $P\in\mathbf P$.

\noindent (ii) (a) $f_{Y|Z}$ is bounded above uniformly in $z, y$ and $P\in\mathbf P$; (b) $f_{Y|Z}(\theta_0(z,u),z)$ is bounded away from zero uniformly in $z, u$ and $P\in\mathbf P$; (c) $y\mapsto D_yf_{Y|Z}(y,z)$ is continuous and bounded in absolute value uniformly in $z\in\mathcal Z$, $y\in\mathcal Y_z$, and $P\in\mathbf P$.

\noindent (iii) $\{h_k\}_{k=1}^\infty$ are functions on $\mathcal Z$ satisfying (a) the eigenvalues of $\Phi_{n,P}$ are bounded above and away from zero uniformly in $n$ and $P\in\mathbf P$; (b) $\|h^{k_n}\|_{P,\infty}\le\xi_n$ uniformly in $P\in\mathbf P$ where $\{\xi_n\}$ is bounded from below; (c) $\sup_{t\in\mathcal T}|a_{n,P}(t)|=O(k_n^{-\varsigma})$ for some absolute constant $\varsigma>0$, uniformly in $n$ and $P\in\mathbf P$; (d) $\|h^{k_n}(z)-h^{k_n}(z')\|\le \varpi_n\|z-z'\|$ for all $z,z'\in\mathcal Z$ and some $\varpi_n$ such that $\{\varpi_n/\xi_n\}$ is bounded away from zero.

\noindent (iv) There is a constant $\delta>0$ satisfying (a) $k_n^3\xi_n^2/n + k_n^{-\varsigma+1}=o(n^{-\delta})$; (b) $\sqrt n k_n^{-\varsigma}\xi_n^{-1}=o(n^{-\delta})$; (c) $(\xi_n^{-1}\varpi_n)^{2d_z}\xi_n^2 =o(n^{1-\delta})$; (d) $k_n^{1/2}l_n+k_n^2\xi_n^2/(nl_n)=o(n^{-\delta})$ for some $l_n\downarrow 0$.

\noindent (v) (a) $\{U_i^*\}_{i=1}^\infty$ is an i.i.d. sequence of $\mathrm{Uniform}(0,1)$ random variables; (b) $\{U_i^*\}_{i=1}^n$ are independent of $\{Z_i\}_{i=1}^n$ for all $n$.
\end{ass}

Assumption \ref{Ass: series npqr} is obtained by tailoring Assumptions S and U in \citet{BelloniChernozhukovChetverikovFernandez2019QR} to our setup and notation. We refer the reader to \citet{BelloniChernozhukovChetverikovFernandez2019QR} for detailed discussions, who also provide more primitive conditions. In particular, Assumption \ref{Ass: series npqr}(v) is imposed to implement their pivotal resampling method. While Example \ref{Ex: NPQR under shape} involves new technical challenges, including coupling a process of increasing dimension, the general strategy to obtain the strong approximation is similar in spirit to the development in Section \ref{Sec: npiv, app}, because it too is based on series estimation. For these reasons, we shall thus keep the treatment concise by relying more on \citet{BelloniChernozhukovChetverikovFernandez2019QR}.

\begin{pro}\label{Pro: npqr, strong}
Assumptions \ref{Ass: series npqr}(i), (ii), (iii)-(a)(b)(c) and (iv) together imply Assumption \ref{Ass: strong approx}(i) with $r_n=\sqrt{n}/\xi_n$, $\hat\theta_n=(h^{k_n})^\transpose\hat\beta_n$, $c_n=n^{-\frac{\delta}{2(2d_z+3)}}$ and
\begin{align}
\mathbb Z_{n,P}=\xi_n^{-1}(h^{k_n})^\transpose J_{n,P}^{-1}  \Phi_{n,P}^{1/2}\mathbb G~,
\end{align}
where $\mathbb G=(\mathbb G_1,\ldots,\mathbb G_m)$ is a vector of independent centered Gaussian processes in $\ell^\infty(\mathcal U)$  such that $E[\mathbb G_j(u)\mathbb G_j(v)]=u\wedge v-uv$ for all $u,v\in\mathcal U$ and all $j=1,\ldots,m$.
\end{pro}
\noindent{\sc Proof:} Define a process $u\mapsto\mathbb U_n(u)$ by: for each $u\in\mathcal U$,
\begin{align}
\mathbb U_n(u)\equiv\frac{1}{\sqrt n}\sum_{i=1}^{n}h^{k_n}(Z_i)(u-1\{U_i\le u\})~,
\end{align}
where recall that $\{U_i\}$ are the i.i.d.\ errors with common distribution $\mathrm{Uniform}(0,1)$. By Assumptions \ref{Ass: series npqr}(i), (ii), (iii)-(a)(c) and (iv)-(a), we may apply Theorem 2 in \citet{BelloniChernozhukovChetverikovFernandez2019QR} to obtain that
\begin{multline}\label{Eqn: npqr, strong, aux1}
\sqrt n \{\hat\beta_n(u)-\beta_{n,P}(u)\}=J_{n,P}^{-1}(u)\mathbb U_n(u)\\ + O_p(\frac{k_n^{3/4}\xi_n^{1/2}\{\log n\}^{1/2}}{n^{1/4}} + \{k_n^{1-\varsigma}\log n\}^{1/2})~,
\end{multline}
uniformly in $u\in\mathcal U$ and $P\in\mathbf P$. By simple algebra, result \eqref{Eqn: npqr, strong, aux1}, and Assumption \ref{Ass: series npqr}(iii)-(c), we in turn have:
\begin{multline}\label{Eqn: npqr, strong, aux2}
r_n \{\hat\theta_n-\theta_0\}=r_n (h^{k_n})^\transpose \{\hat\beta_n-\beta_{n,P}\}-r_na_{n,P}\\
=\xi_n^{-1}(h^{k_n})^\transpose J_{n,P}^{-1}\mathbb U_n
+ O_p(\frac{k_n^{3/4}\xi_n^{1/2}\{\log n\}^{1/2}}{n^{1/4}} + \{k_n^{1-\varsigma}\log n\}^{1/2} + \frac{\sqrt n k_n^{-\varsigma}}{\xi_n})
\end{multline}
in $\ell^\infty(\mathcal T)$, uniformly in $P\in\mathbf P$. By Assumptions \ref{Ass: series npqr}(i)-(a), (iii)-(a)(b) and (iv)-(a)(c), we may invoke Lemma 36 in \citet{BelloniChernozhukovChetverikovFernandez2019QR} to conclude that there exists a zero-mean process $(u,z)\mapsto\hat{\mathbb Z}_{n,P}(u,z)$ satisfying, {\it conditional on $\{Z_i\}_{i=1}^n$}, (a) it has uniformly continuous sample paths almost surely, (b) its covariance functional is
\begin{multline}\label{Eqn: npqr, strong, aux4}
E[\hat{\mathbb Z}_{n,P}(u,z_1)\hat{\mathbb Z}_{n,P}(v,z_2)|\{Z_i\}_{i=1}^n]\\
=\{\frac{1}{n\xi_n^2}\sum_{i=1}^{n}h^{k_n}(z_1)^\transpose J_{n,P}^{-1}h^{k_n}(Z_i)h^{k_n}(Z_i)^\transpose J_{n,P}^{-1} h^{k_n}(z_2)\}(u\wedge v-uv)~,
\end{multline}
for any $u,v\in\mathcal U$ and $z_1,z_2\in\mathcal Z$, and (c), for any absolute constant $\delta'\in(0,\frac{\delta}{2d_z})$,
\begin{align}\label{Eqn: npqr, strong, aux3a}
\xi_n^{-1}(h^{k_n})^\transpose J_{n,P}^{-1}\mathbb U_n=\hat{\mathbb Z}_{n,P}+ o_p(n^{-\frac{\delta'}{2}} + n^{\frac{d_z\delta'}{3}-\frac{\delta}{6}}) \text{ in } \ell^\infty(\mathcal T)~,
\end{align}
uniformly in $P\in\mathbf P$, where the order in \eqref{Eqn: npqr, strong, aux3a} is obtained by simply combining the orders of $r_1$, $r_2$ and $r_3$ in the proof of Lemma 36 in \citet{BelloniChernozhukovChetverikovFernandez2019QR}. In particular, setting $\delta'=\delta/(2d_z+3)$ in \eqref{Eqn: npqr, strong, aux3a} yields: uniformly in $P\in\mathbf P$,
\begin{align}\label{Eqn: npqr, strong, aux3}
\xi_n^{-1}(h^{k_n})^\transpose J_{n,P}^{-1}\mathbb U_n=\hat{\mathbb Z}_{n,P}+ o_p( n^{-\frac{\delta}{2(2d_z+3)}}) \text{ in } \ell^\infty(\mathcal T)~.
\end{align}

Given Assumptions \ref{Ass: series npqr}(ii)-(b) and (iii)-(a)(b), applying Lemma \ref{Lem: Gaussian coupling} with $\mathbf B=\prod_{i=1}^{n}\mathbf R^{d_z}$, $\mathbb X=(Z_1,\ldots,Z_n)$, $\mathbf D=\prod_{j=1}^m\ell^\infty(\mathcal U)$, $\mathbf E=\ell^\infty(\mathcal T)$ with $\mathcal T\equiv\mathcal U\times\mathcal Z$, $\mathbb G_0=(\mathbb G_{0,1},\ldots,\mathbb G_{0,m})$ a vector of independent centered Gaussian variables with common covariance functional $E[\mathbb G_{0,j}(u)\mathbb G_{0,j}(v)]=u\wedge v-uv$ for any $u,v\in\mathcal U$ (so each $\mathbb G_{0,j}$ is a tight Brownian bridge), $\mathbf D_0=\prod_{j=1}^{m}C_{\mathrm u}(\mathcal U)$ with $C_{\mathrm u}(\mathcal U)$ the Banach space of uniformly continuous functions on $\mathcal U$, the map $\hat\psi:\mathbf D\to\mathbf E$ given by $\hat\psi(g)=\xi_n^{-1}(h^{k_n})^\transpose J_{n,P}^{-1}\hat\Phi_n^{1/2}g$ for any $g\in\mathbf D$, and $\mathbf E_0$ the Banach space of uniformly continuously functions on $\mathcal T$ yields that there exists a copy $\mathbb G$ of $\mathbb G_0$ that is independent of $\{Z_i\}_{i=1}^n$ and satisfies
\begin{align}
\hat{\mathbb Z}_{n,P}=\xi_n^{-1}(h^{k_n})^\transpose J_{n,P}^{-1}\hat\Phi_n^{1/2}\mathbb G
\end{align}
almost surely. Now we may set the desired coupling variable $\mathbb Z_{n,P}$ as
\begin{align}
\mathbb Z_{n,P} = \xi_n^{-1}(h^{k_n})^\transpose J_{n,P}^{-1} \Phi_{n,P}^{1/2}\mathbb G~.
\end{align}
By arguments analogous to those leading to \eqref{Eqn: bootstrap, series npiv, aux17}, we may obtain that
\begin{multline}\label{Eqn: npqr, strong, aux5}
E[\|\hat{\mathbb Z}_{n,P}-\mathbb Z_{n,P}\|_{\mathbf H}|\{Z_i\}_{i=1}^n] \lesssim \sup_{u\in\mathcal U}\|J_{n,P}^{-1}(u)\|_o\|\hat\Phi_n^{1/2}-\Phi_{n,P}^{1/2}\|_o \\
\lesssim \frac{1}{\{\lambda_{\min}(\Phi_{n,P})\}^{1/2}}\|\hat\Phi_n-\Phi_{n,P}\|_o=O_p(\sqrt{\frac{\xi_n^2\log k_n}{n}})~,
\end{multline}
uniformly in $P\in\mathbf P$, where the second inequality exploited the fact that the eigenvalues of $J_{n,P}(u)$ are bounded away from zero uniformly in $u\in\mathcal U$, $n$ and $P\in\mathbf P$ (by Assumption \ref{Ass: series npqr}(ii)-(b) and (iii)-(a)) and Proposition 3.2 in \citet{HemmenAndo1980Ideal} (see also Problem X.5.5 in \citet{Bhatia1997Matrix}), and the last step follows by Assumptions \ref{Ass: series npqr}(i)-(a)(b), (iii)-(a)(b) and (iv)-(a) and Theorem E.1 in \citet{Kato2013QuasiB}. It follows from Fubini's theorem, Markov's inequality and result \eqref{Eqn: npqr, strong, aux5} that
\begin{align}\label{Eqn: npqr, strong, aux6}
\|\xi_n^{-1}(h^{k_n})^\transpose J_{n,P}^{-1}\hat{\mathbb Z}_{n,P}-\mathbb Z_{n,P}\|_{\mathbf H}=O_p(\sqrt{\frac{\xi_n^2\log k_n}{n}})~,
\end{align}
uniformly in $P\in\mathbf P$. The proposition then follows from combining \eqref{Eqn: npqr, strong, aux2}, \eqref{Eqn: npqr, strong, aux3}, \eqref{Eqn: npqr, strong, aux6}, the triangle inequality and Assumption \ref{Ass: series npqr}(iv)-(a)(b)(c). \qed

To verify Assumption \ref{Ass: strong approx}(ii), we employ the pivotal method proposed by \citet{BelloniChernozhukovChetverikovFernandez2019QR}. First, for $l_n\downarrow 0$ a suitable bandwidth, we follow \citet{Powell1984CLAD} and estimate the matrix-valued map $u\mapsto J_{n,P}(u)$ by: for any $u\in\mathcal U$,
\begin{align}
\hat J_n(u)=\frac{1}{2nl_n}\sum_{i=1}^{n}1\{|Y_i-h^{k_n}(Z_i)^\transpose \hat\beta_n(u)|\le l_n\} h^{k_n}(Z_i)h^{k_n}(Z_i)^\transpose~.
\end{align}
Given the $\mathrm{Uniform}(0,1)$ random variables $\{U_i^*\}_{i=1}^n$, we may then obtain $\hat{\mathbb G}_n$ as
\begin{align}\label{Eqn: npqr, pivotal}
\hat{\mathbb G}_n=\xi_n^{-1}(h^{k_n})^\transpose \hat J_n^{-}  \frac{1}{\sqrt n}\sum_{i=1}^{n}Z_i(u-1\{U_i^*\le u\})~.
\end{align}

\begin{pro}\label{Pro: npqr, boot}
Assumptions \ref{Ass: series npqr}(i), (ii), (iii), (iv)-(a)(c)(d) and (v) together imply Assumption \ref{Ass: strong approx}(ii) with $\hat{\mathbb G}_n$ given by \eqref{Eqn: npqr, pivotal} and $c_n=n^{-\frac{\delta}{2(2d_z+3)}}$.
\end{pro}
\begin{rem}
Once again, for Assumption \ref{Ass: strong approx} overall, one should take the maximum of the coupling rates in Propositions \ref{Pro: npqr, boot} and \ref{Pro: npqr, strong}. \qed
\end{rem}
\noindent{\sc Proof:} By Assumptions \ref{Ass: series npqr}(i), (ii), (iii)-(a)(c) and (iv)-(d), we may invoke Theorem 3 in \citet{BelloniChernozhukovChetverikovFernandez2019QR} to conclude that
\begin{multline}\label{Eqn: npqr, boot, aux1}
\hat J_n^-\frac{1}{\sqrt n}\sum_{i=1}^{n}Z_i(u-1\{U_i^*\le u\})=J_{n,P}^{-1}\frac{1}{\sqrt n}\sum_{i=1}^{n}Z_i(u-1\{U_i^*\le u\})\\
+O_p(\sqrt{\frac{\xi_n^2k_n^2\log n}{nl_n}} + k_n^{-\varsigma+1/2}+l_n\sqrt{k_n})
\end{multline}
in $\ell^\infty(\mathcal U)$, uniformly in $P\in\mathbf P$. In view of Assumption \ref{Ass: series npqr}(v), we may combine analogs of results \eqref{Eqn: npqr, strong, aux3} and \eqref{Eqn: npqr, strong, aux6} with the triangle inequality to obtain that
\begin{multline}\label{Eqn: npqr, boot, aux2}
\xi_n^{-1}(h^{k_n})^\transpose J_{n,P}^{-1}\frac{1}{\sqrt n}\sum_{i=1}^{n}Z_i(u-1\{U_i^*\le u\})\\
=\bar{\mathbb Z}_{n,P} + o_p( n^{-\frac{\delta}{2(2d_z+3)}}) + O_p(\sqrt{\frac{\xi_n^2\log k_n}{n}})
\end{multline}
in $\ell^\infty(\mathcal T)$, uniformly in $P\in\mathbf P$, where $\bar{\mathbb Z}_{n,P} = \xi_n^{-1}(h^{k_n})^\transpose J_{n,P}^{-1} \Phi_{n,P}^{1/2}\bar{\mathbb G}$ with $\bar{\mathbb G}$ a copy of $\mathbb G$ that is independent of $\{X_i\}_{i=1}^n$. By Assumption \ref{Ass: series npqr}(v)-(d), we note that
\begin{align}\label{Eqn: npqr, boot, aux3}
\sqrt{\frac{\xi_n^2k_n^2\log n}{nl_n}} + k_n^{-\varsigma+1/2}+l_n\sqrt{k_n} + \sqrt{\frac{\xi_n^2\log k_n}{n}}= o(n^{-\delta'/2})~,
\end{align}
for any $\delta'\in(0,\delta)$ and in particular for $\delta'=\frac{\delta}{2d_z+3}$. The proposition then follows by \eqref{Eqn: npqr, boot, aux1}, \eqref{Eqn: npqr, boot, aux2}, \eqref{Eqn: npqr, boot, aux3}, Assumption \ref{Ass: series npqr}(iii)-(b) and the triangle inequality. \qed

\subsection{Rationality and Slutsky Restrictions}\label{Sec: Slutsky, appendix}

The Slutsky restriction is essentially equivalent to the weak axiom (and in fact also the strong axiom if symmetry is present) of revealed preferences \citep{KihlstromMasColellSonnenchein1976Demand}. Hence, rationality of consumer behaviors may be verified by studying the Slutsky restriction, as pursued in both economic theory \citep{JerisonJerison1992Approx,JerisonJerison1993Approx,AguiarSerrano2017Slutsky} and econometrics \citep{Hoderlein2011Rational,ChernozhukovNeweySantos2019CCMM,DetteHoderleinNeumeyer2016Slutsky,HorowitzLee2017Shape}.


\subsubsection{The Model}

For the reader's convenience, we restate the model introduced in Section \ref{Sec: examples}. Let $Q\in\mathbf R^{d_q}$ be a vector of budget shares for ${d_q}$ number of categories of goods, $P\in\mathbf R^{d_q}$ the vector of associated log-prices, $Y\in\mathbf R$ the total expenditure in logarithm, and $Z\in\mathbf R^{d_z}$ a vector of additional observable demographic characteristics. Consider the following system of demand equations
\begin{align}\label{Eqn: ex, Slutsky}
Q  = g_0(P,Y)+\Gamma_0^\transpose Z + U~,
\end{align}
where $g_0: \mathbf R_+^{d_q+1}\to\mathbf R^{d_q}$ is differentiable, $\Gamma_0\in\mathbf M^{d_z\times d_q}$, and $U\in\mathbf R^{d_q}$ is the error term. The semiparametric structure in \eqref{Eqn: ex, Slutsky} is also employed by \citet{BlundellHorowitzParey2012Measuring} to circumvent the curse of dimensionality. For notational simplicity, let $T\equiv (P^\transpose, Y)^\transpose$. The Slutsky matrix of $g_0$ is a mapping $t\equiv(p^\transpose,y)^\transpose\mapsto\theta_0(t)\in\mathbf M^{d_q\times d_q}$ defined by:
\begin{align}\label{Eqn: Slutsky matrix, defn}
\theta_0(t)\equiv D_pg_0(t)+(D_y g_0(t))g_0(t)^\transpose + g_0(t)g_0(t)^\transpose -\mathrm{diag}(g_0(t))~,
\end{align}
where, for a generic vector $a\equiv(a_1,\ldots,a_k)^\transpose$, we denote by $\mathrm{diag}(a)$ or $\mathrm{diag}(a_1,\ldots,a_k)$ the diagonal matrix whose diagonal entries are $a_1,\ldots, a_k$, $D_pg (t)\equiv\partial g(t)/\partial p^\transpose$ and $D_yg (t)\equiv\partial g(t)/\partial y$ for a generic function $t\equiv(p^\transpose,y)^\transpose\mapsto g(t)$. Compared to \eqref{Eqn: Slutsky matrix in ex}, the last two terms in \eqref{Eqn: Slutsky matrix, defn} appear because $Q$ is measured in shares, and $P$ and $Y$ are in logarithm (which is common practice in applied work). 

There are two notable features of the model \eqref{Eqn: ex, Slutsky}. First, endogeneity is a generic concern in the literature. In particular, since total expenditure is largely determined by unobserved preferences \citep{BlundellChenKristensen2007Engel,Hoderlein2011Rational} and is often contaminated by measurement errors \citep{HausmanNeweyIchimuraPowell1991EIV,Newey2001Flexible,DetteHoderleinNeumeyer2016Slutsky}, it is important to allow $Y$ to be correlated with the error $U$. Second, as forcefully argued in \citet{BrownWalker1989Random} and \citet{Lewbel2001Demand}, the additive error $U$ is (inherently) conditionally heteroskedastic through its dependence on (at least) price, in many interesting settings. Our treatment below shall accommodate both features.

\subsubsection{Verification of Main Assumptions: Overview}

The model \eqref{Eqn: ex, Slutsky} with $d_q=1$ may be viewed as a special case of Example \ref{Ex: NPIV under shape}, but is more complicated when $d_q>1$. To make our discussions manageable, we shall thus simplify the arguments that are analogous to those in Section \ref{Sec: npiv, app}.

We commence with estimation of the primitive, $g_0$. Since entries of the function $g_0$ have the same arguments and similar smoothness, we employ the same sequence of basis functions for them as in \citet{BlundellChenKristensen2007Engel}. In order to account for endogeneity, suppose that $V^{(1)}\in\mathbf R^{d_{v_1}}$ and $V^{(2)}\in\mathbf R^{d_{v_2}}$ are vectors of instrumental variables for $T$ and $Z$ respectively. Let $\{h_k\}$ and $\{b_m\}$ be basis functions of $T$ and $V$ respectively, and set $\bar h^{k_n}(t,z)\equiv (h^{k_n}(t)^\transpose, z^\transpose)^\transpose$ and $\bar b^{m_n}(v)\equiv (b^{m_n}(v_1)^\transpose, v_2^\transpose)^\transpose$. Given these basis functions, we may then implement the series 2SLS estimation \citep{AiChen2003Efficient}. To this end, we need to introduce additional notation. Define
\begin{align}
\bar H_n\equiv\begin{bmatrix}
\bar h^{k_n}(T_1,Z_1)^\transpose\\
\vdots\\
\bar h^{k_n}(T_n,Z_n)^\transpose
\end{bmatrix},\, \bar B_n\equiv\begin{bmatrix}
\bar b^{m_n}(V_1)^\transpose\\
\vdots\\
\bar b^{m_n}(V_n)^\transpose
\end{bmatrix},\,\underline Q_n\equiv\begin{bmatrix}
Q_1^\transpose\\
\vdots\\
Q_n^\transpose
\end{bmatrix},\,\underline U_n\equiv\begin{bmatrix}
U_1^\transpose\\
\vdots\\
U_n^\transpose
\end{bmatrix}~.
\end{align}
In turn, we define some sample moments:
\begin{align}
\bar \Phi_n=\frac{1}{n}\bar H_n^\transpose \bar H_n~, \bar \Psi_n=\frac{1}{n}\bar B_n^\transpose \bar B_n~, \bar \Pi_n=\frac{1}{n}\bar B_n^\transpose \bar H_n~.
\end{align}
We may then estimate $g_0$ by $\hat g_n\equiv\hat\Lambda_n^\transpose h^{k_n}$, where $\hat\Lambda_n\in\mathbf M^{k_n\times d_q}$ together with $\hat\Gamma_n\in\mathbf M^{d_z\times d_q}$ are matrices of 2SLS estimators defined by
\begin{align}
\begin{bmatrix}
\hat\Lambda_n\\
\hat\Gamma_n
\end{bmatrix}=(\bar\Pi_n^\transpose\bar\Psi_n^-\bar\Pi_n)^-\bar\Pi_n^\transpose\bar\Psi_n^-\frac{1}{n}\bar B_n^\transpose \underline Q_n~.
\end{align}
Finally, a natural estimator of $\theta_0$ at this point is the plug-in estimator $\hat\theta_n$ given by
\begin{align}
\hat\theta_n=D_p\hat g_n+(D_y \hat g_n)\hat g_n^\transpose + \hat g_n\hat g_n^\transpose -\mathrm{diag}(\hat g_n)~,
\end{align}
where the basis functions $\{h_k\}$ are assumed to be differentiable.

With these notation and definitions in hand, we note that $\theta_0$ is a nonlinear functional of $g_0$. Specifically, let $C_{\mathrm b}^1(\mathcal T)$ be the space of $\mathbf R^{d_q}$-valued functions given by
\begin{align}
C_{\mathrm b}^1(\mathcal T)\equiv \{g:\mathcal T\to\mathbf R^{d_q}: \|g\|_{1,\infty}<\infty\}~,\, \|g\|_{1,\infty}\equiv \sup_{t\in\mathcal T}\{\|g(t)\|+\|\frac{\partial g(t)}{\partial t^\transpose}\|\}~,
\end{align}
and define a functional $\psi: C_{\mathrm b}^1(\mathcal T)\to\mathbf H$ as, for any $g\in C_{\mathrm b}^1(\mathcal T)$,
\begin{align}
\psi(g)\equiv D_pg+(D_y g)g^\transpose+ gg^\transpose-\mathrm{diag}(g)~.
\end{align}
Then we may write $\theta_0=\psi(g_0)$, and estimate it by the plug-in estimator
\begin{align}\label{Eqn: Slutsky matrix, estimator}
\hat\theta_n \equiv \psi(\hat g_n) ~.
\end{align}
The plug-in structure in \eqref{Eqn: Slutsky matrix, estimator} suggests that construction of the strong approximations depends on analytic natures of $\psi$ and statistical properties of $\hat g_n$.

Starting with $\psi$, we tackle the nonlinearity of $\psi$ through linearization following \citet{Newey1997series} and \citet{ChenChristensen2018SupNormOptimal}. In particular, simple algebra reveals that $\psi$ is Fr\'{e}chet differentiable at $g\in C_{\mathrm b}^1(\mathcal T)$ such that, for all $h\in C_{\mathrm b}^1(\mathcal T)$,
\begin{align}\label{Eqn: Slutsky derivative}
\psi_{g}'(h)=D_ph+(D_yg)h^\transpose+(D_yh)g^\transpose+gh^\transpose +hg^\transpose-\mathrm{diag}(h)~.
\end{align}
By leveraging the order of $\|\hat g_n-g_0\|_{1,\infty}$, we may then turn the nonlinear problem into a linear one through the approximation: for some $r_n\uparrow\infty$,
\begin{align}\label{Eqn: Slutsky, linearization}
r_n\{\hat\theta_n-\theta_0\}=r_n\{\psi(\hat g_n)-\psi(g_0)\}=\psi_{g_0}'(r_n\{\hat g_n-g_0\}) + o_p(1) \text{ in }\mathbf H~.
\end{align}
This naturally leads to the study of $\hat g_n$. By appealing to Yurinskii's coupling \citep[Theorem 10.10]{Pollard2002User}, we may obtain that, for some $\varpi_n\downarrow 0$,
\begin{align}\label{Eqn: Slutsky, strong approx of g}
\|r_n\{\hat g_n-g_0\} - \mathbb W_{n,P}\|_{1,\infty} = O_p(\varpi_n) ~,
\end{align}
where $\mathbb W_{n,P} $ is some Gaussian process in $\mathbf H$---see Lemma \ref{Lem: Slutsky coupling} for more details. Combining \eqref{Eqn: Slutsky, linearization} and \eqref{Eqn: Slutsky, strong approx of g}, we may then verify Assumption \ref{Ass: strong approx}(i) for
\begin{align}\label{Eqn: Slutsky, strong approx-i}
\mathbb Z_{n,P}=\psi_{g_0}'(\mathbb W_{n,P})~.
\end{align}
With regard to Assumption \ref{Ass: strong approx}(ii), we employ the sieve score bootstrap following \citet{ChenPouzo2015SieveWald} and \citet{ChenChristensen2018SupNormOptimal}. Towards this end, define $s_n$ and $\xi_n$ as the orders of $\sigma_{\min}(E_P[\bar\Pi_{n}])$ and $\|h^{k_n}\|_{1,\infty}$ respectively as in Example \ref{Ex: NPIV under shape}, set $\hat U_i\equiv Q_i-\hat g_n(T_i)-\hat\Gamma_n^\transpose Z_i$ to be the residuals, and let $\{W_i\}$ be i.i.d.\ bootstrap weights with zero mean and unit variance. Then we may construct
\begin{align}\label{Eqn: Slutsky, strong approx-ii}
\hat{\mathbb G}_n=\psi_{\hat g_n}'(\hat{\mathbb W}_n)~,\quad \hat{\mathbb W}_n\equiv\big(\hat \Omega_n \frac{1}{\sqrt n}\sum_{i=1}^{n}W_i\bar b^{m_n}(V_i)\hat U_i^\transpose\big)^\transpose h^{k_n}~,
\end{align}
with $\hat\Omega_n\in\mathbf M^{k_n\times (m_n+d_{v_2})}$ being the upper block of $ s_n\xi_n^{-1}(\bar\Pi_n^\transpose\bar\Psi_n^-\bar\Pi_n)^- \bar\Pi_n^\transpose\bar\Psi_n^-$.

The next proposition formalizes the discussions above under conditions that are characterized by Assumption \ref{Ass: Slutsky} stated in the next subsection.

\begin{pro}\label{Pro: Slutsky, strong approx}
If \Cref{Ass: Slutsky} holds, then $\hat\theta_n$ in \eqref{Eqn: Slutsky matrix, estimator}, $\mathbb Z_{n,P}$ in \eqref{Eqn: Slutsky, strong approx-i} and $\hat{\mathbb G}_n$ in \eqref{Eqn: Slutsky, strong approx-ii} satisfy \Cref{Ass: strong approx} with $r_n=\sqrt n s_n/\xi_n$ and some $c_n>0$.
\end{pro}

\subsubsection{Verification of Main Assumptions: Details}

We proceed with some additional notation. First, define the ``infeasible'' matrix estimators $\tilde\Lambda_n\in\mathbf M^{k_n\times d_q}$ and $\tilde\Gamma_n\in\mathbf M^{d_z\times d_q}$ by
\begin{align}
\begin{bmatrix}
\tilde\Lambda_n\\
\tilde\Gamma_n
\end{bmatrix}\equiv (\bar\Pi_n^\transpose\bar\Psi_n^-\bar\Pi_n)^-\bar\Pi_n^\transpose\bar\Psi_n^-\frac{1}{n}\bar B_n^\transpose (G_{n,P}+\underline{Z}_n\Gamma_P)~,
\end{align}
where $G_{n,P}\equiv (g_P(T_1),\ldots,g_P(T_n))^\transpose$ and $\underline{Z}_n\equiv[Z_1,\ldots,Z_n]^\transpose$. In turn, we set
\begin{align}
\tilde g_n\equiv \tilde\Lambda_n^\transpose h^{k_n}~,
\end{align}
which plays a role analogous to that of $\tilde h$ in \citet{ChenChristensen2018SupNormOptimal}. Thus, $\tilde g_n-g_P$ may be interpreted as the bias, while $\hat g_n-\tilde g_n$ the (standard) variance. Next, we define some matrices of population moments:
\begin{multline}
\bar \Phi_{n,P}\equiv E_P[\bar h^{k_n}(T,Z)\bar h^{k_n}(T,Z)^\transpose]~,\, \bar \Psi_{n,P}\equiv E_P[\bar b^{m_n}(V)\bar b^{m_n}(V)^\transpose]~,\\ \bar \Pi_{n,P}\equiv E_P[\bar b^{m_n}(V)\bar h^{k_n}(T)^\transpose]~,
\end{multline}
and sieve 2SLS projection matrices $\Lambda_{n,P}\in\mathbf M^{k_n\times d_q}$ and $\Gamma_{n,P}\in\mathbf M^{d_z\times d_q}$:
\begin{align}\label{Eqn: Slutsky 2SLS coefficient}
\begin{bmatrix}
\Lambda_{n,P}\\
\Gamma_{n,P}
\end{bmatrix}  \equiv (\bar \Psi_{n,P}^{-1/2}\bar\Pi_{n,P})_l^{-}\bar\Psi_{n,P}^{-1/2} E_P[\bar b^{m_n}(V)Q^\transpose]~.
\end{align}
In turn, we denote the 2SLS population residual function by $a_{n,P}(t,z)\equiv g_P(t)+
\Gamma_P^\transpose z-\Lambda_{n,P}^\transpose h^{k_n}(t)-\Gamma_{n,P}^\transpose z$ for all $t\in\mathcal T$ and $z\in\mathcal Z$, and set
\begin{align}
A_{n,P}\equiv [a_{n,P}(T_1,Z_1),\ldots,a_{n,P}(T_n,Z_n)]^\transpose~.
\end{align}
To control the sieve approximation error, let $\Lambda_{n,P}^{\mathrm{ols}}\in\mathbf M^{k_n\times d_q}$ and $\Gamma_{n,P}^{\mathrm{ols}}\in\mathbf M^{d_z\times d_q}$ be matrices of the sieve OLS population coefficients, i.e.,
\begin{align}
\begin{bmatrix}
\Lambda_{n,P}^{\mathrm{ols}}\\
\Gamma_{n,P}^{\mathrm{ols}}
\end{bmatrix}  \equiv\bar \Phi_{n,P}^{-1}E_P[\bar h^{k_n}(T,Z)Q^\transpose]~.
\end{align}
Let $\mathcal H_{k_n}$ be the subspace spanned by $h_1,\ldots,h^{k_n}$, and $\mathrm{Proj}_{m,k}: \prod_{j=1}^{d_q}L^2(X)\to\prod_{j=1}^{d_q}\mathcal H_{k_n}$ be the 2SLS projection operator defined by: for any $f\in \prod_{j=1}^{d_q}L^2(X)$,
\begin{align}
\mathrm{Proj}_{m,k}(f) = \Lambda_{n,P}^\transpose h^{k_n}~,
\end{align}
where $\Lambda_{n,P}$ is defined as in \eqref{Eqn: Slutsky 2SLS coefficient} but with $Q$ replaced by $f$. Let $\Upsilon_P: \prod_{j=1}^{d_q}L^2(X)\to \prod_{j=1}^{d_q}L^2(V)$ be the conditional expectation operator, i.e., $\Upsilon_P(f)=E[f|V]$ for any $f\in \prod_{j=1}^{d_q}L^2(X)$. For $\hat U_i\equiv Q_i-\hat\Lambda_n^\transpose h^{k_n}(T_i)-\hat\Gamma_n^\transpose Z_i$, let
\begin{gather}
\hat\Sigma_n   \equiv \frac{1}{n}\sum_{i=1}^{n} (\bar b^{m_n}(V_i)\bar b^{m_n}(V_i)^\transpose)\otimes(\hat U_i\hat U_i^\transpose)~, \\
\Sigma_{n,P}  \equiv E_P[(\bar b^{m_n}(V)\bar b^{m_n}(V)^\transpose)\otimes(U U^\transpose)]~.
\end{gather}
Finally, recall that $X_i=(Q_i,P_i,Y_i,Z_i,V_i)$ in this example.

Having introduced the notation, we now impose the following assumption.

\begin{ass}\label{Ass: Slutsky}
(i) (a) The sample $\{X_i\}_{i=1}^n$ are i.i.d., generated according to \eqref{Eqn: ex, Slutsky} and governed by $P\in\mathbf P$; (b) The support $\mathcal T$ of $T$ is bounded uniformly in $P\in\mathbf P$; (c) $\sup_{P\in\mathbf P}\|g_P\|_{1,\infty}<\infty$; (d) $\Upsilon_P: \prod_{j=1}^{d_q}L^2(X)\to \prod_{j=1}^{d_q}L^2(V)$ is injective.

\noindent (ii) $\|\hat\Gamma_n-\Gamma_P\|_o=O_p(n^{-1/2})$ uniformly in $P\in\mathbf P$.


\noindent (iii) (a) The eigenvalues of $E_P[h^{k_n}(T)h^{k_n}(T)^\transpose]$ are bounded from above uniformly in $n$ and $\mathbf P$; (b) $\|h^{k_n}\|_{1,\infty}\le\xi_n$ with $\xi_n\ge 1$; (c) $\sup_{P\in\mathbf P}\|g_P-(\Lambda_{n,P}^{\mathrm{ols}})^\transpose h^{k_n}\|_{1,\infty}= O(\delta_{n})$ for some $\delta_n=o(1)$; (d) $\|\Upsilon_P(g_0-(\Lambda_{n,P}^{\mathrm{ols}})^\transpose h^{k_n})\|_{L^2(V)}\lesssim s_n\|g_0-(\Lambda_{n,P}^{\mathrm{ols}})^\transpose h^{k_n}\|_{L^2(T)}$.

\noindent (iv) (a) The eigenvalues of $\bar\Psi_{n,P}$ are bounded from above and away from zero uniformly in $n$ and $\mathbf P$; (b) $\sup_{P\in\mathbf P}\|b^{m_n}\|_{P,\infty}\le\xi_n$; (c) $\inf_{P\in\mathbf P}\sigma_{\min}(\bar\Pi_{n,P})\gtrsim s_n>0$ for each $n$; (d) $\|\mathrm{Proj}_{m,k}(g_P-(\Lambda_{n,P}^{\mathrm{ols}})^\transpose h^{k_n})\|_{1,\infty}\lesssim \|g_P-(\Lambda_{n,P}^{\mathrm{ols}})^\transpose h^{k_n}\|_{1,\infty}$.

\noindent (v) (a) There is an absolute constant $\varsigma>0$ satisfying $\sup_{P\in\mathbf P}E_P[\|Z\|^{2+\varsigma}+\|V^{(2)}\|^{2+\varsigma}]<\infty$; (b) $\sup_{P\in\mathbf P}\|E_P[\|U\|^3|V]\|_{P,\infty} <\infty$ and $\sup_{P\in\mathbf P}E_P[\|V^{(2)}\|^3]<\infty$; (c) There is some $\underline\sigma>0$ such that of $\lambda_{\min}(E_P[UU^\transpose|V])\ge\underline\sigma$ almost surely for all $P\in\mathbf P$.

\noindent (vi) (a) $2\le k_n\le m_n\le c_0k_n$ for some $c_0\ge 1$, and the number $d_{v_2}$ of instruments for $Z$ is fixed and larger than $d_z$; (b) $\varpi_n=o(1)$ with $\varpi_n$ defined as
\begin{multline}
\varpi_n=(\frac{\xi_nm_n^2}{\sqrt n})^{1/3} + \frac{\sqrt n s_n\delta_n}{\xi_n}+ \delta_n \sqrt{((\xi_n^2+n^{\frac{2}{2+\varsigma}})\log m_n)\vee m_n} \\ +s_n^{-1}\sqrt{\frac{(\xi_n^2+n^{2/(2+\varsigma)})m_n\log(m_n)}{n}}+\frac{\xi_n m_n}{\sqrt n s_n}~;
\end{multline}
(c) $(\xi_n^{3}\sqrt{\frac{\log m_n}{n}}\big)^{\frac{1}{2}}=o(1)$.

\noindent (vii) (a) $\{W_i\}_{i=1}^\infty$ is an i.i.d. sequence of random variables; (b) $\{W_i\}_{i=1}^n$ are independent of $\{X_i\}_{i=1}^n$ for all $n$; (c) $E[W_1]=0$, $\mathrm{Var}(W_1)=1$ and $E[|W_1|^{3}]<\infty$.

\end{ass}

Assumptions \ref{Ass: Slutsky}(i)-(a)(b)(d) are standard simplifying restrictions on the data generating process, while Assumption \ref{Ass: Slutsky}(i)-(c) imposes a uniform bound on $\|g_P\|_{1,\infty}$ that arises naturally from the consideration of the Slutsky matrix (which involves the derivatives of $g_P$). The $\sqrt n$-consistency of $\hat\Gamma_n$ required by Assumption \ref{Ass: Slutsky} is well-known in the literature \citep{DonaldNewey1994Semilinear,AiChen2003Efficient,ChenChristensen2018SupNormOptimal}, and is imposed to simplify the proof. The uniform boundedness of the $(2+\varsigma)$-th moment of $Z$ and $V^{(2)}$ is required to apply a uniform law of large number for matrices. The remaining assumptions are naturally adapted from Assumption \ref{Ass: series 2SLS estimation}.

Given Assumption \ref{Ass: Slutsky}, we may formalize the strong approximations in Proposition \ref{Pro: Slutsky, strong approx} as follows. First, the coupling variable $\mathbb W_{n,P}$ that appears in \eqref{Eqn: Slutsky, strong approx-i} is $(\Omega_{n,P}G_{n,P})^\transpose h^{k_n}$ with $\Omega_{n,P}\in\mathbf M^{k_n\times (m_n+d_{v_2})}$ the upper block of $s_n\xi_n^{-1}(\bar\Psi_{n,P}^{-1/2}\bar\Pi_{n,P})_l^-\bar\Psi_{n,P}^{-1/2}$ and $G_{n,P}\in\mathbf M^{(m_n+d_{v_2})\times d_q}$ a centered Gaussian matrix that has the same covariance functional as the random matrix $\bar b^{m_n}(V)U^\transpose$. Second, the coupling rate $c_n$ for Assumption \ref{Ass: strong approx}(i) can be taken to be $\varpi_n\ell_n$ for any $\ell_n\to\infty$ (slowly). Third, the coupling rate for Assumption \ref{Ass: strong approx}(ii) can be taken to be $\varpi_n'\ell_n$ with
\begin{align}
\varpi_n'\equiv\varpi_n +  \frac{n^{\frac{1}{2+\varsigma}}}{\sqrt n} + (\xi_n^{3}\sqrt{\frac{\log m_n}{n}}\big)^{\frac{1}{2}}~.
\end{align}
To meet Assumptions \ref{Ass: strong approx}(i) and (ii) simultaneously, we thus take $c_n=\varpi_n'\ell_n$. Finally, the copy $\bar{\mathbb Z}_{n,P}$ for Assumption \ref{Ass: strong approx}(ii) is of the form $(\Omega_{n,P}\bar G_{n,P})^\transpose h^{k_n}$ with $\bar G_{n,P}$ a copy of $G_{n,P}$ that is independent of $\{X_i\}_{i=1}^n$. In what follows, these configurations are understood to be part of Proposition \ref{Pro: Slutsky, strong approx}.

\noindent{\sc Proof of Proposition \ref{Pro: Slutsky, strong approx}:} By linearity of the differential operators $D_p$ and $D_y$, with the help of simple algebra, we may obtain that
\begin{align}\label{Eqn: Slutsky strong approx, aux1}
\psi(\hat g_n)-\psi(g_0)-&\psi_{g_0}'(\hat g_n-g_0)=[D_y(\hat g_n-g_0)](\hat g_n-g_0)^\transpose + (\hat g_n-g_0)(\hat g_n-g_0)^\transpose~.
\end{align}
Since $\|\theta\|_{\mathbf H}\le \|\theta\|_\infty$ for any $\theta\in\mathbf H$, we in turn have from \eqref{Eqn: Slutsky strong approx, aux1} that
\begin{multline}\label{Eqn: Slutsky strong approx, aux2}
\|\psi(\hat g_n)-\psi(g_0)-\psi_{g_0}'(\hat g_n-g_0)\|_{\mathbf H}\\
\lesssim  \|\hat g_n -g_0\|_\infty\|D_y\hat g_n -D_yg_0\|_\infty+\|\hat g_n-g_0\|_\infty^2
\lesssim \|\hat g_n-g_0\|_{1,\infty}^2 ~,
\end{multline}
uniformly in $P\in\mathbf P$. By Assumption \ref{Ass: Slutsky}(vi)-(b), Lemma \ref{Lem: Slutsky rate} and the triangle inequality, it follows from \eqref{Eqn: Slutsky strong approx, aux2} that, uniformly in $P\in\mathbf P$,
\begin{align}\label{Eqn: Slutsky strong approx, aux3}
\|r_n\{\psi(\hat g_n)-\psi(g_0)-\psi_{g_0}'(\hat g_n-g_0)\}\|_{\mathbf H}=O_p(\varpi_n)~.
\end{align}
By simple algebra and Assumption \ref{Ass: Slutsky}(i)-(c), $\|\psi_{g_0}'(h)\|_{\mathbf H} \lesssim \|h\|_{1,\infty}$ for all $h\in C_{\mathrm b}^1(\mathcal T)$. Linearity of $h\mapsto \psi_{g_0}'(h)$, Assumption \ref{Ass: Slutsky}(vi)-(b) and Lemma \ref{Lem: Slutsky coupling} then imply that
\begin{align}\label{Eqn: Slutsky strong approx, aux4}
\|\psi_{g_0}'(r_n\{\hat g_n-g_0\})-\psi_{g_0}'(\mathbb W_{n,P})\|_{\mathbf H} \lesssim \|r_n\{\hat g_n-g_0\} - \mathbb W_{n,P} \|_{1,\infty}=O_p(\varpi_n)~,
\end{align}
uniformly in $P\in\mathbf P$. The first claim of the proposition then follows from combining results \eqref{Eqn: Slutsky strong approx, aux2} and \eqref{Eqn: Slutsky strong approx, aux4} with the triangle inequality.

The proof of the second claim consists of several steps as in the proof of Proposition \ref{Pro: bootstrap, series npiv}. First, by Assumption \ref{Ass: Slutsky}(vii) and the triangle inequality, we have
\begin{multline}\label{Eqn: Slutsky strong approx, aux5}
\hat\Delta_n\equiv  \sum_{i=1}^{n} E[\|W_i\bar b^{m_n}(V_i)\hat U_i^\transpose/\sqrt n\|^3|\{X_i\}_{i=1}^n ]\lesssim n^{-3/2}\sum_{i=1}^{n}\|\bar b^{m_n}(V_i)\|^3\|\hat U_i\|^3\\
\lesssim n^{-3/2}\sum_{i=1}^{n}\|\bar b^{m_n}(V_i)\|^3\|\hat U_i-U_i\|^3+n^{-3/2}\sum_{i=1}^{n}\|\bar b^{m_n}(V_i)\|^3\|U_i\|^3
\end{multline}
By Lemmas \ref{Lem: basis norm} and \ref{Lem: Slutsky residuals} and Assumptions \ref{Ass: Slutsky}(i)-(a), (iv)-(a)(b), (v)-(b) and (vi)-(a)(b), it follows from \eqref{Eqn: Slutsky strong approx, aux5} that, uniformly in $P\in\mathbf P$,
\begin{align}\label{Eqn: Slutsky strong approx, aux6}
\hat\Delta_n= n^{-1/2} \xi_n  O_p(m_n)o_p(1) + n^{-1/2}\xi_n O_p(m_n) = O_p(\frac{\xi_nm_n}{\sqrt n})~.
\end{align}
For notational simplicity, define $S_n\equiv \frac{1}{\sqrt n}\sum_{i=1}^{n}W_i\bar b^{m_n}(V_i)\hat U_i^\transpose$. Then by Assumption \ref{Ass: Slutsky}(vii), we may invoke Theorem 10.10 in \citet{Pollard2002User} to conclude that, for any $\epsilon>0$, there is some $\hat G_n\in\mathbf M^{(m_n+d_{v_2})\times d_q}$ such that $\hat G_n$ shares the same covariance functional as $S_n$ conditional on $\{X_i\}_{i=1}^n$ and
\begin{align}\label{Eqn: Slutsky strong approx, aux7}
P(\|S_n-\hat G_n\|>3\epsilon|\{X_i\}_{i=1}^n)\lesssim \hat\eta_n(1+\frac{|\log(1/\hat\eta_n)|}{(m_n+d_{v_2})d_q})~,
\end{align}
where $\hat\eta_n\equiv \hat\Delta_n (m_n+d_{v_2})d_q\epsilon^{-3}$. By arguments analogous to those leading to \eqref{Eqn: bootstrap, series npiv, aux4}, we may in turn conclude from \eqref{Eqn: Slutsky strong approx, aux6} and \eqref{Eqn: Slutsky strong approx, aux7} that, uniformly in $P\in\mathbf P$,
\begin{align}\label{Eqn: Slutsky strong approx, aux8}
\|S_n-\hat G_n\|=O_p((\frac{\xi_nm_n}{\sqrt n})^{1/3})~.
\end{align}
Define $\tilde{\mathbb W}_n\equiv(\hat\Omega_n\hat G_n)^\transpose h^{k_n}$, and let $\Omega_{n,P}\in\mathbf M^{k_n\times (m_n+d_{v_2})}$ be the upper block of
\begin{align}
s_n\xi_n^{-1}(\bar\Psi_{n,P}^{-1/2}\bar\Pi_{n,P})_l^-\bar\Psi_{n,P}^{-1/2}~.
\end{align}
By the triangle inequality, Fact 11.16.9 in \citet{Bernstein2018Matrix}, results \eqref{Eqn: Slutsky Bahadur, aux10} and \eqref{Eqn: Slutsky Bahadur, aux12}, and Assumptions \ref{Ass: Slutsky}(iv)-(a) and (vi)-(b), we note that, uniformly in $P\in\mathbf P$,
\begin{multline}\label{Eqn: Slutsky strong approx, aux9}
\|\hat\Omega_n\|_o\le \|\hat\Omega_n-\Omega_{n,P}\|_o+\|\Omega_{n,P}\|_o\\
\le s_n\xi_n^{-1} O_p(s_n^{-2}\sqrt{\frac{(\xi_n^2+n^{2/(2+\varsigma)})\log(m_n)}{n}}) + s_n\xi_n^{-1} O(s_n^{-1})=O_p(\xi_n^{-1})~.
\end{multline}
Since $\hat{\mathbb W}_n=(\hat\Omega_n S_n)^\transpose h^{k_n}$ by definition, it follows from \eqref{Eqn: Slutsky strong approx, aux8}, \eqref{Eqn: Slutsky strong approx, aux9} and Assumption \ref{Ass: Slutsky}(iii)-(b) that, uniformly in $P\in\mathbf P$,
\begin{multline}\label{Eqn: Slutsky strong approx, aux10}
\|\hat{\mathbb W}_n-\tilde{\mathbb W}_n\|_{1,\infty} \le \|\hat\Omega_n\|_o\|S_n-\hat G_n\|\|h^{k_n}\|_{1,\infty}\\
\le O_p(\xi_n^{-1}) O_p((\frac{\xi_nm_n}{\sqrt n})^{1/3}) \xi_n = O_p((\frac{\xi_nm_n}{\sqrt n})^{1/3})~.
\end{multline}
Since $\|\psi_g'(h)\|_{\mathbf H}\lesssim (\|g\|_{1,\infty}\vee 1)\|h\|_{1,\infty}$ for all $g,h\in C_{\mathrm b}^1(\mathcal T)$ by \eqref{Eqn: Slutsky derivative} and the triangle inequality, we may obtain by linearity of $h\mapsto\psi_{\hat g_n}'(h)$, the triangle inequality, Lemma \ref{Lem: Slutsky rate}, Assumptions \ref{Ass: Slutsky}(i)-(c) and (vi)-(b), and \eqref{Eqn: Slutsky strong approx, aux10} that, uniformly in $P\in\mathbf P$,
\begin{multline}\label{Eqn: Slutsky strong approx, aux11}
\|\psi_{\hat g_n}'(\hat{\mathbb W}_n)-\psi_{\hat g_n}'(\tilde{\mathbb W}_n)\|_{\mathbf H}\lesssim (\|\hat g_n\|_{1,\infty}\vee 1)\|\hat{\mathbb W}_n-\tilde{\mathbb W}_n\|_{1,\infty}\\
\lesssim \big((\|\hat g_n-g_P\|_{1,\infty}+\|g_P\|_{1,\infty})\vee 1\big)\|\hat{\mathbb W}_n-\tilde{\mathbb W}_n\|_{1,\infty}=O_p((\frac{\xi_nm_n}{\sqrt n})^{1/3})~.
\end{multline}

Next, by Assumption \ref{Ass: Slutsky}(vii) and the elementary formula $\mathrm{vec}(ab^\transpose)=b\otimes a$ for generic vectors $a$ and $b$, we may compute the conditional variance matrix of $\mathrm{vec}(S_n^\transpose)$ as
\begin{align}\label{Eqn: Slutsky strong approx, aux12}
\mathrm{Var}\left(\mathrm{vec}(S_n^\transpose) |\{X_i\}_{i=1}^n\right)
=\frac{1}{n}\sum_{i=1}^{n}(\bar b^{m_n}(V_i)\otimes\hat U_i)(\bar b^{m_n}(V_i)\otimes\hat U_i)^\transpose=\hat\Sigma_n ~.
\end{align}
Thus, \eqref{Eqn: Slutsky strong approx, aux12} implies $\mathrm{vec}(\hat G_n^\transpose)\sim N(0,\hat\Sigma_n)$ conditional on $\{X_i\}_{i=1}^n$. By Lemma \ref{Lem: Gaussian coupling}, there exists some $N_{p_n}\sim N(0,I_{p_n})$ with $p_n\equiv(m_n+d_{v_2})d_q$ such that $N_{p_n}$ is independent of $\{X_i\}_{i=1}^n$ and $\mathrm{vec}(\hat G_n^\transpose)=\hat\Sigma_n^{1/2} N_{p_n}$ almost surely. Let $\bar G_{n,P}\in \mathbf M^{(m_n+d_{v_2})\times d_q}$ be such that $\mathrm{vec}(\bar G_{n,P}^\transpose)=\Sigma_{n,P}^{1/2}N_{p_n}$, and set $\bar{\mathbb W}_{n,P}=(\Omega_{n,P}\bar G_{n,P})^\transpose h^{k_n}$. By construction, $\bar{\mathbb W}_{n,P}$ is centered Gaussian in $\mathbf H$, independent of $\{X_i\}_{i=1}^n$, and a copy of $\mathbb W_{n,P}$.

With $\bar{\mathbb W}_{n,P}$ in hand, we note by the triangle inequality that
\begin{multline}\label{Eqn: Slutsky strong approx, aux13}
\|\psi_{\hat g_n}'(\tilde{\mathbb W}_n) - \psi_{g_0}'(\bar{\mathbb W}_{n,P})\|_{\mathbf H}\le \|\psi_{\hat g_n}'(\tilde{\mathbb W}_n) - \psi_{\hat g_n}'(\bar{\mathbb W}_{n,P})\|_{\mathbf H}\\
+\|\psi_{\hat g_n}'(\bar{\mathbb W}_{n,P}) - \psi_{g_0}'(\bar{\mathbb W}_{n,P})\|_{\mathbf H}~.
\end{multline}
For the first term on the right side of \eqref{Eqn: Slutsky strong approx, aux13}, the triangle inequality again gives
\begin{align}\label{Eqn: Slutsky strong approx, aux14}
\|\psi_{\hat g_n}'(\tilde{\mathbb W}_n) - \psi_{\hat g_n}'(\bar{\mathbb W}_{n,P})\|_{\mathbf H} & \le \|D_p(\tilde{\mathbb W}_n - \bar{\mathbb W}_{n,P})\|_{\mathbf H} +\|(D_y\hat g_n)(\tilde{\mathbb W}_n - \bar{\mathbb W}_{n,P})^\transpose\|_{\mathbf H}\notag \\
 &\quad +\|(D_y(\tilde{\mathbb W}_n - \bar{\mathbb W}_{n,P}))\hat g_n^\transpose\|_{\mathbf H} + \|\hat g_n (\tilde{\mathbb W}_n - \bar{\mathbb W}_{n,P})^\transpose\|_{\mathbf H} \notag\\
 &\quad + \|(\tilde{\mathbb W}_n - \bar{\mathbb W}_{n,P})\hat g_n^\transpose \|_{\mathbf H} +\|\mathrm{diag}(\tilde{\mathbb W}_n - \bar{\mathbb W}_{n,P})\|_{\mathbf H}~.
\end{align}
Consider $\|D_p(\tilde{\mathbb W}_n - \bar{\mathbb W}_{n,P})\|_{\mathbf H}$ first. By definition, we note that
\begin{align}\label{Eqn: Slutsky strong approx, aux15}
\|D_p(\tilde{\mathbb W}_n - \bar{\mathbb W}_{n,P})\|_{\mathbf H} & =\{\int_{\mathcal T}\| \hat G_n^\transpose\hat\Omega_n^\transpose D_ph^{k_n}(t)-\bar G_{n,P}^\transpose\Omega_{n,P}^\transpose D_ph^{k_n}(t)\|^2\, \mathrm dt\}^{1/2}~.
\end{align}
By the simple fact $\|A\|^2=\mathrm{tr}(\mathrm{vec}(A)\mathrm{vec}(A)^\transpose)$ for any generic matrix $A$, Fact 9.4.7 in \citet{Bernstein2018Matrix}, $\mathrm{vec}(\hat G_n^\transpose)=\hat\Sigma_n^{1/2} N_{p_n}$ almost surely, and the definition of $\bar G_{n,P}$, we may in turn obtain that, almost surely,
\begin{align}\label{Eqn: Slutsky strong approx, aux16}
\| \hat G_n^\transpose\hat\Omega_n^\transpose & D_ph^{k_n}(t)-\bar G_{n,P}^\transpose\Omega_{n,P}^\transpose D_ph^{k_n}(t)\|^2\notag \\
&=\mathrm{tr}\Big(\left[((\hat\Omega_n^\transpose D_ph^{k_n}(t))^\transpose \otimes I_{d_q} )\hat\Sigma_n^{1/2} - ((\Omega_{n,P}^\transpose D_ph^{k_n}(t))^\transpose\otimes I_{d_q} )\Sigma_{n,P}^{1/2}\right]N_{p_n}N_{p_n}^\transpose \notag\\
&\quad \cdot \left[((\hat\Omega_n^\transpose D_ph^{k_n}(t))^\transpose \otimes I_{d_q} )\hat\Sigma_n^{1/2} - ((\Omega_{n,P}^\transpose D_ph^{k_n}(t))^\transpose\otimes I_{d_q} )\Sigma_{n,P}^{1/2}\right]^\transpose\Big)~.
\end{align}
Given results \eqref{Eqn: Slutsky strong approx, aux15} and \eqref{Eqn: Slutsky strong approx, aux16}, we may obtain by Jensen's inequality, $N_{p_n}\sim N(0,I_{p_n})$ being independent of $\{X_i\}_{i=1}^n$, and Assumption  \ref{Ass: Slutsky}(i)-(b) that
\begin{multline}\label{Eqn: Slutsky strong approx, aux17}
E[\|D_p(\tilde{\mathbb W}_n - \bar{\mathbb W}_{n,P})\|_{\mathbf H}|\{X_i\}_{i=1}^n]\\
 \lesssim  \sup_{t\in\mathcal T}\|((\hat\Omega_n^\transpose D_ph^{k_n}(t))^\transpose \otimes I_{d_q} )\hat\Sigma_n^{1/2} - ((\Omega_{n,P}^\transpose  D_ph^{k_n}(t))^\transpose\otimes I_{d_q} )\Sigma_{n,P}^{1/2}\|~.
\end{multline}
By the triangle inequality and the simple fact $\|AB\|\le \|A\|\|B\|_o$ for generic matrices $A$ and $B$ such that $AB$ is defined, we obtain that: for each $t\in\mathcal T$,
\begin{align}\label{Eqn: Slutsky strong approx, aux18}
\|((\hat\Omega_n^\transpose D_ph^{k_n}(t))^\transpose & \otimes I_{d_q} )\hat\Sigma_n^{1/2} - ((\Omega_{n,P}^\transpose  D_ph^{k_n}(t))^\transpose\otimes I_{d_q} )\Sigma_{n,P}^{1/2}\| \notag\\
& \le \|(\hat\Omega_n^\transpose D_ph^{k_n}(t))^\transpose \otimes I_{d_q} \|\|\hat\Sigma_n^{1/2} - \Sigma_{n,P}^{1/2}\|_o\notag\\
&\quad + \|(\hat\Omega_n^\transpose D_ph^{k_n}(t))^\transpose \otimes I_{d_q}  - (\Omega_{n,P}^\transpose  D_ph^{k_n}(t))^\transpose\otimes I_{d_q} \| \|\Sigma_{n,P}^{1/2}\|_o~.
\end{align}
To evaluate the upper bound in \eqref{Eqn: Slutsky strong approx, aux18}, we need several facts. First, by Fact 11.10.95 in \citet{Bernstein2018Matrix}, result \eqref{Eqn: Slutsky strong approx, aux9} and Assumption \ref{Ass: Slutsky}(iii)-(b), we note that
\begin{multline}\label{Eqn: Slutsky strong approx, aux19}
\sup_{t\in\mathcal T}\|(\hat\Omega_nD_ph^{k_n}(t))^\transpose \otimes I_{d_q} \| \le\sqrt{d_q}\sup_{t\in\mathcal T}\|(\hat\Omega_nD_ph^{k_n}(t))^\transpose \|\\ \lesssim \|\hat\Omega_n\|_o \|h^{k_n}\|_{1,\infty} \le\frac{s_n}{\xi_n}O_p(s_n^{-1}) O_p(1) \xi_n= O_p(1)~,
\end{multline}
uniformly in $P\in\mathbf P$. Second, by Facts 11.10.95 and 11.16.9 in \citet{Bernstein2018Matrix}, result \eqref{Eqn: Slutsky Bahadur, aux12} and Assumption \ref{Ass: Slutsky}(vi)-(b), we also have: uniformly in $P\in\mathbf P$,
\begin{multline}\label{Eqn: Slutsky strong approx, aux20}
\sup_{t\in\mathcal T}\|(\hat\Omega_nD_ph^{k_n}(t))^\transpose \otimes I_{d_q}  - (\Omega_{n,P} D_ph^{k_n}(t))^\transpose\otimes I_{d_q} \|\\
\le \|\hat\Omega_n-\Omega_{n,P}\|_o \|h^{k_n}\|_{1,\infty}\sqrt{d_q} \le O_p(s_n^{-1}\sqrt{\frac{(\xi_n^2+n^{2/(2+\varsigma)})\log(m_n)}{n}})~.
\end{multline}
Third, by Jensen's inequality \citep[p.40]{Tropp2015Introduction}, result \eqref{Eqn: npiv, variance, aux4} and Assumption \ref{Ass: Slutsky}(v)-(b), we note that, for some absolute constant $\bar\sigma>0$,
\begin{align}\label{Eqn: Slutsky strong approx, aux21}
\|E_P[UU^\transpose|V]\|_o\le E_P[\|UU^\transpose\|_o|V]=E_P[\|U\|^2|V]<\bar\sigma~,
\end{align}
almost surely. Result \eqref{Eqn: Slutsky strong approx, aux21} and Lemma 10.4.1 in \citet{Bernstein2018Matrix} imply that $0\le E_P[UU^\transpose|V]\le \bar\sigma I_{d_q}$ almost surely, and hence, by Fact 10.25.38 in \citet{Bernstein2018Matrix},
\begin{align}\label{Eqn: Slutsky strong approx, aux22}
0\le\Sigma_{n,P}\le \bar\Psi_{n,P}\otimes(\bar\sigma I_{d_q})~.
\end{align}
By Theorem 10.4.9 and Fact 11.10.95 in \citet{Bernstein2018Matrix} and Assumption \ref{Ass: Slutsky}(iv)-(a), we may in turn deduce from result \eqref{Eqn: Slutsky strong approx, aux22} that
\begin{align}\label{Eqn: Slutsky strong approx, aux23}
\sup_{n}\sup_{P\in\mathbf P}\|\Sigma_{n,P}^{1/2}\|_o= \sup_{n}\sup_{P\in\mathbf P}\|\Sigma_{n,P}\|_o^{1/2}\le \sup_{n}\sup_{P\in\mathbf P}\|\bar\Psi_{n,P}\|_o^{1/2}\bar\sigma^{1/2}<\infty~.
\end{align}
Fourth, by Assumption \ref{Ass: Slutsky}(v)-(b), $\sup_{P\in\mathbf P}E_P[\|U\|^3]<\infty$ and hence, by Lemma \ref{Lem: Slutsky variance estimation} and Assumption \ref{Ass: Slutsky}(vi)-(b), we obtain that, uniformly in $P\in\mathbf P$,
\begin{align}\label{Eqn: Slutsky strong approx, aux23a}
\|\hat\Sigma_n^{1/2}- \Sigma_{n,P}^{1/2}\|_o=O_p(\varpi_n + \frac{n^{\frac{1}{2+\varsigma}}}{\sqrt n} + (\xi_n^{3}\sqrt{\frac{\log m_n}{n}}\big)^{\frac{1}{2}})~.
\end{align}
Now, combining results \eqref{Eqn: Slutsky strong approx, aux18}, \eqref{Eqn: Slutsky strong approx, aux19}, \eqref{Eqn: Slutsky strong approx, aux20}, \eqref{Eqn: Slutsky strong approx, aux23} and \eqref{Eqn: Slutsky strong approx, aux23a} with Assumption \ref{Ass: Slutsky}(vi)-(b), we may therefore conclude from \eqref{Eqn: Slutsky strong approx, aux17} that
\begin{align}\label{Eqn: Slutsky strong approx, aux24}
E[\|D_p(\tilde{\mathbb W}_n - \bar{\mathbb W}_{n,P})\|_{\mathbf H}|\{X_i\}_{i=1}^n]= O_p(\varpi_n + \frac{n^{\frac{1}{2+\varsigma}}}{\sqrt n} + (\xi_n^{3}\sqrt{\frac{\log m_n}{n}}\big)^{\frac{1}{2}})~,
\end{align}
uniformly in $P\in\mathbf P$. By Fubini's theorem and Markov's inequality, we may in turn deduce from result \eqref{Eqn: Slutsky strong approx, aux24} that, uniformly in $P\in\mathbf P$,
\begin{align}\label{Eqn: Slutsky strong approx, aux25}
\|D_p(\tilde{\mathbb W}_n - \bar{\mathbb W}_{n,P})\|_{\mathbf H}=O_p(\varpi_n + \frac{n^{\frac{1}{2+\varsigma}}}{\sqrt n} + (\xi_n^{3}\sqrt{\frac{\log m_n}{n}}\big)^{\frac{1}{2}})~.
\end{align}

Treatment of the remaining terms on the right side of \eqref{Eqn: Slutsky strong approx, aux13} is similar. In particular, by arguments analogous to those leading to \eqref{Eqn: Slutsky strong approx, aux17}, we have
\begin{multline}\label{Eqn: Slutsky strong approx, aux26}
E[\|(D_y\hat g_n)(\tilde{\mathbb W}_n - \bar{\mathbb W}_{n,P})^\transpose\|_{\mathbf H}|\{X_i\}_{i=1}^n]\\
 \lesssim  \sup_{t\in\mathcal T}\|\big(I_{d_q}\otimes (D_y\hat g_n(t)h^{k_n}(t)^\transpose \hat\Omega_n \big)\hat\Sigma_n^{1/2} - \big(I_{d_q}\otimes (D_y\hat g_n(t)h^{k_n}(t)^\transpose  \Omega_{n,P} \big)\Sigma_{n,P}^{1/2}\|~.
\end{multline}
Since $\|\hat g_n\|_{1,\infty}=O_p(1)$ by Lemma \ref{Lem: Slutsky rate}, the triangle inequality and Assumptions \ref{Ass: Slutsky}(i)-(c) and (vi)-(b), arguments analogous to those leading to \eqref{Eqn: Slutsky strong approx, aux25} yield
\begin{align}\label{Eqn: Slutsky strong approx, aux27}
\|(D_y\hat g_n)(\tilde{\mathbb W}_n - \bar{\mathbb W}_{n,P})^\transpose\|_{\mathbf H}=O_p(\varpi_n + \frac{n^{\frac{1}{2+\varsigma}}}{\sqrt n} + (\xi_n^{3}\sqrt{\frac{\log m_n}{n}}\big)^{\frac{1}{2}})~,
\end{align}
uniformly in $P\in\mathbf P$. Similar arguments show that other terms on the right side of \eqref{Eqn: Slutsky strong approx, aux13} have the same order as in \eqref{Eqn: Slutsky strong approx, aux27}. This, together with results \eqref{Eqn: Slutsky strong approx, aux11}, \eqref{Eqn: Slutsky strong approx, aux22} and \eqref{Eqn: Slutsky strong approx, aux24}, allows us to obtain that, uniformly in $P\in\mathbf P$,
\begin{align}\label{Eqn: Slutsky strong approx, aux28}
\|\psi_{\hat g_n}'(\tilde{\mathbb W}_n) - \psi_{\hat g_n}'(\bar{\mathbb W}_{n,P})\|_{\mathbf H} = O_p(\varpi_n + \frac{n^{\frac{1}{2+\varsigma}}}{\sqrt n} + (\xi_n^{3}\sqrt{\frac{\log m_n}{n}}\big)^{\frac{1}{2}})~.
\end{align}

Turning to the second term on the right side of \eqref{Eqn: Slutsky strong approx, aux10}, we have by \eqref{Eqn: Slutsky derivative} that
\begin{multline}\label{Eqn: Slutsky strong approx, aux29}
\|\psi_{\hat g_n}'(\bar{\mathbb W}_{n,P}) - \psi_{g_0}'(\bar{\mathbb W}_{n,P})\|_{\mathbf H} \le \|(\hat g_n-g_P) \bar{\mathbb W}_{n,P}^\transpose\|_{\mathbf H} + \|\bar{\mathbb W}_{n,P}(\hat g_n-g_P)^\transpose\|_{\mathbf H}\\
+ \|(D_y\hat g_n-D_yg_P) \bar{\mathbb W}_{n,P}^\transpose\|_{\mathbf H} + \|(D_y\bar{\mathbb W}_{n,P} ) (\hat g_n-g_P)^\transpose \|_{\mathbf H}~.
\end{multline}
By arguments analogous to those leading to \eqref{Eqn: Slutsky strong approx, aux14}, we note that
\begin{multline}\label{Eqn: Slutsky strong approx, aux30}
E[\|(\hat g_n-g_P) \bar{\mathbb W}_{n,P}^\transpose\|_{\mathbf H}|\{X_i\}_{i=1}^n]\le\sup_{t\in\mathcal T}\|\big(I_{d_q}\otimes ((\hat g_n(t)-g_P(t))h^{k_n}(t)^\transpose\Omega_{n,P})\big)\Sigma_{n,P}^{1/2}\| \\
\le \|\hat g_n-g_P\|_\infty \|h^{k_n}\|_\infty \|\Omega_{n,P}\|_o\|\Sigma_{n,P}^{1/2}\|_o \lesssim \|\hat g_n-g_P\|_\infty~,
\end{multline}
uniformly in $P\in\mathbf P$, where we exploited Assumptions \ref{Ass: Slutsky}(iii)-(b) and (iv)-(a), \eqref{Eqn: Slutsky Bahadur, aux10} and \eqref{Eqn: Slutsky strong approx, aux20} for the last step. By Lemma \ref{Lem: Slutsky rate}, Fubini's theorem and Markov's inequality, we thus obtain from \eqref{Eqn: Slutsky strong approx, aux30} that, uniformly in $P\in\mathbf P$,
\begin{multline}\label{Eqn: Slutsky strong approx, aux31}
E[\|(\hat g_n-g_P) \bar{\mathbb W}_{n,P}^\transpose\|_{\mathbf H}] = O_p(s_n^{-1}\xi_n\sqrt{\frac{ m_n}{n}}) \\
+ O_p(s_n^{-1}\delta_n\xi_n\sqrt{\frac{((\xi_n^2+n^{\frac{2}{2+\varsigma}})\log m_n)\vee m_n}{n}}+\delta_n)~.
\end{multline}
Similar arguments show that the remaining terms on the right side of \eqref{Eqn: Slutsky strong approx, aux26} are of the same order as in \eqref{Eqn: Slutsky strong approx, aux31}. Together with Assumption \ref{Ass: Slutsky}(vi)-(b), these imply that
\begin{align}\label{Eqn: Slutsky strong approx, aux32}
\|\psi_{\hat g_n}'(\bar{\mathbb W}_{n,P}) - \psi_{g_0}'(\bar{\mathbb W}_{n,P})\|_{\mathbf H}  = O_p(\varpi_n)~,
\end{align}
uniformly in $P\in\mathbf P$. Combining results \eqref{Eqn: Slutsky strong approx, aux11}, \eqref{Eqn: Slutsky strong approx, aux13}, \eqref{Eqn: Slutsky strong approx, aux28} and \eqref{Eqn: Slutsky strong approx, aux32} with the triangle inequality and Assumption \ref{Ass: Slutsky}(vi)-(b) then yields
\begin{align}\label{Eqn: Slutsky strong approx, aux33}
\|\psi_{\hat g_n}'(\hat{\mathbb W}_n) - \psi_{g_0}'(\bar{\mathbb W}_{n,P})\|_{\mathbf H}  = O_p(\varpi_n + \frac{n^{\frac{1}{2+\varsigma}}}{\sqrt n} + (\xi_n^{3}\sqrt{\frac{\log m_n}{n}}\big)^{\frac{1}{2}})~,
\end{align}
uniformly in $P\in\mathbf P$. This proves the second claim of the proposition. \qed

\begin{lem}\label{Lem: Slutsky rate}
If Assumptions \ref{Ass: Slutsky}(i), (iii), (iv), (v)-(a)(b) and (vi)-(a)(b) hold, then it follows that, uniformly in $P\in\mathbf P$,
\begin{multline}
\|\hat g_n-g_0\|_{1,\infty}  = O_p(s_n^{-1}\xi_n\sqrt{\frac{ m_n}{n}}) \\
+ O_p(s_n^{-1}\delta_n\xi_n\sqrt{\frac{((\xi_n^2+n^{\frac{2}{2+\varsigma}})\log m_n)\vee m_n}{n}}+\delta_n) ~.
\end{multline}
\end{lem}
\noindent{\sc Proof:} This is an immediate consequence of Lemmas \ref{Lem: Slutsky, tilde g} and \ref{Lem: Slutsky variance}. \qed

\begin{lem}\label{Lem: Slutsky Bahadur}
If Assumptions \ref{Ass: Slutsky}(i), (iii), (iv), (v)-(a)(b) and (vi)-(a)(b)(c) hold, then it follows that, uniformly in $P\in\mathbf P$,
\begin{multline}
\begin{bmatrix}
\hat\Lambda_n\\
\hat\Gamma_n
\end{bmatrix}-\begin{bmatrix}
\Lambda_{n,P}\\
\Gamma_{n,P}
\end{bmatrix}=(\bar\Psi_{n,P}^{-1/2}\bar\Pi_{n,P})_l^-\bar\Psi_{n,P}^{-1/2}\frac{1}{n}\bar B_n^\transpose \underline{U}_n \\
+ O_p(s_n^{-1}\delta_n\sqrt{\frac{((\xi_n^2+n^{\frac{2}{2+\varsigma}})\log m_n)\vee m_n}{n}}+s_n^{-2}\sqrt{\frac{(\xi_n^2+n^{2/(2+\varsigma)})m_n\log(m_n)}{n^2}})~.
\end{multline}
\end{lem}
\noindent{\sc Proof:} By Assumptions \ref{Ass: Slutsky}(i)-(a), (iv)-(b) and (v)-(a), and a simple maximal inequality \citep[p.96]{Vaart1996}, we obtain
\begin{align}\label{Eqn: Slutsky Bahadur, aux1}
E_P[\max_{i=1}^n\|\bar b^{m_n}(V_i)\|^2]\le E_P[\max_{i=1}^n\|b^{m_n}(V_i)\|^2]+E_P[\max_{i=1}^n\|V_{2i}\|^2]\lesssim\xi_n^2+ n^{\frac{2}{2+\varsigma}}~.
\end{align}
Given result \eqref{Eqn: Slutsky Bahadur, aux1} and Assumptions \ref{Ass: Slutsky}(i)-(a), (iv)-(a) and (vi)-(a)(b), we may invoke Theorem E.1 in \citet{Kato2013QuasiB} and Markov's inequality to conclude that
\begin{align}\label{Eqn: Slutsky Bahadur, aux2}
\|\bar \Psi_n-\bar\Psi_{n,P}\|_o=O_p(\sqrt{\frac{(\xi_n^2+n^{2/(2+\varsigma)})\log(m_n)}{n}})~,
\end{align}
uniformly in $P\in\mathbf P$. In turn, by result \eqref{Eqn: Slutsky Bahadur, aux2} and Assumption \ref{Ass: Slutsky}(iv)-(a), we obtain by Lemma \ref{Lem: inverse matrix} that, uniformly in $P\in\mathbf P$,
\begin{align}\label{Eqn: Slutsky Bahadur, aux3}
\|\bar \Psi_n^--\bar\Psi_{n,P}^{-1}\|_o=O_p(\sqrt{\frac{(\xi_n^2+n^{2/(2+\varsigma)})\log(m_n)}{n}})~.
\end{align}
Assumption \ref{Ass: Slutsky}(iv)-(a) and result \eqref{Eqn: Slutsky Bahadur, aux3} then allow us to conclude by Proposition 3.2 in \citet{HemmenAndo1980Ideal} that
\begin{multline}\label{Eqn: Slutsky Bahadur, aux4}
\|\bar\Psi_n^{-1/2}-\bar\Psi_{n,P}^{-1/2}\|_o
\le \frac{1}{\{\lambda_{\min}(\bar\Psi_{n,P}^{-1})\}^{1/2}}\|\bar\Psi_n^--\bar\Psi_{n,P}^{-1}\|_o\\ \lesssim O_p(\sqrt{\frac{(\xi_n^2+n^{2/(2+\varsigma)})\log(m_n)}{n}})~,
\end{multline}
uniformly in $P\in\mathbf P$. Next, by Assumptions \ref{Ass: Slutsky}(iii)-(a) and (v)-(a), Theorem III.2.9 in \citet{Bhatia1997Matrix}, Jensen's inequality \citep[p.40]{Tropp2015Introduction} and result \eqref{Eqn: npiv, variance, aux4}, we note that, uniformly in $n$ and $P\in\mathbf P$,
\begin{multline}\label{Eqn: Slutsky Bahadur, aux5}
\|\bar\Phi_{n,P}\|_o\le \|E_P[h^{k_n}(T)h^{k_n}(T)^\transpose]\|_o + \|E_P[ZZ^\transpose]\|_o \\
\le \|E_P[h^{k_n}(T)h^{k_n}(T)^\transpose]\|_o + E_P[\|Z\|^2] <\infty~.
\end{multline}
Given Assumptions \ref{Ass: Slutsky}(i)-(a), (iii)-(b), (v)-(a) and (vi)-(a)(b) and results \eqref{Eqn: Slutsky Bahadur, aux1} and \eqref{Eqn: Slutsky Bahadur, aux5}, it follows by arguments analogous to those leading to \eqref{Eqn: Slutsky Bahadur, aux2} but this time using Corollary E.1 in \citet{Kato2013QuasiB} that, uniformly in $P\in\mathbf P$,
\begin{align}\label{Eqn: Slutsky Bahadur, aux6}
\|\bar \Pi_n-\bar\Pi_{n,P}\|_o=O_p(\sqrt{\frac{(\xi_n^2+n^{2/(2+\varsigma)})\log(m_n)}{n}})~.
\end{align}
By result \eqref{Eqn: Slutsky Bahadur, aux5}, Assumption \ref{Ass: Slutsky}(iv)-(a) and Lemma \ref{Lem: covariance bound}, we have
\begin{align}\label{Eqn: Slutsky Bahadur, aux6a}
\sup_{n\in\mathbf N}\sup_{P\in\mathbf P}\|\bar\Pi_{n,P}\|_o<\infty~.
\end{align}
By Assumption \ref{Ass: Slutsky}(iv)-(a) and results \eqref{Eqn: Slutsky Bahadur, aux4}, \eqref{Eqn: Slutsky Bahadur, aux6} and \eqref{Eqn: Slutsky Bahadur, aux6a}, we may thus apply Lemma \ref{Lem: matrix product} to conclude that, uniformly in $P\in\mathbf P$,
\begin{align}\label{Eqn: Slutsky Bahadur, aux7}
\|\bar\Psi_n^{-1/2}\bar\Pi_n-\bar\Psi_{n,P}^{-1/2}\bar\Pi_{n,P}\|_o=O_p(\sqrt{\frac{(\xi_n^2+n^{2/(2+\varsigma)})\log(m_n)}{n}})~.
\end{align}

By Assumption (vi)-(a) and Corollary 11.6.5 in \citet{Bernstein2018Matrix}, we have
\begin{align}\label{Eqn: Slutsky Bahadur, aux8}
\sigma_{\min} (\bar\Psi_{n,P}^{-1/2}\bar\Pi_{n,P}) \ge \sigma_{\min}(\bar\Psi_{n,P}^{-1/2})\sigma_{\min}(\bar\Pi_{n,P})\gtrsim s_n~,
\end{align}
uniformly in $P\in\mathbf P$, where the second inequality follows by Assumption \ref{Ass: Slutsky}(iv)-(a)(c). Results \eqref{Eqn: Slutsky Bahadur, aux7} and \eqref{Eqn: Slutsky Bahadur, aux8}, Assumption \ref{Ass: Slutsky}(vi)-(b) and Lemma \ref{Lem: inverse matrix} then imply
\begin{align}\label{Eqn: Slutsky Bahadur, aux9}
\|(\bar\Psi_n^{-1/2}\bar\Pi_n)_l^--(\bar\Psi_{n,P}^{-1/2}\bar\Pi_{n,P})_l^-\|_o=O_p(s_n^{-2}\sqrt{\frac{(\xi_n^2+n^{2/(2+\varsigma)})\log(m_n)}{n}})~,
\end{align}
uniformly in $P\in\mathbf P$. By Fact 8.3.33 in \citet{Bernstein2018Matrix} and result \eqref{Eqn: Slutsky Bahadur, aux8}, we obtain that, uniformly in $P\in\mathbf P$,
\begin{align}\label{Eqn: Slutsky Bahadur, aux10}
\|(\bar\Psi_{n,P}^{-1/2}\bar\Pi_{n,P})_l^-\|_o\le \sigma_{\min}(\bar\Psi_{n,P}^{-1/2}\bar\Pi_{n,P})^{-1}=O(s_n^{-1})~.
\end{align}
Moreover, by result \eqref{Eqn: Slutsky Bahadur, aux6a}, we note that, uniformly in $n\in\mathbf N$ and $P\in\mathbf P$,
\begin{align}\label{Eqn: Slutsky Bahadur, aux11}
s_n\le \sigma_{\min}(\bar\Pi_{n,P})\le \sigma_{\max}(\bar\Pi_{n,P})<\infty~.
\end{align}
By results \eqref{Eqn: Slutsky Bahadur, aux4}, \eqref{Eqn: Slutsky Bahadur, aux9}, \eqref{Eqn: Slutsky Bahadur, aux10} and \eqref{Eqn: Slutsky Bahadur, aux11} and Assumption \ref{Ass: Slutsky}(iv)-(a)(b), we may then obtain by Lemma \ref{Lem: matrix product} that, uniformly in $P\in\mathbf P$,
\begin{multline}\label{Eqn: Slutsky Bahadur, aux12}
\|(\bar\Psi_n^{-1/2}\bar\Pi_n)_l^-\bar\Psi_n^{-1/2}-(\bar\Psi_{n,P}^{-1/2}\bar\Pi_{n,P})_l^-\bar\Psi_{n,P}^{-1/2}\|_o\\
=O_p(s_n^{-2}\sqrt{\frac{(\xi_n^2+n^{2/(2+\varsigma)})\log(m_n)}{n}})~.
\end{multline}

Next, define the event $\mathcal E_n $ as
\begin{align}\label{Eqn: Slutsky Bahadur, aux13}
\mathcal E_n \equiv\{\bar\Psi_n^{-1/2}\bar\Pi_n\text{ has full column rank}\}~.
\end{align}
Then, \eqref{Eqn: Slutsky Bahadur, aux7}, \eqref{Eqn: Slutsky Bahadur, aux8}, Assumption \ref{Ass: Slutsky}(vi)-(b) and Lemma \ref{Lem: inverse matrix} imply that
\begin{align}\label{Eqn: Slutsky Bahadur, aux14}
\limsup_{n\to\infty}\sup_{P\in\mathbf P}P(\mathcal E_n ^c)=0~.
\end{align}
Under $\mathcal E_n $, we have $(\bar\Psi_n^{-1/2}\bar\Pi_n)_l^- \bar\Psi_n^{-1/2}\bar\Pi_n = I_{k_n+d_z}$ and hence, by simple algebra,
\begin{align}\label{Eqn: Slutsky Bahadur, aux15}
\begin{bmatrix}
\hat\Lambda_n\\
\hat\Gamma_n
\end{bmatrix}-\begin{bmatrix}
\Lambda_{n,P}\\
\Gamma_{n,P}
\end{bmatrix}=(\bar\Psi_n^{-1/2}\bar\Pi_n)_l^-\bar\Psi_n^{-1/2}\frac{1}{n}\bar B_n^\transpose \underline{U}_n + (\bar\Psi_n^{-1/2}\bar\Pi_n)_l^-\bar\Psi_n^{-1/2}\frac{1}{n}\bar B_n^\transpose A_{n,P}~.
\end{align}
By Jensen's inequality and Assumption \ref{Ass: Slutsky}(i)-(a), we have
\begin{multline} \label{Eqn: Slutsky Bahadur, aux16}
E_P[\|\frac{\bar B_n^\transpose \underline U_n}{n}\|]\le \{\frac{1}{n}\mathrm{tr}(E_P[\bar b^{m_n}(V)\bar b^{m_n}(V)^\transpose \|U\|^2 ])\}^{1/2}\\
 \lesssim \{\frac{1}{n}\mathrm{tr}(E_P[\bar b^{m_n}(V)\bar b^{m_n}(V)^\transpose ])\}^{1/2}
\lesssim  \sqrt{\frac{m_n}{n}}~,
\end{multline}
where the second inequality follows by Assumption \ref{Ass: Slutsky}(v)-(b) and Corollary 10.4.10-(i) in \citet{Bernstein2018Matrix}, and the third one by Assumptions \ref{Ass: Slutsky}(iv)-(a). It follows from results \eqref{Eqn: Slutsky Bahadur, aux12} and \eqref{Eqn: Slutsky Bahadur, aux16} that, uniformly in $P\in\mathbf P$,
\begin{multline}\label{Eqn: Slutsky Bahadur, aux17}
\|[(\bar\Psi_n^{-1/2}\bar\Pi_n)_l^-\bar\Psi_n^{-1/2}-(\bar\Psi_{n,P}^{-1/2}\bar\Pi_{n,P})_l^-\bar\Psi_{n,P}^{-1/2}]\frac{\bar B_n^\transpose U_n}{n}\|_o\\
=O_p(s_n^{-2}\sqrt{\frac{(\xi_n^2+n^{2/(2+\varsigma)})m_n\log(m_n)}{n^2}})~.
\end{multline}
Moreover, simple algebra reveals that, under $\mathcal E_n$,
\begin{align}\label{Eqn: Slutsky Bahadur, aux18}
(\bar\Psi_n^{-1/2}\bar\Pi_n)_l^-\bar\Psi_n^{-1/2}\frac{1}{n}\bar B_n^\transpose A_{n,P} = \begin{bmatrix}
                                                                                            \tilde\Lambda_n\\
                                                                                            \tilde\Gamma_n
                                                                                          \end{bmatrix}-\begin{bmatrix}
                                                                                                          \Lambda_{n,P}\\
                                                                                                          \Gamma_{n,P}
                                                                                                        \end{bmatrix}~.
\end{align}
It follows from results \eqref{Eqn: Slutsky Bahadur, aux14}, \eqref{Eqn: Slutsky Bahadur, aux18} and \eqref{Eqn: Slutsky bias, aux3} that
\begin{align}\label{Eqn: Slutsky Bahadur, aux19}
\|(\bar\Psi_n^{-1/2}\bar\Pi_n)_l^-\bar\Psi_n^{-1/2}\frac{1}{n}\bar B_n^\transpose A_{n,P}\|_o=O_p(s_n^{-1}\delta_n\sqrt{\frac{((\xi_n^2+n^{\frac{2}{2+\varsigma}})\log m_n)\vee m_n}{n}})~,
\end{align}
uniformly in $P\in\mathbf P$. The conclusion of the lemma then follows by combining results \eqref{Eqn: Slutsky Bahadur, aux14}, \eqref{Eqn: Slutsky Bahadur, aux15}, \eqref{Eqn: Slutsky Bahadur, aux17} and \eqref{Eqn: Slutsky Bahadur, aux19} with the triangle inequality. \qed

\begin{lem}\label{Lem: Slutsky coupling}
If Assumptions \ref{Ass: Slutsky}(i)-(a)(b), (iii), (iv), (v)-(a)(b) and (vi)-(a)(b)(c) hold, then it follows that, for $r_n\equiv\sqrt n s_n/\xi_n$,
\begin{multline}
\|r_n\{\hat g_n-g_P\}- \mathbb W_{n,P}\|_{1,\infty}= O_p( (\frac{\xi_nm_n^2}{\sqrt n})^{1/3} + \frac{\sqrt n s_n\delta_n}{\xi_n}) \\
+ O_p( \delta_n \sqrt{((\xi_n^2+n^{\frac{2}{2+\varsigma}})\log m_n)\vee m_n}+s_n^{-1}\sqrt{\frac{(\xi_n^2+n^{2/(2+\varsigma)})m_n\log(m_n)}{n}})~,
\end{multline}
uniformly in $P\in\mathbf P$, where $\mathbb W_{n,P}\equiv (\Omega_{n,P}G_{n,P})^\transpose h^{k_n}$ with $\Omega_{n,P}\in\mathbf M^{k_n\times (m_n+d_{v_2})}$ the upper block of $s_n\xi_n^{-1}(\bar\Psi_{n,P}^{-1/2}\bar\Pi_{n,P})_l^-\bar\Psi_{n,P}^{-1/2}$ and $G_{n,P}\in\mathbf M^{(m_n+d_{v_2})\times d_q}$ a Gaussian matrix that has zero mean and the same covariance structure as $\bar b^{m_n}(V)U^\transpose$.
\end{lem}
\noindent{\sc Proof:} Let $\Delta_{n,P}\equiv\sum_{i=1}^{n} E_P[\| \bar b^{m_n}(V_i) U_i^\transpose/\sqrt n\|^3]$. By Assumption \ref{Ass: Slutsky}(i)-(a), (iv)-(a)(b) and (v)-(b), we may obtain that
\begin{multline}\label{Eqn: Slutsky coupling, aux1}
\Delta_{n,P}= E_P[\frac{\|\bar b^{m_n}(V_1)U_1^\transpose\|^3}{\sqrt n}]\lesssim E_P[\frac{\|\bar b^{m_n}(V_1) \|^3}{\sqrt n}]\\
\le  E_P[\frac{\xi_n\| b^{m_n}(V_1)\|^2+\|V^{(2)}\|^3}{\sqrt n}]\lesssim \frac{\xi_nm_n}{\sqrt n}~,
\end{multline}
where the last step follows by $\xi_n\ge 1$ (by Assumption \ref{Ass: Slutsky}(iii)-(b)) and Lemma \ref{Lem: basis norm}. By result \eqref{Eqn: Slutsky coupling, aux1} and Assumption \ref{Ass: Slutsky}(i)-(a), Theorem 10.10 in \citet{Pollard2002User} implies that, for any $\epsilon>0$, there is some Gaussian $G_{n,P}\in \mathbf M^{(m_n+d_{v_2})\times d_q}$ that has the same covariance functional as the random matrix $\bar b^{m_n}(V) U^\transpose$ and satisfies
\begin{align}\label{Eqn: Slutsky coupling, aux2}
P(\|\frac{1}{\sqrt n}\sum_{i=1}^{n}\bar b^{m_n}(V_i) U_i^\transpose-G_{n,P}\|>3  \epsilon)\lesssim \eta_{n,P}(1+\frac{|\log (1/\eta_{n,P})|}{(m_n+d_{v_2})d_q}) ~,
\end{align}
with $\eta_{n,P}\equiv \Delta_{n,P}(m_n+d_{v_2})d_q\epsilon^{-3}$. By result \eqref{Eqn: Slutsky coupling, aux1}, we note that
\begin{align}\label{Eqn: Slutsky coupling, aux3}
\eta_{n,P}\lesssim \frac{\xi_nm_n^2}{\sqrt n \epsilon^3}~.
\end{align}
By \eqref{Eqn: Slutsky coupling, aux2}, \eqref{Eqn: Slutsky coupling, aux3} and $\xi_nm_n^2/\sqrt n=o(1)$ (by Assumption \ref{Ass: Slutsky}(vi)-(b)), we may therefore conclude as in the proof of Proposition \ref{Pro: strong approx, series npiv} that, uniformly in $P\in\mathbf P$,
\begin{align}\label{Eqn: Slutsky coupling, aux4}
\|\frac{1}{\sqrt n}\sum_{i=1}^{n}\bar b^{m_n}(V_i) U_i^\transpose-G_{n,P}\| = O_p((\frac{\xi_nm_n^2}{\sqrt n})^{1/3})~.
\end{align}

By Lemma \ref{Lem: Slutsky Bahadur}, Assumption \ref{Ass: Slutsky}(iv)-(a), \eqref{Eqn: Slutsky Bahadur, aux10} and \eqref{Eqn: Slutsky coupling, aux4}, we obtain that
\begin{multline}\label{Eqn: Slutsky coupling, aux5}
\sqrt n \{\begin{bmatrix}
\hat\Lambda_n\\
\hat\Gamma_n
\end{bmatrix}-\begin{bmatrix}
\Lambda_{n,P}\\
\Gamma_{n,P}
\end{bmatrix}\}=(\bar\Psi_{n,P}^{-1/2}\bar\Pi_{n,P})_l^-\bar\Psi_{n,P}^{-1/2} G_{n,P} + O_p(s_n^{-1}(\frac{\xi_nm_n^2}{\sqrt n})^{1/3}) \\
+ O_p(s_n^{-1}\delta_n\sqrt{((\xi_n^2+n^{\frac{2}{2+\varsigma}})\log m_n)\vee m_n}+s_n^{-2}\sqrt{\frac{(\xi_n^2+n^{2/(2+\varsigma)})m_n\log(m_n)}{n}})~,
\end{multline}
uniformly in $P\in\mathbf P$. Let $\mathbb W_{n,P}\equiv (\Omega_{n,P}G_{n,P})^\transpose h^{k_n}$. Then it follows from \eqref{Eqn: Slutsky coupling, aux5}, Assumption \ref{Ass: Slutsky}(iii)-(b) and Fact 11.16.9 in \citet{Bernstein2018Matrix} that
\begin{multline}\label{Eqn: Slutsky coupling, aux6}
\|r_n \{\hat\Lambda_n-\Lambda_{n,P}\}^\transpose h^{k_n} - \mathbb W_{n,P}\|_{1,\infty}  = O_p( (\frac{\xi_nm_n^2}{\sqrt n})^{1/3}) \\
+ O_p( \delta_n \sqrt{((\xi_n^2+n^{\frac{2}{2+\varsigma}})\log m_n)\vee m_n}+s_n^{-1}\sqrt{\frac{(\xi_n^2+n^{2/(2+\varsigma)})m_n\log(m_n)}{n}})~,
\end{multline}
uniformly in $P\in\mathbf P$. Next, by Assumption \ref{Ass: Slutsky}(iii)-(c) and (iv)-(d), we may employ arguments similar to those leading to result \eqref{Eqn: npiv, residual, aux8} to conclude that
\begin{multline}\label{Eqn: Slutsky coupling, aux7}
\|r_n\{\hat g_n-g_P\} -r_n \{\hat\Lambda_n-\Lambda_{n,P}\}^\transpose h^{k_n}\|_{1,\infty}=\|r_n \{g_P-\Lambda_{n,P}^\transpose h^{k_n}\}\|_{1,\infty}\\
\lesssim \|r_n \{g_P-(\Lambda_{n,P}^{\mathrm{ols}})^\transpose h^{k_n}\}\|_{1,\infty}=O(\frac{\sqrt n s_n\delta_n}{\xi_n})~,
\end{multline}
uniformly in $P\in\mathbf P$. The conclusion of the lemma then follows from combining results \eqref{Eqn: Slutsky coupling, aux6} and \eqref{Eqn: Slutsky coupling, aux7} with the triangle inequality.\qed

\begin{lem}\label{Lem: Slutsky variance estimation}
Let Assumptions \ref{Ass: Slutsky}(i)-(a)(b), (iii), (iv), (v) and (vi)-(a)(b)(c) hold. If $\sup_{P\in\mathbf P}E_P[\|U\|^{2+\delta}]<\infty$ for some $\delta>0$, then, uniformly in $P\in\mathbf P$,
\begin{multline}
\|\hat\Sigma_n^{1/2}- \Sigma_{n,P}^{1/2}\|_o = O_p(\big(\xi_n^{1+2/\delta}\sqrt{\frac{\log m_n}{n}}\big)^{\frac{\delta}{\delta+1}})\\
+O_p(s_n^{-1}\xi_n\sqrt{\frac{ m_n}{n}}+s_n^{-1}\delta_n\xi_n\sqrt{\frac{((\xi_n^2+n^{\frac{2}{2+\varsigma}})\log m_n)\vee m_n}{n}}+\delta_n+\frac{n^{\frac{1}{2+\varsigma}}}{\sqrt n})~.
\end{multline}
\end{lem}
\noindent{\sc Proof:} Analogous to what we have done in the proof of Lemma \ref{Pro: bootstrap, series npiv}, define
\begin{multline}\label{Eqn: Slutsky variance estimation, aux1}
\tilde\Sigma_n\equiv\frac{1}{n}\sum_{i=1}^{n}(\bar b^{m_n}(V_i)\bar b^{m_n}(V_i)^\transpose)\otimes  (U_i U_i^\transpose)\\ =\frac{1}{n}\sum_{i=1}^{n}(\bar b^{m_n}(V_i)\otimes U_i)(\bar b^{m_n}(V_i) \otimes   U_i)^\transpose~.
\end{multline}
By simple algebra, we may then obtain the identity:
\begin{multline}\label{Eqn: Slutsky variance estimation, aux2}
\hat\Sigma_n-\tilde\Sigma_n  = \frac{1}{n}\sum_{i=1}^{n}  (\bar b^{m_n}(V_i)\bar b^{m_n}(V_i)^\transpose)\otimes  ((\hat U_i-U_i)(\hat U_i-U_i)^\transpose) \\
 + \frac{1}{n}\sum_{i=1}^{n}  (\bar b^{m_n}(V_i)\bar b^{m_n}(V_i)^\transpose)\otimes ((\hat U_i-U_i)U_i^\transpose + U_i(\hat U_i-U_i)^\transpose)~.
\end{multline}
By the simple fact \eqref{Eqn: npiv, variance, aux4} and the triangle inequality, we note that
\begin{align}
\|(\hat U_i-U_i)(\hat U_i-U_i)^\transpose\|_o & \le \|\hat U_i-U_i\|^2~, \label{Eqn: Slutsky variance estimation, aux3} \\
 \|(\hat U_i-U_i)U_i^\transpose + U_i(\hat U_i-U_i)^\transpose\|_o & \le 2\|\hat U_i-U_i\|\|U_i\| ~. \label{Eqn: Slutsky variance estimation, aux4}
\end{align}
By Fact 7.12.9, Corollary 10.4.2 and Fact 10.25.37 in \citet{Bernstein2018Matrix}, it follows from results \eqref{Eqn: Slutsky variance estimation, aux3} and \eqref{Eqn: Slutsky variance estimation, aux4} that
\begin{align}
0\le (\bar b^{m_n}(V_i)\bar b^{m_n}(V_i)^\transpose)\otimes & ((\hat U_i-U_i)(\hat U_i-U_i)^\transpose) \notag\\
  & \le \max_{i=1}^n\|\hat U_i-U_i\|^2 \{ (\bar b^{m_n}(V_i)\bar b^{m_n}(V_i)^\transpose)\otimes I_{d_q}\}~,\label{Eqn: Slutsky variance estimation, aux6}
\end{align}
and that, for $\Xi_{n,i}\equiv 2\max_{i=1}^n\|\hat U_i-U_i\|\{(\bar b^{m_n}(V_i)\bar b^{m_n}(V_i)^\transpose)\otimes( \|U_i\| I_{d_q})\}$,
\begin{align}
-\Xi_{n,i}\le (\bar b^{m_n}(V_i)\bar b^{m_n}(V_i)^\transpose)\otimes ((\hat U_i-U_i)U_i^\transpose + U_i(\hat U_i-U_i)^\transpose) \le \Xi_{n,i}~.\label{Eqn: Slutsky variance estimation, aux7}
\end{align}
By \eqref{Eqn: Slutsky variance estimation, aux6} and \eqref{Eqn: Slutsky variance estimation, aux7}, we may apply Lemma \ref{Lem: matrix operator norm} and Theorem 10.4.9 in \citet{Bernstein2018Matrix} (combined with Fact 7.12.9 in \citet{Bernstein2018Matrix}) to conclude from \eqref{Eqn: Slutsky variance estimation, aux2} that
\begin{multline}\label{Eqn: Slutsky variance estimation, aux8}
\|\hat\Sigma_n-\tilde\Sigma_n\|_o  \le \max_{i=1}^n \|\hat U_i-U_i\|^2 \cdot \|\frac{1}{n}\sum_{i=1}^{n}(\bar b^{m_n}(V_i)\bar b^{m_n}(V_i)^\transpose)\otimes I_{d_q}\|_o  \\
 + 2 \max_{i=1}^n \|\hat U_i-U_i\|  \cdot  \|\frac{1}{n}\sum_{i=1}^{n}(\bar b^{m_n}(V_i)\bar b^{m_n}(V_i)^\transpose)\otimes( \|U_i\| I_{d_q})\|_o ~.
\end{multline}
By Lemma \ref{Lem: Slutsky residuals}, Assumptions \ref{Ass: Slutsky}(i)-(a), (iv)-(a), (v)-(b) and (vi)-(b), and Fact 11.10.95 in \citet{Bernstein2018Matrix}, we obtain from \eqref{Eqn: Slutsky variance estimation, aux8} that, uniformly in $P\in\mathbf P$,
\begin{multline}\label{Eqn: Slutsky variance estimation, aux9}
\|\hat\Sigma_n-\tilde\Sigma_n\|_o\\
= O_p(s_n^{-1}\xi_n\sqrt{\frac{ m_n}{n}}+s_n^{-1}\delta_n\xi_n\sqrt{\frac{((\xi_n^2+n^{\frac{2}{2+\varsigma}})\log m_n)\vee m_n}{n}}+\delta_n+\frac{n^{\frac{1}{2+\varsigma}}}{\sqrt n})~.
\end{multline}

Next, since $\sup_{P\in\mathbf P}E_P[\|U\|^{2+\delta}]<\infty$ as given, we may employ arguments analogous to those in the proof of Lemma \ref{Lem: npiv, variance} to obtain that, uniformly in $P\in\mathbf P$,
\begin{align}\label{Eqn: Slutsky variance estimation, aux10}
\|\tilde\Sigma_n-\Sigma_{n,P}\|_o=O_p(\big(\xi_n^{1+2/\delta}\sqrt{\frac{\log m_n}{n}}\big)^{\frac{\delta}{\delta+1}})~.
\end{align}
It follows from \eqref{Eqn: Slutsky variance estimation, aux9}, \eqref{Eqn: Slutsky variance estimation, aux10} and the triangle inequality that
\begin{multline}\label{Eqn: Slutsky variance estimation, aux11}
\|\hat\Sigma_n- \Sigma_{n,P}\|_o = O_p(\big(\xi_n^{1+2/\delta}\sqrt{\frac{\log m_n}{n}}\big)^{\frac{\delta}{\delta+1}})\\
+O_p(s_n^{-1}\xi_n\sqrt{\frac{ m_n}{n}}+s_n^{-1}\delta_n\xi_n\sqrt{\frac{((\xi_n^2+n^{\frac{2}{2+\varsigma}})\log m_n)\vee m_n}{n}}+\delta_n+\frac{n^{\frac{1}{2+\varsigma}}}{\sqrt n})~,
\end{multline}
uniformly in $P\in\mathbf P$. By the law of iterated expectations, we note that
\begin{align}\label{Eqn: Slutsky variance estimation, aux12}
\Sigma_{n,P}=E_P\big[(\bar b^{m_n}(V)\bar b^{m_n}(V)^\transpose)\otimes E_P[UU^\transpose|V]\big]~.
\end{align}
By result \eqref{Eqn: Slutsky variance estimation, aux12} and Assumptions \ref{Ass: Slutsky}(iv)-(a) and (v)-(d), we may apply Fact 10.25.37 and Proposition 9.1.10 in \citet{Bernstein2018Matrix} to deduce that $\inf_{P\in\mathbf P}\lambda_{\min}(\Sigma_{n,P})$ is bounded away from zero uniformly in $n$. In turn, we may then obtain by Proposition 3.2 in \citet{HemmenAndo1980Ideal} that
\begin{align}\label{Eqn: Slutsky variance estimation, aux13}
\|\hat\Sigma_n^{1/2}-\Sigma_{n,P}^{1/2}\|_o\le\frac{1}{\{\lambda_{\min}(\Sigma_{n,P})\}^{1/2}}\|\hat\Sigma_n- \Sigma_{n,P}\|_o\lesssim \|\hat\Sigma_n- \Sigma_{n,P}\|_o~.
\end{align}
The lemma then follows from combining results \eqref{Eqn: Slutsky variance estimation, aux11} and \eqref{Eqn: Slutsky variance estimation, aux13}. \qed

\begin{lem}\label{Lem: Slutsky, tilde g}
If Assumptions \ref{Ass: Slutsky}(i), (iii), (iv), (v)-(a) and (vi)-(a)(b) hold, then it follows that, uniformly in $P\in\mathbf P$,
\begin{align}
\|\tilde g_n-g_0\|_{1,\infty}  =O_p(s_n^{-1}\delta_n\xi_n\sqrt{\frac{((\xi_n^2+n^{\frac{2}{2+\varsigma}})\log m_n)\vee m_n}{n}}+\delta_n)~.
\end{align}
\end{lem}
\noindent{\sc Proof:} By definition, we note that
\begin{align}\label{Eqn: Slutsky bias, aux1}
\tilde g_n-g_0 = (\tilde\Lambda_n-\Lambda_{n,P})^\transpose h^{k_n} + \Lambda_{n,P}^\transpose h^{k_n}-g_0~.
\end{align}
Therefore, we commence by controlling $\tilde\Lambda_n-\Lambda_{n,P}$, which is essentially what Lemma \ref{Lem: npiv, bias} is concerned with in the context of Example \ref{Ex: NPIV under shape}. Let $\mathcal E_n $ be defined as in \eqref{Eqn: Slutsky Bahadur, aux13}. Since $(\bar\Psi_n^{-1/2}\bar\Pi_n)_l^- \bar\Psi_n^{-1/2}\bar\Pi_n = I_{k_n+d_z}$ under $\mathcal E_n $, simple algebra reveals that, under $\mathcal E_n$,
\begin{multline}\label{Eqn: Slutsky bias, aux2}
\begin{bmatrix}
\tilde\Lambda_n\\
\tilde\Gamma_n
\end{bmatrix}  - \begin{bmatrix}
\Lambda_{n,P}\\
\Gamma_{n,P}
\end{bmatrix}
 =  (\bar\Psi_{n,P}^{-1/2}\bar\Pi_{n,P})_l^-\bar\Psi_{n,P}^{-1/2}\{\frac{\bar B_n^\transpose D_{n,P}}{n}-E_P[\bar b^{m_n}(V)d_{n,P}(T)^\transpose]\} \\
 +  \{(\bar\Psi_n^{-1/2}\bar\Pi_n)_l^-\bar\Psi_n^{-1/2}\bar\Psi_{n,P}^{1/2}-(\bar\Psi_{n,P}^{-1/2}\bar\Pi_{n,P})_l^-\}\bar\Psi_{n,P}^{-1/2}\frac{\bar B_n^\transpose D_{n,P}}{n}~,
\end{multline}
where $D_{n,P}\equiv(d_{n,P}(T_1),\ldots,d_{n,P}(T_n))^\transpose$ and $d_{n,P}(t)\equiv g_P(t)-(\Lambda_{n,P}^{\mathrm{ols}})^\transpose h^{k_n}(t)$ for $t\in\mathbf R^{d_q+1}$. Then, by Assumptions \ref{Ass: Slutsky}(i)-(a)(b)(d), (iii), (iv)-(a)(b)(c), (v)-(a), and (vi)-(a)(b), together with Jensen's inequality \citep[p.40]{Tropp2015Introduction}, we may argue as in the proof of Lemma \ref{Lem: npiv, bias} to conclude from \eqref{Eqn: Slutsky bias, aux2} that
\begin{align}\label{Eqn: Slutsky bias, aux3}
\|\begin{bmatrix}
\tilde\Lambda_n\\
\tilde\Gamma_n
\end{bmatrix}  - \begin{bmatrix}
\Lambda_{n,P}\\
\Gamma_{n,P}
\end{bmatrix}\|_o= O_p(s_n^{-1}\delta_n\sqrt{\frac{((\xi_n^2+n^{\frac{2}{2+\varsigma}})\log m_n)\vee m_n}{n}})~,
\end{align}
uniformly in $P\in\mathbf P$. By result \eqref{Eqn: Slutsky bias, aux3}, Fact 11.16.9 in \citet{Bernstein2018Matrix} and Assumption \ref{Ass: Slutsky}(iii)-(b), we in turn have that, uniformly in $P\in\mathbf P$,
\begin{align}\label{Eqn: Slutsky bias, aux4}
\|(\tilde\Lambda_n-\Lambda_{n,P})^\transpose h^{k_n}\|_{1,\infty}= O_p(s_n^{-1}\delta_n\xi_n\sqrt{\frac{((\xi_n^2+n^{\frac{2}{2+\varsigma}})\log m_n)\vee m_n}{n}})~.
\end{align}
Next, by Assumptions \ref{Ass: Slutsky}(iii)-(c) and (iv)-(d), we may conclude by arguments analogous to those leading to \eqref{Eqn: npiv, residual, aux8} that, uniformly in $P\in\mathbf P$,
\begin{align}\label{Eqn: Slutsky bias, aux5}
\|\mathrm{Proj}_{m,k}(g_P)-g_P\|_{1,\infty}=O(\delta_n)~.
\end{align}
The conclusion of the lemma then follows by combining results \eqref{Eqn: Slutsky Bahadur, aux14}, \eqref{Eqn: Slutsky bias, aux1}, \eqref{Eqn: Slutsky bias, aux4} and \eqref{Eqn: Slutsky bias, aux5} with the triangle inequality. \qed

\begin{lem}\label{Lem: Slutsky variance}
If Assumptions \ref{Ass: Slutsky}(i)-(a)(d), (iii)-(a)(b), (iv)-(a)(b)(c), (v)-(a)(b) and (vi)-(a)(b) hold, then it follows that, uniformly in $P\in\mathbf P$,
\begin{align}
\|\hat g_n-\tilde g_n\|_{1,\infty}  = O_p(s_n^{-1}\xi_n\sqrt{\frac{ m_n}{n}})~.
\end{align}
\end{lem}
\noindent{\sc Proof:} By definition, we note that
\begin{align}
\hat g_n-\tilde g_n & =(\hat\Lambda_n-\tilde\Lambda_n)^\transpose h^{k_n}~, \label{Eqn: Slutsky variance, aux1} \\
\begin{bmatrix}
\hat\Lambda_n\\
\hat\Gamma_n
\end{bmatrix}-\begin{bmatrix}
\tilde\Lambda_n\\
\tilde\Gamma_n
\end{bmatrix} & = [(\bar\Psi_n^{-1/2}\bar\Pi_n)_l^-\bar\Psi_n^{-1/2}\frac{\bar B_n^\transpose U_n}{n}~.  \label{Eqn: Slutsky variance, aux2}
\end{align}
By result \eqref{Eqn: Slutsky Bahadur, aux17} and the triangle inequality, we have: uniformly in $P\in\mathbf P$,
\begin{multline}\label{Eqn: Slutsky variance, aux3}
\|\begin{bmatrix}
\hat\Lambda_n\\
\hat\Gamma_n
\end{bmatrix}-\begin{bmatrix}
\tilde\Lambda_n\\
\tilde\Gamma_n
\end{bmatrix}\|_o\le O_p(s_n^{-2}\sqrt{\frac{(\xi_n^2+n^{2/(2+\varsigma)})m_n\log(m_n)}{n^2}})\\ + (\bar\Psi_{n,P}^{-1/2}\bar\Pi_{n,P})_l^-\bar\Psi_{n,P}^{-1/2}\frac{\bar B_n^\transpose U_n}{n}~.
\end{multline}
By Assumption \ref{Ass: Slutsky}(iv)-(a), \eqref{Eqn: Slutsky Bahadur, aux10} and \eqref{Eqn: Slutsky Bahadur, aux16}, we note that
\begin{align}\label{Eqn: Slutsky variance, aux4}
\|(\bar\Psi_{n,P}^{-1/2}\bar\Pi_{n,P})_l^-\bar\Psi_{n,P}^{-1/2}\frac{\bar B_n^\transpose U_n}{n}\|_o\le O_p(s_n^{-1}\sqrt{\frac{m_n}{n}})~,
\end{align}
uniformly in $P\in\mathbf P$. By results \eqref{Eqn: Slutsky variance, aux3} and \eqref{Eqn: Slutsky variance, aux4}, Assumption \ref{Ass: Slutsky}(vi)-(b), and Fact 9.14.10 in \citet{Bernstein2018Matrix}, we then have: uniformly in $P\in\mathbf P$,
\begin{align}\label{Eqn: Slutsky variance, aux5}
\|\hat\Lambda_n-\tilde\Lambda_n\|_o = O_p(s_n^{-1}\sqrt{\frac{m_n}{n}})~.
\end{align}
The lemma then follows from \eqref{Eqn: Slutsky variance, aux1}, \eqref{Eqn: Slutsky variance, aux5} and Assumption \ref{Ass: Slutsky}(iii)-(b). \qed

\begin{lem}\label{Lem: Slutsky residuals}
If Assumptions \ref{Ass: Slutsky}(i), (ii), (iii), (iv), (v)-(a)(b) and (vi)-(a)(b) hold, then it follows that, uniformly in $P\in\mathbf P$,
\begin{multline}
\max_{i=1}^n\|\hat U_i-U_i\|=O_p(s_n^{-1}\xi_n\sqrt{\frac{ m_n}{n}}) \\
+ O_p(s_n^{-1}\delta_n\xi_n\sqrt{\frac{((\xi_n^2+n^{\frac{2}{2+\varsigma}})\log m_n)\vee m_n}{n}}+\delta_n+\frac{n^{\frac{1}{2+\varsigma}}}{\sqrt n}) ~.
\end{multline}
\end{lem}
\noindent{\sc Proof:} By the triangle inequality and simple algebra, we have
\begin{multline}\label{Eqn: Slutsky residuals, aux1}
\max_{i=1}^n\|\hat U_i-U_i\|=\max_{i=1}^n\|g_P(P_i,Y_i)-\hat g_n(P_i,Y_i)+(\Gamma_P-\hat\Gamma_n)^\transpose Z_i\|\\
\le\|\hat g_n-g_P\|_\infty + \|\Gamma_P-\hat\Gamma_n\|\max_{i=1}^n\|Z_i\|~.
\end{multline}
By Assumptions \ref{Ass: Slutsky}(i)-(a) and (v)-(a), we may obtain by a maximal inequality \citep[p.98]{Vaart1996} that
\begin{align}\label{Eqn: Slutsky residuals, aux2}
\sup_{P\in\mathbf P}E_P[\max_{i=1}^n\|Z_i\|]\le n^{\frac{1}{2+\varsigma}} \sup_{P\in\mathbf P} \{E_P[\|Z_1\|^{2+\varsigma}]\}^{\frac{1}{2+\varsigma}}\lesssim n^{\frac{1}{2+\varsigma}}~.
\end{align}
The lemma then follows from combining Lemma \ref{Lem: Slutsky rate}, \eqref{Eqn: Slutsky residuals, aux1} and \eqref{Eqn: Slutsky residuals, aux2}. \qed

\subsection{Supporting Lemmas}

\begin{lem}\label{Lem: matrix operator norm}
Let $A_1,\ldots,A_n$ be positive semidefinite matrices in $\mathbf M^{d\times d}$ and $c_1,\ldots,c_n\in\mathbf R$ be arbitrary scalars. Then it follows that
\begin{align}
\|\sum_{i=1}^{n}c_i A_i\|_{o}\le \|\sum_{i=1}^{n}|c_i| A_i\|_{o}\le \max_{i=1}^n|c_i| \|\sum_{i=1}^{n}A_i\|_{o}~.
\end{align}
\end{lem}
\noindent{\sc Proof:} By Proposition 10.1.2-(iv) in \citet{Bernstein2018Matrix} and simple algebra, we note
\begin{align}\label{Eqn: matrix operator norm, aux1}
-\sum_{i=1}^{n}|c_i| A_i\le \sum_{i=1}^{n}c_i A_i\le \sum_{i=1}^{n}|c_i| A_i~.
\end{align}
By result \eqref{Eqn: matrix operator norm, aux1} and Theorem 10.4.9 in \citet{Bernstein2018Matrix}, we in turn have
\begin{multline}\label{Eqn: matrix operator norm, aux2}
-\lambda_{\max}(\sum_{i=1}^{n}|c_i| A_i)=\lambda_{\min}(-\sum_{i=1}^{n}|c_i| A_i)\le \lambda_{\min}(\sum_{i=1}^{n}c_i A_i)\\
\le\lambda_{\max}(\sum_{i=1}^{n}c_i A_i)\le \lambda_{\max}(\sum_{i=1}^{n}|c_i| A_i)~.
\end{multline}
By Fact 7.12.9 in \citet{Bernstein2018Matrix} and the fact that the singular values and eigenvalues of any positive semidefinite matrix coincide, we thus obtain from \eqref{Eqn: matrix operator norm, aux2} that
\begin{align}
\sigma_{\max}(\sum_{i=1}^{n}c_i A_i)\le \sigma_{\max}(\sum_{i=1}^{n}|c_i| A_i)~,
\end{align}
as desired for the first inequality. Next, we note that
\begin{align}\label{Eqn: matrix operator norm, aux3}
0\le \sum_{i=1}^{n}|c_i| A_i\le \max_{i=1}^n |c_i| \sum_{i=1}^{n} A_i~.
\end{align}
Since the singular values and eigenvalues of any positive semidefinite matrix coincide, the second inequality then follows by \eqref{Eqn: matrix operator norm, aux3} and Theorem 10.4.9 in \citet{Bernstein2018Matrix}. \qed

\begin{lem}\label{Lem: inverse matrix}
Let $\mathbf P$ be a family of probability measures, $\Sigma_{n,P}\in\mathbf M^{m_n\times k_n}$ possibly dependent on $P\in\mathbf P$ and $n\in\mathbf N$, and $\hat\Sigma_n$ be an estimator of $\Sigma_{n,P}$ with $\|\hat\Sigma_n-\Sigma_{n,P}\|_{o}=O_p(a_n)$ uniformly in $P\in\mathbf P$ for some $a_n\ge 0$. If $\inf_{P\in\mathbf P} \sigma_{\min}(\Sigma_{n,P})\ge s_n>0$ for each $n$ and $a_n/s_n=o(1)$, then uniformly in $P\in\mathbf P$ it holds that $P(\sigma_{\min}(\hat\Sigma_n)<s_n/2)=o(1)$, $\|(\hat\Sigma_n)_l^--(\Sigma_{n,P})_l^{-}\|_{o}=O_p(a_n/s_n^2)$, and  $\|(\hat\Sigma_n)_l^-\|_{o}=O_p(1/s_n)$.
\end{lem}
\begin{rem}
From the proof, we easily see that the same lemma in fact also holds with $(\cdot)_l^-$ replaced by the Moore-Penrose inverses.\qed
\end{rem}
\noindent{\sc Proof:} Since $\|\hat\Sigma_n-\Sigma_{n,P}\|_{o}=O_p(a_n)$ uniformly in $P\in\mathbf P$ by assumption, we may conclude by Fact 11.16.40 in \citet{Bernstein2018Matrix} that
\begin{align}\label{Eqn: inverse matrix, aux1}
|\sigma_{\min}(\hat\Sigma_n)-\sigma_{\min}(\Sigma_{n,P})|\le \|\hat\Sigma_n-\Sigma_{n,P}\|_{o} =O_p(a_n)~,
\end{align}
uniformly in $P\in\mathbf P$. Since $\inf_{P\in\mathbf P} \sigma_{\min}(\Sigma_{n,P})\ge s_n>0$, we in turn have
\begin{multline}\label{Eqn: inverse matrix, aux2}
\limsup_{n\to\infty}\sup_{P\in\mathbf P}P(\sigma_{\min}(\hat\Sigma_n)< \frac{s_n}{2})\le \limsup_{n\to\infty}\sup_{P\in\mathbf P}P(|\sigma_{\min}(\hat\Sigma_n)-\sigma_{\min}(\Sigma_{n,P})|> \frac{s_n}{2})\\
\le \limsup_{n\to\infty}\sup_{P\in\mathbf P}P(\frac{|\sigma_{\min}(\hat\Sigma_n)-\sigma_{\min}(\Sigma_{n,P})|}{s_n}> \frac{1}{2}) \to 0~,
\end{multline}
where the final step follows from result \eqref{Eqn: inverse matrix, aux1} and $a_n/s_n=o(1)$. This establishes the first claim of the lemma.

For the second claim, we note that $\inf_{P\in\mathbf P} \sigma_{\min}(\Sigma_{n,P})\ge s_n>0$ implies
\begin{multline}\label{Eqn: inverse matrix, aux3}
\limsup_{n\to\infty}\sup_{P\in\mathbf P}P(\|\hat\Sigma_n-\Sigma_{n,P}\|_{o}> \frac{1}{2}\sigma_{\min}(\Sigma_{n,P}))\\
\le \limsup_{n\to\infty}\sup_{P\in\mathbf P}P(\|\hat\Sigma_n-\Sigma_{n,P}\|_{o}\ge  \frac{1}{2}\frac{s_n}{a_n}a_n)=0~,
\end{multline}
where we exploited $s_n/a_n\to\infty$ and $\|\hat\Sigma_n-\Sigma_{n,P}\|_{o}=O_p(a_n)$ uniformly. Next, define
\begin{align}
\mathcal A_{n,P}\equiv \{\|\hat\Sigma_n-\Sigma_{n,P}\|_{o}\le \frac{1}{2}\sigma_{\min}(\Sigma_{n,P}), \text{ and } \hat\Sigma_n\text{ has full column rank}\}~.
\end{align}
Results \eqref{Eqn: inverse matrix, aux2} and \eqref{Eqn: inverse matrix, aux3} then together imply that
\begin{align}\label{Eqn: inverse matrix, aux4}
\limsup_{n\to\infty}\sup_{P\in\mathbf P}P(\mathcal A_{n,P}^c)=0~.
\end{align}
Fix $M>0$. It follows from result \eqref{Eqn: inverse matrix, aux4}, Lemma F.4 in \citet{ChenChristensen2018SupNormOptimal} and Lemma 8.3.33 in \citet{Bernstein2018Matrix} that, uniformly in $P\in\mathbf P$,
\begin{multline}\label{Eqn: inverse matrix, aux5}
P(\|(\hat\Sigma_n)_l^- - (\Sigma_{n,P})_l^{-}\|_{o}>\frac{a_n}{s_n^2}M)\le P(\|(\hat\Sigma_n)_l^- - (\Sigma_{n,P})_l^{-}\|_{o}>\frac{a_n}{s_n^2}M, \mathcal A_{n,P})+P(\mathcal A_{n,P}^c) \\
\le P(s_n^{-2}\|\hat\Sigma_n - \Sigma_{n,P}\|_{o}\gtrsim\frac{a_n}{s_n^2}M)+ o(1)=o(1)
\end{multline}
as $n\to\infty$ followed by $M\to\infty$. This shows the second claim.

For the last one, note by the triangle inequality that, uniformly in $P\in\mathbf P$,
\begin{align}
\|(\hat\Sigma_n)_l^-\|_{o}\le \|(\hat\Sigma_n)_l^--(\Sigma_{n,P})_l^{-}\|_{o}+\|(\Sigma_{n,P})_l^{-}\|_{o}\le O_p(\frac{a_n}{s_n^2})+\frac{1}{s_n}=O_p(\frac{1}{s_n})~,
\end{align}
where the second inequality follows by the second claim of the lemma and Fact 8.3.33 in \citet{Bernstein2018Matrix}, and the last step is due to the assumption $a_n/s_n=o(1)$. This completes the proof of the third claim. \qed

\begin{lem}\label{Lem: matrix product}
Let $\mathbf P$ be a family of probability measures, and $A_{n,P}\in\mathbf M^{m_n\times k_n}$ and $B_{n,P}\in\mathbf M^{k_n\times l_n}$, with $m_n$, $k_n$ and $l_n$ possibly depending on $n$. Let $\hat A_n$ and $\hat B_n$ be estimators such that $\|\hat A_n-A_{n,P}\|_o=O_p(a_n)$ and $\|\hat B_n-B_{n,P}\|_o=O_p(b_n)$ both uniformly in $P\in\mathbf P$ where $a_n=O(1)$ and $b_n=O(1)$. Suppose $\|A_{n,P}\|_o=O(c_n)$ and $\|B_{n,P}\|_o=O(d_n)$ with $b_n/d_n=O(1)$, both uniformly in $P\in\mathbf P$. Then, uniformly in $P\in\mathbf P$,
\begin{align}
\|\hat A_n\hat B_n-A_{n,P}B_{n,P}\|_o= O_p(a_nd_n+ b_nc_n)~.
\end{align}
\end{lem}
\noindent{\sc Proof:} By the triangle inequality, we have
\begin{align}\label{Eqn: matrix product, aux1}
\|\hat A_n\hat B_n-A_{n,P}B_{n,P}\|_o\le \|\hat A_n-A_{n,P}\|_o\|\hat B_n\|_o+\|A_{n,P}\|_o\|\hat B_n-B_{n,P}\|_o~.
\end{align}
Again by the triangle inequality and $b_n=O(d_n)$, we also have: uniformly in $P\in\mathbf P$,
\begin{align}\label{Eqn: matrix product, aux2}
\|\hat B_n\|_o\le \|\hat B_n-B_{n,P}\|_o+\|B_{n,P}\|_o=O_p(b_n)+ O(d_n)=O_p(d_n)~.
\end{align}
By result \eqref{Eqn: matrix product, aux2} and the assumptions, we obtain from \eqref{Eqn: matrix product, aux1} that
\begin{align}\label{Eqn: matrix product, aux3}
\|\hat A_n\hat B_n-A_{n,P}B_{n,P}\|_o\le O_p(a_n)O_p(d_n)+O(c_n)O_p(b_n)=O_p(a_nd_n + b_nc_n)~,
\end{align}
uniformly in $P\in\mathbf P$. This proves the claim of the lemma. \qed

\begin{lem}\label{Lem: covariance bound}
Let $Z\in\mathbf R^k$ and $V\in\mathbf R^m$ be random vectors with $m\ge k$. Then
\begin{align}
\sigma_{\max}(E[VZ^\transpose])\le \sigma_{\max} (E[VV^\transpose])+\sigma_{\max} (E[ZZ^\transpose])~.
\end{align}
\end{lem}
\noindent{\sc Proof:} Define $X\equiv (V^\transpose, Z^\transpose)^\transpose$ and note that
\begin{align}\label{Eqn: covariance bound, aux1}
E[XX^\transpose]=\begin{bmatrix}
                   E[VV^\transpose] & E[VZ^\transpose] \\
                   E[ZV^\transpose] & E[ZZ^\transpose]
                 \end{bmatrix}~.
\end{align}
By Fact 11.16.9 in \citet{Bernstein2018Matrix}, we have in view of \eqref{Eqn: covariance bound, aux1} that
\begin{align}\label{Eqn: covariance bound, aux2}
\sigma_{\max}(E[VZ^\transpose])\le \sigma_{\max}(E[XX^\transpose])~.
\end{align}
Since eigenvalues and singular values of any positive semidefinite matrix coincide, we obtain by Theorem III.2.9 in \citet{Bhatia1997Matrix} that
\begin{align}\label{Eqn: covariance bound, aux3}
\sigma_{\max}(E[XX^\transpose])\le \sigma_{\max} (E[VV^\transpose])+ \sigma_{\max} (E[ZZ^\transpose])~.
\end{align}
The conclusion of the lemma then follows from \eqref{Eqn: covariance bound, aux2} and \eqref{Eqn: covariance bound, aux3}. \qed

The following lemma is nearly trivial, and is recorded here to simplify the proofs as it is routinely needed in Appendices \ref{Sec: npiv, app} and \ref{Sec: Slutsky, appendix}.

\begin{lem}\label{Lem: basis norm}
Let $X\in\mathcal X$ be random and let $h$ be a $k\times 1$ vector of functions on $\mathcal X$. Then $E[\|h(X)\|]\le\sqrt{k\lambda_{\max}}$ and $E[\|h(X)\|^2]\le k\lambda_{\max}$ with $\lambda_{\max}\equiv \lambda_{\max}(E[h(X)h(X)^\transpose])$. \qed
\end{lem}
\noindent{\sc Proof:} By Jensen's inequality, we have
\begin{align}\label{Eqn: basis norm, aux1}
E[\|h(X)\|]\le \{E[\|h(X)\|^2]\}^{1/2}=\{\mathrm{tr}(E[h(X)h(X)^\transpose])\}^{1/2}\le \sqrt{k\lambda_{\max}}~,
\end{align}
where the last step exploited the fact that $\mathrm{tr}(E[h(X)h(X)^\transpose])$ equals the sum of the eigenvalues of $E[h(X)h(X)^\transpose]$. Both claims of the lemma then follow from \eqref{Eqn: basis norm, aux1}. \qed

\begin{lem}\label{Lem: Gaussian coupling}
Let $(\Omega,\mathcal A,P)$ be a probability space, $\mathbb G_0:\Omega\to\mathbf D$ a Gaussian variable that is tight and centered in a Banach space $\mathbf D$, and $\mathbb X:\Omega\to\mathbf B$ a Borel map that is independent of $\mathbb G_0$ with $\mathbf B$ a separable Banach space. Further, for $\mathbf E$ a Banach space, let $\hat\psi: \mathbf D\to\mathbf E$ be a map that may depend on $\omega\in\Omega$ but only through $\mathbb X$ and is continuous and linear for almost all realizations of $\mathbb X$. If $\mathbb Z: \Omega\to\mathbf E$ is centered Gaussian conditional on $\mathbb X$ such that $P(\mathbb Z\in\mathbf E_0|\mathbb X)=1$ almost surely for some complete and separable subspace $\mathbf E_0\subset\mathbf E$, and that it has the same covariance operator as $\hat\psi(\mathbb G_0)$ conditional on $\mathbb X$ almost surely, then it follows that there exists a copy $\mathbb G: \Omega\to\mathbf D$ of $\mathbb G_0$ such that $\mathbb Z=\hat\psi(\mathbb G)$ almost surely and $\mathbb G$ is independent of $\mathbb X$.
\end{lem}
\noindent{\sc Proof:} We shall make the dependence of $\hat\psi$ on $\mathbb X$ explicit by writing $\hat\psi(\mathbb X)$, $\hat\psi(\mathbb X(\omega))$ for $\mathbb X$ evaluated at $\mathbb X(\omega)$, or $\hat\psi(x)$ for $\mathbb X=x$, which are all maps from $\mathbf D$ to $\mathbf E$. Define $\mathbb Z_0\equiv\hat\psi(\mathbb G_0)$. Further, let $P_{\mathbb Z_0|\mathbb X}$ be (a version of) the conditional distribution of $\mathbb Z_0$ given $\mathbb X$, and denote by $P_{\mathbb Z_0|\mathbb X}(\cdot,\omega)$ the conditional probability measure of $\mathbb Z_0$ given $\mathbb X=\mathbb X(\omega)$. The notation $P_{\mathbb Z|\mathbb X}$ and $P_{\mathbb Z|\mathbb X}(\cdot,\omega)$ are analogously defined. By assumption, we may pick a set $\Omega_0\in\mathcal A$ of full probability measure (i.e., $P(\Omega_0)=1$) such that, for each $\omega\in\Omega_0$, (i) $\hat\psi(\mathbb X(\omega)):\mathbf D\to\mathbf E$ is continuous and linear when $\mathbb X=\mathbb X(\omega)$, (ii) $\mathbb Z$ conditional on $\mathbb X=\mathbb X(\omega)$ is centered Gaussian, (iii) $P_{\mathbb Z_0|\mathbb X}(\cdot,\omega)$ and $P_{\mathbb Z|\mathbb X}(\cdot,\omega)$ share the same covariance operator, and (iv) $P(\mathbb Z\in\mathbf E_0|\mathbb X=\mathbb X(\omega))=1$.

Fix $\omega\in\Omega_0$. Since $\mathbb G_0$ is independent of $\mathbb X$ and $\hat\psi: \mathbf D\to\mathbf E$ depends on $\omega$ only through $\mathbb X$, it follows that $P_{\mathbb Z_0|\mathbb X}(\cdot,\omega)$ is centered Gaussian by Lemma 2.2.2 in \citet{Bogavcev1998gaussian} and continuity and linearity of $\hat\psi(\mathbb X(\omega)): \mathbf D\to\mathbf E$. Since $\mathbb G_0$ is tight, it follows by Lemma 1.3.2 in \citet{Vaart1996} and the corollary to Theorem I.3.1 in \citet{Vakhania_Tarieladze_Chobanyan1987probability} that the law of $\mathbb G_0$ is Radon so that $P_{\mathbb Z_0|\mathbb X}(\cdot,\omega)$ is Radon by continuity of $\hat\psi(\mathbb X(\omega))$. In addition, since $P(\mathbb Z\in\mathbf E_0|\mathbb X=\mathbb X(\omega))=1$, the corollary to Theorem I.3.1 in \citet{Vakhania_Tarieladze_Chobanyan1987probability} also implies that $P_{\mathbb Z|\mathbb X}(\cdot,\omega)$ is Radon. Therefore, since $P_{\mathbb Z_0|\mathbb X}(\cdot,\omega)$ and $P_{\mathbb Z|\mathbb X}(\cdot,\omega)$ are centered Gaussian and share the same covariance operator by assumption, Proposition IV.2.7 in \citet{Vakhania_Tarieladze_Chobanyan1987probability} implies that $P_{\mathbb Z_0|\mathbb X}(\cdot,\omega)$ and $P_{\mathbb Z|\mathbb X}(\cdot,\omega)$ have the same characteristic functional. Since they are also Radon, $P_{\mathbb Z_0|\mathbb X}(\cdot,\omega)$ and $P_{\mathbb Z|\mathbb X}(\cdot,\omega)$ are equal by Lemma 7.13.5 in \citet{Bogachev2007II}. In turn, Theorem 10.2.1 in \citet{Dudley2002real} then implies that $(\mathbb X,\mathbb Z_0)$ and $(\mathbb X,\mathbb Z)$ are equal in distribution. Since $P(\mathbb Z\in\mathbf E_0|\mathbb X)=1$ almost surely, we obtain by the law of iterated expectations that $P(\mathbb Z\in\mathbf E_0)=1$, and hence also $P(\mathbb Z_0\in\mathbf E_0)=1$.

Let $L$ be the joint law of $(\mathbb G_0,\mathbb X,\mathbb Z_0)$. Since $\mathbb G_0$ is Radon as noted above and centered Gaussian by assumption, it follows by the remark to Proposition 7.4 in \citet{Davydov1998local} that the topological support of $\mathbb G_0$, denoted $\mathbf D_0$, is a separable Banach space. Let $\Pi_{\mathbf D_0}: \mathbf D\to\mathbf D_0$ and $\Pi_{\mathbf E_0}: \mathbf E\to\mathbf E_0$ be the orthogonal projections onto $\mathbf D_0$ and $\mathbf E_0$ respectively. Then $\Pi_{\mathbf D_0}\mathbb G_0=\mathbb G_0$ and $\Pi_{\mathbf E_0}\mathbb Z_0=\mathbb Z_0$ almost surely. Thus, $\Pi_{\mathbf D_0}\mathbb G_0$ is equal in law to $\mathbb G_0$, $(\mathbb X,\Pi_{\mathbf E_0}\mathbb Z_0)$ is equal in law to $(\mathbb X,\mathbb Z)$, and the joint law of $(\Pi_{\mathbf D_0}\mathbb G_0,\mathbb X,\Pi_{\mathbf E_0}\mathbb Z_0)$ is $L$. By Lemma 2.11 in \citet{DudleyPhilipp1983Invariance}, we may then conclude that there is a copy $\mathbb G:\Omega\to\mathbf D_0$ of $\Pi_{\mathbf D_0}\mathbb G_0$ and hence also of $\mathbb G_0$ such that the joint law of $(\mathbb G, \mathbb X, \mathbb Z)$ is $L$. It follows that $\mathbb G$ and $\mathbb X$ are independent. In addition, since the law $L$ is induced by the map $(g,x)\mapsto (g,x,(\hat\psi(x))(g))$ under the measure of $(\mathbb G_0,\mathbb X)$, we must have $\mathbb Z=\hat\psi(\mathbb G)$ almost surely. This completes the proof of the lemma. \qed

\section{Proofs for Results in Appendix \ref{Sec: special case}}\label{App: proofs for the special case}

\noindent{\sc Proof of Proposition \ref{Pro: strong approximation, special}:} Let $L_{n,P}$ and $L_P$ be the laws of $r_n\{\hat\theta_n-\theta_P\}$ and $\mathbb G_P$ respectively (evaluated under $P$), which exist by measurability of $\hat\theta_n$ as an estimator---note that measurability in separable spaces is a rather weak restriction \citep[p.37-8]{Ledoux_Talagrand1991probability}. Moreover, let $\rho$ be the Prohorov distance of laws on $\mathbf H$. By Lemma 3.29 in \citet{Dudley2014UCLT} and Assumption \ref{Ass: strong approx, regular}(i), we have:
\begin{align}\label{Eqn: strong approximation, special, aux1}
\limsup_{n\to\infty}\sup_{P\in\mathbf P}\rho(L_{n,P},L_p)= 0~.
\end{align}
Since $\mathbf H$ is a separable Hilbert space, $L_{n,P}$ and $L_P$ are tight by Lemma 1.3.2 in \citet{Vaart1996}. By Theorem 10.8 in \citet{Pollard2002User}, we may conclude from result \eqref{Eqn: strong approximation, special, aux1} that there exist random variables $\tilde{\mathbb G}_{n,P}$ and $\tilde{\mathbb Z}_{n,P}$ in $\mathbf H$ such that
\begin{align}\label{Eqn: strong approximation, special, aux2}
\|\tilde{\mathbb G}_{n,P}-\tilde{\mathbb Z}_{n,P}\|_{\mathbf H}=o_p(1)
\end{align}
uniformly in $P\in\mathbf P$, where $\tilde{\mathbb G}_{n,P}\overset{d}{=}r_n\{\hat\theta_n-\theta_P\}$ and $\tilde{\mathbb Z}_{n,P}\overset{d}{=}\mathbb G_P$. By Lemma 2.11 in \citet{DudleyPhilipp1983Invariance}, there exist $\{\mathbb Z_{n,P}\}$ satisfying $(r_n\{\hat\theta_n-\theta_P\},\mathbb Z_{n,P})\overset{d}{=}(\tilde{\mathbb G}_{n,P},\tilde{\mathbb Z}_{n,P})$. This, together with result \eqref{Eqn: strong approximation, special, aux2}, verifies Assumption \ref{Ass: strong approx}(i) with $c_n=1$.

Next, by Fubini's theorem and Jensen's inequality, we obtain that
\begin{align}\label{Eqn: strong approximation, special, aux4}
\sup_{f\in\mathrm{BL}_1(\mathbf H)} \big|E_P[f(\hat{\mathbb G}_n)]-E[f&(\mathbb G_P)]\big|=\sup_{f\in\mathrm{BL}_1(\mathbf H)}\big|E_P\big[E[f(\hat{\mathbb G}_n)|\{X_i\}_{i=1}^n]\big]-E[f(\mathbb G_P)]\big|\notag\\
&\le E_P\big[\sup_{f\in\mathrm{BL}_1(\mathbf H)}\big|E[f(\hat{\mathbb G}_n)|\{X_i\}_{i=1}^n]-E[f(\mathbb G_P)]\big|\big]~.
\end{align}
Fix $\epsilon>0$. Noting that each $f\in\mathrm{BL}_1(\mathbf H)$ is bounded by 1, we then obtain:
\begin{multline}\label{Eqn: strong approximation, special, aux5}
E_P\big[\sup_{f\in\mathrm{BL}_1(\mathbf H)}\big|E[f(\hat{\mathbb G}_n)|\{X_i\}_{i=1}^n]-E[f(\mathbb G_P)]\big|\big]\\
\le \epsilon+ 2 P(\sup_{f\in\mathrm{BL}_1(\mathbf H)}\big|E[f(\hat{\mathbb G}_n)|\{X_i\}_{i=1}^n]-E[f(\mathbb G_P)]\big|>\epsilon)~,
\end{multline}
uniformly in $P\in\mathbf P$. Since $\epsilon$ is arbitrary, it follows by results \eqref{Eqn: strong approximation, special, aux4} and \eqref{Eqn: strong approximation, special, aux5} that
\begin{align}\label{Eqn: strong approximation, special, aux6}
\sup_{P\in\mathbf P}\sup_{f\in\mathrm{BL}_1(\mathbf H)} \big|E_P[f(\hat{\mathbb G}_n)]-E[f(\mathbb G_P)]\big|=o(1)~.
\end{align}
We may then employ analogous arguments as before to conclude from \eqref{Eqn: strong approximation, special, aux6} that, there exist copies $\{\bar{\mathbb Z}_{n,P}\}$ of $\mathbb G_P$ for each $P\in\mathbf P$ satisfying: uniformly in $P\in\mathbf P$,
\begin{align}\label{Eqn: strong approximation, special, aux7}
\|\hat{\mathbb G}_n-\bar{\mathbb Z}_{n,P}\|_{\mathbf H}=o_p(1)~.
\end{align}
Clearly, $\bar{\mathbb Z}_{n,P}$ is also a copy of $\mathbb Z_{n,P}$ for each $P\in\mathbf P$ and $n$.

For the first claim of the proposition, it remains to show that $\bar{\mathbb Z}_{n,P}$ is asymptotically independent of the data $\{X_i\}_{i=1}^n$. Observing that $f\in \mathrm{BL}_1(\mathbf H)$ is bounded by 1 and Lipschitz continuous, we have
\begin{multline}\label{Eqn: strong approximation, special, aux8}
\sup_{f\in\mathrm{BL}_1(\mathbf H)} |E[f(\hat{\mathbb G}_n)|\{X_i\}_{i=1}^n]-E[f(\bar{\mathbb Z}_{n,P})|\{X_i\}_{i=1}^n]|\\
\le \epsilon+ 2 P(\|\hat{\mathbb G}_n-\bar{\mathbb Z}_{n,P}\|_{\mathbf H}>\epsilon|\{X_i\}_{i=1}^n)=\epsilon+o_p(1)~,
\end{multline}
uniformly in $P\in\mathbf P$, where the last step follows from Markov's inequality and result \eqref{Eqn: strong approximation, special, aux7}. Since $\epsilon$ is arbitrary, result \eqref{Eqn: strong approximation, special, aux8} implies that
\begin{align}\label{Eqn: strong approximation, special, aux9}
\sup_{f\in\mathrm{BL}_1(\mathbf H)} |E[f(\hat{\mathbb G}_n)|\{X_i\}_{i=1}^n]-E[f(\bar{\mathbb Z}_{n,P})|\{X_i\}_{i=1}^n]| = o_p(1)~,
\end{align}
uniformly in $P\in\mathbf P$. Since each $\bar{\mathbb Z}_{n,P}$ is a copy of $\mathbb G_P$, we must have $E[f(\bar{\mathbb Z}_{n,P})]=E[f(\mathbb G_P)]$ for all $f\in\mathrm{BL}_1(\mathbf H)$. This, together with Assumption \ref{Ass: strong approx, regular}(ii), yields
\begin{align}\label{Eqn: strong approximation, special, aux10}
\sup_{f\in\mathrm{BL}_1(\mathbf H)} |E[f(\hat{\mathbb G}_n)|\{X_i\}_{i=1}^n]-E[f(\bar{\mathbb Z}_{n,P})]| = o_p(1)
\end{align}
uniformly in $P\in\mathbf P$. By the triangle inequality, we note that
\begin{align}\label{Eqn: strong approximation, special, aux11}
\sup_{f\in\mathrm{BL}_1(\mathbf H)} |E[f(\bar{\mathbb Z}_{n,P})|& \{X_i\}_{i=1}^n]-E[f(\bar{\mathbb Z}_{n,P})]|\notag\\
 & \le \sup_{f\in\mathrm{BL}_1(\mathbf H)} |E[f(\bar{\mathbb Z}_{n,P})|\{X_i\}_{i=1}^n]-E[f(\hat{\mathbb G}_n)|\{X_i\}_{i=1}^n]|\notag\\
 & \qquad + \sup_{f\in\mathrm{BL}_1(\mathbf H)} |E[f(\hat{\mathbb G}_n)|\{X_i\}_{i=1}^n]-E[f(\bar{\mathbb Z}_{n,P})]|~.
\end{align}
Combining results \eqref{Eqn: strong approximation, special, aux9}, \eqref{Eqn: strong approximation, special, aux10} and \eqref{Eqn: strong approximation, special, aux11}, we may thus conclude that
\begin{align}
\sup_{f\in\mathrm{BL}_1(\mathbf H)} |E[f(\bar{\mathbb Z}_{n,P})| \{X_i\}_{i=1}^n]-E[f(\bar{\mathbb Z}_{n,P})]| = o_p(1)~,
\end{align}
uniformly in $P\in\mathbf P$. This completes the proof of the first claim.

For the second claim, fix $P\in\mathbf P_0$. Then we have by the proof of Lemma \ref{Lem: projection monotonicity} that
\begin{align}\label{Eqn: strong approximation, special, aux12}
\psi_{\kappa_n,P}(h)=\min_{|a|\le \kappa_n}\|h+a\theta_P-\Pi_\Lambda(h+a\theta_P)\|_{\mathbf H}
\end{align}
for all $h\in\mathbf H$. It follows from results \eqref{Eqn: strong approximation, special, aux12} and \eqref{Eqn: derivative estimator, aux5} (ahead) that
\begin{align}\label{Eqn: strong approximation, special, aux13}
\psi_{\kappa_n,P}(h)\to\phi_{\theta_P}'(h)
\end{align}
for all $h\in\mathbf H$. Therefore, we may conclude by \eqref{Eqn: strong approximation, special, aux13} that
\begin{align}\label{Eqn: strong approximation, special, aux14}
\psi_{\kappa_n,P}(\mathbb G_P)\to\phi_{\theta_P}'(\mathbb G_P)
\end{align}
(almost) surely. Since $\mathbb Z_{n,P}$ are copies of $\mathbb G_P$, it follows that each $\psi_{\kappa_n,P}(\mathbb G_P)$ is equal in distribution to $\psi_{\kappa_n,P}(\mathbb Z_{n,P})$. The second claim then follows from \eqref{Eqn: strong approximation, special, aux14}. \qed

\noindent{\sc Proof of Proposition \ref{Pro: derivative estimator}:} Fix $h\in\mathbf H$ and $P\in\mathbf P_0$. We proceed in two steps. First, we show that $\bar\phi_n'(h)\to\phi_{\theta_P}'(h)$ where $\bar\phi_n': \mathbf H\to\mathbf R$ is defined by
\begin{align}\label{Eqn: derivative estimator, aux0}
\bar\phi_n'(h)\equiv \min_{|\alpha|\le \kappa_n} \|h+\alpha\theta_P-\Pi_\Lambda(h+\alpha\theta_P)\|_{\mathbf H}~.
\end{align}
To this end, fix $\epsilon>0$. Then, we may pick some $\alpha^*\in \mathbf R$ such that
\begin{align}\label{Eqn: derivative estimator, aux2}
\inf_{\alpha\in\mathbf R}\|h+\alpha\theta_P-\Pi_\Lambda(h+\alpha\theta_P)\|_{\mathbf H}\ge \|h+\alpha^*\theta_P-\Pi_\Lambda(h+\alpha^*\theta_P)\|_{\mathbf H}-\epsilon~.
\end{align}
Since $\kappa_n\to\infty$ by assumption, we have: for all $n$ sufficient large so that $|\alpha^*|\le \kappa_n$,
\begin{align}\label{Eqn: derivative estimator, aux3}
\|h+\alpha^*\theta_P-\Pi_\Lambda(h+\alpha^*\theta_P)\|_{\mathbf H}\ge \inf_{|\alpha|\le \kappa_n} \|h+\alpha\theta_P-\Pi_\Lambda(h+\alpha\theta_P)\|_{\mathbf H}~.
\end{align}
Combining results \eqref{Eqn: derivative estimator, aux2} and \eqref{Eqn: derivative estimator, aux3}, together with the simple fact that the right side of \eqref{Eqn: derivative estimator, aux3} is no less than $\phi_{\theta_P}'(h)$ for all $n$ by Lemma \ref{Lem: derivative, alternative form}, we thus obtain
\begin{align}\label{Eqn: derivative estimator, aux4}
\phi_{\theta_P}'(h)\le \lim_{n\to\infty} \inf_{|\alpha|\le \kappa_n} \|h+\alpha\theta_P-\Pi_\Lambda(h+\alpha\theta_P)\|_{\mathbf H}\le \phi_{\theta_P}'(h)+\epsilon~.
\end{align}
Since $\epsilon$ is arbitrary, it follows from equation \eqref{Eqn: derivative estimator, aux0} and result \eqref{Eqn: derivative estimator, aux4} that
\begin{align}\label{Eqn: derivative estimator, aux5}
\lim_{n\to\infty} \bar\phi_n'(h)=\phi_{\theta_P}'(h)~.
\end{align}

Next, by Assumption \ref{Ass: convex setup} and the proof of Lemma \ref{Lem: projection monotonicity}, we have
\begin{align}\label{Eqn: derivative estimator, aux51}
\hat\phi_n'(h)=\min_{|\alpha|\le \kappa_n} \|h+\alpha\Pi_\Lambda\hat\theta_n-\Pi_\Lambda (h+\alpha\Pi_\Lambda\hat\theta_n)\|_{\mathbf H}~.
\end{align}
In turn, we obtain from result \eqref{Eqn: derivative estimator, aux51} that
\begin{align}\label{Eqn: derivative estimator, aux6}
&|\hat\phi_n'(h)-\bar\phi_n'(h)| \notag\\
&= \big |\min_{|\alpha|\le \kappa_n} \|h+\alpha\Pi_\Lambda\hat\theta_n-\Pi_\Lambda (h+\alpha\Pi_\Lambda\hat\theta_n)\|_{\mathbf H}- \min_{|\alpha|\le \kappa_n} \|h+\alpha\theta_P-\Pi_\Lambda(h+\alpha\theta_P)\|_{\mathbf H}\big |\notag\\
&\le \max_{|\alpha|\le \kappa_n} \|\alpha\Pi_\Lambda\hat\theta_n -\alpha\theta_P\|_{\mathbf H}\le \frac{\kappa_n}{ r_n}\| r_n \{\hat\theta_n-\theta_P\}\|_{\mathbf H}=o(1)O_p(1)=o_p(1)~,
\end{align}
where the first inequality follows from the Lipschitz continuity of the min operator and Theorem 3.16 in \citet{AliprantisandBorder2006}, and the second inequality is due to $\theta_P=\Pi_\Lambda\theta_P$ and Lemma 6.54-d in \citet{AliprantisandBorder2006}. Combination of results \eqref{Eqn: derivative estimator, aux5} and \eqref{Eqn: derivative estimator, aux6} then leads to the desired conclusion. \qed

\begin{lem}\label{Lem: subadditivity}
Let Assumption \ref{Ass: convex setup}(i) hold and $\phi(h)\equiv \|h-\Pi_\Lambda h\|_{\mathbf H}$ for all $h\in\mathbf H$. Then $h\mapsto\phi(h)$ is convex. If in addition Assumption \ref{Ass: convex setup}(ii) holds, then $h\mapsto\phi(h)$ is also positively homogeneous of degree one and subadditive.
\end{lem}
\noindent{\sc Proof:} Given Assumption \ref{Ass: convex setup}(i),  $\phi$ is convex by Corollary 12.12 in \citet{BauschkeCombettes2017Convex}. If Assumption \ref{Ass: convex setup}(ii) also holds, then $\phi$ is positively homogeneous of degree one by Proposition 29.29 in \citet{BauschkeCombettes2017Convex}, and thus subadditive by Proposition 10.3 in \citet{BauschkeCombettes2017Convex}. \qed

\begin{lem}\label{Lem: asymptotic independence}
Let $\tilde c_{n,P}(1-\alpha)$ be the $(1-\alpha)$ conditional quantile of $\psi_{\kappa_n,P}(\bar{\mathbb Z}_{n,P})$ given $\{X_i\}_{i=1}^n$. Then \eqref{Eqn: asymptotic independence} implies that, for some $\epsilon_n\downarrow 0$ and any $\alpha_n\in(0,1-\epsilon_n)$,
\begin{align}
\liminf_{n\to\infty}\inf_{P\in\mathbf P}P(\tilde c_{n,P}(1-\alpha_n)\ge c_{n,P}(1-\alpha_n-\epsilon_n)-\epsilon_n)=1~.
\end{align}
\end{lem}
\begin{rem}
As explained in Section \ref{Sec: special case}, the coupling order $o_p(1)$ is sufficient in settings where $\hat\theta_n$ converges in distribution. Then, in the proof of Theorem \ref{Thm: size control}, the quantile $c_{n,P}(1-\alpha-\eta_n)$ may be replaced by $\tilde c_{n,P}(1-\alpha-\eta_n)$, which may still be justified by Lemma 11 in \citet{ChernozhukovLeeRosen2013Intersection}. Given the condition \eqref{Eqn: asymptotic independence}, Lemma \ref{Lem: asymptotic independence} in turn allows us to bound $\tilde c_{n,P}(1-\alpha-\eta_n)$ from below by $c_{n,P}(1-\alpha-\eta_n-\epsilon_n)-\epsilon_n$ asymptotically (by making $\epsilon_n$ larger if necessary), so that the remaining arguments there may still apply. This explains that, in the setting of Section \ref{Sec: special case}, the independence condition in Assumption \ref{Ass: strong approx}(ii) may be replaced with \eqref{Eqn: asymptotic independence}. \qed
\end{rem}
\noindent{\sc Proof of Lemma \ref{Lem: asymptotic independence}:} Since $f\circ\psi_{\kappa_n,P}\in\mathrm{BL}_1(\mathbf H)$ whenever $f\in\mathrm{BL}_1(\mathbf R)$ by Theorem 3.16 in \citet{AliprantisandBorder2006}, we may obtain from \eqref{Eqn: asymptotic independence} that,
\begin{align}\label{Eqn: asymptotic independence, aux1}
\sup_{f\in\mathrm{BL}_1(\mathbf R)} |E[f(\psi_{\kappa_n,P}(\bar{\mathbb Z}_{n,P}))|\{X_i\}_{i=1}^n]-E[f(\psi_{\kappa_n,P}(\bar{\mathbb Z}_{n,P}))]|=o_p(1)~,
\end{align}
uniformly in $P\in\mathbf P$. Let $\tilde F_{n,P}$ be the conditional cdf of $\psi_{\kappa_n,P}(\bar{\mathbb Z}_{n,P})$ given $\{X_i\}_{i=1}^n$ and $F_{n,P}$ be the unconditional cdf of $\psi_{\kappa_n,P}(\bar{\mathbb Z}_{n,P})$. By Lemma 3.29 in \citet{Dudley2014UCLT}, we may then obtain from result \eqref{Eqn: asymptotic independence, aux1} that
\begin{align}\label{Eqn: asymptotic independence, aux2}
\liminf_{n\to\infty}\inf_{P\in\mathbf P}P(\tilde F_{n,P}(x)\le F_{n,P}(x+\epsilon_n)+\epsilon_n\,\forall x\in\mathbf R)=1~,
\end{align}
for some $\epsilon_n\downarrow 0$ slowly. If the event that $\tilde F_{n,P}(x)\le F_{n,P}(x+\epsilon_n)+\epsilon_n$ for $x\in\mathbf R$ occurs, then by the definition of quantile, we must have for all $\alpha_n\in(0,1-\epsilon_n)$ that
\begin{align}
F_{n,P}(\tilde c_{n,P}(1-\alpha_n)+\epsilon_n)\ge  \tilde F_{n,P}(\tilde c_{n,P}(1-\alpha_n))-\epsilon_n\ge 1-\alpha_n-\epsilon_n~,
\end{align}
which implies by the definition of quantile that
\begin{align}\label{Eqn: asymptotic independence, aux3}
\tilde c_{n,P}(1-\alpha_n)+\epsilon_n\ge c_{n,P}(1-\alpha_n-\epsilon_n)~.
\end{align}
The lemma thus follows from combining results \eqref{Eqn: asymptotic independence, aux2} and \eqref{Eqn: asymptotic independence, aux3}. \qed

\begin{lem}\label{Lem: derivative, alternative form}
If Assumption \ref{Ass: convex setup} holds and $\theta_0\in\Lambda$, then, for any $h\in\mathbf H$,
\begin{align}\label{Eqn: derivative, alternative form}
\phi_{\theta_0}'(h)\equiv\|h-\Pi_{T_{\theta_0}}h\|_{\mathbf H}=\inf_{\alpha\in\mathbf R} \|h+\alpha\theta_0-\Pi_\Lambda(h+\alpha\theta_0)\|_{\mathbf H}~.
\end{align}
\end{lem}

\noindent{\sc Proof:} Fix $h\in\mathbf H$. By Assumption \ref{Ass: convex setup} and Lemma 4.2.5 in \citet{Aubin_Frankowska2009set}, $T_{\theta_0}=\overline{\Lambda+\mathbf R\theta_0}$. By the definition of projection, we thus have
\begin{multline}\label{Eqn: derivative, alternative form, aux}
\phi_{\theta_0}'(h)=\|h-\Pi_{T_{\theta_0}}h\|_{\mathbf H}=\inf_{\lambda'\in \overline{\Lambda+\mathbf R \theta_0}}\|h-\lambda'\|_{\mathbf H}=\inf_{\lambda'\in \Lambda+\mathbf R \theta_0}\|h-\lambda'\|_{\mathbf H} \\
=\inf_{\alpha\in\mathbf R}\inf_{\lambda\in\Lambda} \|h-\lambda-\alpha\theta_0\|_{\mathbf H}=\inf_{\alpha\in\mathbf R} \|h-\alpha\theta_0-\Pi_\Lambda(h-\alpha\theta_0)\|_{\mathbf H}~,
\end{multline} 
where the third equality follows by continuity of $\lambda'\mapsto\|h-\lambda'\|_{\mathbf H}$ and Proposition 11.1(i) in \citet{BauschkeCombettes2017Convex}, and the last one is by applying the definition of projection for each $\alpha\in\mathbf R$. The lemma then follows from \eqref{Eqn: derivative, alternative form, aux} by noting that the minimization in the last step is taken over $\mathbf R$ which is symmetric around zero. \qed

\begin{figure}[t]
\centering\footnotesize
\begin{tikzpicture}[>=latex',scale=1.1,dot/.style={circle,inner sep=1pt,fill,name=#1},
                    extended line/.style={shorten >=-#1,shorten <=-#1},
                    extended line/.default=1cm]
\draw (-4,-2.5)--(-4,3)--(2,3);
\draw (-4,-2.5)--(-5/3,-2.5);
\draw[loosely dotted,->] (-4,0)--(4,0);
\draw (4,0) node[anchor=west] {$0$};
\draw[loosely dotted,->] (0,-2.5)--(0,3);
\draw (0,3) node[anchor=south] {$0$};
\fill[red] (0,0) circle (0.8pt);
\draw (0,0) node[anchor=north west] {$0$};
\draw[->,NavyBlue] (0,0) -- +({atan(1.5)}:0.5);
\draw[->,NavyBlue] (0,0) -- +({atan(-2/3)}:0.5);
\draw[->,NavyBlue] (0,0) -- +({180+atan(1.5)}:0.5);
\fill[fill=Honeydew3,semitransparent] (0,0)--(2,3)--(4,3)--(4,0)--(0,0);
\draw (-5/3,-2.5)--(2,3);
\draw[dashdotted] (0,0)--(4,0);
\draw (2,1) node[anchor=west] {$\Lambda$};
\node[dot=theta,label={east: $\theta_P$}] at (4/5,6/5) {};
\draw[->,NavyBlue] (4/5,6/5) -- +({atan(1.5)}:0.5);
\draw[->,NavyBlue] (4/5,6/5) -- +({atan(-2/3)}:0.5);
\draw[->,NavyBlue] (4/5,6/5) -- +({180+atan(1.5)}:0.5);
\fill[very nearly transparent] (2,3)--(4,3)--(4,-2.5)--(-5/3,-2.5)--(2,3);
\draw (2,-1) node[anchor=west] {$T_{\theta_P}$};
\node[dot=h,label={west:{$h$}}] at (-2.5,-0.5)  {};
\node (zero) at (-5/3,-2.5)  {};
\coordinate[label=right: $\Pi_{T_{\theta_P}}(h)$] (h') at ($(theta)!(h)!(zero)$);
\fill (h') circle (0.8pt);
\draw[densely dotted]  (h') -- (h);
\draw[extended line=2cm] (h)--+($(theta)-(0,0)$);
\draw (-0.9,1.8) node[anchor=south] {$h+\alpha\theta_P:\alpha\ge 0$};
\draw (-3,-1.55) node[anchor=north] {$h+\alpha\theta_P:\alpha\le 0$};
\draw[dashed] (h)--(0,0);
\draw[dashed] (-37/18,1/6)--(0,0);
\draw[dashed] (-3/2,1)--(0,0);
\fill[red] (-3/2,1) circle (0.8pt);
\draw (-3/2,1)  node[anchor=south east] {$h+\alpha^*\theta_P$};
\end{tikzpicture}
\caption{Lemma \ref{Lem: derivative, alternative form} shows that the distance from $h$ to the tangent cone (i.e., $\phi_{\theta_0}'(h)$) is simply the shortest distance from points on the line $h+\alpha\theta_0$ to $\Lambda$ as $\alpha$ ranges over $\mathbf R$.}
\label{Fig: distance to tangent cone}
\end{figure}
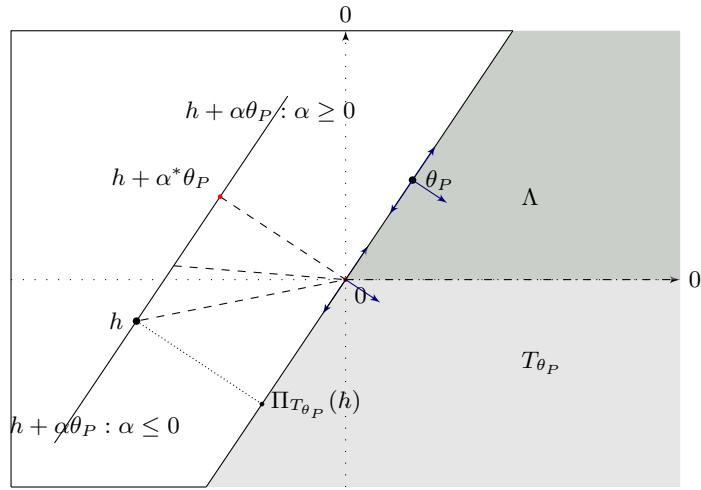

\section{Additional Simulation Results}\label{App: full simulations}

The section collects the complete set of results for the simulation designs in Section \ref{Sec: simulations} and Appendix \ref{App: more simulations}.
{
\setlength{\tabcolsep}{6.5pt}
\renewcommand{\arraystretch}{1.2}
\begin{table}[!ht]
\caption{Empirical Size of Monotonicity Tests for \eqref{Eqn: MC1,aux1} at $\alpha=5\%$}
\centering\footnotesize
\begin{threeparttable}
\sisetup{table-number-alignment = center, table-format = 1.3} 
\begin{tabularx}{\linewidth}{@{}cc *{3}{S[round-mode = places,round-precision = 3]} c *{3}{S[round-mode = places,round-precision = 3]} c *{3}{S[round-mode = places,round-precision = 3]}@{}} 
\hline
\hline
\multirow{2}{*}{$n$} & \multirow{2}{*}{$\gamma_n$} & \multicolumn{3}{c}{FS-C3: $k_n=7$} & & \multicolumn{3}{c}{FS-C5: $k_n=9$} & & \multicolumn{3}{c}{FS-C7: $k_n=11$}\\
\cline{3-5} \cline{7-9} \cline{11-13}
& & {D1} & {D2} & {D3}  & & {D1} & {D2} & {D3} & & {D1} & {D2} & {D3}\\
\hline
\multirow{9}{*}{$500$}   & $1/n^{1/2}$            & 0.0537 & 0.0157 & 0.0030 & & 0.0577 & 0.0207 & 0.0033 & & 0.0590 & 0.0197 & 0.0030\\
                         & $1/n^{3/4}$            & 0.0530 & 0.0157 & 0.0030 & & 0.0577 & 0.0207 & 0.0033 & & 0.0583 & 0.0193 & 0.0027\\
                         & $1/n$                  & 0.0530 & 0.0157 & 0.0030 & & 0.0577 & 0.0207 & 0.0033 & & 0.0583 & 0.0190 & 0.0027\\
                         & $0.1/\log n$           & 0.0530 & 0.0157 & 0.0030 & & 0.0577 & 0.0207 & 0.0033 & & 0.0583 & 0.0193 & 0.0027\\
                         & $0.05/\log n$          & 0.0530 & 0.0157 & 0.0030 & & 0.0577 & 0.0207 & 0.0033 & & 0.0583 & 0.0193 & 0.0027\\
                         & $0.01/\log n$          & 0.0530 & 0.0157 & 0.0030 & & 0.0577 & 0.0207 & 0.0033 & & 0.0583 & 0.0190 & 0.0027\\
                         & $0.1$                  & 0.0540 & 0.0160 & 0.0033 & & 0.0580 & 0.0210 & 0.0037 & & 0.0590 & 0.0197 & 0.0033\\
                         & $0.05$                 & 0.0537 & 0.0157 & 0.0030 & & 0.0577 & 0.0207 & 0.0033 & & 0.0590 & 0.0197 & 0.0030\\
                         & $0.01$                 & 0.0530 & 0.0157 & 0.0030 & & 0.0577 & 0.0207 & 0.0033 & & 0.0583 & 0.0193 & 0.0027\\
\hline
\multirow{9}{*}{$750$}   & $1/n^{1/2}$            & 0.0527 & 0.0097 & 0.0017 & & 0.0567 & 0.0143 & 0.0023 & & 0.0590 & 0.0173 & 0.0030\\
                         & $1/n^{3/4}$            & 0.0523 & 0.0097 & 0.0013 & & 0.0560 & 0.0140 & 0.0017 & & 0.0590 & 0.0173 & 0.0030\\
                         & $1/n$                  & 0.0523 & 0.0097 & 0.0013 & & 0.0560 & 0.0137 & 0.0017 & & 0.0590 & 0.0173 & 0.0030\\
                         & $0.1/\log n$           & 0.0523 & 0.0097 & 0.0013 & & 0.0563 & 0.0140 & 0.0020 & & 0.0590 & 0.0173 & 0.0030\\
                         & $0.05/\log n$          & 0.0523 & 0.0097 & 0.0013 & & 0.0560 & 0.0140 & 0.0017 & & 0.0590 & 0.0173 & 0.0030\\
                         & $0.01/\log n$          & 0.0523 & 0.0097 & 0.0013 & & 0.0560 & 0.0137 & 0.0017 & & 0.0590 & 0.0173 & 0.0030\\
                         & $0.1$                  & 0.0527 & 0.0097 & 0.0017 & & 0.0567 & 0.0143 & 0.0030 & & 0.0597 & 0.0173 & 0.0033\\
                         & $0.05$                 & 0.0527 & 0.0097 & 0.0017 & & 0.0567 & 0.0143 & 0.0027 & & 0.0590 & 0.0173 & 0.0030\\
                         & $0.01$                 & 0.0523 & 0.0097 & 0.0013 & & 0.0560 & 0.0140 & 0.0020 & & 0.0590 & 0.0173 & 0.0030\\
\hline
\multirow{9}{*}{$1000$}  & $1/n^{1/2}$            & 0.0570 & 0.0110 & 0.0013 & & 0.0563 & 0.0117 & 0.0003 & & 0.0560 & 0.0130 & 0.0007\\
                         & $1/n^{3/4}$            & 0.0557 & 0.0107 & 0.0007 & & 0.0560 & 0.0113 & 0.0003 & & 0.0560 & 0.0130 & 0.0007\\
                         & $1/n$                  & 0.0557 & 0.0107 & 0.0007 & & 0.0560 & 0.0113 & 0.0003 & & 0.0560 & 0.0127 & 0.0007\\
                         & $0.1/\log n$           & 0.0567 & 0.0110 & 0.0010 & & 0.0560 & 0.0117 & 0.0003 & & 0.0560 & 0.0130 & 0.0007\\
                         & $0.05/\log n$          & 0.0560 & 0.0107 & 0.0010 & & 0.0560 & 0.0117 & 0.0003 & & 0.0560 & 0.0130 & 0.0007\\
                         & $0.01/\log n$          & 0.0557 & 0.0107 & 0.0007 & & 0.0560 & 0.0113 & 0.0003 & & 0.0560 & 0.0127 & 0.0007\\
                         & $0.1$                  & 0.0570 & 0.0113 & 0.0017 & & 0.0570 & 0.0120 & 0.0007 & & 0.0573 & 0.0133 & 0.0007\\
                         & $0.05$                 & 0.0570 & 0.0110 & 0.0013 & & 0.0567 & 0.0117 & 0.0003 & & 0.0567 & 0.0130 & 0.0007\\
                         & $0.01$                 & 0.0560 & 0.0107 & 0.0010 & & 0.0560 & 0.0117 & 0.0003 & & 0.0560 & 0.0130 & 0.0007\\
\hline
\multicolumn{2}{c}{\multirow{2}{*}{$n$}}  & \multicolumn{3}{c}{LSW-S} & & \multicolumn{3}{c}{LSW-L} & & \multicolumn{3}{c}{C-OS}\\
\cline{3-5} \cline{7-9} \cline{11-13}
\multicolumn{2}{c}{}     & {D1} & {D2} & {D3}  & & {D1} & {D2} & {D3} & & {D1} & {D2} & {D3}\\
\hline
\multicolumn{2}{c}{500}                          & 0.0600 & 0.0413 & 0.0083 & & 0.0660 & 0.0350 & 0.0037 & & 0.0597  & 0.0413  & 0.0120\\
\multicolumn{2}{c}{750}                          & 0.0567 & 0.0363 & 0.0050 & & 0.0587 & 0.0303 & 0.0057 & & 0.0540  & 0.0337  & 0.0080\\
\multicolumn{2}{c}{1000}                         & 0.0610 & 0.0347 & 0.0047 & & 0.0653 & 0.0347 & 0.0033 & & 0.0493  & 0.0357  & 0.0090\\
\hline
\hline
\end{tabularx}
\begin{tablenotes}[flushleft]
\item {\it Note:} The parameter $\gamma_n$ determines $\hat\kappa_n$ proposed in Section \ref{Sec: tuning parameter} with $c_n=1/\log n$ and $r_n=(n/k_n)^{1/2}$.
\end{tablenotes}
\end{threeparttable}
\label{Tab: MC1Mon, size1, app} 
\end{table}
}

{
\setlength{\tabcolsep}{6.5pt}
\renewcommand{\arraystretch}{1.2}
\begin{table}[!ht]
\caption{Empirical Size of Convexity Tests for \eqref{Eqn: MC1,aux1} at $\alpha=5\%$}
\centering\footnotesize
\begin{threeparttable}
\sisetup{table-number-alignment = center, table-format = 1.3} 
\begin{tabularx}{\linewidth}{@{}cc *{3}{S[round-mode = places,round-precision = 3]} c *{3}{S[round-mode = places,round-precision = 3]} c *{3}{S[round-mode = places,round-precision = 3]}@{}} 
\hline
\hline
\multirow{2}{*}{$n$} & \multirow{2}{*}{$\gamma_n$} & \multicolumn{3}{c}{FS-C3: $k_n=7$} & & \multicolumn{3}{c}{FS-C5: $k_n=9$} & & \multicolumn{3}{c}{FS-C7: $k_n=11$}\\
\cline{3-5} \cline{7-9} \cline{11-13}
& & {D1} & {D2} & {D3}  & & {D1} & {D2} & {D3} & & {D1} & {D2} & {D3}\\
\hline
\multirow{9}{*}{$500$}   & $1/n^{1/2}$            & 0.0487 & 0.0423 & 0.0100 & & 0.0567 & 0.0480 & 0.0160 & & 0.0537 & 0.0457 & 0.0177\\
                         & $1/n^{3/4}$            & 0.0487 & 0.0423 & 0.0093 & & 0.0563 & 0.0477 & 0.0157 & & 0.0533 & 0.0453 & 0.0173\\
                         & $1/n$                  & 0.0480 & 0.0423 & 0.0093 & & 0.0563 & 0.0473 & 0.0157 & & 0.0533 & 0.0453 & 0.0173\\
                         & $0.1/\log n$           & 0.0487 & 0.0423 & 0.0100 & & 0.0563 & 0.0477 & 0.0157 & & 0.0533 & 0.0453 & 0.0173\\
                         & $0.05/\log n$          & 0.0487 & 0.0423 & 0.0093 & & 0.0563 & 0.0477 & 0.0157 & & 0.0533 & 0.0453 & 0.0173\\
                         & $0.01/\log n$          & 0.0480 & 0.0423 & 0.0093 & & 0.0563 & 0.0473 & 0.0157 & & 0.0533 & 0.0453 & 0.0173\\
                         & $0.1$                  & 0.0490 & 0.0427 & 0.0100 & & 0.0567 & 0.0480 & 0.0163 & & 0.0543 & 0.0460 & 0.0180\\
                         & $0.05$                 & 0.0487 & 0.0427 & 0.0100 & & 0.0567 & 0.0480 & 0.0163 & & 0.0540 & 0.0457 & 0.0177\\
                         & $0.01$                 & 0.0487 & 0.0423 & 0.0093 & & 0.0563 & 0.0477 & 0.0157 & & 0.0533 & 0.0453 & 0.0173\\
\hline
\multirow{9}{*}{$750$}   & $1/n^{1/2}$            & 0.0577 & 0.0463 & 0.0077 & & 0.0627 & 0.0547 & 0.0120 & & 0.0597 & 0.0563 & 0.0207\\
                         & $1/n^{3/4}$            & 0.0577 & 0.0460 & 0.0073 & & 0.0623 & 0.0547 & 0.0110 & & 0.0597 & 0.0553 & 0.0197\\
                         & $1/n$                  & 0.0577 & 0.0460 & 0.0070 & & 0.0623 & 0.0547 & 0.0107 & & 0.0593 & 0.0553 & 0.0190\\
                         & $0.1/\log n$           & 0.0577 & 0.0463 & 0.0077 & & 0.0627 & 0.0547 & 0.0113 & & 0.0597 & 0.0557 & 0.0203\\
                         & $0.05/\log n$          & 0.0577 & 0.0460 & 0.0077 & & 0.0623 & 0.0547 & 0.0110 & & 0.0597 & 0.0557 & 0.0200\\
                         & $0.01/\log n$          & 0.0577 & 0.0460 & 0.0070 & & 0.0623 & 0.0547 & 0.0107 & & 0.0593 & 0.0553 & 0.0190\\
                         & $0.1$                  & 0.0577 & 0.0467 & 0.0077 & & 0.0630 & 0.0547 & 0.0133 & & 0.0603 & 0.0563 & 0.0213\\
                         & $0.05$                 & 0.0577 & 0.0463 & 0.0077 & & 0.0627 & 0.0547 & 0.0123 & & 0.0597 & 0.0563 & 0.0207\\
                         & $0.01$                 & 0.0577 & 0.0460 & 0.0077 & & 0.0623 & 0.0547 & 0.0110 & & 0.0593 & 0.0557 & 0.0203\\
\hline
\multirow{9}{*}{$1000$}  & $1/n^{1/2}$            & 0.0523 & 0.0450 & 0.0053 & & 0.0553 & 0.0470 & 0.0097 & & 0.0540 & 0.0453 & 0.0130\\
                         & $1/n^{3/4}$            & 0.0523 & 0.0440 & 0.0053 & & 0.0550 & 0.0467 & 0.0097 & & 0.0540 & 0.0447 & 0.0127\\
                         & $1/n$                  & 0.0523 & 0.0440 & 0.0053 & & 0.0550 & 0.0467 & 0.0097 & & 0.0540 & 0.0443 & 0.0127\\
                         & $0.1/\log n$           & 0.0523 & 0.0443 & 0.0053 & & 0.0550 & 0.0470 & 0.0097 & & 0.0540 & 0.0450 & 0.0127\\
                         & $0.05/\log n$          & 0.0523 & 0.0443 & 0.0053 & & 0.0550 & 0.0470 & 0.0097 & & 0.0540 & 0.0447 & 0.0127\\
                         & $0.01/\log n$          & 0.0523 & 0.0440 & 0.0053 & & 0.0550 & 0.0467 & 0.0097 & & 0.0540 & 0.0443 & 0.0127\\
                         & $0.1$                  & 0.0523 & 0.0450 & 0.0057 & & 0.0553 & 0.0473 & 0.0103 & & 0.0540 & 0.0460 & 0.0137\\
                         & $0.05$                 & 0.0523 & 0.0450 & 0.0053 & & 0.0553 & 0.0470 & 0.0097 & & 0.0540 & 0.0457 & 0.0133\\
                         & $0.01$                 & 0.0523 & 0.0443 & 0.0053 & & 0.0550 & 0.0470 & 0.0097 & & 0.0540 & 0.0450 & 0.0127\\
\hline
\multicolumn{2}{c}{\multirow{2}{*}{$n$}}  & \multicolumn{3}{c}{LSW-S} & & \multicolumn{3}{c}{LSW-L} & & \multicolumn{3}{c}{ }\\
\cline{3-5} \cline{7-9}
\multicolumn{2}{c}{}     & {D1} & {D2} & {D3}  & & {D1} & {D2} & {D3} & &   &   &  \\
\hline
\multicolumn{2}{c}{500}                          & 0.0593 & 0.0583 & 0.0480 & & 0.0633  & 0.0657  & 0.0497  & &   &   & \\
\multicolumn{2}{c}{750}                          & 0.0633 & 0.0577 & 0.0487 & & 0.0643  & 0.0640  & 0.0467  & &   &   & \\
\multicolumn{2}{c}{1000}                         & 0.0567 & 0.0547 & 0.0457 & & 0.0580  & 0.0583  & 0.0460  & &   &   & \\
\hline
\hline
\end{tabularx}
\begin{tablenotes}[flushleft]
\item {\it Note:} The parameter $\gamma_n$ determines $\hat\kappa_n$ proposed in Section \ref{Sec: tuning parameter} with $c_n=1/\log n$ and $r_n=(n/k_n)^{1/2}$.
\end{tablenotes}
\end{threeparttable}
\label{Tab: MC1Con, size1, app} 
\end{table}
}

{
\setlength{\tabcolsep}{6.5pt}
\renewcommand{\arraystretch}{1.2}
\begin{table}[!ht]
\caption{Empirical Size of Monotonicity-Convexity Tests for \eqref{Eqn: MC1,aux1} at $\alpha=5\%$}
\centering\footnotesize
\begin{threeparttable}
\sisetup{table-number-alignment = center, table-format = 1.3} 
\begin{tabularx}{\linewidth}{@{}cc *{3}{S[round-mode = places,round-precision = 3]} c *{3}{S[round-mode = places,round-precision = 3]} c *{3}{S[round-mode = places,round-precision = 3]}@{}} 
\hline
\hline
\multirow{2}{*}{$n$} & \multirow{2}{*}{$\gamma_n$} & \multicolumn{3}{c}{FS-C3: $k_n=7$} & & \multicolumn{3}{c}{FS-C5: $k_n=9$} & & \multicolumn{3}{c}{FS-C7: $k_n=11$}\\
\cline{3-5} \cline{7-9} \cline{11-13}
& & {D1} & {D2} & {D3}  & & {D1} & {D2} & {D3} & & {D1} & {D2} & {D3}\\
\hline
\multirow{9}{*}{$500$}   & $1/n^{1/2}$            & 0.0503 & 0.0273 & 0.0083 & & 0.0550 & 0.0333 & 0.0117 & & 0.0550 & 0.0330 & 0.0143\\
                         & $1/n^{3/4}$            & 0.0503 & 0.0263 & 0.0073 & & 0.0543 & 0.0323 & 0.0113 & & 0.0543 & 0.0320 & 0.0140\\
                         & $1/n$                  & 0.0500 & 0.0263 & 0.0070 & & 0.0540 & 0.0323 & 0.0110 & & 0.0543 & 0.0317 & 0.0133\\
                         & $0.1/\log n$           & 0.0503 & 0.0263 & 0.0077 & & 0.0547 & 0.0327 & 0.0113 & & 0.0543 & 0.0323 & 0.0143\\
                         & $0.05/\log n$          & 0.0503 & 0.0263 & 0.0073 & & 0.0543 & 0.0323 & 0.0113 & & 0.0543 & 0.0320 & 0.0140\\
                         & $0.01/\log n$          & 0.0500 & 0.0263 & 0.0070 & & 0.0540 & 0.0323 & 0.0110 & & 0.0543 & 0.0317 & 0.0133\\
                         & $0.1$                  & 0.0503 & 0.0277 & 0.0090 & & 0.0557 & 0.0337 & 0.0120 & & 0.0553 & 0.0337 & 0.0147\\
                         & $0.05$                 & 0.0503 & 0.0273 & 0.0083 & & 0.0550 & 0.0333 & 0.0117 & & 0.0550 & 0.0330 & 0.0143\\
                         & $0.01$                 & 0.0503 & 0.0263 & 0.0073 & & 0.0543 & 0.0327 & 0.0113 & & 0.0543 & 0.0320 & 0.0140\\
\hline
\multirow{9}{*}{$750$}   & $1/n^{1/2}$            & 0.0563 & 0.0263 & 0.0057 & & 0.0597 & 0.0353 & 0.0090 & & 0.0577 & 0.0347 & 0.0180\\
                         & $1/n^{3/4}$            & 0.0563 & 0.0257 & 0.0053 & & 0.0590 & 0.0347 & 0.0080 & & 0.0573 & 0.0343 & 0.0177\\
                         & $1/n$                  & 0.0563 & 0.0257 & 0.0053 & & 0.0590 & 0.0340 & 0.0080 & & 0.0573 & 0.0343 & 0.0173\\
                         & $0.1/\log n$           & 0.0563 & 0.0257 & 0.0057 & & 0.0590 & 0.0347 & 0.0083 & & 0.0573 & 0.0343 & 0.0180\\
                         & $0.05/\log n$          & 0.0563 & 0.0257 & 0.0053 & & 0.0590 & 0.0347 & 0.0080 & & 0.0573 & 0.0343 & 0.0177\\
                         & $0.01/\log n$          & 0.0563 & 0.0257 & 0.0053 & & 0.0590 & 0.0340 & 0.0080 & & 0.0573 & 0.0343 & 0.0173\\
                         & $0.1$                  & 0.0570 & 0.0263 & 0.0060 & & 0.0603 & 0.0353 & 0.0100 & & 0.0580 & 0.0360 & 0.0187\\
                         & $0.05$                 & 0.0570 & 0.0263 & 0.0057 & & 0.0597 & 0.0353 & 0.0093 & & 0.0577 & 0.0350 & 0.0183\\
                         & $0.01$                 & 0.0563 & 0.0257 & 0.0053 & & 0.0590 & 0.0347 & 0.0080 & & 0.0573 & 0.0343 & 0.0180\\
\hline
\multirow{9}{*}{$1000$}  & $1/n^{1/2}$            & 0.0550 & 0.0227 & 0.0037 & & 0.0557 & 0.0297 & 0.0067 & & 0.0533 & 0.0300 & 0.0103\\
                         & $1/n^{3/4}$            & 0.0550 & 0.0223 & 0.0037 & & 0.0553 & 0.0290 & 0.0060 & & 0.0533 & 0.0300 & 0.0103\\
                         & $1/n$                  & 0.0550 & 0.0220 & 0.0037 & & 0.0553 & 0.0287 & 0.0057 & & 0.0533 & 0.0300 & 0.0103\\
                         & $0.1/\log n$           & 0.0550 & 0.0227 & 0.0037 & & 0.0557 & 0.0293 & 0.0063 & & 0.0533 & 0.0300 & 0.0103\\
                         & $0.05/\log n$          & 0.0550 & 0.0223 & 0.0037 & & 0.0553 & 0.0290 & 0.0060 & & 0.0533 & 0.0300 & 0.0103\\
                         & $0.01/\log n$          & 0.0550 & 0.0220 & 0.0037 & & 0.0553 & 0.0287 & 0.0057 & & 0.0533 & 0.0300 & 0.0103\\
                         & $0.1$                  & 0.0550 & 0.0237 & 0.0040 & & 0.0557 & 0.0303 & 0.0073 & & 0.0537 & 0.0303 & 0.0103\\
                         & $0.05$                 & 0.0550 & 0.0237 & 0.0037 & & 0.0557 & 0.0300 & 0.0067 & & 0.0533 & 0.0300 & 0.0103\\
                         & $0.01$                 & 0.0550 & 0.0227 & 0.0037 & & 0.0557 & 0.0293 & 0.0060 & & 0.0533 & 0.0300 & 0.0103\\
\hline
\multicolumn{2}{c}{\multirow{2}{*}{$n$}}  & \multicolumn{3}{c}{LSW-S} & & \multicolumn{3}{c}{LSW-S} & & \multicolumn{3}{c}{ }\\
\cline{3-5} \cline{7-9}
\multicolumn{2}{c}{}     & {D1} & {D2} & {D3}  & & {D1} & {D2} & {D3} & &   &   &  \\
\hline
\multicolumn{2}{c}{500}                          & 0.0653 & 0.0570 & 0.0297 & & 0.0680  & 0.0570  & 0.0303  & &   &   & \\
\multicolumn{2}{c}{750}                          & 0.0653 & 0.0517 & 0.0320 & & 0.0687  & 0.0530  & 0.0310  & &   &   & \\
\multicolumn{2}{c}{1000}                         & 0.0597 & 0.0483 & 0.0263 & & 0.0647  & 0.0543  & 0.0263  & &   &   & \\
\hline
\hline
\end{tabularx}
\begin{tablenotes}[flushleft]
\item {\it Note:} The parameter $\gamma_n$ determines $\hat\kappa_n$ proposed in Section \ref{Sec: tuning parameter} with $c_n=1/\log n$ and $r_n=(n/k_n)^{1/2}$.
\end{tablenotes}
\end{threeparttable}
\label{Tab: MC1MonCon, size1, app} 
\end{table}
}

\if AllResultsYes\choice

{
\setlength{\tabcolsep}{5.6pt}
\renewcommand{\arraystretch}{1.2}
\begin{table}[!ht]
\caption{Empirical Size of Monotonicity Tests for \eqref{Eqn: MC2,aux1} at $\alpha=5\%$}
\vspace{-0.1 in}
\centering\small
\sisetup{table-number-alignment = center, table-format = 1.3} 
\begin{tabularx}{\linewidth}{@{}cc *{3}{S[round-mode = places,round-precision = 3]} c *{3}{S[round-mode = places,round-precision = 3]} c *{3}{S[round-mode = places,round-precision = 3]}@{}} 
\hline
\hline
\multirow{3}{*}{$n$} & \multirow{3}{*}{$\gamma_n$} & \multicolumn{11}{c}{Our test with quadratic B-spline}  \\
\cline{3-13}
& & \multicolumn{3}{c}{0 interior knots} & & \multicolumn{3}{c}{1 interior knot} & & \multicolumn{3}{c}{2 interior knots}\\
\cline{3-5} \cline{7-9} \cline{11-13}
& & {D1} & {D2} & {D3}  & & {D1} & {D2} & {D3} & & {D1} & {D2} & {D3}\\
\hline
\multirow{9}{*}{$500$}   & $1/n^{1/2}$            & 0.0643 & 0.0203 & 0.0000 & & 0.0690 & 0.0303 & 0.0020 & & 0.0813 & 0.0460 & 0.0057\\
                         & $1/n^{3/4}$            & 0.0627 & 0.0190 & 0.0000 & & 0.0687 & 0.0300 & 0.0020 & & 0.0803 & 0.0447 & 0.0053\\
                         & $1/n$                  & 0.0613 & 0.0183 & 0.0000 & & 0.0680 & 0.0300 & 0.0017 & & 0.0797 & 0.0437 & 0.0050\\
                         & $0.1/\log n$           & 0.0633 & 0.0197 & 0.0000 & & 0.0690 & 0.0300 & 0.0020 & & 0.0810 & 0.0450 & 0.0053\\
                         & $0.05/\log n$          & 0.0623 & 0.0187 & 0.0000 & & 0.0687 & 0.0300 & 0.0020 & & 0.0803 & 0.0443 & 0.0053\\
                         & $0.01/\log n$          & 0.0613 & 0.0183 & 0.0000 & & 0.0680 & 0.0300 & 0.0017 & & 0.0797 & 0.0437 & 0.0050\\
                         & $0.1$                  & 0.0657 & 0.0203 & 0.0000 & & 0.0693 & 0.0307 & 0.0023 & & 0.0820 & 0.0473 & 0.0063\\
                         & $0.05$                 & 0.0643 & 0.0203 & 0.0000 & & 0.0690 & 0.0303 & 0.0023 & & 0.0813 & 0.0463 & 0.0057\\
                         & $0.01$                 & 0.0627 & 0.0193 & 0.0000 & & 0.0687 & 0.0300 & 0.0020 & & 0.0807 & 0.0447 & 0.0053\\
\hline
\multirow{9}{*}{$750$}   & $1/n^{1/2}$            & 0.0607 & 0.0113 & 0.0000 & & 0.0673 & 0.0253 & 0.0003 & & 0.0777 & 0.0360 & 0.0007\\
                         & $1/n^{3/4}$            & 0.0597 & 0.0103 & 0.0000 & & 0.0660 & 0.0250 & 0.0003 & & 0.0767 & 0.0347 & 0.0007\\
                         & $1/n$                  & 0.0583 & 0.0103 & 0.0000 & & 0.0653 & 0.0247 & 0.0003 & & 0.0763 & 0.0340 & 0.0007\\
                         & $0.1/\log n$           & 0.0597 & 0.0107 & 0.0000 & & 0.0663 & 0.0250 & 0.0003 & & 0.0770 & 0.0357 & 0.0007\\
                         & $0.05/\log n$          & 0.0597 & 0.0103 & 0.0000 & & 0.0660 & 0.0250 & 0.0003 & & 0.0767 & 0.0350 & 0.0007\\
                         & $0.01/\log n$          & 0.0583 & 0.0103 & 0.0000 & & 0.0653 & 0.0247 & 0.0003 & & 0.0763 & 0.0340 & 0.0007\\
                         & $0.1$                  & 0.0617 & 0.0120 & 0.0000 & & 0.0693 & 0.0257 & 0.0003 & & 0.0793 & 0.0367 & 0.0007\\
                         & $0.05$                 & 0.0607 & 0.0113 & 0.0000 & & 0.0673 & 0.0257 & 0.0003 & & 0.0783 & 0.0360 & 0.0007\\
                         & $0.01$                 & 0.0597 & 0.0107 & 0.0000 & & 0.0660 & 0.0250 & 0.0003 & & 0.0767 & 0.0353 & 0.0007\\
\hline
\multirow{9}{*}{$1000$}  & $1/n^{1/2}$            & 0.0503 & 0.0110 & 0.0000 & & 0.0567 & 0.0217 & 0.0000 & & 0.0610 & 0.0257 & 0.0010\\
                         & $1/n^{3/4}$            & 0.0500 & 0.0107 & 0.0000 & & 0.0557 & 0.0210 & 0.0000 & & 0.0603 & 0.0253 & 0.0007\\
                         & $1/n$                  & 0.0500 & 0.0107 & 0.0000 & & 0.0553 & 0.0210 & 0.0000 & & 0.0600 & 0.0250 & 0.0007\\
                         & $0.1/\log n$           & 0.0503 & 0.0110 & 0.0000 & & 0.0560 & 0.0213 & 0.0000 & & 0.0607 & 0.0257 & 0.0010\\
                         & $0.05/\log n$          & 0.0500 & 0.0107 & 0.0000 & & 0.0560 & 0.0213 & 0.0000 & & 0.0607 & 0.0253 & 0.0007\\
                         & $0.01/\log n$          & 0.0500 & 0.0107 & 0.0000 & & 0.0553 & 0.0210 & 0.0000 & & 0.0600 & 0.0250 & 0.0007\\
                         & $0.1$                  & 0.0507 & 0.0110 & 0.0000 & & 0.0580 & 0.0223 & 0.0000 & & 0.0627 & 0.0277 & 0.0010\\
                         & $0.05$                 & 0.0503 & 0.0110 & 0.0000 & & 0.0567 & 0.0217 & 0.0000 & & 0.0617 & 0.0270 & 0.0010\\
                         & $0.01$                 & 0.0500 & 0.0107 & 0.0000 & & 0.0560 & 0.0213 & 0.0000 & & 0.0607 & 0.0257 & 0.0010\\
\hline
\multirow{9}{*}{$5000$}  & $1/n^{1/2}$            & 0.0513 & 0.0007 & 0.0000 & & 0.0530 & 0.0040 & 0.0000 & & 0.0573 & 0.0080 & 0.0000\\
                         & $1/n^{3/4}$            & 0.0510 & 0.0007 & 0.0000 & & 0.0517 & 0.0040 & 0.0000 & & 0.0573 & 0.0080 & 0.0000\\
                         & $1/n$                  & 0.0510 & 0.0007 & 0.0000 & & 0.0517 & 0.0040 & 0.0000 & & 0.0573 & 0.0080 & 0.0000\\
                         & $0.1/\log n$           & 0.0513 & 0.0007 & 0.0000 & & 0.0530 & 0.0040 & 0.0000 & & 0.0573 & 0.0080 & 0.0000\\
                         & $0.05/\log n$          & 0.0510 & 0.0007 & 0.0000 & & 0.0520 & 0.0040 & 0.0000 & & 0.0573 & 0.0080 & 0.0000\\
                         & $0.01/\log n$          & 0.0510 & 0.0007 & 0.0000 & & 0.0517 & 0.0040 & 0.0000 & & 0.0573 & 0.0080 & 0.0000\\
                         & $0.1$                  & 0.0527 & 0.0007 & 0.0000 & & 0.0533 & 0.0040 & 0.0000 & & 0.0577 & 0.0087 & 0.0000\\
                         & $0.05$                 & 0.0520 & 0.0007 & 0.0000 & & 0.0533 & 0.0040 & 0.0000 & & 0.0577 & 0.0087 & 0.0000\\
                         & $0.01$                 & 0.0513 & 0.0007 & 0.0000 & & 0.0527 & 0.0040 & 0.0000 & & 0.0573 & 0.0080 & 0.0000\\
\hline
\multicolumn{2}{c}{\multirow{2}{*}{$n$}}  & \multicolumn{3}{c}{LSW-S} & & \multicolumn{3}{c}{LSW-L} & & \multicolumn{3}{c}{C-OS}\\
\cline{3-5} \cline{7-9} \cline{11-13}
\multicolumn{2}{c}{}     & {D1} & {D2} & {D3}  & & {D1} & {D2} & {D3} & & {D1} & {D2} & {D3}\\
\hline
\multicolumn{2}{c}{500}                          & 0.0430 & 0.0200 & 0.0000 & & 0.0610  & 0.0260  & 0.0020  & & 0.0683  & 0.0607  & 0.0367\\
\multicolumn{2}{c}{750}                          & 0.0540 & 0.0170 & 0.0000 & & 0.0660  & 0.0310  & 0.0000  & & 0.0570  & 0.0453  & 0.0230\\
\multicolumn{2}{c}{1000}                         & 0.0440 & 0.0210 & 0.0000 & & 0.0540  & 0.0190  & 0.0000  & & 0.0553  & 0.0423  & 0.0187\\
\multicolumn{2}{c}{5000}                         &  &  &  & &   &   &   & & 0.0460  & 0.0383  & 0.0170\\
\hline
\hline
\end{tabularx}
\label{Tab: MC2Mon, size1, app} 
\end{table}
}

{
\setlength{\tabcolsep}{5.6pt}
\renewcommand{\arraystretch}{1.2}
\begin{table}[!ht]
\caption{MC2: Empirical Size of Monotonicity Test at $\alpha=5\%$}
\vspace{-0.1 in}
\centering\small
\sisetup{table-number-alignment = center, table-format = 1.3} 
\begin{tabularx}{\linewidth}{@{}cc *{3}{S[round-mode = places,round-precision = 3]} c *{3}{S[round-mode = places,round-precision = 3]} c *{3}{S[round-mode = places,round-precision = 3]}@{}} 
\hline
\hline
\multirow{3}{*}{$n$} & \multirow{3}{*}{$\gamma_n$} & \multicolumn{11}{c}{Our test with cubic B-spline}  \\
\cline{3-13}
& & \multicolumn{3}{c}{0 interior knots} & & \multicolumn{3}{c}{1 interior knot} & & \multicolumn{3}{c}{2 interior knots}\\
\cline{3-5} \cline{7-9} \cline{11-13}
& & {D1} & {D2} & {D3}  & & {D1} & {D2} & {D3} & & {D1} & {D2} & {D3}\\
\hline
\multirow{9}{*}{$500$}   & $1/n^{1/2}$            & 0.0680 & 0.0307 & 0.0010 & & 0.0843 & 0.0463 & 0.0043 & & 0.0930 & 0.0560 & 0.0060\\
                         & $1/n^{3/4}$            & 0.0673 & 0.0307 & 0.0007 & & 0.0837 & 0.0447 & 0.0043 & & 0.0920 & 0.0540 & 0.0057\\
                         & $1/n$                  & 0.0670 & 0.0303 & 0.0007 & & 0.0830 & 0.0437 & 0.0040 & & 0.0903 & 0.0530 & 0.0050\\
                         & $0.1/\log n$           & 0.0673 & 0.0307 & 0.0007 & & 0.0840 & 0.0450 & 0.0043 & & 0.0920 & 0.0550 & 0.0060\\
                         & $0.05/\log n$          & 0.0673 & 0.0307 & 0.0007 & & 0.0833 & 0.0443 & 0.0043 & & 0.0920 & 0.0537 & 0.0057\\
                         & $0.01/\log n$          & 0.0670 & 0.0303 & 0.0007 & & 0.0830 & 0.0437 & 0.0040 & & 0.0903 & 0.0530 & 0.0050\\
                         & $0.1$                  & 0.0690 & 0.0310 & 0.0010 & & 0.0853 & 0.0473 & 0.0053 & & 0.0943 & 0.0573 & 0.0070\\
                         & $0.05$                 & 0.0680 & 0.0310 & 0.0010 & & 0.0843 & 0.0470 & 0.0043 & & 0.0930 & 0.0563 & 0.0060\\
                         & $0.01$                 & 0.0673 & 0.0307 & 0.0007 & & 0.0837 & 0.0447 & 0.0043 & & 0.0920 & 0.0540 & 0.0057\\
\hline
\multirow{9}{*}{$750$}   & $1/n^{1/2}$            & 0.0657 & 0.0227 & 0.0000 & & 0.0753 & 0.0327 & 0.0007 & & 0.0883 & 0.0413 & 0.0030\\
                         & $1/n^{3/4}$            & 0.0650 & 0.0223 & 0.0000 & & 0.0740 & 0.0327 & 0.0007 & & 0.0873 & 0.0393 & 0.0027\\
                         & $1/n$                  & 0.0640 & 0.0220 & 0.0000 & & 0.0737 & 0.0323 & 0.0003 & & 0.0867 & 0.0387 & 0.0027\\
                         & $0.1/\log n$           & 0.0650 & 0.0223 & 0.0000 & & 0.0743 & 0.0327 & 0.0007 & & 0.0880 & 0.0400 & 0.0027\\
                         & $0.05/\log n$          & 0.0650 & 0.0223 & 0.0000 & & 0.0740 & 0.0327 & 0.0007 & & 0.0873 & 0.0393 & 0.0027\\
                         & $0.01/\log n$          & 0.0640 & 0.0220 & 0.0000 & & 0.0737 & 0.0323 & 0.0003 & & 0.0867 & 0.0387 & 0.0027\\
                         & $0.1$                  & 0.0663 & 0.0233 & 0.0000 & & 0.0770 & 0.0330 & 0.0007 & & 0.0893 & 0.0427 & 0.0030\\
                         & $0.05$                 & 0.0657 & 0.0230 & 0.0000 & & 0.0757 & 0.0330 & 0.0007 & & 0.0883 & 0.0420 & 0.0030\\
                         & $0.01$                 & 0.0650 & 0.0223 & 0.0000 & & 0.0740 & 0.0327 & 0.0007 & & 0.0873 & 0.0397 & 0.0027\\
\hline
\multirow{9}{*}{$1000$}  & $1/n^{1/2}$            & 0.0537 & 0.0207 & 0.0000 & & 0.0593 & 0.0243 & 0.0007 & & 0.0740 & 0.0310 & 0.0017\\
                         & $1/n^{3/4}$            & 0.0527 & 0.0203 & 0.0000 & & 0.0590 & 0.0230 & 0.0003 & & 0.0727 & 0.0300 & 0.0017\\
                         & $1/n$                  & 0.0527 & 0.0203 & 0.0000 & & 0.0590 & 0.0227 & 0.0003 & & 0.0723 & 0.0300 & 0.0017\\
                         & $0.1/\log n$           & 0.0530 & 0.0203 & 0.0000 & & 0.0590 & 0.0230 & 0.0007 & & 0.0733 & 0.0307 & 0.0017\\
                         & $0.05/\log n$          & 0.0527 & 0.0203 & 0.0000 & & 0.0590 & 0.0230 & 0.0003 & & 0.0730 & 0.0303 & 0.0017\\
                         & $0.01/\log n$          & 0.0527 & 0.0203 & 0.0000 & & 0.0590 & 0.0227 & 0.0003 & & 0.0723 & 0.0300 & 0.0017\\
                         & $0.1$                  & 0.0543 & 0.0210 & 0.0000 & & 0.0620 & 0.0247 & 0.0007 & & 0.0740 & 0.0323 & 0.0017\\
                         & $0.05$                 & 0.0537 & 0.0210 & 0.0000 & & 0.0607 & 0.0247 & 0.0007 & & 0.0740 & 0.0313 & 0.0017\\
                         & $0.01$                 & 0.0527 & 0.0203 & 0.0000 & & 0.0590 & 0.0230 & 0.0003 & & 0.0730 & 0.0303 & 0.0017\\
\hline
\multirow{9}{*}{$5000$}  & $1/n^{1/2}$            & 0.0527 & 0.0020 & 0.0000 & & 0.0597 & 0.0047 & 0.0000 & & 0.0570 & 0.0057 & 0.0000\\
                         & $1/n^{3/4}$            & 0.0520 & 0.0020 & 0.0000 & & 0.0577 & 0.0043 & 0.0000 & & 0.0563 & 0.0057 & 0.0000\\
                         & $1/n$                  & 0.0523 & 0.0020 & 0.0000 & & 0.0577 & 0.0043 & 0.0000 & & 0.0563 & 0.0057 & 0.0000\\
                         & $0.1/\log n$           & 0.0527 & 0.0020 & 0.0000 & & 0.0593 & 0.0047 & 0.0000 & & 0.0570 & 0.0057 & 0.0000\\
                         & $0.05/\log n$          & 0.0520 & 0.0020 & 0.0000 & & 0.0583 & 0.0047 & 0.0000 & & 0.0563 & 0.0057 & 0.0000\\
                         & $0.01/\log n$          & 0.0520 & 0.0020 & 0.0000 & & 0.0577 & 0.0043 & 0.0000 & & 0.0563 & 0.0057 & 0.0000\\
                         & $0.1$                  & 0.0537 & 0.0023 & 0.0000 & & 0.0607 & 0.0060 & 0.0000 & & 0.0580 & 0.0060 & 0.0000\\
                         & $0.05$                 & 0.0533 & 0.0023 & 0.0000 & & 0.0600 & 0.0053 & 0.0000 & & 0.0577 & 0.0060 & 0.0000\\
                         & $0.01$                 & 0.0523 & 0.0020 & 0.0000 & & 0.0590 & 0.0047 & 0.0000 & & 0.0570 & 0.0057 & 0.0000\\
\hline
\multicolumn{2}{c}{\multirow{2}{*}{$n$}}  & \multicolumn{3}{c}{LSW-S} & & \multicolumn{3}{c}{LSW-S} & & \multicolumn{3}{c}{???}\\
\cline{3-5} \cline{7-9} \cline{11-13}
\multicolumn{2}{c}{}     & {D1} & {D2} & {D3}  & & {D1} & {D2} & {D3} & & {D1} & {D2} & {D3}\\
\hline
\multicolumn{2}{c}{500}                          & 0.0430 & 0.0200 & 0.0000 & & 0.0610  & 0.0260  & 0.0020  & & 0.0683  & 0.0607  & 0.0367\\
\multicolumn{2}{c}{750}                          & 0.0540 & 0.0170 & 0.0000 & & 0.0660  & 0.0310  & 0.0000  & & 0.0570  & 0.0453  & 0.0230\\
\multicolumn{2}{c}{1000}                         & 0.0440 & 0.0210 & 0.0000 & & 0.0540  & 0.0190  & 0.0000  & & 0.0553  & 0.0423  & 0.0187\\
\multicolumn{2}{c}{5000}                         &  &  &  & &   &   &   & & 0.0460  & 0.0383  & 0.0170\\
\hline
\hline
\end{tabularx}
\label{Tab: MC2Mon, size2, app} 
\end{table}
}

\else 

{
\setlength{\tabcolsep}{3.2pt}
\renewcommand{\arraystretch}{1.2}
\begin{table}[!ht]
\caption{Empirical Size of Monotonicity Tests for \eqref{Eqn: MC2,aux1} at $\alpha=5\%$}
\centering\footnotesize
\begin{threeparttable}
\sisetup{table-number-alignment = center, table-format = 1.3} 
\begin{tabularx}{\linewidth}{@{}cc *{3}{S[round-mode = places,round-precision = 3]} c *{3}{S[round-mode = places,round-precision = 3]} c *{3}{S[round-mode = places,round-precision = 3]} c *{3}{S[round-mode = places,round-precision = 3]}@{}} 
\hline
\hline
\multirow{2}{*}{$n$} & \multirow{2}{*}{$\gamma_n$} & \multicolumn{3}{c}{FS-Q0: $k_n=9$} & & \multicolumn{3}{c}{FS-Q1: $k_n=16$} & & \multicolumn{3}{c}{FS-C0: $k_n=16$} & & \multicolumn{3}{c}{FS-C1: $k_n=25$}\\
\cline{3-5} \cline{7-9} \cline{11-13} \cline{15-17}
& & {D1} & {D2} & {D3}  & & {D1} & {D2} & {D3} & & {D1} & {D2} & {D3} & & {D1} & {D2} & {D3}\\  
\hline
\multirow{9}{*}{$500$}& $1/n^{1/2}$    & 0.0643 & 0.0203 & 0.0000 & & 0.0690 & 0.0303 & 0.0020 & & 0.0680 & 0.0307 & 0.0010 & & 0.0843 & 0.0463 & 0.0043\\
                       & $1/n^{3/4}$   & 0.0627 & 0.0190 & 0.0000 & & 0.0687 & 0.0300 & 0.0020 & & 0.0673 & 0.0307 & 0.0007 & & 0.0837 & 0.0447 & 0.0043\\
                       & $1/n$         & 0.0613 & 0.0183 & 0.0000 & & 0.0680 & 0.0300 & 0.0017 & & 0.0670 & 0.0303 & 0.0007 & & 0.0830 & 0.0437 & 0.0040\\
                       & $0.1/\log n$  & 0.0633 & 0.0197 & 0.0000 & & 0.0690 & 0.0300 & 0.0020 & & 0.0673 & 0.0307 & 0.0007 & & 0.0840 & 0.0450 & 0.0043\\
                       & $0.05/\log n$ & 0.0623 & 0.0187 & 0.0000 & & 0.0687 & 0.0300 & 0.0020 & & 0.0673 & 0.0307 & 0.0007 & & 0.0833 & 0.0443 & 0.0043\\
                       & $0.01/\log n$ & 0.0613 & 0.0183 & 0.0000 & & 0.0680 & 0.0300 & 0.0017 & & 0.0670 & 0.0303 & 0.0007 & & 0.0830 & 0.0437 & 0.0040\\
                       & $0.1$         & 0.0657 & 0.0203 & 0.0000 & & 0.0693 & 0.0307 & 0.0023 & & 0.0690 & 0.0310 & 0.0010 & & 0.0853 & 0.0473 & 0.0053\\
                       & $0.05$        & 0.0643 & 0.0203 & 0.0000 & & 0.0690 & 0.0303 & 0.0023 & & 0.0680 & 0.0310 & 0.0010 & & 0.0843 & 0.0470 & 0.0043\\
                       & $0.01$        & 0.0627 & 0.0193 & 0.0000 & & 0.0687 & 0.0300 & 0.0020 & & 0.0673 & 0.0307 & 0.0007 & & 0.0837 & 0.0447 & 0.0043\\
\hline
\multirow{9}{*}{$750$}& $1/n^{1/2}$    & 0.0607 & 0.0113 & 0.0000 & & 0.0673 & 0.0253 & 0.0003 & & 0.0657 & 0.0227 & 0.0000 & & 0.0753 & 0.0327 & 0.0007\\
                       & $1/n^{3/4}$   & 0.0597 & 0.0103 & 0.0000 & & 0.0660 & 0.0250 & 0.0003 & & 0.0650 & 0.0223 & 0.0000 & & 0.0740 & 0.0327 & 0.0007\\
                       & $1/n$         & 0.0583 & 0.0103 & 0.0000 & & 0.0653 & 0.0247 & 0.0003 & & 0.0640 & 0.0220 & 0.0000 & & 0.0737 & 0.0323 & 0.0003\\
                       & $0.1/\log n$  & 0.0597 & 0.0107 & 0.0000 & & 0.0663 & 0.0250 & 0.0003 & & 0.0650 & 0.0223 & 0.0000 & & 0.0743 & 0.0327 & 0.0007\\
                       & $0.05/\log n$ & 0.0597 & 0.0103 & 0.0000 & & 0.0660 & 0.0250 & 0.0003 & & 0.0650 & 0.0223 & 0.0000 & & 0.0740 & 0.0327 & 0.0007\\
                       & $0.01/\log n$ & 0.0583 & 0.0103 & 0.0000 & & 0.0653 & 0.0247 & 0.0003 & & 0.0640 & 0.0220 & 0.0000 & & 0.0737 & 0.0323 & 0.0003\\
                       & $0.1$         & 0.0617 & 0.0120 & 0.0000 & & 0.0693 & 0.0257 & 0.0003 & & 0.0663 & 0.0233 & 0.0000 & & 0.0770 & 0.0330 & 0.0007\\
                       & $0.05$        & 0.0607 & 0.0113 & 0.0000 & & 0.0673 & 0.0257 & 0.0003 & & 0.0657 & 0.0230 & 0.0000 & & 0.0757 & 0.0330 & 0.0007\\
                       & $0.01$        & 0.0597 & 0.0107 & 0.0000 & & 0.0660 & 0.0250 & 0.0003 & & 0.0650 & 0.0223 & 0.0000 & & 0.0740 & 0.0327 & 0.0007\\
\hline
\multirow{9}{*}{$1000$}& $1/n^{1/2}$   & 0.0503 & 0.0110 & 0.0000 & & 0.0567 & 0.0217 & 0.0000 & & 0.0537 & 0.0207 & 0.0000 & & 0.0593 & 0.0243 & 0.0007\\
                       & $1/n^{3/4}$   & 0.0500 & 0.0107 & 0.0000 & & 0.0557 & 0.0210 & 0.0000 & & 0.0527 & 0.0203 & 0.0000 & & 0.0590 & 0.0230 & 0.0003\\
                       & $1/n$         & 0.0500 & 0.0107 & 0.0000 & & 0.0553 & 0.0210 & 0.0000 & & 0.0527 & 0.0203 & 0.0000 & & 0.0590 & 0.0227 & 0.0003\\
                       & $0.1/\log n$  & 0.0503 & 0.0110 & 0.0000 & & 0.0560 & 0.0213 & 0.0000 & & 0.0530 & 0.0203 & 0.0000 & & 0.0590 & 0.0230 & 0.0007\\
                       & $0.05/\log n$ & 0.0500 & 0.0107 & 0.0000 & & 0.0560 & 0.0213 & 0.0000 & & 0.0527 & 0.0203 & 0.0000 & & 0.0590 & 0.0230 & 0.0003\\
                       & $0.01/\log n$ & 0.0500 & 0.0107 & 0.0000 & & 0.0553 & 0.0210 & 0.0000 & & 0.0527 & 0.0203 & 0.0000 & & 0.0590 & 0.0227 & 0.0003\\
                       & $0.1$         & 0.0507 & 0.0110 & 0.0000 & & 0.0580 & 0.0223 & 0.0000 & & 0.0543 & 0.0210 & 0.0000 & & 0.0620 & 0.0247 & 0.0007\\
                       & $0.05$        & 0.0503 & 0.0110 & 0.0000 & & 0.0567 & 0.0217 & 0.0000 & & 0.0537 & 0.0210 & 0.0000 & & 0.0607 & 0.0247 & 0.0007\\
                       & $0.01$        & 0.0500 & 0.0107 & 0.0000 & & 0.0560 & 0.0213 & 0.0000 & & 0.0527 & 0.0203 & 0.0000 & & 0.0590 & 0.0230 & 0.0003\\
\hline
\multicolumn{2}{c}{\multirow{2}{*}{$n$}}  & \multicolumn{3}{c}{LSW-S} & & \multicolumn{3}{c}{LSW-S} & & \multicolumn{3}{c}{C-OS} & & \multicolumn{3}{c}{ }\\
\cline{3-5} \cline{7-9} \cline{11-13}
\multicolumn{2}{c}{}     & {D1} & {D2} & {D3}  & & {D1} & {D2} & {D3} & & {D1} & {D2} & {D3} & &   &   &  \\
\hline
\multicolumn{2}{c}{500}                & 0.0430 & 0.0200 & 0.0000 & & 0.0610  & 0.0260  & 0.0020 & & 0.0683  & 0.0607  & 0.0367\\
\multicolumn{2}{c}{750}                & 0.0540 & 0.0170 & 0.0000 & & 0.0660  & 0.0310  & 0.0000 & & 0.0570  & 0.0453  & 0.0230\\
\multicolumn{2}{c}{1000}               & 0.0440 & 0.0210 & 0.0000 & & 0.0540  & 0.0190  & 0.0000 & & 0.0553  & 0.0423  & 0.0187\\
\hline
\hline
\end{tabularx}
\begin{tablenotes}[flushleft]
\item {\it Note:} The parameter $\gamma_n$ determines $\hat\kappa_n$ proposed in Section \ref{Sec: tuning parameter} with $c_n=1/\log n$ and $r_n=(n/k_n)^{1/2}$.
\end{tablenotes}
\end{threeparttable}
\label{Tab: MC2Mon, size, app}
\end{table}
}

\fi

\if AllResultsYes\choice
{
\setlength{\tabcolsep}{5.6pt}
\renewcommand{\arraystretch}{1.2}
\begin{table}[!ht]
\caption{MC2: Empirical Size of Convexity Test at $\alpha=5\%$}
\vspace{-0.1 in}
\centering\small
\sisetup{table-number-alignment = center, table-format = 1.3} 
\begin{tabularx}{\linewidth}{@{}cc *{3}{S[round-mode = places,round-precision = 3]} c *{3}{S[round-mode = places,round-precision = 3]} c *{3}{S[round-mode = places,round-precision = 3]}@{}} 
\hline
\hline
\multirow{3}{*}{$n$} & \multirow{3}{*}{$\gamma_n$} & \multicolumn{11}{c}{Our test with quadratic B-spline}  \\
\cline{3-13}
& & \multicolumn{3}{c}{0 interior knots} & & \multicolumn{3}{c}{1 interior knot} & & \multicolumn{3}{c}{2 interior knots}\\
\cline{3-5} \cline{7-9} \cline{11-13}
& & {D1} & {D2} & {D3}  & & {D1} & {D2} & {D3} & & {D1} & {D2} & {D3}\\
\hline
\multirow{9}{*}{$500$}   & $1/n^{1/2}$            & 0.0633 & 0.0620 & 0.0153 & & 0.0707 & 0.0690 & 0.0303 & & 0.0840 & 0.0847 & 0.0487\\
                         & $1/n^{3/4}$            & 0.0627 & 0.0610 & 0.0147 & & 0.0690 & 0.0673 & 0.0293 & & 0.0837 & 0.0843 & 0.0470\\
                         & $1/n$                  & 0.0620 & 0.0607 & 0.0147 & & 0.0687 & 0.0667 & 0.0287 & & 0.0837 & 0.0837 & 0.0470\\
                         & $0.1/\log n$           & 0.0633 & 0.0613 & 0.0150 & & 0.0693 & 0.0680 & 0.0300 & & 0.0837 & 0.0843 & 0.0477\\
                         & $0.05/\log n$          & 0.0627 & 0.0610 & 0.0147 & & 0.0690 & 0.0673 & 0.0293 & & 0.0837 & 0.0843 & 0.0470\\
                         & $0.01/\log n$          & 0.0620 & 0.0607 & 0.0147 & & 0.0687 & 0.0667 & 0.0287 & & 0.0837 & 0.0837 & 0.0470\\
                         & $0.1$                  & 0.0650 & 0.0623 & 0.0153 & & 0.0723 & 0.0703 & 0.0310 & & 0.0847 & 0.0863 & 0.0493\\
                         & $0.05$                 & 0.0633 & 0.0620 & 0.0153 & & 0.0707 & 0.0690 & 0.0307 & & 0.0840 & 0.0847 & 0.0487\\
                         & $0.01$                 & 0.0630 & 0.0610 & 0.0147 & & 0.0690 & 0.0673 & 0.0293 & & 0.0837 & 0.0843 & 0.0470\\
\hline
\multirow{9}{*}{$750$}   & $1/n^{1/2}$            & 0.0657 & 0.0650 & 0.0110 & & 0.0747 & 0.0753 & 0.0283 & & 0.0757 & 0.0780 & 0.0413\\
                         & $1/n^{3/4}$            & 0.0650 & 0.0630 & 0.0107 & & 0.0737 & 0.0737 & 0.0277 & & 0.0740 & 0.0770 & 0.0403\\
                         & $1/n$                  & 0.0643 & 0.0627 & 0.0107 & & 0.0730 & 0.0733 & 0.0267 & & 0.0727 & 0.0760 & 0.0403\\
                         & $0.1/\log n$           & 0.0650 & 0.0633 & 0.0107 & & 0.0740 & 0.0747 & 0.0280 & & 0.0750 & 0.0773 & 0.0403\\
                         & $0.05/\log n$          & 0.0650 & 0.0630 & 0.0107 & & 0.0737 & 0.0737 & 0.0277 & & 0.0743 & 0.0770 & 0.0403\\
                         & $0.01/\log n$          & 0.0643 & 0.0627 & 0.0107 & & 0.0730 & 0.0733 & 0.0267 & & 0.0727 & 0.0760 & 0.0403\\
                         & $0.1$                  & 0.0660 & 0.0663 & 0.0110 & & 0.0767 & 0.0767 & 0.0293 & & 0.0760 & 0.0790 & 0.0430\\
                         & $0.05$                 & 0.0660 & 0.0650 & 0.0110 & & 0.0753 & 0.0757 & 0.0287 & & 0.0757 & 0.0783 & 0.0420\\
                         & $0.01$                 & 0.0650 & 0.0630 & 0.0107 & & 0.0737 & 0.0737 & 0.0280 & & 0.0747 & 0.0773 & 0.0403\\
\hline
\multirow{9}{*}{$1000$}  & $1/n^{1/2}$            & 0.0577 & 0.0613 & 0.0037 & & 0.0683 & 0.0680 & 0.0180 & & 0.0663 & 0.0647 & 0.0270\\
                         & $1/n^{3/4}$            & 0.0567 & 0.0590 & 0.0037 & & 0.0673 & 0.0667 & 0.0177 & & 0.0647 & 0.0623 & 0.0263\\
                         & $1/n$                  & 0.0567 & 0.0590 & 0.0037 & & 0.0667 & 0.0663 & 0.0177 & & 0.0647 & 0.0620 & 0.0253\\
                         & $0.1/\log n$           & 0.0573 & 0.0597 & 0.0037 & & 0.0673 & 0.0670 & 0.0180 & & 0.0657 & 0.0637 & 0.0267\\
                         & $0.05/\log n$          & 0.0573 & 0.0590 & 0.0037 & & 0.0673 & 0.0670 & 0.0180 & & 0.0650 & 0.0630 & 0.0263\\
                         & $0.01/\log n$          & 0.0567 & 0.0590 & 0.0037 & & 0.0667 & 0.0663 & 0.0177 & & 0.0647 & 0.0620 & 0.0253\\
                         & $0.1$                  & 0.0577 & 0.0627 & 0.0040 & & 0.0693 & 0.0713 & 0.0190 & & 0.0673 & 0.0657 & 0.0287\\
                         & $0.05$                 & 0.0577 & 0.0617 & 0.0040 & & 0.0687 & 0.0697 & 0.0183 & & 0.0667 & 0.0650 & 0.0273\\
                         & $0.01$                 & 0.0573 & 0.0593 & 0.0037 & & 0.0673 & 0.0670 & 0.0180 & & 0.0653 & 0.0633 & 0.0263\\
\hline
\multirow{9}{*}{$5000$}  & $1/n^{1/2}$            & 0.0560 & 0.0587 & 0.0007 & & 0.0617 & 0.0583 & 0.0053 & & 0.0727 & 0.0683 & 0.0103\\
                         & $1/n^{3/4}$            & 0.0550 & 0.0583 & 0.0003 & & 0.0613 & 0.0583 & 0.0053 & & 0.0723 & 0.0680 & 0.0103\\
                         & $1/n$                  & 0.0550 & 0.0583 & 0.0003 & & 0.0613 & 0.0583 & 0.0053 & & 0.0723 & 0.0680 & 0.0103\\
                         & $0.1/\log n$           & 0.0557 & 0.0583 & 0.0007 & & 0.0617 & 0.0583 & 0.0053 & & 0.0727 & 0.0680 & 0.0103\\
                         & $0.05/\log n$          & 0.0553 & 0.0583 & 0.0007 & & 0.0613 & 0.0583 & 0.0053 & & 0.0727 & 0.0680 & 0.0103\\
                         & $0.01/\log n$          & 0.0550 & 0.0583 & 0.0003 & & 0.0613 & 0.0583 & 0.0053 & & 0.0723 & 0.0680 & 0.0103\\
                         & $0.1$                  & 0.0580 & 0.0607 & 0.0007 & & 0.0627 & 0.0590 & 0.0063 & & 0.0730 & 0.0700 & 0.0113\\
                         & $0.05$                 & 0.0570 & 0.0600 & 0.0007 & & 0.0620 & 0.0583 & 0.0057 & & 0.0727 & 0.0693 & 0.0110\\
                         & $0.01$                 & 0.0557 & 0.0583 & 0.0007 & & 0.0617 & 0.0583 & 0.0053 & & 0.0727 & 0.0680 & 0.0103\\
\hline
\multicolumn{2}{c}{\multirow{2}{*}{$n$}}  & \multicolumn{3}{c}{LSW-S} & & \multicolumn{3}{c}{LSW-S} & & \multicolumn{3}{c}{???}\\
\cline{3-5} \cline{7-9} \cline{11-13}
\multicolumn{2}{c}{}     & {D1} & {D2} & {D3}  & & {D1} & {D2} & {D3} & & {D1} & {D2} & {D3}\\
\hline
\multicolumn{2}{c}{500}                          & 0.0460 & 0.0490 & 0.0430 & & 0.0590  & 0.0550  & 0.0490  & &   &   & \\
\multicolumn{2}{c}{750}                          & 0.0680 & 0.0560 & 0.0490 & & 0.0710  & 0.0740  & 0.0480  & &   &   & \\
\multicolumn{2}{c}{1000}                         & 0.0530 & 0.0530 & 0.0430 & & 0.0620  & 0.0510  & 0.0370  & &   &   & \\
\multicolumn{2}{c}{5000}                         &  &  &  & &   &   &   & &   &   & \\
\hline
\hline
\end{tabularx}
\label{Tab: MC2Con, size1, app} 
\end{table}
}

{
\setlength{\tabcolsep}{5.6pt}
\renewcommand{\arraystretch}{1.2}
\begin{table}[!ht]
\caption{MC2: Empirical Size of Convexity Test at $\alpha=5\%$}
\vspace{-0.1 in}
\centering\small
\sisetup{table-number-alignment = center, table-format = 1.3} 
\begin{tabularx}{\linewidth}{@{}cc *{3}{S[round-mode = places,round-precision = 3]} c *{3}{S[round-mode = places,round-precision = 3]} c *{3}{S[round-mode = places,round-precision = 3]}@{}} 
\hline
\hline
\multirow{3}{*}{$n$} & \multirow{3}{*}{$\gamma_n$} & \multicolumn{11}{c}{Our test with cubic B-spline}  \\
\cline{3-13}
& & \multicolumn{3}{c}{0 interior knots} & & \multicolumn{3}{c}{1 interior knot} & & \multicolumn{3}{c}{2 interior knots}\\
\cline{3-5} \cline{7-9} \cline{11-13}
& & {D1} & {D2} & {D3}  & & {D1} & {D2} & {D3} & & {D1} & {D2} & {D3}\\
\hline
\multirow{9}{*}{$500$}   & $1/n^{1/2}$            & 0.0710 & 0.0687 & 0.0300 & & 0.0860 & 0.0833 & 0.0460 & & 0.0843 & 0.0910 & 0.0573\\
                         & $1/n^{3/4}$            & 0.0707 & 0.0680 & 0.0290 & & 0.0843 & 0.0823 & 0.0447 & & 0.0813 & 0.0890 & 0.0560\\
                         & $1/n$                  & 0.0697 & 0.0667 & 0.0283 & & 0.0830 & 0.0810 & 0.0443 & & 0.0807 & 0.0887 & 0.0557\\
                         & $0.1/\log n$           & 0.0707 & 0.0683 & 0.0297 & & 0.0847 & 0.0823 & 0.0453 & & 0.0827 & 0.0903 & 0.0563\\
                         & $0.05/\log n$          & 0.0703 & 0.0677 & 0.0290 & & 0.0843 & 0.0823 & 0.0443 & & 0.0813 & 0.0890 & 0.0560\\
                         & $0.01/\log n$          & 0.0697 & 0.0667 & 0.0283 & & 0.0830 & 0.0810 & 0.0443 & & 0.0807 & 0.0887 & 0.0557\\
                         & $0.1$                  & 0.0723 & 0.0700 & 0.0307 & & 0.0867 & 0.0843 & 0.0467 & & 0.0853 & 0.0923 & 0.0577\\
                         & $0.05$                 & 0.0710 & 0.0690 & 0.0300 & & 0.0860 & 0.0833 & 0.0463 & & 0.0847 & 0.0913 & 0.0573\\
                         & $0.01$                 & 0.0707 & 0.0680 & 0.0293 & & 0.0843 & 0.0823 & 0.0447 & & 0.0813 & 0.0890 & 0.0560\\
\hline
\multirow{9}{*}{$750$}   & $1/n^{1/2}$            & 0.0757 & 0.0753 & 0.0250 & & 0.0710 & 0.0733 & 0.0360 & & 0.0797 & 0.0803 & 0.0450\\
                         & $1/n^{3/4}$            & 0.0730 & 0.0740 & 0.0243 & & 0.0693 & 0.0710 & 0.0350 & & 0.0783 & 0.0797 & 0.0440\\
                         & $1/n$                  & 0.0720 & 0.0733 & 0.0237 & & 0.0687 & 0.0707 & 0.0347 & & 0.0770 & 0.0790 & 0.0430\\
                         & $0.1/\log n$           & 0.0747 & 0.0750 & 0.0243 & & 0.0700 & 0.0723 & 0.0357 & & 0.0790 & 0.0800 & 0.0447\\
                         & $0.05/\log n$          & 0.0733 & 0.0740 & 0.0243 & & 0.0693 & 0.0710 & 0.0353 & & 0.0783 & 0.0797 & 0.0443\\
                         & $0.01/\log n$          & 0.0720 & 0.0733 & 0.0237 & & 0.0687 & 0.0707 & 0.0347 & & 0.0770 & 0.0790 & 0.0430\\
                         & $0.1$                  & 0.0777 & 0.0760 & 0.0253 & & 0.0727 & 0.0760 & 0.0367 & & 0.0803 & 0.0813 & 0.0460\\
                         & $0.05$                 & 0.0760 & 0.0753 & 0.0250 & & 0.0720 & 0.0753 & 0.0367 & & 0.0800 & 0.0810 & 0.0453\\
                         & $0.01$                 & 0.0740 & 0.0743 & 0.0243 & & 0.0693 & 0.0713 & 0.0357 & & 0.0783 & 0.0797 & 0.0443\\
\hline
\multirow{9}{*}{$1000$}  & $1/n^{1/2}$            & 0.0703 & 0.0693 & 0.0143 & & 0.0673 & 0.0663 & 0.0283 & & 0.0743 & 0.0733 & 0.0350\\
                         & $1/n^{3/4}$            & 0.0697 & 0.0673 & 0.0137 & & 0.0667 & 0.0647 & 0.0267 & & 0.0733 & 0.0717 & 0.0343\\
                         & $1/n$                  & 0.0693 & 0.0670 & 0.0137 & & 0.0660 & 0.0647 & 0.0267 & & 0.0733 & 0.0713 & 0.0343\\
                         & $0.1/\log n$           & 0.0703 & 0.0683 & 0.0140 & & 0.0667 & 0.0653 & 0.0273 & & 0.0737 & 0.0730 & 0.0350\\
                         & $0.05/\log n$          & 0.0697 & 0.0680 & 0.0137 & & 0.0667 & 0.0647 & 0.0273 & & 0.0733 & 0.0723 & 0.0347\\
                         & $0.01/\log n$          & 0.0693 & 0.0670 & 0.0137 & & 0.0660 & 0.0647 & 0.0267 & & 0.0733 & 0.0713 & 0.0343\\
                         & $0.1$                  & 0.0717 & 0.0700 & 0.0147 & & 0.0683 & 0.0667 & 0.0300 & & 0.0753 & 0.0743 & 0.0350\\
                         & $0.05$                 & 0.0713 & 0.0693 & 0.0143 & & 0.0673 & 0.0663 & 0.0293 & & 0.0743 & 0.0740 & 0.0350\\
                         & $0.01$                 & 0.0703 & 0.0680 & 0.0140 & & 0.0667 & 0.0650 & 0.0273 & & 0.0733 & 0.0727 & 0.0347\\
\hline
\multirow{9}{*}{$5000$}  & $1/n^{1/2}$            & 0.0627 & 0.0560 & 0.0020 & & 0.0710 & 0.0683 & 0.0077 & & 0.0680 & 0.0667 & 0.0110\\
                         & $1/n^{3/4}$            & 0.0623 & 0.0560 & 0.0020 & & 0.0703 & 0.0677 & 0.0073 & & 0.0670 & 0.0667 & 0.0110\\
                         & $1/n$                  & 0.0623 & 0.0560 & 0.0020 & & 0.0703 & 0.0677 & 0.0073 & & 0.0670 & 0.0667 & 0.0110\\
                         & $0.1/\log n$           & 0.0627 & 0.0560 & 0.0020 & & 0.0707 & 0.0683 & 0.0077 & & 0.0680 & 0.0667 & 0.0110\\
                         & $0.05/\log n$          & 0.0623 & 0.0560 & 0.0020 & & 0.0703 & 0.0680 & 0.0073 & & 0.0670 & 0.0667 & 0.0110\\
                         & $0.01/\log n$          & 0.0623 & 0.0560 & 0.0020 & & 0.0703 & 0.0677 & 0.0073 & & 0.0670 & 0.0667 & 0.0110\\
                         & $0.1$                  & 0.0637 & 0.0577 & 0.0027 & & 0.0727 & 0.0693 & 0.0080 & & 0.0690 & 0.0677 & 0.0113\\
                         & $0.05$                 & 0.0630 & 0.0570 & 0.0027 & & 0.0723 & 0.0690 & 0.0077 & & 0.0683 & 0.0677 & 0.0110\\
                         & $0.01$                 & 0.0627 & 0.0560 & 0.0020 & & 0.0707 & 0.0680 & 0.0077 & & 0.0677 & 0.0667 & 0.0110\\
\hline
\multicolumn{2}{c}{\multirow{2}{*}{$n$}}  & \multicolumn{3}{c}{LSW-S} & & \multicolumn{3}{c}{LSW-S} & & \multicolumn{3}{c}{???}\\
\cline{3-5} \cline{7-9} \cline{11-13}
\multicolumn{2}{c}{}     & {D1} & {D2} & {D3}  & & {D1} & {D2} & {D3} & & {D1} & {D2} & {D3}\\
\hline
\multicolumn{2}{c}{500}                          & 0.0460 & 0.0490 & 0.0430 & & 0.0590  & 0.0550  & 0.0490  & &   &   & \\
\multicolumn{2}{c}{750}                          & 0.0680 & 0.0560 & 0.0490 & & 0.0710  & 0.0740  & 0.0480  & &   &   & \\
\multicolumn{2}{c}{1000}                         & 0.0530 & 0.0530 & 0.0430 & & 0.0620  & 0.0510  & 0.0370  & &   &   & \\
\multicolumn{2}{c}{5000}                         &  &  &  & &   &   &   & &   &   & \\
\hline
\hline
\end{tabularx}
\label{Tab: MC2Con, size2, app} 
\end{table}
}

\else 

{
\setlength{\tabcolsep}{3.2pt}
\renewcommand{\arraystretch}{1.2}
\begin{table}[!ht]
\caption{Empirical Size of Concavity Tests for \eqref{Eqn: MC2 supp,aux1} at $\alpha=5\%$}
\centering\footnotesize
\begin{threeparttable}
\sisetup{table-number-alignment = center, table-format = 1.3} 
\begin{tabularx}{\linewidth}{@{}cc *{3}{S[round-mode = places,round-precision = 3]} c *{3}{S[round-mode = places,round-precision = 3]} c *{3}{S[round-mode = places,round-precision = 3]} c *{3}{S[round-mode = places,round-precision = 3]}@{}} 
\hline
\hline
\multirow{3}{*}{$n$} & \multirow{3}{*}{$\gamma_n$} & \multicolumn{3}{c}{FS-Q0: $k_n=9$} & & \multicolumn{3}{c}{FS-Q1: $k_n=16$} & & \multicolumn{3}{c}{FS-C0: $k_n=16$} & & \multicolumn{3}{c}{FS-C1: $k_n=25$}\\
\cline{3-5} \cline{7-9} \cline{11-13} \cline{15-17}
& & {D1} & {D2} & {D3}  & & {D1} & {D2} & {D3} & & {D1} & {D2} & {D3} & & {D1} & {D2} & {D3}\\  
\hline
\multirow{9}{*}{$500$}& $1/n^{1/2}$    & 0.0633 & 0.0620 & 0.0153 & & 0.0707 & 0.0690 & 0.0303 & & 0.0710 & 0.0687 & 0.0300 & & 0.0860 & 0.0833 & 0.0460\\
                       & $1/n^{3/4}$   & 0.0627 & 0.0610 & 0.0147 & & 0.0690 & 0.0673 & 0.0293 & & 0.0707 & 0.0680 & 0.0290 & & 0.0843 & 0.0823 & 0.0447\\
                       & $1/n$         & 0.0620 & 0.0607 & 0.0147 & & 0.0687 & 0.0667 & 0.0287 & & 0.0697 & 0.0667 & 0.0283 & & 0.0830 & 0.0810 & 0.0443\\
                       & $0.1/\log n$  & 0.0633 & 0.0613 & 0.0150 & & 0.0693 & 0.0680 & 0.0300 & & 0.0707 & 0.0683 & 0.0297 & & 0.0847 & 0.0823 & 0.0453\\
                       & $0.05/\log n$ & 0.0627 & 0.0610 & 0.0147 & & 0.0690 & 0.0673 & 0.0293 & & 0.0703 & 0.0677 & 0.0290 & & 0.0843 & 0.0823 & 0.0443\\
                       & $0.01/\log n$ & 0.0620 & 0.0607 & 0.0147 & & 0.0687 & 0.0667 & 0.0287 & & 0.0697 & 0.0667 & 0.0283 & & 0.0830 & 0.0810 & 0.0443\\
                       & $0.1$         & 0.0650 & 0.0623 & 0.0153 & & 0.0723 & 0.0703 & 0.0310 & & 0.0723 & 0.0700 & 0.0307 & & 0.0867 & 0.0843 & 0.0467\\
                       & $0.05$        & 0.0633 & 0.0620 & 0.0153 & & 0.0707 & 0.0690 & 0.0307 & & 0.0710 & 0.0690 & 0.0300 & & 0.0860 & 0.0833 & 0.0463\\
                       & $0.01$        & 0.0630 & 0.0610 & 0.0147 & & 0.0690 & 0.0673 & 0.0293 & & 0.0707 & 0.0680 & 0.0293 & & 0.0843 & 0.0823 & 0.0447\\
\hline
\multirow{9}{*}{$750$}& $1/n^{1/2}$    & 0.0657 & 0.0650 & 0.0110 & & 0.0747 & 0.0753 & 0.0283 & & 0.0757 & 0.0753 & 0.0250 & & 0.0710 & 0.0733 & 0.0360\\
                       & $1/n^{3/4}$   & 0.0650 & 0.0630 & 0.0107 & & 0.0737 & 0.0737 & 0.0277 & & 0.0730 & 0.0740 & 0.0243 & & 0.0693 & 0.0710 & 0.0350\\
                       & $1/n$         & 0.0643 & 0.0627 & 0.0107 & & 0.0730 & 0.0733 & 0.0267 & & 0.0720 & 0.0733 & 0.0237 & & 0.0687 & 0.0707 & 0.0347\\
                       & $0.1/\log n$  & 0.0650 & 0.0633 & 0.0107 & & 0.0740 & 0.0747 & 0.0280 & & 0.0747 & 0.0750 & 0.0243 & & 0.0700 & 0.0723 & 0.0357\\
                       & $0.05/\log n$ & 0.0650 & 0.0630 & 0.0107 & & 0.0737 & 0.0737 & 0.0277 & & 0.0733 & 0.0740 & 0.0243 & & 0.0693 & 0.0710 & 0.0353\\
                       & $0.01/\log n$ & 0.0643 & 0.0627 & 0.0107 & & 0.0730 & 0.0733 & 0.0267 & & 0.0720 & 0.0733 & 0.0237 & & 0.0687 & 0.0707 & 0.0347\\
                       & $0.1$         & 0.0660 & 0.0663 & 0.0110 & & 0.0767 & 0.0767 & 0.0293 & & 0.0777 & 0.0760 & 0.0253 & & 0.0727 & 0.0760 & 0.0367\\
                       & $0.05$        & 0.0660 & 0.0650 & 0.0110 & & 0.0753 & 0.0757 & 0.0287 & & 0.0760 & 0.0753 & 0.0250 & & 0.0720 & 0.0753 & 0.0367\\
                       & $0.01$        & 0.0650 & 0.0630 & 0.0107 & & 0.0737 & 0.0737 & 0.0280 & & 0.0740 & 0.0743 & 0.0243 & & 0.0693 & 0.0713 & 0.0357\\
\hline
\multirow{9}{*}{$1000$}& $1/n^{1/2}$   & 0.0577 & 0.0613 & 0.0037 & & 0.0683 & 0.0680 & 0.0180 & & 0.0703 & 0.0693 & 0.0143 & & 0.0673 & 0.0663 & 0.0283\\
                       & $1/n^{3/4}$   & 0.0567 & 0.0590 & 0.0037 & & 0.0673 & 0.0667 & 0.0177 & & 0.0697 & 0.0673 & 0.0137 & & 0.0667 & 0.0647 & 0.0267\\
                       & $1/n$         & 0.0567 & 0.0590 & 0.0037 & & 0.0667 & 0.0663 & 0.0177 & & 0.0693 & 0.0670 & 0.0137 & & 0.0660 & 0.0647 & 0.0267\\
                       & $0.1/\log n$  & 0.0573 & 0.0597 & 0.0037 & & 0.0673 & 0.0670 & 0.0180 & & 0.0703 & 0.0683 & 0.0140 & & 0.0667 & 0.0653 & 0.0273\\
                       & $0.05/\log n$ & 0.0573 & 0.0590 & 0.0037 & & 0.0673 & 0.0670 & 0.0180 & & 0.0697 & 0.0680 & 0.0137 & & 0.0667 & 0.0647 & 0.0273\\
                       & $0.01/\log n$ & 0.0567 & 0.0590 & 0.0037 & & 0.0667 & 0.0663 & 0.0177 & & 0.0693 & 0.0670 & 0.0137 & & 0.0660 & 0.0647 & 0.0267\\
                       & $0.1$         & 0.0577 & 0.0627 & 0.0040 & & 0.0693 & 0.0713 & 0.0190 & & 0.0717 & 0.0700 & 0.0147 & & 0.0683 & 0.0667 & 0.0300\\
                       & $0.05$        & 0.0577 & 0.0617 & 0.0040 & & 0.0687 & 0.0697 & 0.0183 & & 0.0713 & 0.0693 & 0.0143 & & 0.0673 & 0.0663 & 0.0293\\
                       & $0.01$        & 0.0573 & 0.0593 & 0.0037 & & 0.0673 & 0.0670 & 0.0180 & & 0.0703 & 0.0680 & 0.0140 & & 0.0667 & 0.0650 & 0.0273\\
\hline
\multicolumn{2}{c}{\multirow{2}{*}{$n$}}  & \multicolumn{3}{c}{LSW-S} & & \multicolumn{3}{c}{LSW-S} & & \multicolumn{3}{c}{ } & & \multicolumn{3}{c}{ }\\
\cline{3-5} \cline{7-9}
\multicolumn{2}{c}{}     & {D1} & {D2} & {D3}  & & {D1} & {D2} & {D3} & &   &   &  & & & &\\
\hline
\multicolumn{2}{c}{500}                & 0.0460 & 0.0490 & 0.0430 & & 0.0590  & 0.0550  & 0.0490 & &  &  &  & &  &  & \\
\multicolumn{2}{c}{750}                & 0.0680 & 0.0560 & 0.0490 & & 0.0710  & 0.0740  & 0.0480 & &  &  &  & &  &  & \\
\multicolumn{2}{c}{1000}               & 0.0530 & 0.0530 & 0.0430 & & 0.0620  & 0.0510  & 0.0370 & &  &  &  & &  &  & \\
\hline
\hline
\end{tabularx}
\begin{tablenotes}[flushleft]
\item {\it Note:} The parameter $\gamma_n$ determines $\hat\kappa_n$ proposed in Section \ref{Sec: tuning parameter} with $c_n=1/\log n$ and $r_n=(n/k_n)^{1/2}$.
\end{tablenotes}
\end{threeparttable}
\label{Tab: MC2Con, size, app}
\end{table}
}

\fi

\if AllResultsYes\choice

{
\setlength{\tabcolsep}{3.2pt}
\renewcommand{\arraystretch}{1.2}
\begin{table}[!ht]
\caption{Empirical Size of Testing Slutsky Restrictions for \eqref{Eqn: MC3, aux1} at $\alpha=5\%$}
\vspace{-0.1 in}
\centering\footnotesize
\sisetup{table-number-alignment = center, table-format = 1.3} 
\begin{tabularx}{\linewidth}{@{}cc *{3}{S[round-mode = places,round-precision = 3]} c *{3}{S[round-mode = places,round-precision = 3]} c *{3}{S[round-mode = places,round-precision = 3]} c *{3}{S[round-mode = places,round-precision = 3]}@{}} 
\hline
\hline
\multirow{3}{*}{$n$} & \multirow{3}{*}{$\gamma_n$} & \multicolumn{3}{c}{FS-Q0} & & \multicolumn{3}{c}{FS-Q1} & & \multicolumn{3}{c}{FS-C0} & & \multicolumn{3}{c}{FS-C1}\\
\cline{3-5} \cline{7-9} \cline{11-13} \cline{15-17}
& & {D1} & {D2} & {D3}  & & {D1} & {D2} & {D3} & & {D1} & {D2} & {D3} & & {D1} & {D2} & {D3}\\  
\hline
\multirow{9}{*}{$1000$}& $1/n^{1/2}$   & 0.0650 & 0.0327 & 0.0207 & & 0.0817 & 0.0503 & 0.0327 & & 0.0883 & 0.0523 & 0.0333 & & 0.1420 & 0.0900 & 0.0657\\
                       & $1/n^{3/4}$   & 0.0650 & 0.0327 & 0.0203 & & 0.0820 & 0.0503 & 0.0327 & & 0.0887 & 0.0523 & 0.0333 & & 0.1420 & 0.0900 & 0.0660\\
                       & $1/n$         & 0.0650 & 0.0327 & 0.0203 & & 0.0820 & 0.0503 & 0.0327 & & 0.0887 & 0.0523 & 0.0333 & & 0.1420 & 0.0900 & 0.0660\\
                       & $0.1/\log n$  & 0.0650 & 0.0327 & 0.0207 & & 0.0820 & 0.0503 & 0.0327 & & 0.0883 & 0.0523 & 0.0333 & & 0.1420 & 0.0900 & 0.0657\\
                       & $0.05/\log n$ & 0.0650 & 0.0327 & 0.0203 & & 0.0820 & 0.0503 & 0.0327 & & 0.0883 & 0.0523 & 0.0333 & & 0.1420 & 0.0900 & 0.0660\\
                       & $0.01/\log n$ & 0.0650 & 0.0327 & 0.0203 & & 0.0813 & 0.0503 & 0.0327 & & 0.0887 & 0.0523 & 0.0333 & & 0.1420 & 0.0900 & 0.0660\\
                       & $0.1$         & 0.0650 & 0.0327 & 0.0200 & & 0.0813 & 0.0503 & 0.0327 & & 0.0883 & 0.0523 & 0.0333 & & 0.1420 & 0.0900 & 0.0657\\
                       & $0.05$        & 0.0650 & 0.0327 & 0.0203 & & 0.0813 & 0.0503 & 0.0327 & & 0.0883 & 0.0523 & 0.0333 & & 0.1420 & 0.0900 & 0.0657\\
                       & $0.01$        & 0.0650 & 0.0327 & 0.0203 & & 0.0820 & 0.0503 & 0.0327 & & 0.0883 & 0.0523 & 0.0333 & & 0.1420 & 0.0900 & 0.0660\\
\hline
\multirow{9}{*}{$2000$}& $1/n^{1/2}$   & 0.0523 & 0.0220 & 0.0110 & & 0.0693 & 0.0297 & 0.0147 & & 0.0713 & 0.0343 & 0.0153 & & 0.0880 & 0.0500 & 0.0290\\
                       & $1/n^{3/4}$   & 0.0523 & 0.0217 & 0.0110 & & 0.0693 & 0.0297 & 0.0147 & & 0.0713 & 0.0343 & 0.0153 & & 0.0883 & 0.0497 & 0.0290\\
                       & $1/n$         & 0.0523 & 0.0217 & 0.0110 & & 0.0693 & 0.0297 & 0.0147 & & 0.0713 & 0.0343 & 0.0153 & & 0.0883 & 0.0497 & 0.0290\\
                       & $0.1/\log n$  & 0.0523 & 0.0220 & 0.0110 & & 0.0693 & 0.0297 & 0.0147 & & 0.0713 & 0.0343 & 0.0153 & & 0.0880 & 0.0500 & 0.0290\\
                       & $0.05/\log n$ & 0.0523 & 0.0217 & 0.0110 & & 0.0693 & 0.0297 & 0.0147 & & 0.0713 & 0.0343 & 0.0153 & & 0.0880 & 0.0497 & 0.0290\\
                       & $0.01/\log n$ & 0.0523 & 0.0217 & 0.0110 & & 0.0693 & 0.0297 & 0.0147 & & 0.0713 & 0.0343 & 0.0153 & & 0.0883 & 0.0497 & 0.0290\\
                       & $0.1$         & 0.0520 & 0.0220 & 0.0110 & & 0.0693 & 0.0297 & 0.0147 & & 0.0713 & 0.0343 & 0.0153 & & 0.0880 & 0.0500 & 0.0293\\
                       & $0.05$        & 0.0520 & 0.0220 & 0.0110 & & 0.0693 & 0.0297 & 0.0147 & & 0.0713 & 0.0343 & 0.0153 & & 0.0880 & 0.0500 & 0.0293\\
                       & $0.01$        & 0.0523 & 0.0220 & 0.0110 & & 0.0693 & 0.0297 & 0.0147 & & 0.0713 & 0.0343 & 0.0153 & & 0.0880 & 0.0500 & 0.0290\\
\hline
\multirow{9}{*}{$3000$}& $1/n^{1/2}$   & 0.0487 & 0.0167 & 0.0070 & & 0.0577 & 0.0237 & 0.0123 & & 0.0577 & 0.0220 & 0.0113 & & 0.0740 & 0.0300 & 0.0140\\
                       & $1/n^{3/4}$   & 0.0490 & 0.0167 & 0.0070 & & 0.0577 & 0.0237 & 0.0123 & & 0.0577 & 0.0220 & 0.0113 & & 0.0740 & 0.0297 & 0.0137\\
                       & $1/n$         & 0.0490 & 0.0167 & 0.0070 & & 0.0577 & 0.0237 & 0.0123 & & 0.0577 & 0.0220 & 0.0113 & & 0.0740 & 0.0297 & 0.0137\\
                       & $0.1/\log n$  & 0.0487 & 0.0167 & 0.0070 & & 0.0577 & 0.0237 & 0.0123 & & 0.0577 & 0.0220 & 0.0113 & & 0.0740 & 0.0300 & 0.0137\\
                       & $0.05/\log n$ & 0.0487 & 0.0167 & 0.0070 & & 0.0577 & 0.0237 & 0.0123 & & 0.0577 & 0.0220 & 0.0113 & & 0.0740 & 0.0300 & 0.0137\\
                       & $0.01/\log n$ & 0.0490 & 0.0167 & 0.0070 & & 0.0577 & 0.0237 & 0.0123 & & 0.0577 & 0.0220 & 0.0113 & & 0.0740 & 0.0297 & 0.0137\\
                       & $0.1$         & 0.0487 & 0.0163 & 0.0070 & & 0.0577 & 0.0237 & 0.0123 & & 0.0573 & 0.0217 & 0.0113 & & 0.0743 & 0.0300 & 0.0137\\
                       & $0.05$        & 0.0487 & 0.0167 & 0.0070 & & 0.0577 & 0.0237 & 0.0123 & & 0.0577 & 0.0220 & 0.0113 & & 0.0743 & 0.0300 & 0.0137\\
                       & $0.01$        & 0.0487 & 0.0167 & 0.0070 & & 0.0577 & 0.0237 & 0.0123 & & 0.0577 & 0.0220 & 0.0113 & & 0.0740 & 0.0300 & 0.0137\\
\hline
\multirow{9}{*}{$5000$}& $1/n^{1/2}$   & 0.0513 & 0.0180 & 0.0073 & & 0.0627 & 0.0203 & 0.0080 & & 0.0603 & 0.0183 & 0.0073 & & 0.0577 & 0.0207 & 0.0083\\
                       & $1/n^{3/4}$   & 0.0513 & 0.0180 & 0.0073 & & 0.0627 & 0.0203 & 0.0080 & & 0.0603 & 0.0183 & 0.0073 & & 0.0580 & 0.0207 & 0.0083\\
                       & $1/n$         & 0.0513 & 0.0180 & 0.0073 & & 0.0627 & 0.0203 & 0.0080 & & 0.0603 & 0.0183 & 0.0073 & & 0.0580 & 0.0207 & 0.0083\\
                       & $0.1/\log n$  & 0.0513 & 0.0180 & 0.0073 & & 0.0627 & 0.0203 & 0.0080 & & 0.0603 & 0.0183 & 0.0073 & & 0.0580 & 0.0207 & 0.0083\\
                       & $0.05/\log n$ & 0.0513 & 0.0180 & 0.0073 & & 0.0627 & 0.0203 & 0.0080 & & 0.0603 & 0.0183 & 0.0073 & & 0.0580 & 0.0207 & 0.0083\\
                       & $0.01/\log n$ & 0.0513 & 0.0180 & 0.0073 & & 0.0627 & 0.0203 & 0.0080 & & 0.0603 & 0.0183 & 0.0073 & & 0.0580 & 0.0207 & 0.0083\\
                       & $0.1$         & 0.0513 & 0.0180 & 0.0073 & & 0.0627 & 0.0203 & 0.0080 & & 0.0603 & 0.0183 & 0.0073 & & 0.0577 & 0.0207 & 0.0083\\
                       & $0.05$        & 0.0513 & 0.0180 & 0.0073 & & 0.0627 & 0.0203 & 0.0080 & & 0.0603 & 0.0183 & 0.0073 & & 0.0577 & 0.0207 & 0.0083\\
                       & $0.01$        & 0.0513 & 0.0180 & 0.0073 & & 0.0627 & 0.0203 & 0.0080 & & 0.0603 & 0.0183 & 0.0073 & & 0.0580 & 0.0207 & 0.0083\\
\hline
\hline
\end{tabularx}
\label{Tab: MC3, size, app} 
\end{table}
}

\else 

{
\setlength{\tabcolsep}{3.2pt}
\renewcommand{\arraystretch}{1.2}
\begin{table}[!ht]
\caption{Empirical Size of Testing Slutsky Restrictions for \eqref{Eqn: MC3, aux1} at $\alpha=5\%$}
\centering\footnotesize
\begin{threeparttable}
\sisetup{table-number-alignment = center, table-format = 1.3} 
\begin{tabularx}{\linewidth}{@{}cc *{3}{S[round-mode = places,round-precision = 3]} c *{3}{S[round-mode = places,round-precision = 3]} c *{3}{S[round-mode = places,round-precision = 3]} c *{3}{S[round-mode = places,round-precision = 3]}@{}} 
\hline
\hline
\multirow{3}{*}{$n$} & \multirow{3}{*}{$\gamma_n$} & \multicolumn{3}{c}{FS-Q0: $k_n=27$} & & \multicolumn{3}{c}{FS-Q1: $k_n=64$} & & \multicolumn{3}{c}{FS-C0: $k_n=64$} & & \multicolumn{3}{c}{FS-C1: $k_n=125$}\\
\cline{3-5} \cline{7-9} \cline{11-13} \cline{15-17}
& & {D1} & {D2} & {D3}  & & {D1} & {D2} & {D3} & & {D1} & {D2} & {D3} & & {D1} & {D2} & {D3}\\  
\hline
\multirow{9}{*}{$1000$}& $1/n^{1/2}$   & 0.0727 & 0.0373 & 0.0223 & & 0.0920 & 0.0570 & 0.0373 & & 0.0983 & 0.0583 & 0.0397 & & 0.1553 & 0.1017 & 0.0723\\
                       & $1/n^{3/4}$   & 0.0723 & 0.0373 & 0.0220 & & 0.0920 & 0.0563 & 0.0373 & & 0.0983 & 0.0580 & 0.0393 & & 0.1540 & 0.1013 & 0.0723\\
                       & $1/n$         & 0.0717 & 0.0373 & 0.0220 & & 0.0920 & 0.0560 & 0.0373 & & 0.0980 & 0.0580 & 0.0393 & & 0.1533 & 0.1013 & 0.0723\\
                       & $0.1/\log n$  & 0.0727 & 0.0373 & 0.0223 & & 0.0920 & 0.0567 & 0.0373 & & 0.0983 & 0.0583 & 0.0397 & & 0.1550 & 0.1013 & 0.0723\\
                       & $0.05/\log n$ & 0.0727 & 0.0373 & 0.0220 & & 0.0920 & 0.0563 & 0.0373 & & 0.0983 & 0.0583 & 0.0393 & & 0.1540 & 0.1013 & 0.0723\\
                       & $0.01/\log n$ & 0.0717 & 0.0373 & 0.0220 & & 0.0920 & 0.0560 & 0.0373 & & 0.0980 & 0.0580 & 0.0393 & & 0.1533 & 0.1013 & 0.0723\\
                       & $0.1$         & 0.0730 & 0.0373 & 0.0223 & & 0.0920 & 0.0570 & 0.0377 & & 0.0990 & 0.0587 & 0.0400 & & 0.1563 & 0.1020 & 0.0727\\
                       & $0.05$        & 0.0727 & 0.0373 & 0.0223 & & 0.0920 & 0.0570 & 0.0373 & & 0.0983 & 0.0587 & 0.0400 & & 0.1557 & 0.1017 & 0.0727\\
                       & $0.01$        & 0.0727 & 0.0373 & 0.0223 & & 0.0920 & 0.0567 & 0.0373 & & 0.0983 & 0.0583 & 0.0393 & & 0.1550 & 0.1013 & 0.0723\\
\hline
\multirow{9}{*}{$3000$}& $1/n^{1/2}$   & 0.0547 & 0.0193 & 0.0087 & & 0.0650 & 0.0260 & 0.0137 & & 0.0657 & 0.0240 & 0.0123 & & 0.0840 & 0.0357 & 0.0167\\
                       & $1/n^{3/4}$   & 0.0540 & 0.0193 & 0.0087 & & 0.0647 & 0.0260 & 0.0137 & & 0.0650 & 0.0237 & 0.0123 & & 0.0830 & 0.0350 & 0.0163\\
                       & $1/n$         & 0.0540 & 0.0193 & 0.0087 & & 0.0647 & 0.0260 & 0.0137 & & 0.0650 & 0.0237 & 0.0123 & & 0.0830 & 0.0350 & 0.0163\\
                       & $0.1/\log n$  & 0.0547 & 0.0193 & 0.0087 & & 0.0650 & 0.0260 & 0.0137 & & 0.0653 & 0.0237 & 0.0123 & & 0.0840 & 0.0357 & 0.0167\\
                       & $0.05/\log n$ & 0.0540 & 0.0193 & 0.0087 & & 0.0647 & 0.0260 & 0.0137 & & 0.0653 & 0.0237 & 0.0123 & & 0.0833 & 0.0353 & 0.0163\\
                       & $0.01/\log n$ & 0.0540 & 0.0193 & 0.0087 & & 0.0647 & 0.0260 & 0.0137 & & 0.0650 & 0.0237 & 0.0123 & & 0.0830 & 0.0350 & 0.0163\\
                       & $0.1$         & 0.0550 & 0.0193 & 0.0090 & & 0.0650 & 0.0263 & 0.0137 & & 0.0663 & 0.0243 & 0.0123 & & 0.0843 & 0.0363 & 0.0170\\
                       & $0.05$        & 0.0547 & 0.0193 & 0.0087 & & 0.0650 & 0.0263 & 0.0137 & & 0.0660 & 0.0240 & 0.0123 & & 0.0843 & 0.0363 & 0.0170\\
                       & $0.01$        & 0.0543 & 0.0193 & 0.0087 & & 0.0650 & 0.0260 & 0.0137 & & 0.0653 & 0.0240 & 0.0123 & & 0.0840 & 0.0357 & 0.0167\\
\hline
\multirow{9}{*}{$5000$}& $1/n^{1/2}$   & 0.0560 & 0.0203 & 0.0083 & & 0.0673 & 0.0223 & 0.0093 & & 0.0667 & 0.0213 & 0.0083 & & 0.0670 & 0.0233 & 0.0090\\
                       & $1/n^{3/4}$   & 0.0560 & 0.0200 & 0.0083 & & 0.0670 & 0.0223 & 0.0093 & & 0.0660 & 0.0210 & 0.0083 & & 0.0667 & 0.0233 & 0.0090\\
                       & $1/n$         & 0.0560 & 0.0200 & 0.0083 & & 0.0670 & 0.0223 & 0.0093 & & 0.0660 & 0.0210 & 0.0083 & & 0.0667 & 0.0233 & 0.0090\\
                       & $0.1/\log n$  & 0.0560 & 0.0203 & 0.0083 & & 0.0673 & 0.0223 & 0.0093 & & 0.0663 & 0.0213 & 0.0083 & & 0.0670 & 0.0233 & 0.0090\\
                       & $0.05/\log n$ & 0.0560 & 0.0200 & 0.0083 & & 0.0670 & 0.0223 & 0.0093 & & 0.0663 & 0.0213 & 0.0083 & & 0.0670 & 0.0233 & 0.0090\\
                       & $0.01/\log n$ & 0.0560 & 0.0200 & 0.0083 & & 0.0670 & 0.0223 & 0.0093 & & 0.0660 & 0.0210 & 0.0083 & & 0.0667 & 0.0233 & 0.0090\\
                       & $0.1$         & 0.0563 & 0.0203 & 0.0083 & & 0.0680 & 0.0223 & 0.0100 & & 0.0667 & 0.0213 & 0.0083 & & 0.0677 & 0.0237 & 0.0090\\
                       & $0.05$        & 0.0563 & 0.0203 & 0.0083 & & 0.0677 & 0.0223 & 0.0097 & & 0.0667 & 0.0213 & 0.0083 & & 0.0670 & 0.0233 & 0.0090\\
                       & $0.01$        & 0.0560 & 0.0203 & 0.0083 & & 0.0673 & 0.0223 & 0.0093 & & 0.0663 & 0.0213 & 0.0083 & & 0.0670 & 0.0233 & 0.0090\\
\hline
\hline
\end{tabularx}
\begin{tablenotes}[flushleft]
\item {\it Note:} The parameter $\gamma_n$ determines $\hat\kappa_n$ proposed in Section \ref{Sec: tuning parameter} with $c_n=1/\log n$ and $r_n=(n/k_n)^{1/2}$.
\end{tablenotes}
\end{threeparttable}
\label{Tab: MC3, size, app} 
\end{table}
}

\fi

\pgfplotstableread{
delta alpha   P1      P2      P3      L1      L2      L3      C1      C2      C3
0     0.05  0.0537  0.0530  0.0530  0.0530  0.0530  0.0530  0.0540  0.0537  0.0530
1     0.05  0.0713  0.0697  0.0697  0.0700  0.0697  0.0697  0.0717  0.0713  0.0697
2     0.05  0.1227  0.1220  0.1203  0.1220  0.1220  0.1203  0.1237  0.1227  0.1220
3     0.05  0.2167  0.2157  0.2143  0.2157  0.2157  0.2143  0.2197  0.2173  0.2157
4     0.05  0.3667  0.3647  0.3633  0.3650  0.3647  0.3633  0.3697  0.3667  0.3647
5     0.05  0.5383  0.5363  0.5357  0.5367  0.5363  0.5357  0.5407  0.5387  0.5363
6     0.05  0.7040  0.7000  0.6977  0.7017  0.7000  0.6977  0.7097  0.7043  0.7000
7     0.05  0.8357  0.8337  0.8323  0.8340  0.8337  0.8323  0.8373  0.8357  0.8337
8     0.05  0.9237  0.9217  0.9213  0.9230  0.9217  0.9213  0.9247  0.9237  0.9220
9     0.05  0.9650  0.9643  0.9640  0.9647  0.9643  0.9640  0.9660  0.9650  0.9643
10    0.05  0.9873  0.9863  0.9860  0.9870  0.9863  0.9860  0.9877  0.9873  0.9863
}\FirstMonNFiveKnotsThree

\pgfplotstableread{
delta alpha   P1      P2      P3      L1      L2      L3      C1      C2      C3
0     0.05  0.0577  0.0577  0.0577  0.0577  0.0577  0.0577  0.0580  0.0577  0.0577
1     0.05  0.0737  0.0733  0.0733  0.0733  0.0733  0.0733  0.0743  0.0737  0.0733
2     0.05  0.1190  0.1183  0.1183  0.1187  0.1183  0.1183  0.1197  0.1190  0.1183
3     0.05  0.2023  0.2013  0.2000  0.2017  0.2013  0.2000  0.2027  0.2023  0.2013
4     0.05  0.3227  0.3203  0.3197  0.3207  0.3197  0.3197  0.3243  0.3227  0.3203
5     0.05  0.5020  0.4980  0.4960  0.4997  0.4977  0.4960  0.5033  0.5023  0.4980
6     0.05  0.6753  0.6723  0.6717  0.6727  0.6723  0.6717  0.6760  0.6757  0.6727
7     0.05  0.8120  0.8100  0.8087  0.8103  0.8100  0.8087  0.8143  0.8120  0.8100
8     0.05  0.9107  0.9093  0.9077  0.9093  0.9093  0.9077  0.9117  0.9110  0.9093
9     0.05  0.9600  0.9583  0.9577  0.9583  0.9583  0.9577  0.9603  0.9600  0.9583
10    0.05  0.9847  0.9847  0.9843  0.9847  0.9847  0.9843  0.9857  0.9850  0.9847
}\FirstMonNFiveKnotsFive

\pgfplotstableread{
delta alpha   P1      P2      P3      L1      L2      L3      C1      C2      C3
0     0.05  0.0590  0.0583  0.0583  0.0583  0.0583  0.0583  0.0590  0.0590  0.0583
1     0.05  0.0720  0.0710  0.0707  0.0713  0.0710  0.0707  0.0720  0.0720  0.0710
2     0.05  0.1123  0.1117  0.1117  0.1120  0.1117  0.1117  0.1137  0.1127  0.1117
3     0.05  0.1820  0.1813  0.1800  0.1817  0.1807  0.1800  0.1823  0.1820  0.1813
4     0.05  0.2973  0.2947  0.2940  0.2957  0.2940  0.2940  0.2990  0.2973  0.2947
5     0.05  0.4493  0.4473  0.4467  0.4480  0.4470  0.4467  0.4517  0.4497  0.4473
6     0.05  0.6173  0.6160  0.6147  0.6167  0.6153  0.6147  0.6213  0.6187  0.6160
7     0.05  0.7697  0.7670  0.7667  0.7677  0.7673  0.7667  0.7720  0.7700  0.7673
8     0.05  0.8830  0.8820  0.8810  0.8820  0.8820  0.8810  0.8837  0.8830  0.8820
9     0.05  0.9453  0.9440  0.9440  0.9443  0.9440  0.9440  0.9457  0.9453  0.9440
10    0.05  0.9777  0.9773  0.9773  0.9777  0.9773  0.9773  0.9780  0.9777  0.9773
}\FirstMonNFiveKnotsSeven

\pgfplotstableread{
delta alpha   P1      P2      P3      L1      L2      L3      C1      C2      C3
0     0.05  0.0487  0.0487  0.0480  0.0487  0.0487  0.0480  0.0490  0.0487  0.0487
1     0.05  0.0393  0.0393  0.0387  0.0397  0.0393  0.0390  0.0407  0.0393  0.0393
2     0.05  0.0533  0.0527  0.0523  0.0527  0.0527  0.0523  0.0540  0.0533  0.0527
3     0.05  0.0793  0.0787  0.0783  0.0790  0.0787  0.0783  0.0803  0.0793  0.0787
4     0.05  0.1177  0.1160  0.1153  0.1167  0.1160  0.1153  0.1207  0.1183  0.1160
5     0.05  0.1890  0.1860  0.1853  0.1870  0.1857  0.1853  0.1920  0.1890  0.1860
6     0.05  0.2780  0.2723  0.2690  0.2733  0.2720  0.2690  0.2807  0.2787  0.2720
7     0.05  0.3900  0.3833  0.3803  0.3847  0.3830  0.3803  0.3930  0.3903  0.3827
8     0.05  0.5007  0.4913  0.4897  0.4930  0.4913  0.4897  0.5063  0.5010  0.4913
9     0.05  0.6200  0.6117  0.6080  0.6123  0.6110  0.6080  0.6243  0.6203  0.6117
10    0.05  0.7313  0.7240  0.7200  0.7257  0.7227  0.7200  0.7357  0.7313  0.7240
}\FirstConNFiveKnotsThree

\pgfplotstableread{
delta alpha   P1      P2      P3      L1      L2      L3      C1      C2      C3
0     0.05  0.0567  0.0563  0.0563  0.0563  0.0563  0.0563  0.0567  0.0567  0.0563
1     0.05  0.0470  0.0467  0.0467  0.0470  0.0467  0.0467  0.0473  0.0470  0.0467
2     0.05  0.0583  0.0583  0.0573  0.0583  0.0577  0.0573  0.0583  0.0583  0.0583
3     0.05  0.0807  0.0787  0.0787  0.0787  0.0787  0.0787  0.0807  0.0807  0.0787
4     0.05  0.1173  0.1147  0.1147  0.1153  0.1147  0.1147  0.1183  0.1177  0.1147
5     0.05  0.1703  0.1670  0.1663  0.1687  0.1670  0.1663  0.1727  0.1707  0.1670
6     0.05  0.2413  0.2390  0.2387  0.2400  0.2390  0.2387  0.2430  0.2420  0.2390
7     0.05  0.3407  0.3333  0.3320  0.3353  0.3333  0.3320  0.3443  0.3417  0.3337
8     0.05  0.4553  0.4490  0.4470  0.4510  0.4487  0.4470  0.4583  0.4560  0.4490
9     0.05  0.5707  0.5630  0.5597  0.5660  0.5620  0.5597  0.5743  0.5710  0.5633
10    0.05  0.6793  0.6740  0.6713  0.6753  0.6733  0.6713  0.6853  0.6797  0.6740
}\FirstConNFiveKnotsFive

\pgfplotstableread{
delta alpha   P1      P2      P3      L1      L2      L3      C1      C2      C3
0     0.05  0.0537  0.0533  0.0533  0.0533  0.0533  0.0533  0.0543  0.0540  0.0533
1     0.05  0.0467  0.0460  0.0453  0.0460  0.0460  0.0453  0.0477  0.0473  0.0460
2     0.05  0.0563  0.0557  0.0550  0.0557  0.0557  0.0550  0.0563  0.0563  0.0557
3     0.05  0.0743  0.0730  0.0733  0.0740  0.0730  0.0733  0.0750  0.0740  0.0730
4     0.05  0.1033  0.1030  0.1017  0.1030  0.1030  0.1017  0.1050  0.1037  0.1030
5     0.05  0.1593  0.1573  0.1560  0.1580  0.1567  0.1560  0.1610  0.1593  0.1570
6     0.05  0.2183  0.2160  0.2143  0.2170  0.2157  0.2143  0.2203  0.2183  0.2160
7     0.05  0.3053  0.3023  0.3003  0.3033  0.3023  0.3003  0.3073  0.3053  0.3023
8     0.05  0.4047  0.4017  0.3997  0.4030  0.4017  0.3997  0.4090  0.4053  0.4017
9     0.05  0.5113  0.5067  0.5040  0.5087  0.5060  0.5040  0.5143  0.5117  0.5070
10    0.05  0.6290  0.6267  0.6240  0.6267  0.6260  0.6240  0.6313  0.6290  0.6267
}\FirstConNFiveKnotsSeven

\pgfplotstableread{
delta alpha   P1      P2      P3      L1      L2      L3      C1      C2      C3
0     0.05  0.0503  0.0503  0.0500  0.0503  0.0503  0.0500  0.0503  0.0503  0.0503
1     0.05  0.0627  0.0623  0.0617  0.0627  0.0623  0.0617  0.0630  0.0627  0.0623
2     0.05  0.1130  0.1127  0.1127  0.1130  0.1127  0.1127  0.1130  0.1130  0.1127
3     0.05  0.2003  0.2000  0.1983  0.2003  0.2000  0.1983  0.2007  0.2003  0.2000
4     0.05  0.3450  0.3420  0.3410  0.3427  0.3420  0.3410  0.3457  0.3450  0.3423
5     0.05  0.5190  0.5157  0.5147  0.5170  0.5157  0.5147  0.5200  0.5190  0.5157
6     0.05  0.6867  0.6843  0.6840  0.6850  0.6843  0.6840  0.6890  0.6867  0.6847
7     0.05  0.8270  0.8250  0.8237  0.8257  0.8243  0.8237  0.8283  0.8273  0.8250
8     0.05  0.9150  0.9140  0.9140  0.9143  0.9140  0.9140  0.9157  0.9150  0.9140
9     0.05  0.9647  0.9640  0.9637  0.9640  0.9640  0.9637  0.9663  0.9650  0.9640
10    0.05  0.9880  0.9880  0.9877  0.9880  0.9880  0.9877  0.9880  0.9880  0.9880
}\FirstMonConNFiveKnotsThree

\pgfplotstableread{
delta alpha   P1      P2      P3      L1      L2      L3      C1      C2      C3
0     0.05  0.0550  0.0543  0.0540  0.0547  0.0543  0.0540  0.0557  0.0550  0.0543
1     0.05  0.0660  0.0657  0.0653  0.0660  0.0657  0.0653  0.0660  0.0660  0.0657
2     0.05  0.1103  0.1097  0.1097  0.1100  0.1097  0.1097  0.1107  0.1103  0.1097
3     0.05  0.1797  0.1787  0.1783  0.1790  0.1787  0.1783  0.1807  0.1797  0.1787
4     0.05  0.2973  0.2960  0.2957  0.2970  0.2960  0.2957  0.2990  0.2973  0.2963
5     0.05  0.4783  0.4753  0.4740  0.4767  0.4753  0.4740  0.4797  0.4783  0.4757
6     0.05  0.6607  0.6583  0.6577  0.6593  0.6583  0.6577  0.6627  0.6607  0.6583
7     0.05  0.7980  0.7973  0.7967  0.7973  0.7973  0.7967  0.8003  0.7983  0.7973
8     0.05  0.9053  0.9040  0.9037  0.9047  0.9040  0.9037  0.9060  0.9053  0.9040
9     0.05  0.9563  0.9563  0.9560  0.9563  0.9563  0.9560  0.9573  0.9563  0.9563
10    0.05  0.9847  0.9840  0.9840  0.9840  0.9840  0.9840  0.9853  0.9847  0.9840
}\FirstMonConNFiveKnotsFive

\pgfplotstableread{
delta alpha   P1      P2      P3      L1      L2      L3      C1      C2      C3
0     0.05  0.0550  0.0543  0.0543  0.0543  0.0543  0.0543  0.0553  0.0550  0.0543
1     0.05  0.0650  0.0650  0.0647  0.0650  0.0650  0.0647  0.0653  0.0650  0.0650
2     0.05  0.1077  0.1077  0.1070  0.1077  0.1077  0.1070  0.1077  0.1077  0.1077
3     0.05  0.1710  0.1693  0.1687  0.1697  0.1693  0.1687  0.1713  0.1710  0.1693
4     0.05  0.2823  0.2813  0.2807  0.2817  0.2813  0.2807  0.2830  0.2823  0.2813
5     0.05  0.4330  0.4313  0.4310  0.4313  0.4313  0.4310  0.4333  0.4330  0.4313
6     0.05  0.6023  0.5997  0.5987  0.6000  0.5997  0.5987  0.6037  0.6027  0.5997
7     0.05  0.7580  0.7560  0.7550  0.7567  0.7560  0.7550  0.7600  0.7587  0.7560
8     0.05  0.8777  0.8763  0.8760  0.8770  0.8763  0.8760  0.8783  0.8780  0.8767
9     0.05  0.9433  0.9420  0.9420  0.9423  0.9420  0.9420  0.9433  0.9433  0.9420
10    0.05  0.9763  0.9757  0.9757  0.9757  0.9757  0.9757  0.9763  0.9763  0.9757
}\FirstMonConNFiveKnotsSeven

\pgfplotstableread{
delta alpha   P1      P2      P3      L1      L2      L3      C1      C2      C3
0     0.05  0.0527  0.0523  0.0523  0.0523  0.0523  0.0523  0.0527  0.0527  0.0523
1     0.05  0.0777  0.0773  0.0773  0.0773  0.0773  0.0773  0.0790  0.0777  0.0773
2     0.05  0.1583  0.1567  0.1563  0.1573  0.1567  0.1563  0.1587  0.1583  0.1567
3     0.05  0.3117  0.3093  0.3090  0.3113  0.3097  0.3090  0.3133  0.3123  0.3097
4     0.05  0.5163  0.5110  0.5107  0.5140  0.5113  0.5107  0.5173  0.5167  0.5127
5     0.05  0.7143  0.7117  0.7113  0.7117  0.7117  0.7113  0.7173  0.7147  0.7117
6     0.05  0.8820  0.8793  0.8787  0.8807  0.8793  0.8787  0.8833  0.8823  0.8800
7     0.05  0.9577  0.9560  0.9560  0.9567  0.9560  0.9560  0.9580  0.9580  0.9563
8     0.05  0.9883  0.9873  0.9873  0.9877  0.9877  0.9873  0.9887  0.9883  0.9877
9     0.05  0.9977  0.9973  0.9973  0.9973  0.9973  0.9973  0.9980  0.9980  0.9973
10    0.05  0.9993  0.9993  0.9993  0.9993  0.9993  0.9993  0.9993  0.9993  0.9993
}\FirstMonNSevenKnotsThree

\pgfplotstableread{
delta alpha   P1      P2      P3      L1      L2      L3      C1      C2      C3
0     0.05  0.0567  0.0560  0.0560  0.0563  0.0560  0.0560  0.0567  0.0567  0.0560
1     0.05  0.0723  0.0720  0.0717  0.0723  0.0720  0.0717  0.0730  0.0730  0.0720
2     0.05  0.1420  0.1407  0.1403  0.1410  0.1407  0.1403  0.1430  0.1423  0.1410
3     0.05  0.2810  0.2800  0.2790  0.2810  0.2800  0.2790  0.2817  0.2813  0.2807
4     0.05  0.4770  0.4753  0.4743  0.4757  0.4753  0.4743  0.4790  0.4777  0.4753
5     0.05  0.6883  0.6853  0.6853  0.6870  0.6857  0.6853  0.6903  0.6890  0.6863
6     0.05  0.8533  0.8520  0.8513  0.8527  0.8520  0.8513  0.8553  0.8540  0.8527
7     0.05  0.9470  0.9460  0.9460  0.9467  0.9460  0.9460  0.9477  0.9470  0.9460
8     0.05  0.9847  0.9843  0.9843  0.9847  0.9843  0.9843  0.9850  0.9847  0.9843
9     0.05  0.9977  0.9977  0.9973  0.9977  0.9977  0.9973  0.9977  0.9977  0.9977
10    0.05  0.9993  0.9993  0.9993  0.9993  0.9993  0.9993  0.9993  0.9993  0.9993
}\FirstMonNSevenKnotsFive

\pgfplotstableread{
delta alpha   P1      P2      P3      L1      L2      L3      C1      C2      C3
0     0.05  0.0590  0.0590  0.0590  0.0590  0.0590  0.0590  0.0597  0.0590  0.0590
1     0.05  0.0740  0.0733  0.0727  0.0733  0.0733  0.0727  0.0743  0.0743  0.0733
2     0.05  0.1350  0.1337  0.1337  0.1343  0.1337  0.1337  0.1353  0.1350  0.1340
3     0.05  0.2520  0.2500  0.2500  0.2507  0.2503  0.2500  0.2530  0.2520  0.2503
4     0.05  0.4357  0.4343  0.4337  0.4350  0.4347  0.4337  0.4363  0.4357  0.4350
5     0.05  0.6393  0.6367  0.6350  0.6387  0.6370  0.6350  0.6420  0.6397  0.6377
6     0.05  0.8203  0.8193  0.8190  0.8197  0.8193  0.8190  0.8223  0.8207  0.8193
7     0.05  0.9307  0.9303  0.9303  0.9303  0.9303  0.9303  0.9307  0.9307  0.9303
8     0.05  0.9777  0.9773  0.9773  0.9777  0.9773  0.9773  0.9783  0.9777  0.9777
9     0.05  0.9950  0.9950  0.9950  0.9950  0.9950  0.9950  0.9950  0.9950  0.9950
10    0.05  0.9993  0.9993  0.9993  0.9993  0.9993  0.9993  0.9993  0.9993  0.9993
}\FirstMonNSevenKnotsSeven

\pgfplotstableread{
delta alpha   P1      P2      P3      L1      L2      L3      C1      C2      C3
0     0.05  0.0577  0.0577  0.0577  0.0577  0.0577  0.0577  0.0577  0.0577  0.0577
1     0.05  0.0437  0.0433  0.0433  0.0437  0.0433  0.0433  0.0443  0.0437  0.0433
2     0.05  0.0577  0.0563  0.0557  0.0570  0.0563  0.0557  0.0583  0.0580  0.0563
3     0.05  0.1023  0.1003  0.0993  0.1013  0.1000  0.0997  0.1053  0.1027  0.1010
4     0.05  0.1763  0.1727  0.1713  0.1743  0.1727  0.1713  0.1783  0.1767  0.1727
5     0.05  0.2873  0.2827  0.2813  0.2837  0.2827  0.2813  0.2933  0.2887  0.2830
6     0.05  0.4267  0.4193  0.4157  0.4233  0.4200  0.4157  0.4333  0.4277  0.4217
7     0.05  0.5803  0.5723  0.5710  0.5743  0.5727  0.5710  0.5893  0.5830  0.5737
8     0.05  0.7210  0.7127  0.7090  0.7157  0.7133  0.7090  0.7267  0.7220  0.7143
9     0.05  0.8263  0.8203  0.8177  0.8237  0.8210  0.8177  0.8320  0.8277  0.8230
10    0.05  0.9083  0.9040  0.9017  0.9053  0.9040  0.9017  0.9137  0.9107  0.9043
}\FirstConNSevenKnotsThree

\pgfplotstableread{
delta alpha   P1      P2      P3      L1      L2      L3      C1      C2      C3
0     0.05  0.0627  0.0623  0.0623  0.0627  0.0623  0.0623  0.0630  0.0627  0.0623
1     0.05  0.0487  0.0480  0.0480  0.0483  0.0480  0.0480  0.0493  0.0487  0.0477
2     0.05  0.0607  0.0600  0.0593  0.0607  0.0603  0.0593  0.0613  0.0607  0.0607
3     0.05  0.1027  0.1013  0.1010  0.1023  0.1017  0.1010  0.1040  0.1033  0.1017
4     0.05  0.1660  0.1637  0.1633  0.1647  0.1637  0.1633  0.1677  0.1670  0.1643
5     0.05  0.2563  0.2517  0.2503  0.2527  0.2517  0.2503  0.2590  0.2570  0.2523
6     0.05  0.3703  0.3623  0.3593  0.3673  0.3633  0.3593  0.3763  0.3713  0.3640
7     0.05  0.5133  0.5077  0.5057  0.5103  0.5080  0.5057  0.5167  0.5143  0.5087
8     0.05  0.6573  0.6510  0.6490  0.6520  0.6510  0.6490  0.6617  0.6587  0.6513
9     0.05  0.7810  0.7763  0.7753  0.7793  0.7763  0.7753  0.7843  0.7817  0.7780
10    0.05  0.8713  0.8683  0.8673  0.8700  0.8687  0.8673  0.8740  0.8717  0.8693
}\FirstConNSevenKnotsFive

\pgfplotstableread{
delta alpha   P1      P2      P3      L1      L2      L3      C1      C2      C3
0     0.05  0.0597  0.0597  0.0593  0.0597  0.0597  0.0593  0.0603  0.0597  0.0593
1     0.05  0.0500  0.0500  0.0500  0.0500  0.0500  0.0500  0.0507  0.0500  0.0500
2     0.05  0.0640  0.0637  0.0633  0.0637  0.0637  0.0633  0.0647  0.0640  0.0637
3     0.05  0.0957  0.0947  0.0940  0.0957  0.0947  0.0940  0.0967  0.0957  0.0950
4     0.05  0.1460  0.1430  0.1423  0.1440  0.1430  0.1423  0.1480  0.1463  0.1433
5     0.05  0.2333  0.2310  0.2290  0.2323  0.2313  0.2290  0.2353  0.2347  0.2317
6     0.05  0.3343  0.3303  0.3300  0.3313  0.3303  0.3300  0.3373  0.3343  0.3303
7     0.05  0.4677  0.4630  0.4610  0.4653  0.4633  0.4610  0.4703  0.4687  0.4643
8     0.05  0.6033  0.5990  0.5960  0.6000  0.5990  0.5960  0.6067  0.6033  0.5997
9     0.05  0.7297  0.7257  0.7227  0.7273  0.7253  0.7227  0.7337  0.7303  0.7260
10    0.05  0.8360  0.8333  0.8320  0.8347  0.8330  0.8320  0.8397  0.8363  0.8337
}\FirstConNSevenKnotsSeven

\pgfplotstableread{
delta alpha   P1      P2      P3      L1      L2      L3      C1      C2      C3
0     0.05  0.0563  0.0563  0.0563  0.0563  0.0563  0.0563  0.0570  0.0570  0.0563
1     0.05  0.0717  0.0717  0.0710  0.0717  0.0717  0.0710  0.0723  0.0720  0.0717
2     0.05  0.1387  0.1383  0.1383  0.1383  0.1383  0.1383  0.1393  0.1387  0.1383
3     0.05  0.2930  0.2903  0.2897  0.2920  0.2910  0.2897  0.2940  0.2933  0.2910
4     0.05  0.4953  0.4930  0.4920  0.4950  0.4933  0.4920  0.4990  0.4957  0.4933
5     0.05  0.7030  0.7003  0.7000  0.7010  0.7003  0.7000  0.7043  0.7030  0.7007
6     0.05  0.8737  0.8723  0.8713  0.8730  0.8727  0.8713  0.8747  0.8740  0.8727
7     0.05  0.9527  0.9523  0.9523  0.9523  0.9523  0.9523  0.9537  0.9533  0.9523
8     0.05  0.9867  0.9863  0.9860  0.9863  0.9863  0.9860  0.9867  0.9867  0.9863
9     0.05  0.9977  0.9973  0.9973  0.9973  0.9973  0.9973  0.9977  0.9977  0.9973
10    0.05  0.9987  0.9987  0.9987  0.9987  0.9987  0.9987  0.9987  0.9987  0.9987
}\FirstMonConNSevenKnotsThree

\pgfplotstableread{
delta alpha   P1      P2      P3      L1      L2      L3      C1      C2      C3
0     0.05  0.0597  0.0590  0.0590  0.0590  0.0590  0.0590  0.0603  0.0597  0.0590
1     0.05  0.0693  0.0687  0.0687  0.0690  0.0687  0.0687  0.0693  0.0690  0.0690
2     0.05  0.1303  0.1300  0.1290  0.1303  0.1300  0.1290  0.1307  0.1303  0.1303
3     0.05  0.2603  0.2593  0.2590  0.2597  0.2593  0.2590  0.2620  0.2603  0.2597
4     0.05  0.4547  0.4533  0.4527  0.4543  0.4537  0.4527  0.4550  0.4550  0.4540
5     0.05  0.6650  0.6637  0.6633  0.6643  0.6637  0.6633  0.6680  0.6660  0.6643
6     0.05  0.8447  0.8437  0.8433  0.8447  0.8437  0.8433  0.8447  0.8447  0.8440
7     0.05  0.9397  0.9397  0.9390  0.9397  0.9397  0.9390  0.9403  0.9397  0.9397
8     0.05  0.9843  0.9840  0.9840  0.9840  0.9840  0.9840  0.9843  0.9843  0.9840
9     0.05  0.9983  0.9980  0.9980  0.9983  0.9983  0.9980  0.9983  0.9983  0.9983
10    0.05  0.9993  0.9993  0.9993  0.9993  0.9993  0.9993  0.9993  0.9993  0.9993
}\FirstMonConNSevenKnotsFive

\pgfplotstableread{
delta alpha   P1      P2      P3      L1      L2      L3      C1      C2      C3
0     0.05  0.0577  0.0573  0.0573  0.0573  0.0573  0.0573  0.0580  0.0577  0.0573
1     0.05  0.0717  0.0713  0.0710  0.0717  0.0713  0.0710  0.0720  0.0717  0.0713
2     0.05  0.1257  0.1250  0.1247  0.1250  0.1250  0.1247  0.1260  0.1260  0.1250
3     0.05  0.2330  0.2320  0.2320  0.2323  0.2320  0.2320  0.2347  0.2333  0.2320
4     0.05  0.4053  0.4030  0.4033  0.4047  0.4030  0.4033  0.4067  0.4057  0.4030
5     0.05  0.6260  0.6233  0.6230  0.6240  0.6237  0.6230  0.6273  0.6263  0.6237
6     0.05  0.8133  0.8117  0.8117  0.8130  0.8117  0.8117  0.8143  0.8140  0.8117
7     0.05  0.9283  0.9283  0.9280  0.9283  0.9283  0.9280  0.9287  0.9283  0.9283
8     0.05  0.9760  0.9757  0.9757  0.9760  0.9757  0.9757  0.9760  0.9760  0.9757
9     0.05  0.9960  0.9960  0.9957  0.9960  0.9960  0.9957  0.9960  0.9960  0.9960
10    0.05  0.9990  0.9990  0.9990  0.9990  0.9990  0.9990  0.9990  0.9990  0.9990
}\FirstMonConNSevenKnotsSeven


\pgfplotstableread{
delta alpha   P1      P2      P3      L1      L2      L3      C1      C2      C3
0     0.05  0.0570  0.0557  0.0557  0.0567  0.0560  0.0557  0.0570  0.0570  0.0560
1     0.05  0.0867  0.0857  0.0857  0.0863  0.0857  0.0857  0.0877  0.0870  0.0860
2     0.05  0.1903  0.1890  0.1890  0.1897  0.1893  0.1890  0.1917  0.1907  0.1897
3     0.05  0.4123  0.4093  0.4087  0.4100  0.4093  0.4087  0.4143  0.4127  0.4097
4     0.05  0.6593  0.6580  0.6580  0.6580  0.6580  0.6580  0.6630  0.6600  0.6580
5     0.05  0.8633  0.8607  0.8597  0.8617  0.8610  0.8597  0.8663  0.8640  0.8613
6     0.05  0.9590  0.9573  0.9573  0.9580  0.9573  0.9573  0.9593  0.9593  0.9577
7     0.05  0.9890  0.9890  0.9890  0.9890  0.9890  0.9890  0.9890  0.9890  0.9890
8     0.05  0.9983  0.9983  0.9983  0.9983  0.9983  0.9983  0.9983  0.9983  0.9983
9     0.05  1.0000  1.0000  1.0000  1.0000  1.0000  1.0000  1.0000  1.0000  1.0000
10    0.05  1.0000  1.0000  1.0000  1.0000  1.0000  1.0000  1.0000  1.0000  1.0000
}\FirstMonNTenKnotsThree

\pgfplotstableread{
delta alpha   P1      P2      P3      L1      L2      L3      C1      C2      C3
0     0.05  0.0563  0.0560  0.0560  0.0560  0.0560  0.0560  0.0570  0.0567  0.0560
1     0.05  0.0830  0.0823  0.0820  0.0823  0.0823  0.0820  0.0830  0.0830  0.0823
2     0.05  0.1690  0.1683  0.1683  0.1687  0.1687  0.1683  0.1710  0.1700  0.1687
3     0.05  0.3630  0.3613  0.3603  0.3613  0.3613  0.3603  0.3650  0.3643  0.3613
4     0.05  0.6210  0.6183  0.6180  0.6203  0.6193  0.6180  0.6233  0.6217  0.6200
5     0.05  0.8283  0.8260  0.8257  0.8280  0.8270  0.8257  0.8310  0.8297  0.8273
6     0.05  0.9493  0.9490  0.9487  0.9493  0.9490  0.9487  0.9500  0.9497  0.9490
7     0.05  0.9887  0.9883  0.9883  0.9887  0.9883  0.9883  0.9887  0.9887  0.9887
8     0.05  0.9980  0.9980  0.9980  0.9980  0.9980  0.9980  0.9980  0.9980  0.9980
9     0.05  1.0000  1.0000  1.0000  1.0000  1.0000  1.0000  1.0000  1.0000  1.0000
10    0.05  1.0000  1.0000  1.0000  1.0000  1.0000  1.0000  1.0000  1.0000  1.0000
}\FirstMonNTenKnotsFive

\pgfplotstableread{
delta alpha   P1      P2      P3      L1      L2      L3      C1      C2      C3
0     0.05  0.0560  0.0560  0.0560  0.0560  0.0560  0.0560  0.0573  0.0567  0.0560
1     0.05  0.0783  0.0777  0.0780  0.0780  0.0780  0.0780  0.0783  0.0783  0.0780
2     0.05  0.1527  0.1513  0.1510  0.1513  0.1513  0.1510  0.1527  0.1527  0.1513
3     0.05  0.3240  0.3223  0.3220  0.3230  0.3227  0.3220  0.3243  0.3240  0.3227
4     0.05  0.5683  0.5667  0.5663  0.5673  0.5673  0.5663  0.5693  0.5687  0.5673
5     0.05  0.7877  0.7843  0.7840  0.7857  0.7853  0.7837  0.7897  0.7883  0.7857
6     0.05  0.9343  0.9337  0.9337  0.9337  0.9337  0.9337  0.9350  0.9347  0.9337
7     0.05  0.9827  0.9823  0.9823  0.9827  0.9827  0.9823  0.9827  0.9827  0.9827
8     0.05  0.9967  0.9963  0.9963  0.9967  0.9963  0.9963  0.9967  0.9967  0.9963
9     0.05  1.0000  1.0000  1.0000  1.0000  1.0000  1.0000  1.0000  1.0000  1.0000
10    0.05  1.0000  1.0000  1.0000  1.0000  1.0000  1.0000  1.0000  1.0000  1.0000
}\FirstMonNTenKnotsSeven

\pgfplotstableread{
delta alpha   P1      P2      P3      L1      L2      L3      C1      C2      C3
0     0.05  0.0523  0.0523  0.0523  0.0523  0.0523  0.0523  0.0523  0.0523  0.0523
1     0.05  0.0440  0.0433  0.0433  0.0443  0.0433  0.0433  0.0450  0.0447  0.0433
2     0.05  0.0733  0.0720  0.0720  0.0730  0.0723  0.0720  0.0747  0.0737  0.0723
3     0.05  0.1403  0.1370  0.1363  0.1377  0.1370  0.1363  0.1430  0.1410  0.1373
4     0.05  0.2423  0.2367  0.2363  0.2390  0.2367  0.2360  0.2477  0.2427  0.2387
5     0.05  0.3817  0.3747  0.3737  0.3783  0.3760  0.3737  0.3910  0.3843  0.3773
6     0.05  0.5657  0.5567  0.5540  0.5613  0.5583  0.5540  0.5750  0.5687  0.5597
7     0.05  0.7227  0.7167  0.7150  0.7190  0.7180  0.7150  0.7293  0.7247  0.7180
8     0.05  0.8487  0.8447  0.8427  0.8460  0.8450  0.8427  0.8537  0.8497  0.8453
9     0.05  0.9223  0.9190  0.9183  0.9210  0.9203  0.9183  0.9260  0.9233  0.9200
10    0.05  0.9677  0.9657  0.9653  0.9667  0.9657  0.9653  0.9703  0.9687  0.9667
}\FirstConNTenKnotsThree

\pgfplotstableread{
delta alpha   P1      P2      P3      L1      L2      L3      C1      C2      C3
0     0.05  0.0553  0.0550  0.0550  0.0550  0.0550  0.0550  0.0553  0.0553  0.0550
1     0.05  0.0430  0.0417  0.0417  0.0423  0.0420  0.0417  0.0433  0.0430  0.0420
2     0.05  0.0650  0.0647  0.0647  0.0650  0.0647  0.0647  0.0660  0.0653  0.0647
3     0.05  0.1260  0.1227  0.1217  0.1230  0.1227  0.1217  0.1273  0.1267  0.1227
4     0.05  0.2057  0.2030  0.2027  0.2040  0.2030  0.2027  0.2087  0.2070  0.2037
5     0.05  0.3310  0.3260  0.3250  0.3290  0.3263  0.3250  0.3350  0.3323  0.3280
6     0.05  0.4940  0.4877  0.4863  0.4907  0.4883  0.4863  0.4987  0.4953  0.4887
7     0.05  0.6577  0.6533  0.6523  0.6553  0.6540  0.6523  0.6610  0.6600  0.6550
8     0.05  0.7993  0.7963  0.7957  0.7983  0.7963  0.7957  0.8027  0.8010  0.7967
9     0.05  0.8983  0.8963  0.8950  0.8970  0.8963  0.8950  0.9020  0.9000  0.8963
10    0.05  0.9567  0.9543  0.9540  0.9557  0.9547  0.9540  0.9570  0.9570  0.9547
}\FirstConNTenKnotsFive

\pgfplotstableread{
delta alpha   P1      P2      P3      L1      L2      L3      C1      C2      C3
0     0.05  0.0540  0.0540  0.0540  0.0540  0.0540  0.0540  0.0540  0.0540  0.0540
1     0.05  0.0450  0.0447  0.0447  0.0447  0.0447  0.0447  0.0457  0.0457  0.0447
2     0.05  0.0687  0.0683  0.0683  0.0683  0.0683  0.0683  0.0697  0.0690  0.0683
3     0.05  0.1120  0.1107  0.1097  0.1120  0.1113  0.1097  0.1137  0.1123  0.1113
4     0.05  0.1837  0.1807  0.1803  0.1823  0.1813  0.1803  0.1847  0.1837  0.1817
5     0.05  0.2880  0.2843  0.2817  0.2857  0.2850  0.2817  0.2927  0.2897  0.2850
6     0.05  0.4343  0.4267  0.4260  0.4290  0.4273  0.4260  0.4393  0.4353  0.4280
7     0.05  0.5963  0.5883  0.5860  0.5910  0.5893  0.5860  0.6017  0.5983  0.5907
8     0.05  0.7440  0.7407  0.7397  0.7417  0.7410  0.7397  0.7480  0.7443  0.7413
9     0.05  0.8643  0.8630  0.8607  0.8633  0.8630  0.8607  0.8673  0.8647  0.8633
10    0.05  0.9377  0.9350  0.9347  0.9367  0.9350  0.9347  0.9390  0.9380  0.9360
}\FirstConNTenKnotsSeven

\pgfplotstableread{
delta alpha   P1      P2      P3      L1      L2      L3      C1      C2      C3
0     0.05  0.0550  0.0550  0.0550  0.0550  0.0550  0.0550  0.0550  0.0550  0.0550
1     0.05  0.0757  0.0757  0.0757  0.0757  0.0757  0.0757  0.0757  0.0757  0.0757
2     0.05  0.1740  0.1730  0.1730  0.1733  0.1730  0.1730  0.1743  0.1743  0.1730
3     0.05  0.3880  0.3853  0.3850  0.3870  0.3857  0.3850  0.3897  0.3887  0.3867
4     0.05  0.6477  0.6463  0.6463  0.6470  0.6467  0.6463  0.6497  0.6490  0.6470
5     0.05  0.8570  0.8550  0.8547  0.8557  0.8550  0.8547  0.8593  0.8573  0.8550
6     0.05  0.9590  0.9583  0.9580  0.9587  0.9583  0.9580  0.9597  0.9590  0.9583
7     0.05  0.9883  0.9883  0.9880  0.9883  0.9883  0.9880  0.9887  0.9883  0.9883
8     0.05  0.9980  0.9980  0.9980  0.9980  0.9980  0.9980  0.9980  0.9980  0.9980
9     0.05  1.0000  0.9997  0.9997  1.0000  0.9997  0.9997  1.0000  1.0000  0.9997
10    0.05  1.0000  1.0000  1.0000  1.0000  1.0000  1.0000  1.0000  1.0000  1.0000
}\FirstMonConNTenKnotsThree

\pgfplotstableread{
delta alpha   P1      P2      P3      L1      L2      L3      C1      C2      C3
0     0.05  0.0557  0.0553  0.0553  0.0557  0.0553  0.0553  0.0557  0.0557  0.0557
1     0.05  0.0697  0.0693  0.0690  0.0697  0.0693  0.0690  0.0700  0.0700  0.0693
2     0.05  0.1523  0.1520  0.1520  0.1520  0.1520  0.1520  0.1530  0.1527  0.1520
3     0.05  0.3403  0.3383  0.3380  0.3390  0.3387  0.3380  0.3447  0.3417  0.3387
4     0.05  0.6027  0.6000  0.5990  0.6020  0.6007  0.5990  0.6050  0.6033  0.6017
5     0.05  0.8220  0.8210  0.8207  0.8217  0.8213  0.8207  0.8227  0.8223  0.8213
6     0.05  0.9463  0.9463  0.9463  0.9463  0.9463  0.9463  0.9470  0.9467  0.9463
7     0.05  0.9870  0.9870  0.9870  0.9870  0.9870  0.9870  0.9873  0.9870  0.9870
8     0.05  0.9983  0.9983  0.9983  0.9983  0.9983  0.9983  0.9983  0.9983  0.9983
9     0.05  1.0000  1.0000  1.0000  1.0000  1.0000  1.0000  1.0000  1.0000  1.0000
10    0.05  1.0000  1.0000  1.0000  1.0000  1.0000  1.0000  1.0000  1.0000  1.0000
}\FirstMonConNTenKnotsFive

\pgfplotstableread{
delta alpha   P1      P2      P3      L1      L2      L3      C1      C2      C3
0     0.05  0.0533  0.0533  0.0533  0.0533  0.0533  0.0533  0.0537  0.0533  0.0533
1     0.05  0.0710  0.0707  0.0707  0.0710  0.0707  0.0707  0.0717  0.0717  0.0707
2     0.05  0.1400  0.1393  0.1390  0.1393  0.1393  0.1390  0.1410  0.1407  0.1393
3     0.05  0.3047  0.3047  0.3040  0.3047  0.3047  0.3040  0.3050  0.3050  0.3047
4     0.05  0.5490  0.5483  0.5483  0.5487  0.5483  0.5483  0.5513  0.5500  0.5487
5     0.05  0.7827  0.7807  0.7807  0.7820  0.7810  0.7807  0.7840  0.7827  0.7813
6     0.05  0.9290  0.9280  0.9280  0.9287  0.9280  0.9280  0.9297  0.9290  0.9287
7     0.05  0.9793  0.9793  0.9793  0.9793  0.9793  0.9793  0.9800  0.9797  0.9793
8     0.05  0.9967  0.9967  0.9967  0.9967  0.9967  0.9967  0.9967  0.9967  0.9967
9     0.05  0.9997  0.9997  0.9997  0.9997  0.9997  0.9997  0.9997  0.9997  0.9997
10    0.05  1.0000  1.0000  1.0000  1.0000  1.0000  1.0000  1.0000  1.0000  1.0000
}\FirstMonConNTenKnotsSeven

\pgfplotstableread{
delta alpha   Five1  Seven1   Ten1   Five2   Seven2   Ten2    Fifty2
0     0.05   0.0597  0.0540  0.0493  0.0683  0.0570  0.0553   0.0460
1     0.05   0.0623  0.0600  0.0620  0.0760  0.0600  0.0583   0.0653
2     0.05   0.0767  0.0873  0.1067  0.0797  0.0693  0.0690   0.1023
3     0.05   0.1173  0.1673  0.2260  0.0923  0.0783  0.0860   0.2077
4     0.05   0.1893  0.3010  0.4423  0.0980  0.0983  0.1077   0.4153
5     0.05   0.3210  0.5050  0.6797  0.1123  0.1190  0.1393   0.6890
6     0.05   0.4780  0.7067  0.8657  0.1283  0.1593  0.1877   0.9043
7     0.05   0.6383  0.8530  0.9613  0.1540  0.2120  0.2513   0.9810
8     0.05   0.7813  0.9470  0.9913  0.1810  0.2693  0.3413   0.9993
9     0.05   0.8807  0.9850  0.9987  0.2250  0.3410  0.4530   1.0000
10    0.05   0.9450  0.9957  1.0000  0.2737  0.4260  0.5753   1.0000
}\Chetverikov

\pgfplotstableread{
delta alpha   FiveOp  FiveUn  SevenOp  SevenUn  TenOp    TenUn
0     0.05    0.0660  0.0600  0.0587   0.0567   0.0653   0.0610
1     0.05    0.0713  0.0607  0.0667   0.0563   0.0790   0.0653
2     0.05    0.0857  0.0660  0.0843   0.0677   0.0973   0.0737
3     0.05    0.1177  0.0787  0.1343   0.0783   0.1467   0.0933
4     0.05    0.1727  0.0993  0.2147   0.1053   0.2473   0.1217
5     0.05    0.2497  0.1293  0.3287   0.1453   0.3817   0.1693
6     0.05    0.3537  0.1717  0.4823   0.2027   0.5753   0.2360
7     0.05    0.4753  0.2280  0.6403   0.2870   0.7690   0.3387
8     0.05    0.6113  0.2973  0.7873   0.4017   0.8967   0.4620
9     0.05    0.7433  0.3777  0.9010   0.5153   0.9647   0.6140
10    0.05    0.8547  0.4803  0.9600   0.6527   0.9897   0.7613
}\FirstLSWMon

\pgfplotstableread{
delta alpha   FiveOp  FiveUn  SevenOp  SevenUn   TenOp   TenUn
0     0.05    0.0633  0.0593   0.0643   0.0633   0.0580  0.0567
1     0.05    0.0643  0.0557   0.0690   0.0630   0.0593  0.0607
2     0.05    0.0717  0.0593   0.0670   0.0647   0.0660  0.0603
3     0.05    0.0757  0.0637   0.0827   0.0687   0.0740  0.0640
4     0.05    0.0847  0.0713   0.0920   0.0783   0.0940  0.0700
5     0.05    0.1000  0.0773   0.1110   0.0860   0.1170  0.0810
6     0.05    0.1160  0.0813   0.1443   0.0947   0.1480  0.0927
7     0.05    0.1390  0.0967   0.1777   0.1080   0.1940  0.1040
8     0.05    0.1707  0.1083   0.2163   0.1290   0.2513  0.1287
9     0.05    0.2137  0.1233   0.2747   0.1487   0.3220  0.1447
10    0.05    0.2707  0.1453   0.3497   0.1867   0.4100  0.1860
}\FirstLSWCon

\pgfplotstableread{
delta alpha   FiveOp  FiveUn  SevenOp   SevenUn  TenOp    TenUn
0     0.05    0.0680  0.0653   0.0687   0.0653   0.0647   0.0597
1     0.05    0.0710  0.0663   0.0710   0.0633   0.0700   0.0633
2     0.05    0.0747  0.0663   0.0723   0.0677   0.0750   0.0697
3     0.05    0.0883  0.0737   0.0880   0.0750   0.0913   0.0750
4     0.05    0.0997  0.0777   0.1117   0.0867   0.1133   0.0903
5     0.05    0.1240  0.0910   0.1437   0.0987   0.1550   0.1043
6     0.05    0.1563  0.1093   0.1920   0.1190   0.2157   0.1253
7     0.05    0.2043  0.1333   0.2650   0.1493   0.3000   0.1577
8     0.05    0.2650  0.1520   0.3467   0.1883   0.4100   0.2083
9     0.05    0.3337  0.1887   0.4543   0.2410   0.5390   0.2683
10    0.05    0.4297  0.2340   0.5940   0.3037   0.6987   0.3603
}\FirstLSWMonCon

\pgfplotstableread{
delta alpha   P1      P2      P3      L1      L2      L3      C1      C2      C3
0     0.05  0.0643  0.0627  0.0613  0.0633  0.0623  0.0613  0.0657  0.0643  0.0627
1     0.05  0.1133  0.1117  0.1110  0.1120  0.1117  0.1110  0.1143  0.1133  0.1117
2     0.05  0.2017  0.1997  0.1987  0.2007  0.1997  0.1987  0.2027  0.2017  0.1997
3     0.05  0.3200  0.3190  0.3173  0.3190  0.3190  0.3173  0.3210  0.3203  0.3190
4     0.05  0.4613  0.4607  0.4597  0.4603  0.4603  0.4597  0.4613  0.4613  0.4607
5     0.05  0.6157  0.6153  0.6147  0.6157  0.6153  0.6147  0.6157  0.6157  0.6157
6     0.05  0.7460  0.7457  0.7457  0.7457  0.7457  0.7457  0.7467  0.7460  0.7457
7     0.05  0.8577  0.8577  0.8573  0.8577  0.8577  0.8573  0.8580  0.8580  0.8577
8     0.05  0.9317  0.9313  0.9313  0.9313  0.9313  0.9313  0.9317  0.9317  0.9313
9     0.05  0.9790  0.9790  0.9790  0.9790  0.9790  0.9790  0.9790  0.9790  0.9790
10    0.05  0.9937  0.9937  0.9937  0.9937  0.9937  0.9937  0.9937  0.9937  0.9937
}\SecondMonNFiveQuadraticKnotsZero

\pgfplotstableread{
delta alpha   P1      P2      P3      L1      L2      L3      C1      C2      C3
0     0.05  0.0690  0.0687  0.0680  0.0690  0.0687  0.0680  0.0693  0.0690  0.0687
1     0.05  0.0993  0.0987  0.0980  0.0987  0.0987  0.0980  0.1000  0.0993  0.0987
2     0.05  0.1473  0.1460  0.1450  0.1463  0.1457  0.1450  0.1477  0.1473  0.1460
3     0.05  0.2030  0.2017  0.2010  0.2020  0.2013  0.2010  0.2037  0.2030  0.2013
4     0.05  0.2863  0.2850  0.2843  0.2853  0.2847  0.2843  0.2867  0.2863  0.2850
5     0.05  0.3903  0.3893  0.3887  0.3900  0.3893  0.3887  0.3907  0.3903  0.3893
6     0.05  0.5060  0.5060  0.5060  0.5060  0.5060  0.5060  0.5067  0.5063  0.5060
7     0.05  0.6207  0.6200  0.6200  0.6200  0.6200  0.6200  0.6207  0.6207  0.6200
8     0.05  0.7243  0.7237  0.7237  0.7240  0.7237  0.7237  0.7243  0.7243  0.7237
9     0.05  0.8223  0.8223  0.8223  0.8223  0.8223  0.8223  0.8223  0.8223  0.8223
10    0.05  0.8980  0.8983  0.8980  0.8980  0.8980  0.8980  0.8980  0.8980  0.8980
}\SecondMonNFiveQuadraticKnotsOne

\pgfplotstableread{
delta alpha   P1      P2      P3      L1      L2      L3      C1      C2      C3
0     0.05  0.0813  0.0803  0.0797  0.0810  0.0803  0.0797  0.0820  0.0813  0.0807
1     0.05  0.1077  0.1067  0.1060  0.1070  0.1067  0.1060  0.1087  0.1077  0.1067
2     0.05  0.1400  0.1380  0.1377  0.1390  0.1380  0.1377  0.1400  0.1400  0.1380
3     0.05  0.1893  0.1883  0.1860  0.1887  0.1880  0.1860  0.1900  0.1893  0.1883
4     0.05  0.2520  0.2500  0.2493  0.2500  0.2500  0.2493  0.2533  0.2520  0.2500
5     0.05  0.3210  0.3203  0.3187  0.3207  0.3197  0.3187  0.3227  0.3210  0.3203
6     0.05  0.4027  0.4013  0.4007  0.4027  0.4010  0.4007  0.4037  0.4030  0.4013
7     0.05  0.5023  0.5020  0.5020  0.5023  0.5020  0.5020  0.5023  0.5023  0.5020
8     0.05  0.6090  0.6080  0.6077  0.6087  0.6077  0.6077  0.6093  0.6090  0.6080
9     0.05  0.7060  0.7060  0.7060  0.7060  0.7060  0.7060  0.7060  0.7060  0.7060
10    0.05  0.7890  0.7890  0.7887  0.7890  0.7890  0.7887  0.7893  0.7890  0.7890
}\SecondMonNFiveQuadraticKnotsTwo

\pgfplotstableread{
delta alpha   P1      P2      P3      L1      L2      L3      C1      C2      C3
0     0.05  0.0680  0.0673  0.0670  0.0673  0.0673  0.0670  0.0690  0.0680  0.0673
1     0.05  0.0977  0.0963  0.0960  0.0967  0.0960  0.0960  0.0997  0.0980  0.0967
2     0.05  0.1420  0.1417  0.1413  0.1420  0.1417  0.1413  0.1443  0.1423  0.1417
3     0.05  0.2083  0.2073  0.2070  0.2080  0.2073  0.2070  0.2097  0.2083  0.2073
4     0.05  0.2887  0.2870  0.2860  0.2880  0.2867  0.2860  0.2887  0.2887  0.2870
5     0.05  0.3950  0.3950  0.3947  0.3950  0.3950  0.3947  0.3953  0.3953  0.3950
6     0.05  0.5093  0.5083  0.5073  0.5090  0.5083  0.5073  0.5097  0.5097  0.5087
7     0.05  0.6237  0.6240  0.6240  0.6240  0.6240  0.6240  0.6237  0.6237  0.6240
8     0.05  0.7313  0.7310  0.7307  0.7313  0.7310  0.7307  0.7317  0.7313  0.7310
9     0.05  0.8307  0.8307  0.8303  0.8307  0.8307  0.8303  0.8307  0.8307  0.8307
10    0.05  0.8970  0.8970  0.8970  0.8970  0.8970  0.8970  0.8970  0.8970  0.8970
}\SecondMonNFiveCubicKnotsZero

\pgfplotstableread{
delta alpha   P1      P2      P3      L1      L2      L3      C1      C2      C3
0     0.05  0.0843  0.0837  0.0830  0.0840  0.0833  0.0830  0.0853  0.0843  0.0837
1     0.05  0.1090  0.1073  0.1060  0.1080  0.1070  0.1060  0.1113  0.1097  0.1073
2     0.05  0.1480  0.1473  0.1467  0.1473  0.1470  0.1467  0.1483  0.1483  0.1473
3     0.05  0.1953  0.1937  0.1930  0.1937  0.1937  0.1930  0.1957  0.1953  0.1937
4     0.05  0.2513  0.2507  0.2503  0.2513  0.2507  0.2503  0.2517  0.2513  0.2507
5     0.05  0.3283  0.3270  0.3263  0.3273  0.3267  0.3263  0.3293  0.3283  0.3270
6     0.05  0.4197  0.4190  0.4183  0.4193  0.4190  0.4183  0.4197  0.4197  0.4190
7     0.05  0.5143  0.5140  0.5137  0.5143  0.5140  0.5137  0.5143  0.5143  0.5140
8     0.05  0.6240  0.6233  0.6230  0.6237  0.6233  0.6230  0.6243  0.6240  0.6233
9     0.05  0.7250  0.7250  0.7247  0.7250  0.7250  0.7247  0.7257  0.7250  0.7250
10    0.05  0.8033  0.8033  0.8033  0.8033  0.8033  0.8033  0.8037  0.8033  0.8033
}\SecondMonNFiveCubicKnotsOne

\pgfplotstableread{
delta alpha   P1      P2      P3      L1      L2      L3      C1      C2      C3
0     0.05  0.0930  0.0920  0.0903  0.0920  0.0920  0.0903  0.0943  0.0930  0.0920
1     0.05  0.1177  0.1160  0.1157  0.1167  0.1157  0.1157  0.1180  0.1177  0.1160
2     0.05  0.1477  0.1470  0.1453  0.1477  0.1470  0.1453  0.1490  0.1477  0.1470
3     0.05  0.1853  0.1833  0.1823  0.1837  0.1827  0.1823  0.1860  0.1853  0.1833
4     0.05  0.2410  0.2407  0.2390  0.2407  0.2403  0.2390  0.2417  0.2410  0.2407
5     0.05  0.3097  0.3090  0.3080  0.3093  0.3087  0.3080  0.3100  0.3097  0.3090
6     0.05  0.3910  0.3910  0.3907  0.3910  0.3910  0.3907  0.3917  0.3910  0.3910
7     0.05  0.4737  0.4733  0.4723  0.4737  0.4730  0.4723  0.4737  0.4737  0.4733
8     0.05  0.5587  0.5580  0.5580  0.5580  0.5580  0.5580  0.5587  0.5587  0.5580
9     0.05  0.6450  0.6450  0.6450  0.6450  0.6450  0.6450  0.6450  0.6450  0.6450
10    0.05  0.7317  0.7313  0.7310  0.7313  0.7313  0.7310  0.7317  0.7317  0.7313
}\SecondMonNFiveCubicKnotsTwo

\pgfplotstableread{
delta alpha   P1      P2      P3      L1      L2      L3      C1      C2      C3
0     0.05  0.0607  0.0597  0.0583  0.0597  0.0597  0.0583  0.0617  0.0607  0.0597
1     0.05  0.1300  0.1277  0.1270  0.1290  0.1280  0.1270  0.1313  0.1307  0.1283
2     0.05  0.2543  0.2530  0.2530  0.2533  0.2533  0.2530  0.2550  0.2543  0.2533
3     0.05  0.4107  0.4103  0.4100  0.4107  0.4103  0.4100  0.4120  0.4113  0.4107
4     0.05  0.5913  0.5907  0.5907  0.5907  0.5907  0.5907  0.5917  0.5917  0.5907
5     0.05  0.7647  0.7643  0.7643  0.7643  0.7643  0.7643  0.7647  0.7647  0.7643
6     0.05  0.8847  0.8847  0.8843  0.8847  0.8847  0.8843  0.8850  0.8847  0.8847
7     0.05  0.9587  0.9587  0.9587  0.9587  0.9587  0.9587  0.9587  0.9587  0.9587
8     0.05  0.9853  0.9853  0.9853  0.9853  0.9853  0.9853  0.9853  0.9853  0.9853
9     0.05  0.9973  0.9973  0.9973  0.9973  0.9973  0.9973  0.9973  0.9973  0.9973
10    0.05  0.9993  0.9993  0.9993  0.9993  0.9993  0.9993  0.9993  0.9993  0.9993
}\SecondMonNSevenQuadraticKnotsZero

\pgfplotstableread{
delta alpha   P1      P2      P3      L1      L2      L3      C1      C2      C3
0     0.05  0.0673  0.0660  0.0653  0.0663  0.0660  0.0653  0.0693  0.0673  0.0660
1     0.05  0.1113  0.1103  0.1097  0.1107  0.1103  0.1097  0.1127  0.1117  0.1103
2     0.05  0.1790  0.1777  0.1773  0.1783  0.1777  0.1773  0.1803  0.1793  0.1780
3     0.05  0.2743  0.2730  0.2730  0.2733  0.2730  0.2730  0.2753  0.2747  0.2730
4     0.05  0.3860  0.3850  0.3850  0.3853  0.3850  0.3850  0.3863  0.3860  0.3850
5     0.05  0.5210  0.5210  0.5207  0.5210  0.5210  0.5207  0.5217  0.5210  0.5210
6     0.05  0.6490  0.6497  0.6493  0.6493  0.6497  0.6493  0.6503  0.6493  0.6493
7     0.05  0.7757  0.7757  0.7757  0.7757  0.7757  0.7757  0.7763  0.7760  0.7757
8     0.05  0.8720  0.8720  0.8720  0.8720  0.8720  0.8720  0.8720  0.8720  0.8720
9     0.05  0.9387  0.9387  0.9387  0.9387  0.9387  0.9387  0.9387  0.9387  0.9387
10    0.05  0.9760  0.9760  0.9760  0.9760  0.9760  0.9760  0.9760  0.9760  0.9760
}\SecondMonNSevenQuadraticKnotsOne

\pgfplotstableread{
delta alpha   P1      P2      P3      L1      L2      L3      C1      C2      C3
0     0.05  0.0777  0.0767  0.0763  0.0770  0.0767  0.0763  0.0793  0.0783  0.0767
1     0.05  0.1130  0.1123  0.1117  0.1123  0.1123  0.1117  0.1137  0.1130  0.1123
2     0.05  0.1633  0.1617  0.1613  0.1620  0.1617  0.1613  0.1650  0.1640  0.1617
3     0.05  0.2280  0.2273  0.2267  0.2273  0.2273  0.2267  0.2283  0.2280  0.2273
4     0.05  0.3157  0.3143  0.3143  0.3147  0.3143  0.3147  0.3163  0.3160  0.3143
5     0.05  0.4250  0.4240  0.4237  0.4247  0.4240  0.4237  0.4260  0.4250  0.4240
6     0.05  0.5500  0.5497  0.5497  0.5500  0.5497  0.5497  0.5500  0.5500  0.5500
7     0.05  0.6647  0.6643  0.6640  0.6647  0.6643  0.6640  0.6650  0.6647  0.6647
8     0.05  0.7663  0.7667  0.7667  0.7663  0.7667  0.7667  0.7667  0.7667  0.7667
9     0.05  0.8530  0.8530  0.8533  0.8530  0.8530  0.8533  0.8530  0.8530  0.8530
10    0.05  0.9210  0.9207  0.9207  0.9207  0.9207  0.9207  0.9207  0.9207  0.9210
}\SecondMonNSevenQuadraticKnotsTwo

\pgfplotstableread{
delta alpha   P1      P2      P3      L1      L2      L3      C1      C2      C3
0     0.05  0.0657  0.0650  0.0640  0.0650  0.0650  0.0640  0.0663  0.0657  0.0650
1     0.05  0.1097  0.1090  0.1090  0.1090  0.1090  0.1090  0.1103  0.1100  0.1090
2     0.05  0.1837  0.1830  0.1827  0.1830  0.1830  0.1827  0.1850  0.1843  0.1830
3     0.05  0.2763  0.2743  0.2737  0.2750  0.2743  0.2737  0.2777  0.2767  0.2750
4     0.05  0.3923  0.3920  0.3917  0.3923  0.3920  0.3917  0.3930  0.3923  0.3923
5     0.05  0.5243  0.5243  0.5243  0.5243  0.5243  0.5243  0.5257  0.5247  0.5243
6     0.05  0.6520  0.6517  0.6517  0.6520  0.6517  0.6517  0.6527  0.6520  0.6520
7     0.05  0.7750  0.7750  0.7750  0.7750  0.7750  0.7750  0.7760  0.7753  0.7750
8     0.05  0.8763  0.8763  0.8763  0.8767  0.8763  0.8763  0.8767  0.8767  0.8767
9     0.05  0.9413  0.9413  0.9413  0.9413  0.9413  0.9413  0.9413  0.9413  0.9413
10    0.05  0.9737  0.9737  0.9737  0.9737  0.9737  0.9737  0.9737  0.9737  0.9737
}\SecondMonNSevenCubicKnotsZero

\pgfplotstableread{
delta alpha   P1      P2      P3      L1      L2      L3      C1      C2      C3
0     0.05  0.0753  0.0740  0.0737  0.0743  0.0740  0.0737  0.0770  0.0757  0.0740
1     0.05  0.1147  0.1123  0.1113  0.1140  0.1127  0.1113  0.1163  0.1153  0.1133
2     0.05  0.1593  0.1587  0.1583  0.1590  0.1587  0.1583  0.1607  0.1593  0.1590
3     0.05  0.2350  0.2327  0.2323  0.2333  0.2330  0.2323  0.2363  0.2353  0.2333
4     0.05  0.3273  0.3270  0.3263  0.3273  0.3270  0.3263  0.3283  0.3273  0.3270
5     0.05  0.4347  0.4340  0.4333  0.4343  0.4340  0.4333  0.4350  0.4347  0.4340
6     0.05  0.5623  0.5620  0.5617  0.5620  0.5620  0.5617  0.5623  0.5627  0.5620
7     0.05  0.6817  0.6820  0.6813  0.6817  0.6820  0.6813  0.6817  0.6817  0.6817
8     0.05  0.7857  0.7857  0.7857  0.7857  0.7857  0.7857  0.7857  0.7857  0.7857
9     0.05  0.8687  0.8687  0.8687  0.8687  0.8687  0.8687  0.8687  0.8687  0.8687
10    0.05  0.9307  0.9307  0.9307  0.9307  0.9307  0.9307  0.9307  0.9307  0.9307
}\SecondMonNSevenCubicKnotsOne

\pgfplotstableread{
delta alpha   P1      P2      P3      L1      L2      L3      C1      C2      C3
0     0.05  0.0883  0.0873  0.0867  0.0880  0.0873  0.0867  0.0893  0.0883  0.0873
1     0.05  0.1220  0.1207  0.1207  0.1210  0.1207  0.1207  0.1233  0.1223  0.1207
2     0.05  0.1663  0.1643  0.1640  0.1647  0.1643  0.1640  0.1667  0.1663  0.1643
3     0.05  0.2283  0.2267  0.2267  0.2273  0.2267  0.2267  0.2290  0.2283  0.2267
4     0.05  0.3080  0.3067  0.3067  0.3067  0.3067  0.3067  0.3087  0.3080  0.3067
5     0.05  0.4037  0.4027  0.4017  0.4030  0.4027  0.4017  0.4050  0.4047  0.4027
6     0.05  0.5080  0.5077  0.5067  0.5077  0.5077  0.5067  0.5100  0.5083  0.5077
7     0.05  0.6100  0.6097  0.6093  0.6097  0.6097  0.6093  0.6103  0.6100  0.6097
8     0.05  0.7073  0.7070  0.7073  0.7070  0.7070  0.7073  0.7070  0.7073  0.7073
9     0.05  0.8100  0.8100  0.8100  0.8100  0.8100  0.8100  0.8100  0.8100  0.8100
10    0.05  0.8770  0.8770  0.8770  0.8770  0.8770  0.8770  0.8770  0.8770  0.8770
}\SecondMonNSevenCubicKnotsTwo

\pgfplotstableread{
delta alpha   P1      P2      P3      L1      L2      L3      C1      C2      C3
0     0.05  0.0503  0.0500  0.0500  0.0503  0.0500  0.0500  0.0507  0.0503  0.0500
1     0.05  0.1283  0.1267  0.1267  0.1273  0.1270  0.1267  0.1290  0.1287  0.1270
2     0.05  0.2620  0.2607  0.2607  0.2607  0.2607  0.2607  0.2643  0.2633  0.2607
3     0.05  0.4577  0.4580  0.4573  0.4577  0.4580  0.4573  0.4587  0.4583  0.4580
4     0.05  0.6730  0.6730  0.6733  0.6730  0.6730  0.6733  0.6733  0.6733  0.6737
5     0.05  0.8480  0.8477  0.8473  0.8480  0.8477  0.8473  0.8483  0.8480  0.8477
6     0.05  0.9493  0.9493  0.9493  0.9493  0.9493  0.9493  0.9497  0.9493  0.9493
7     0.05  0.9823  0.9823  0.9823  0.9823  0.9823  0.9823  0.9823  0.9823  0.9823
8     0.05  0.9963  0.9963  0.9963  0.9963  0.9963  0.9963  0.9963  0.9963  0.9963
9     0.05  0.9993  0.9993  0.9993  0.9993  0.9993  0.9993  0.9993  0.9993  0.9993
10    0.05  1.0000  1.0000  1.0000  1.0000  1.0000  1.0000  1.0000  1.0000  1.0000
}\SecondMonNTenQuadraticKnotsZero

\pgfplotstableread{
delta alpha   P1      P2      P3      L1      L2      L3      C1      C2      C3
0     0.05  0.0567  0.0557  0.0553  0.0560  0.0560  0.0553  0.0580  0.0567  0.0560
1     0.05  0.1000  0.0993  0.0990  0.1000  0.0997  0.0990  0.1020  0.1007  0.0997
2     0.05  0.1723  0.1720  0.1720  0.1720  0.1720  0.1720  0.1727  0.1723  0.1720
3     0.05  0.2863  0.2857  0.2857  0.2863  0.2860  0.2857  0.2870  0.2867  0.2860
4     0.05  0.4303  0.4300  0.4300  0.4303  0.4300  0.4300  0.4303  0.4310  0.4300
5     0.05  0.5907  0.5900  0.5900  0.5903  0.5903  0.5900  0.5910  0.5907  0.5907
6     0.05  0.7630  0.7627  0.7627  0.7627  0.7627  0.7627  0.7630  0.7630  0.7627
7     0.05  0.8780  0.8780  0.8780  0.8780  0.8780  0.8780  0.8780  0.8780  0.8780
8     0.05  0.9593  0.9593  0.9593  0.9593  0.9593  0.9593  0.9593  0.9593  0.9593
9     0.05  0.9837  0.9837  0.9837  0.9837  0.9837  0.9837  0.9837  0.9837  0.9837
10    0.05  0.9957  0.9957  0.9957  0.9957  0.9957  0.9957  0.9957  0.9957  0.9957
}\SecondMonNTenQuadraticKnotsOne

\pgfplotstableread{
delta alpha   P1      P2      P3      L1      L2      L3      C1      C2      C3
0     0.05  0.0610  0.0603  0.0600  0.0607  0.0607  0.0600  0.0627  0.0617  0.0607
1     0.05  0.0963  0.0947  0.0947  0.0953  0.0950  0.0947  0.0980  0.0963  0.0953
2     0.05  0.1500  0.1497  0.1493  0.1497  0.1497  0.1497  0.1507  0.1500  0.1497
3     0.05  0.2317  0.2310  0.2310  0.2313  0.2310  0.2310  0.2320  0.2317  0.2310
4     0.05  0.3427  0.3410  0.3410  0.3417  0.3410  0.3410  0.3427  0.3427  0.3413
5     0.05  0.4713  0.4710  0.4710  0.4713  0.4710  0.4710  0.4717  0.4713  0.4713
6     0.05  0.6207  0.6207  0.6207  0.6207  0.6207  0.6207  0.6207  0.6207  0.6207
7     0.05  0.7630  0.7630  0.7630  0.7630  0.7630  0.7630  0.7630  0.7630  0.7630
8     0.05  0.8727  0.8727  0.8727  0.8727  0.8727  0.8727  0.8727  0.8727  0.8727
9     0.05  0.9470  0.9470  0.9470  0.9470  0.9470  0.9470  0.9470  0.9470  0.9470
10    0.05  0.9747  0.9743  0.9743  0.9743  0.9743  0.9743  0.9747  0.9747  0.9743
}\SecondMonNTenQuadraticKnotsTwo

\pgfplotstableread{
delta alpha   P1      P2      P3      L1      L2      L3      C1      C2      C3
0     0.05  0.0537  0.0527  0.0527  0.0530  0.0527  0.0527  0.0543  0.0537  0.0527
1     0.05  0.1043  0.1037  0.1027  0.1037  0.1037  0.1027  0.1057  0.1053  0.1037
2     0.05  0.1803  0.1800  0.1800  0.1803  0.1800  0.1800  0.1803  0.1803  0.1800
3     0.05  0.2930  0.2917  0.2917  0.2923  0.2920  0.2917  0.2937  0.2930  0.2920
4     0.05  0.4403  0.4403  0.4400  0.4403  0.4403  0.4400  0.4410  0.4407  0.4403
5     0.05  0.6027  0.6020  0.6017  0.6023  0.6020  0.6017  0.6033  0.6027  0.6020
6     0.05  0.7633  0.7633  0.7630  0.7633  0.7633  0.7630  0.7633  0.7633  0.7633
7     0.05  0.8787  0.8787  0.8787  0.8787  0.8787  0.8787  0.8787  0.8787  0.8787
8     0.05  0.9557  0.9557  0.9557  0.9557  0.9557  0.9557  0.9553  0.9557  0.9557
9     0.05  0.9833  0.9837  0.9837  0.9833  0.9837  0.9837  0.9833  0.9833  0.9837
10    0.05  0.9940  0.9940  0.9940  0.9940  0.9940  0.9940  0.9940  0.9940  0.9940
}\SecondMonNTenCubicKnotsZero

\pgfplotstableread{
delta alpha   P1      P2      P3      L1      L2      L3      C1      C2      C3
0     0.05  0.0593  0.0590  0.0590  0.0590  0.0590  0.0590  0.0620  0.0607  0.0590
1     0.05  0.0973  0.0970  0.0970  0.0970  0.0970  0.0970  0.0977  0.0977  0.0970
2     0.05  0.1537  0.1530  0.1527  0.1533  0.1533  0.1527  0.1550  0.1537  0.1533
3     0.05  0.2353  0.2347  0.2347  0.2350  0.2347  0.2347  0.2370  0.2360  0.2347
4     0.05  0.3480  0.3477  0.3477  0.3480  0.3480  0.3477  0.3487  0.3480  0.3480
5     0.05  0.4860  0.4860  0.4857  0.4860  0.4860  0.4857  0.4867  0.4860  0.4860
6     0.05  0.6300  0.6300  0.6297  0.6300  0.6300  0.6297  0.6307  0.6303  0.6297
7     0.05  0.7793  0.7787  0.7787  0.7790  0.7787  0.7787  0.7793  0.7793  0.7787
8     0.05  0.8813  0.8813  0.8813  0.8813  0.8813  0.8813  0.8813  0.8817  0.8813
9     0.05  0.9510  0.9510  0.9510  0.9510  0.9510  0.9510  0.9510  0.9510  0.9510
10    0.05  0.9800  0.9800  0.9800  0.9800  0.9800  0.9800  0.9800  0.9800  0.9800
}\SecondMonNTenCubicKnotsOne

\pgfplotstableread{
delta alpha   P1      P2      P3      L1      L2      L3      C1      C2      C3
0     0.05  0.0740  0.0727  0.0723  0.0733  0.0730  0.0723  0.0740  0.0740  0.0730
1     0.05  0.1013  0.1007  0.1003  0.1010  0.1010  0.1003  0.1017  0.1017  0.1010
2     0.05  0.1597  0.1573  0.1570  0.1587  0.1577  0.1570  0.1617  0.1597  0.1583
3     0.05  0.2323  0.2317  0.2313  0.2317  0.2317  0.2313  0.2330  0.2327  0.2317
4     0.05  0.3290  0.3273  0.3273  0.3283  0.3273  0.3273  0.3293  0.3293  0.3280
5     0.05  0.4467  0.4467  0.4467  0.4467  0.4467  0.4467  0.4467  0.4467  0.4467
6     0.05  0.5760  0.5753  0.5753  0.5760  0.5753  0.5753  0.5763  0.5763  0.5753
7     0.05  0.7007  0.7007  0.7003  0.7010  0.7007  0.7003  0.7013  0.7007  0.7010
8     0.05  0.8120  0.8120  0.8120  0.8120  0.8117  0.8120  0.8120  0.8120  0.8120
9     0.05  0.8990  0.8990  0.8990  0.8990  0.8990  0.8990  0.8987  0.8987  0.8990
10    0.05  0.9497  0.9497  0.9497  0.9497  0.9497  0.9497  0.9500  0.9497  0.9497
}\SecondMonNTenCubicKnotsTwo

\pgfplotstableread{
delta alpha   P1      P2      P3      L1      L2      L3      C1      C2      C3
0     0.05  0.0513  0.0510  0.0510  0.0513  0.0510  0.0510  0.0527  0.0520  0.0513
1     0.05  0.2777  0.2770  0.2770  0.2770  0.2767  0.2770  0.2790  0.2787  0.2773
2     0.05  0.7353  0.7360  0.7360  0.7350  0.7360  0.7360  0.7337  0.7343  0.7353
3     0.05  0.9750  0.9750  0.9750  0.9750  0.9750  0.9750  0.9743  0.9747  0.9750
4     0.05  1.0000  1.0000  1.0000  1.0000  1.0000  1.0000  1.0000  1.0000  1.0000
5     0.05  1.0000  1.0000  1.0000  1.0000  1.0000  1.0000  1.0000  1.0000  1.0000
6     0.05  1.0000  1.0000  1.0000  1.0000  1.0000  1.0000  1.0000  1.0000  1.0000
7     0.05  1.0000  1.0000  1.0000  1.0000  1.0000  1.0000  1.0000  1.0000  1.0000
8     0.05  1.0000  1.0000  1.0000  1.0000  1.0000  1.0000  1.0000  1.0000  1.0000
9     0.05  1.0000  1.0000  1.0000  1.0000  1.0000  1.0000  1.0000  1.0000  1.0000
10    0.05  1.0000  1.0000  1.0000  1.0000  1.0000  1.0000  1.0000  1.0000  1.0000
}\SecondMonNFiftyQuadraticKnotsZero

\pgfplotstableread{
delta alpha   P1      P2      P3      L1      L2      L3      C1      C2      C3
0     0.05  0.0530  0.0517  0.0517  0.0530  0.0520  0.0517  0.0533  0.0533  0.0527
1     0.05  0.1773  0.1767  0.1767  0.1773  0.1770  0.1767  0.1777  0.1773  0.1773
2     0.05  0.4620  0.4623  0.4623  0.4620  0.4620  0.4623  0.4623  0.4620  0.4620
3     0.05  0.8080  0.8073  0.8070  0.8073  0.8073  0.8073  0.8060  0.8063  0.8077
4     0.05  0.9770  0.9770  0.9770  0.9770  0.9770  0.9770  0.9767  0.9770  0.9770
5     0.05  0.9987  0.9987  0.9987  0.9987  0.9987  0.9987  0.9987  0.9987  0.9987
6     0.05  1.0000  1.0000  1.0000  1.0000  1.0000  1.0000  1.0000  1.0000  1.0000
7     0.05  1.0000  1.0000  1.0000  1.0000  1.0000  1.0000  1.0000  1.0000  1.0000
8     0.05  1.0000  1.0000  1.0000  1.0000  1.0000  1.0000  1.0000  1.0000  1.0000
9     0.05  1.0000  1.0000  1.0000  1.0000  1.0000  1.0000  1.0000  1.0000  1.0000
10    0.05  1.0000  1.0000  1.0000  1.0000  1.0000  1.0000  1.0000  1.0000  1.0000
}\SecondMonNFiftyQuadraticKnotsOne

\pgfplotstableread{
delta alpha   P1      P2      P3      L1      L2      L3      C1      C2      C3
0     0.05  0.0573  0.0573  0.0573  0.0573  0.0573  0.0573  0.0577  0.0577  0.0573
1     0.05  0.1560  0.1553  0.1553  0.1560  0.1553  0.1553  0.1577  0.1570  0.1557
2     0.05  0.3837  0.3833  0.3833  0.3837  0.3833  0.3833  0.3840  0.3833  0.3837
3     0.05  0.6837  0.6837  0.6837  0.6837  0.6837  0.6837  0.6837  0.6833  0.6837
4     0.05  0.9110  0.9117  0.9117  0.9110  0.9117  0.9117  0.9107  0.9110  0.9110
5     0.05  0.9933  0.9933  0.9933  0.9933  0.9933  0.9933  0.9933  0.9933  0.9933
6     0.05  1.0000  1.0000  1.0000  1.0000  1.0000  1.0000  1.0000  1.0000  1.0000
7     0.05  1.0000  1.0000  1.0000  1.0000  1.0000  1.0000  1.0000  1.0000  1.0000
8     0.05  1.0000  1.0000  1.0000  1.0000  1.0000  1.0000  1.0000  1.0000  1.0000
9     0.05  1.0000  1.0000  1.0000  1.0000  1.0000  1.0000  1.0000  1.0000  1.0000
10    0.05  1.0000  1.0000  1.0000  1.0000  1.0000  1.0000  1.0000  1.0000  1.0000
}\SecondMonNFiftyQuadraticKnotsTwo

\pgfplotstableread{
delta alpha   P1      P2      P3      L1      L2      L3      C1      C2      C3
0     0.05  0.0527  0.0520  0.0523  0.0527  0.0520  0.0520  0.0537  0.0533  0.0523
1     0.05  0.1840  0.1843  0.1843  0.1840  0.1843  0.1843  0.1847  0.1850  0.1840
2     0.05  0.4703  0.4703  0.4703  0.4710  0.4703  0.4703  0.4700  0.4700  0.4707
3     0.05  0.8057  0.8053  0.8053  0.8053  0.8053  0.8053  0.8047  0.8053  0.8053
4     0.05  0.9753  0.9750  0.9750  0.9750  0.9750  0.9750  0.9747  0.9750  0.9750
5     0.05  0.9990  0.9990  0.9990  0.9990  0.9990  0.9990  0.9990  0.9990  0.9990
6     0.05  1.0000  1.0000  1.0000  1.0000  1.0000  1.0000  1.0000  1.0000  1.0000
7     0.05  1.0000  1.0000  1.0000  1.0000  1.0000  1.0000  1.0000  1.0000  1.0000
8     0.05  1.0000  1.0000  1.0000  1.0000  1.0000  1.0000  1.0000  1.0000  1.0000
9     0.05  1.0000  1.0000  1.0000  1.0000  1.0000  1.0000  1.0000  1.0000  1.0000
10    0.05  1.0000  1.0000  1.0000  1.0000  1.0000  1.0000  1.0000  1.0000  1.0000
}\SecondMonNFiftyCubicKnotsZero

\pgfplotstableread{
delta alpha   P1      P2      P3      L1      L2      L3      C1      C2      C3
0     0.05  0.0597  0.0577  0.0577  0.0593  0.0583  0.0577  0.0607  0.0600  0.0590
1     0.05  0.1573  0.1570  0.1570  0.1573  0.1570  0.1570  0.1587  0.1580  0.1573
2     0.05  0.3987  0.4000  0.4000  0.3987  0.3990  0.4000  0.3980  0.3983  0.3987
3     0.05  0.6987  0.6990  0.6990  0.6987  0.6987  0.6990  0.6993  0.6990  0.6987
4     0.05  0.9187  0.9187  0.9187  0.9187  0.9187  0.9187  0.9183  0.9187  0.9187
5     0.05  0.9953  0.9953  0.9953  0.9953  0.9953  0.9953  0.9953  0.9953  0.9953
6     0.05  1.0000  1.0000  1.0000  1.0000  1.0000  1.0000  1.0000  1.0000  1.0000
7     0.05  1.0000  1.0000  1.0000  1.0000  1.0000  1.0000  1.0000  1.0000  1.0000
8     0.05  1.0000  1.0000  1.0000  1.0000  1.0000  1.0000  1.0000  1.0000  1.0000
9     0.05  1.0000  1.0000  1.0000  1.0000  1.0000  1.0000  1.0000  1.0000  1.0000
10    0.05  1.0000  1.0000  1.0000  1.0000  1.0000  1.0000  1.0000  1.0000  1.0000
}\SecondMonNFiftyCubicKnotsOne

\pgfplotstableread{
delta alpha   P1      P2      P3      L1      L2      L3      C1      C2      C3
0     0.05  0.0570  0.0563  0.0563  0.0570  0.0563  0.0563  0.0580  0.0577  0.0570
1     0.05  0.1580  0.1573  0.1573  0.1580  0.1580  0.1573  0.1590  0.1587  0.1580
2     0.05  0.3623  0.3613  0.3613  0.3620  0.3620  0.3613  0.3637  0.3630  0.3623
3     0.05  0.6480  0.6480  0.6480  0.6480  0.6480  0.6480  0.6480  0.6480  0.6477
4     0.05  0.8833  0.8833  0.8833  0.8833  0.8833  0.8833  0.8833  0.8833  0.8833
5     0.05  0.9850  0.9850  0.9850  0.9850  0.9850  0.9850  0.9850  0.9850  0.9850
6     0.05  0.9993  0.9993  0.9993  0.9993  0.9993  0.9993  0.9993  0.9993  0.9993
7     0.05  1.0000  1.0000  1.0000  1.0000  1.0000  1.0000  1.0000  1.0000  1.0000
8     0.05  1.0000  1.0000  1.0000  1.0000  1.0000  1.0000  1.0000  1.0000  1.0000
9     0.05  1.0000  1.0000  1.0000  1.0000  1.0000  1.0000  1.0000  1.0000  1.0000
10    0.05  1.0000  1.0000  1.0000  1.0000  1.0000  1.0000  1.0000  1.0000  1.0000
}\SecondMonNFiftyCubicKnotsTwo

\pgfplotstableread{
delta alpha   P1      P2      P3      L1      L2      L3      C1      C2      C3
0     0.05  0.0633  0.0627  0.0620  0.0633  0.0627  0.0620  0.0650  0.0633  0.0630
1     0.05  0.1043  0.1037  0.1033  0.1037  0.1037  0.1033  0.1057  0.1043  0.1037
2     0.05  0.1677  0.1673  0.1657  0.1677  0.1667  0.1657  0.1693  0.1683  0.1673
3     0.05  0.2587  0.2557  0.2553  0.2570  0.2557  0.2553  0.2587  0.2587  0.2563
4     0.05  0.3590  0.3573  0.3563  0.3577  0.3573  0.3563  0.3603  0.3593  0.3573
5     0.05  0.4863  0.4823  0.4813  0.4827  0.4823  0.4817  0.4873  0.4867  0.4823
6     0.05  0.6170  0.6157  0.6150  0.6163  0.6157  0.6153  0.6187  0.6170  0.6160
7     0.05  0.7433  0.7430  0.7413  0.7423  0.7423  0.7413  0.7440  0.7433  0.7423
8     0.05  0.8397  0.8387  0.8383  0.8390  0.8387  0.8383  0.8410  0.8403  0.8387
9     0.05  0.9087  0.9083  0.9080  0.9087  0.9080  0.9080  0.9087  0.9087  0.9083
10    0.05  0.9533  0.9533  0.9527  0.9533  0.9533  0.9527  0.9533  0.9533  0.9533
}\SecondConNFiveQuadraticKnotsZero

\pgfplotstableread{
delta alpha   P1      P2      P3      L1      L2      L3      C1      C2      C3
0     0.05  0.0707  0.0690  0.0687  0.0693  0.0690  0.0687  0.0723  0.0707  0.0690
1     0.05  0.0947  0.0930  0.0923  0.0937  0.0930  0.0923  0.0957  0.0950  0.0930
2     0.05  0.1447  0.1420  0.1417  0.1427  0.1420  0.1417  0.1457  0.1447  0.1420
3     0.05  0.2073  0.2040  0.2030  0.2060  0.2040  0.2027  0.2100  0.2080  0.2043
4     0.05  0.2930  0.2913  0.2913  0.2917  0.2913  0.2913  0.2950  0.2930  0.2913
5     0.05  0.3837  0.3833  0.3827  0.3837  0.3833  0.3827  0.3850  0.3840  0.3833
6     0.05  0.4907  0.4883  0.4873  0.4887  0.4877  0.4873  0.4910  0.4907  0.4883
7     0.05  0.6020  0.6017  0.6017  0.6017  0.6017  0.6017  0.6027  0.6020  0.6017
8     0.05  0.7147  0.7140  0.7130  0.7133  0.7133  0.7123  0.7150  0.7137  0.7137
9     0.05  0.8193  0.8173  0.8173  0.8180  0.8173  0.8170  0.8197  0.8193  0.8177
10    0.05  0.8823  0.8813  0.8813  0.8820  0.8813  0.8817  0.8823  0.8823  0.8813
}\SecondConNFiveQuadraticKnotsOne

\pgfplotstableread{
delta alpha   P1      P2      P3      L1      L2      L3      C1      C2      C3
0     0.05  0.0840  0.0837  0.0837  0.0837  0.0837  0.0837  0.0847  0.0840  0.0837
1     0.05  0.1037  0.1027  0.1017  0.1030  0.1027  0.1017  0.1040  0.1037  0.1027
2     0.05  0.1313  0.1283  0.1280  0.1290  0.1280  0.1280  0.1320  0.1317  0.1283
3     0.05  0.1637  0.1617  0.1603  0.1620  0.1617  0.1603  0.1640  0.1637  0.1617
4     0.05  0.1987  0.1977  0.1973  0.1980  0.1973  0.1973  0.2007  0.1990  0.1977
5     0.05  0.2517  0.2493  0.2473  0.2507  0.2493  0.2470  0.2530  0.2517  0.2493
6     0.05  0.3047  0.3040  0.3027  0.3040  0.3037  0.3023  0.3060  0.3047  0.3037
7     0.05  0.3763  0.3747  0.3743  0.3750  0.3747  0.3743  0.3770  0.3763  0.3750
8     0.05  0.4527  0.4510  0.4500  0.4510  0.4507  0.4503  0.4533  0.4530  0.4507
9     0.05  0.5320  0.5317  0.5310  0.5317  0.5313  0.5310  0.5330  0.5320  0.5317
10    0.05  0.6097  0.6077  0.6070  0.6083  0.6077  0.6070  0.6097  0.6097  0.6077
}\SecondConNFiveQuadraticKnotsTwo

\pgfplotstableread{
delta alpha   P1      P2      P3      L1      L2      L3      C1      C2      C3
0     0.05  0.0710  0.0707  0.0697  0.0707  0.0703  0.0697  0.0723  0.0710  0.0707
1     0.05  0.0993  0.0990  0.0977  0.0990  0.0987  0.0977  0.1023  0.0993  0.0990
2     0.05  0.1530  0.1507  0.1500  0.1517  0.1507  0.1500  0.1570  0.1537  0.1507
3     0.05  0.2297  0.2267  0.2240  0.2280  0.2263  0.2240  0.2317  0.2297  0.2267
4     0.05  0.3197  0.3170  0.3150  0.3180  0.3170  0.3150  0.3220  0.3203  0.3170
5     0.05  0.4150  0.4123  0.4113  0.4127  0.4120  0.4113  0.4163  0.4150  0.4123
6     0.05  0.5313  0.5287  0.5280  0.5287  0.5283  0.5280  0.5323  0.5313  0.5287
7     0.05  0.6513  0.6493  0.6480  0.6507  0.6490  0.6480  0.6527  0.6513  0.6497
8     0.05  0.7563  0.7540  0.7537  0.7543  0.7540  0.7537  0.7573  0.7563  0.7540
9     0.05  0.8500  0.8497  0.8480  0.8497  0.8493  0.8480  0.8507  0.8497  0.8497
10    0.05  0.9070  0.9070  0.9067  0.9067  0.9073  0.9067  0.9077  0.9070  0.9073
}\SecondConNFiveCubicKnotsZero

\pgfplotstableread{
delta alpha   P1      P2      P3      L1      L2      L3      C1      C2      C3
0     0.05  0.0860  0.0843  0.0830  0.0847  0.0843  0.0830  0.0867  0.0860  0.0843
1     0.05  0.1017  0.1003  0.1000  0.1010  0.1003  0.1000  0.1027  0.1017  0.1003
2     0.05  0.1320  0.1307  0.1300  0.1307  0.1303  0.1300  0.1320  0.1320  0.1307
3     0.05  0.1653  0.1640  0.1640  0.1647  0.1640  0.1640  0.1663  0.1653  0.1640
4     0.05  0.2070  0.2047  0.2033  0.2053  0.2040  0.2033  0.2090  0.2070  0.2043
5     0.05  0.2570  0.2553  0.2540  0.2560  0.2550  0.2540  0.2583  0.2573  0.2553
6     0.05  0.3153  0.3130  0.3127  0.3133  0.3130  0.3127  0.3160  0.3153  0.3130
7     0.05  0.3880  0.3870  0.3850  0.3880  0.3870  0.3850  0.3887  0.3880  0.3877
8     0.05  0.4607  0.4600  0.4597  0.4603  0.4600  0.4593  0.4617  0.4607  0.4597
9     0.05  0.5423  0.5403  0.5397  0.5410  0.5400  0.5397  0.5427  0.5423  0.5403
10    0.05  0.6257  0.6230  0.6223  0.6240  0.6233  0.6223  0.6260  0.6257  0.6240
}\SecondConNFiveCubicKnotsOne

\pgfplotstableread{
delta alpha   P1      P2      P3      L1      L2      L3      C1      C2      C3
0     0.05  0.0843  0.0813  0.0807  0.0827  0.0813  0.0807  0.0853  0.0847  0.0813
1     0.05  0.1087  0.1070  0.1057  0.1070  0.1070  0.1057  0.1097  0.1090  0.1070
2     0.05  0.1237  0.1230  0.1210  0.1230  0.1230  0.1210  0.1243  0.1237  0.1230
3     0.05  0.1540  0.1523  0.1517  0.1523  0.1523  0.1517  0.1550  0.1540  0.1523
4     0.05  0.1823  0.1813  0.1813  0.1813  0.1813  0.1813  0.1837  0.1823  0.1813
5     0.05  0.2240  0.2223  0.2213  0.2227  0.2223  0.2213  0.2247  0.2240  0.2223
6     0.05  0.2637  0.2620  0.2620  0.2620  0.2620  0.2620  0.2650  0.2640  0.2620
7     0.05  0.3180  0.3167  0.3163  0.3173  0.3167  0.3163  0.3183  0.3183  0.3167
8     0.05  0.3830  0.3807  0.3810  0.3813  0.3807  0.3810  0.3837  0.3827  0.3810
9     0.05  0.4447  0.4437  0.4427  0.4437  0.4430  0.4427  0.4460  0.4450  0.4437
10    0.05  0.5207  0.5197  0.5200  0.5200  0.5200  0.5197  0.5223  0.5213  0.5193
}\SecondConNFiveCubicKnotsTwo

\pgfplotstableread{
delta alpha   P1      P2      P3      L1      L2      L3      C1      C2      C3
0     0.05  0.0657  0.0650  0.0643  0.0650  0.0650  0.0643  0.0660  0.0660  0.0650
1     0.05  0.1237  0.1230  0.1227  0.1233  0.1230  0.1227  0.1253  0.1237  0.1230
2     0.05  0.2073  0.2047  0.2037  0.2063  0.2053  0.2033  0.2107  0.2083  0.2057
3     0.05  0.3357  0.3347  0.3347  0.3350  0.3347  0.3347  0.3383  0.3363  0.3347
4     0.05  0.4790  0.4770  0.4767  0.4773  0.4770  0.4767  0.4810  0.4797  0.4770
5     0.05  0.6287  0.6280  0.6270  0.6280  0.6280  0.6270  0.6317  0.6287  0.6280
6     0.05  0.7777  0.7770  0.7767  0.7773  0.7770  0.7770  0.7780  0.7780  0.7773
7     0.05  0.8857  0.8850  0.8850  0.8853  0.8853  0.8850  0.8857  0.8853  0.8857
8     0.05  0.9470  0.9467  0.9467  0.9467  0.9470  0.9467  0.9470  0.9470  0.9470
9     0.05  0.9787  0.9787  0.9783  0.9783  0.9787  0.9787  0.9790  0.9790  0.9783
10    0.05  0.9943  0.9947  0.9943  0.9947  0.9943  0.9943  0.9947  0.9943  0.9947
}\SecondConNSevenQuadraticKnotsZero

\pgfplotstableread{
delta alpha   P1      P2      P3      L1      L2      L3      C1      C2      C3
0     0.05  0.0747  0.0737  0.0730  0.0740  0.0737  0.0730  0.0767  0.0753  0.0737
1     0.05  0.1207  0.1187  0.1173  0.1190  0.1187  0.1173  0.1220  0.1210  0.1187
2     0.05  0.1780  0.1770  0.1770  0.1773  0.1770  0.1770  0.1793  0.1783  0.1773
3     0.05  0.2767  0.2747  0.2730  0.2763  0.2747  0.2730  0.2787  0.2773  0.2753
4     0.05  0.3817  0.3803  0.3800  0.3810  0.3807  0.3800  0.3827  0.3823  0.3807
5     0.05  0.5133  0.5110  0.5103  0.5117  0.5113  0.5103  0.5140  0.5137  0.5113
6     0.05  0.6483  0.6467  0.6463  0.6480  0.6470  0.6463  0.6490  0.6483  0.6470
7     0.05  0.7713  0.7700  0.7700  0.7707  0.7703  0.7700  0.7720  0.7707  0.7703
8     0.05  0.8760  0.8753  0.8750  0.8763  0.8760  0.8750  0.8757  0.8760  0.8760
9     0.05  0.9350  0.9343  0.9343  0.9347  0.9343  0.9343  0.9350  0.9353  0.9343
10    0.05  0.9737  0.9737  0.9737  0.9737  0.9737  0.9737  0.9737  0.9737  0.9737
}\SecondConNSevenQuadraticKnotsOne

\pgfplotstableread{
delta alpha   P1      P2      P3      L1      L2      L3      C1      C2      C3
0     0.05  0.0757  0.0740  0.0727  0.0750  0.0743  0.0727  0.0760  0.0757  0.0747
1     0.05  0.0993  0.0980  0.0977  0.0987  0.0980  0.0977  0.1017  0.1000  0.0980
2     0.05  0.1317  0.1287  0.1273  0.1300  0.1290  0.1273  0.1327  0.1317  0.1290
3     0.05  0.1760  0.1753  0.1750  0.1757  0.1753  0.1750  0.1780  0.1760  0.1757
4     0.05  0.2427  0.2423  0.2420  0.2423  0.2423  0.2420  0.2443  0.2427  0.2423
5     0.05  0.3140  0.3130  0.3117  0.3133  0.3130  0.3117  0.3173  0.3157  0.3133
6     0.05  0.3933  0.3927  0.3917  0.3923  0.3923  0.3917  0.3943  0.3933  0.3923
7     0.05  0.4973  0.4960  0.4950  0.4960  0.4960  0.4957  0.4980  0.4973  0.4960
8     0.05  0.5933  0.5923  0.5923  0.5927  0.5927  0.5920  0.5950  0.5933  0.5923
9     0.05  0.7037  0.7033  0.7023  0.7027  0.7033  0.7017  0.7040  0.7037  0.7030
10    0.05  0.7907  0.7900  0.7900  0.7907  0.7903  0.7900  0.7913  0.7907  0.7903
}\SecondConNSevenQuadraticKnotsTwo

\pgfplotstableread{
delta alpha   P1      P2      P3      L1      L2      L3      C1      C2      C3
0     0.05  0.0757  0.0730  0.0720  0.0747  0.0733  0.0720  0.0777  0.0760  0.0740
1     0.05  0.1267  0.1237  0.1217  0.1247  0.1237  0.1217  0.1277  0.1270  0.1237
2     0.05  0.1913  0.1883  0.1877  0.1890  0.1887  0.1877  0.1950  0.1917  0.1890
3     0.05  0.2953  0.2950  0.2943  0.2953  0.2950  0.2943  0.2967  0.2953  0.2953
4     0.05  0.4113  0.4087  0.4080  0.4097  0.4087  0.4077  0.4137  0.4123  0.4090
5     0.05  0.5463  0.5457  0.5450  0.5467  0.5460  0.5450  0.5473  0.5467  0.5460
6     0.05  0.6877  0.6863  0.6857  0.6863  0.6863  0.6857  0.6877  0.6877  0.6867
7     0.05  0.8070  0.8063  0.8060  0.8070  0.8067  0.8063  0.8090  0.8070  0.8063
8     0.05  0.8947  0.8940  0.8937  0.8947  0.8943  0.8940  0.8947  0.8943  0.8940
9     0.05  0.9503  0.9503  0.9503  0.9503  0.9503  0.9503  0.9503  0.9503  0.9503
10    0.05  0.9837  0.9837  0.9837  0.9837  0.9840  0.9840  0.9840  0.9837  0.9840
}\SecondConNSevenCubicKnotsZero

\pgfplotstableread{
delta alpha   P1      P2      P3      L1      L2      L3      C1      C2      C3
0     0.05  0.0710  0.0693  0.0687  0.0700  0.0693  0.0687  0.0727  0.0720  0.0693
1     0.05  0.0993  0.0980  0.0980  0.0993  0.0983  0.0980  0.1007  0.1000  0.0990
2     0.05  0.1297  0.1277  0.1270  0.1283  0.1277  0.1270  0.1310  0.1297  0.1280
3     0.05  0.1900  0.1887  0.1887  0.1890  0.1887  0.1887  0.1933  0.1917  0.1887
4     0.05  0.2530  0.2513  0.2497  0.2520  0.2517  0.2497  0.2553  0.2537  0.2517
5     0.05  0.3277  0.3263  0.3250  0.3273  0.3263  0.3250  0.3290  0.3287  0.3270
6     0.05  0.4150  0.4140  0.4137  0.4147  0.4143  0.4137  0.4153  0.4150  0.4147
7     0.05  0.5140  0.5133  0.5133  0.5137  0.5137  0.5130  0.5153  0.5140  0.5137
8     0.05  0.6117  0.6113  0.6113  0.6117  0.6117  0.6113  0.6127  0.6117  0.6113
9     0.05  0.7107  0.7097  0.7090  0.7097  0.7097  0.7093  0.7117  0.7107  0.7100
10    0.05  0.7997  0.7987  0.7990  0.7993  0.7990  0.7990  0.7997  0.7997  0.7990
}\SecondConNSevenCubicKnotsOne

\pgfplotstableread{
delta alpha   P1      P2      P3      L1      L2      L3      C1      C2      C3
0     0.05  0.0797  0.0783  0.0770  0.0790  0.0783  0.0770  0.0803  0.0800  0.0783
1     0.05  0.0973  0.0960  0.0957  0.0963  0.0960  0.0957  0.0990  0.0973  0.0960
2     0.05  0.1280  0.1263  0.1257  0.1267  0.1263  0.1257  0.1297  0.1283  0.1263
3     0.05  0.1673  0.1660  0.1650  0.1667  0.1660  0.1650  0.1687  0.1677  0.1660
4     0.05  0.2157  0.2143  0.2133  0.2157  0.2143  0.2133  0.2173  0.2163  0.2143
5     0.05  0.2763  0.2757  0.2753  0.2763  0.2760  0.2753  0.2780  0.2767  0.2760
6     0.05  0.3400  0.3390  0.3377  0.3390  0.3390  0.3377  0.3407  0.3400  0.3390
7     0.05  0.4187  0.4177  0.4170  0.4183  0.4177  0.4170  0.4193  0.4190  0.4180
8     0.05  0.5007  0.4990  0.4990  0.5007  0.4990  0.4990  0.5010  0.5007  0.5000
9     0.05  0.5897  0.5887  0.5880  0.5890  0.5887  0.5880  0.5900  0.5897  0.5887
10    0.05  0.6793  0.6787  0.6783  0.6793  0.6787  0.6787  0.6803  0.6793  0.6790
}\SecondConNSevenCubicKnotsTwo

\pgfplotstableread{
delta alpha   P1      P2      P3      L1      L2      L3      C1      C2      C3
0     0.05  0.0577  0.0567  0.0567  0.0573  0.0573  0.0567  0.0577  0.0577  0.0573
1     0.05  0.1293  0.1283  0.1277  0.1283  0.1283  0.1277  0.1307  0.1297  0.1283
2     0.05  0.2500  0.2483  0.2480  0.2493  0.2490  0.2480  0.2523  0.2513  0.2490
3     0.05  0.4020  0.4010  0.4007  0.4017  0.4007  0.4007  0.4033  0.4020  0.4013
4     0.05  0.5830  0.5817  0.5810  0.5827  0.5823  0.5810  0.5840  0.5833  0.5827
5     0.05  0.7557  0.7547  0.7543  0.7550  0.7547  0.7543  0.7557  0.7557  0.7547
6     0.05  0.8857  0.8860  0.8853  0.8857  0.8853  0.8857  0.8863  0.8860  0.8853
7     0.05  0.9510  0.9507  0.9507  0.9507  0.9510  0.9507  0.9510  0.9510  0.9507
8     0.05  0.9817  0.9817  0.9817  0.9817  0.9817  0.9817  0.9813  0.9817  0.9817
9     0.05  0.9950  0.9950  0.9953  0.9953  0.9950  0.9953  0.9950  0.9950  0.9950
10    0.05  0.9990  0.9990  0.9990  0.9990  0.9990  0.9990  0.9990  0.9990  0.9990
}\SecondConNTenQuadraticKnotsZero

\pgfplotstableread{
delta alpha   P1      P2      P3      L1      L2      L3      C1      C2      C3
0     0.05  0.0683  0.0673  0.0667  0.0673  0.0673  0.0667  0.0693  0.0687  0.0673
1     0.05  0.1093  0.1087  0.1087  0.1090  0.1090  0.1087  0.1107  0.1103  0.1090
2     0.05  0.1943  0.1920  0.1920  0.1933  0.1927  0.1920  0.1970  0.1953  0.1927
3     0.05  0.3290  0.3273  0.3263  0.3283  0.3277  0.3263  0.3313  0.3303  0.3277
4     0.05  0.4630  0.4617  0.4617  0.4623  0.4620  0.4613  0.4640  0.4640  0.4617
5     0.05  0.6253  0.6243  0.6243  0.6253  0.6243  0.6243  0.6263  0.6253  0.6250
6     0.05  0.7703  0.7703  0.7700  0.7703  0.7703  0.7700  0.7707  0.7703  0.7703
7     0.05  0.8763  0.8757  0.8753  0.8760  0.8753  0.8757  0.8760  0.8760  0.8760
8     0.05  0.9527  0.9523  0.9530  0.9527  0.9527  0.9523  0.9527  0.9527  0.9527
9     0.05  0.9810  0.9810  0.9810  0.9810  0.9810  0.9807  0.9810  0.9810  0.9810
10    0.05  0.9920  0.9923  0.9920  0.9920  0.9920  0.9920  0.9920  0.9920  0.9920
}\SecondConNTenQuadraticKnotsOne

\pgfplotstableread{
delta alpha   P1      P2      P3      L1      L2      L3      C1      C2      C3
0     0.05  0.0663  0.0647  0.0647  0.0657  0.0650  0.0647  0.0673  0.0667  0.0653
1     0.05  0.0890  0.0883  0.0883  0.0890  0.0887  0.0883  0.0897  0.0893  0.0887
2     0.05  0.1350  0.1323  0.1320  0.1347  0.1330  0.1320  0.1373  0.1357  0.1343
3     0.05  0.1943  0.1933  0.1930  0.1943  0.1933  0.1930  0.1957  0.1947  0.1933
4     0.05  0.2647  0.2627  0.2620  0.2637  0.2630  0.2623  0.2670  0.2650  0.2630
5     0.05  0.3593  0.3580  0.3577  0.3583  0.3580  0.3583  0.3603  0.3590  0.3583
6     0.05  0.4777  0.4770  0.4767  0.4767  0.4770  0.4767  0.4787  0.4777  0.4767
7     0.05  0.6067  0.6060  0.6057  0.6060  0.6057  0.6057  0.6070  0.6063  0.6063
8     0.05  0.7220  0.7210  0.7210  0.7213  0.7210  0.7210  0.7230  0.7223  0.7213
9     0.05  0.8227  0.8223  0.8227  0.8227  0.8227  0.8217  0.8230  0.8230  0.8227
10    0.05  0.9033  0.9030  0.9033  0.9033  0.9033  0.9030  0.9033  0.9033  0.9033
}\SecondConNTenQuadraticKnotsTwo

\pgfplotstableread{
delta alpha   P1      P2      P3      L1      L2      L3      C1      C2      C3
0     0.05  0.0703  0.0697  0.0693  0.0703  0.0697  0.0693  0.0717  0.0713  0.0703
1     0.05  0.1243  0.1217  0.1210  0.1220  0.1220  0.1210  0.1260  0.1243  0.1220
2     0.05  0.2163  0.2143  0.2137  0.2157  0.2147  0.2137  0.2193  0.2167  0.2150
3     0.05  0.3447  0.3437  0.3430  0.3440  0.3440  0.3433  0.3470  0.3463  0.3440
4     0.05  0.5017  0.5000  0.4990  0.5010  0.5003  0.4990  0.5027  0.5020  0.5010
5     0.05  0.6690  0.6677  0.6673  0.6680  0.6677  0.6673  0.6703  0.6693  0.6677
6     0.05  0.8040  0.8030  0.8030  0.8033  0.8030  0.8033  0.8053  0.8037  0.8033
7     0.05  0.9043  0.9037  0.9033  0.9040  0.9040  0.9033  0.9047  0.9043  0.9040
8     0.05  0.9633  0.9633  0.9633  0.9633  0.9633  0.9633  0.9633  0.9633  0.9633
9     0.05  0.9843  0.9843  0.9843  0.9843  0.9843  0.9843  0.9847  0.9843  0.9843
10    0.05  0.9953  0.9953  0.9953  0.9957  0.9957  0.9953  0.9957  0.9953  0.9953
}\SecondConNTenCubicKnotsZero

\pgfplotstableread{
delta alpha   P1      P2      P3      L1      L2      L3      C1      C2      C3
0     0.05  0.0673  0.0667  0.0660  0.0667  0.0667  0.0660  0.0683  0.0673  0.0667
1     0.05  0.0937  0.0930  0.0930  0.0933  0.0930  0.0930  0.0950  0.0940  0.0930
2     0.05  0.1383  0.1353  0.1350  0.1370  0.1360  0.1350  0.1410  0.1397  0.1367
3     0.05  0.2003  0.1993  0.1987  0.1997  0.1993  0.1987  0.2013  0.2007  0.1993
4     0.05  0.2740  0.2730  0.2717  0.2740  0.2737  0.2717  0.2753  0.2743  0.2740
5     0.05  0.3627  0.3617  0.3613  0.3623  0.3620  0.3613  0.3640  0.3630  0.3620
6     0.05  0.4910  0.4897  0.4897  0.4907  0.4900  0.4897  0.4913  0.4910  0.4903
7     0.05  0.6147  0.6140  0.6140  0.6147  0.6140  0.6137  0.6157  0.6150  0.6143
8     0.05  0.7323  0.7320  0.7317  0.7320  0.7320  0.7317  0.7327  0.7327  0.7320
9     0.05  0.8293  0.8290  0.8290  0.8290  0.8290  0.8287  0.8297  0.8297  0.8290
10    0.05  0.9057  0.9053  0.9053  0.9060  0.9057  0.9050  0.9060  0.9063  0.9057
}\SecondConNTenCubicKnotsOne

\pgfplotstableread{
delta alpha   P1      P2      P3      L1      L2      L3      C1      C2      C3
0     0.05  0.0743  0.0733  0.0733  0.0737  0.0733  0.0733  0.0753  0.0743  0.0733
1     0.05  0.0990  0.0970  0.0967  0.0977  0.0973  0.0967  0.1003  0.0993  0.0977
2     0.05  0.1343  0.1323  0.1317  0.1333  0.1323  0.1317  0.1353  0.1347  0.1330
3     0.05  0.1757  0.1753  0.1750  0.1753  0.1753  0.1750  0.1757  0.1757  0.1753
4     0.05  0.2250  0.2237  0.2233  0.2240  0.2237  0.2233  0.2257  0.2250  0.2237
5     0.05  0.3023  0.3010  0.3007  0.3020  0.3013  0.3007  0.3040  0.3030  0.3017
6     0.05  0.3993  0.3983  0.3980  0.3990  0.3987  0.3980  0.4003  0.3997  0.3990
7     0.05  0.4963  0.4953  0.4950  0.4950  0.4953  0.4953  0.4973  0.4967  0.4953
8     0.05  0.6020  0.6000  0.6003  0.6013  0.6000  0.6000  0.6030  0.6023  0.6013
9     0.05  0.7033  0.7037  0.7037  0.7033  0.7037  0.7037  0.7037  0.7033  0.7033
10    0.05  0.7927  0.7920  0.7923  0.7927  0.7923  0.7923  0.7923  0.7923  0.7923
}\SecondConNTenCubicKnotsTwo

\pgfplotstableread{
delta alpha   P1      P2      P3      L1      L2      L3      C1      C2      C3
0     0.05  0.0560  0.0550  0.0550  0.0557  0.0553  0.0550  0.0580  0.0570  0.0557
1     0.05  0.2783  0.2770  0.2770  0.2783  0.2780  0.2770  0.2813  0.2803  0.2783
2     0.05  0.6600  0.6600  0.6597  0.6600  0.6600  0.6600  0.6633  0.6620  0.6603
3     0.05  0.9337  0.9333  0.9330  0.9333  0.9333  0.9333  0.9333  0.9333  0.9337
4     0.05  0.9957  0.9957  0.9957  0.9957  0.9957  0.9957  0.9960  0.9960  0.9957
5     0.05  1.0000  1.0000  1.0000  1.0000  1.0000  1.0000  1.0000  1.0000  1.0000
6     0.05  1.0000  1.0000  1.0000  1.0000  1.0000  1.0000  1.0000  1.0000  1.0000
7     0.05  1.0000  1.0000  1.0000  1.0000  1.0000  1.0000  1.0000  1.0000  1.0000
8     0.05  1.0000  1.0000  1.0000  1.0000  1.0000  1.0000  1.0000  1.0000  1.0000
9     0.05  1.0000  1.0000  1.0000  1.0000  1.0000  1.0000  1.0000  1.0000  1.0000
10    0.05  1.0000  1.0000  1.0000  1.0000  1.0000  1.0000  1.0000  1.0000  1.0000
}\SecondConNFiftyQuadraticKnotsZero

\pgfplotstableread{
delta alpha   P1      P2      P3      L1      L2      L3      C1      C2      C3
0     0.05  0.0617  0.0613  0.0613  0.0617  0.0613  0.0613  0.0627  0.0620  0.0617
1     0.05  0.2127  0.2117  0.2117  0.2123  0.2120  0.2117  0.2153  0.2143  0.2123
2     0.05  0.5293  0.5280  0.5283  0.5293  0.5287  0.5283  0.5303  0.5290  0.5293
3     0.05  0.8503  0.8507  0.8503  0.8507  0.8503  0.8507  0.8510  0.8510  0.8503
4     0.05  0.9840  0.9843  0.9840  0.9840  0.9843  0.9843  0.9843  0.9843  0.9843
5     0.05  0.9993  0.9993  0.9993  0.9993  0.9993  0.9993  0.9993  0.9993  0.9993
6     0.05  1.0000  1.0000  1.0000  1.0000  1.0000  1.0000  1.0000  1.0000  1.0000
7     0.05  1.0000  1.0000  1.0000  1.0000  1.0000  1.0000  1.0000  1.0000  1.0000
8     0.05  1.0000  1.0000  1.0000  1.0000  1.0000  1.0000  1.0000  1.0000  1.0000
9     0.05  1.0000  1.0000  1.0000  1.0000  1.0000  1.0000  1.0000  1.0000  1.0000
10    0.05  1.0000  1.0000  1.0000  1.0000  1.0000  1.0000  1.0000  1.0000  1.0000
}\SecondConNFiftyQuadraticKnotsOne

\pgfplotstableread{
delta alpha   P1      P2      P3      L1      L2      L3      C1      C2      C3
0     0.05  0.0727  0.0723  0.0723  0.0727  0.0727  0.0723  0.0730  0.0727  0.0727
1     0.05  0.1530  0.1527  0.1527  0.1530  0.1527  0.1527  0.1543  0.1530  0.1530
2     0.05  0.3030  0.3030  0.3030  0.3030  0.3030  0.3030  0.3047  0.3037  0.3030
3     0.05  0.5553  0.5550  0.5547  0.5550  0.5547  0.5550  0.5563  0.5550  0.5553
4     0.05  0.8217  0.8220  0.8217  0.8217  0.8213  0.8213  0.8223  0.8223  0.8217
5     0.05  0.9597  0.9600  0.9600  0.9600  0.9597  0.9600  0.9597  0.9600  0.9597
6     0.05  0.9950  0.9957  0.9957  0.9953  0.9953  0.9953  0.9953  0.9953  0.9953
7     0.05  1.0000  1.0000  1.0000  1.0000  1.0000  1.0000  1.0000  1.0000  1.0000
8     0.05  1.0000  1.0000  1.0000  1.0000  1.0000  1.0000  1.0000  1.0000  1.0000
9     0.05  1.0000  1.0000  1.0000  1.0000  1.0000  1.0000  1.0000  1.0000  1.0000
10    0.05  1.0000  1.0000  1.0000  1.0000  1.0000  1.0000  1.0000  1.0000  1.0000
}\SecondConNFiftyQuadraticKnotsTwo

\pgfplotstableread{
delta alpha   P1      P2      P3      L1      L2      L3      C1      C2      C3
0     0.05  0.0627  0.0623  0.0623  0.0627  0.0623  0.0623  0.0637  0.0630  0.0627
1     0.05  0.2290  0.2277  0.2273  0.2290  0.2280  0.2273  0.2340  0.2317  0.2287
2     0.05  0.5703  0.5687  0.5693  0.5710  0.5693  0.5690  0.5750  0.5727  0.5700
3     0.05  0.8790  0.8790  0.8783  0.8797  0.8793  0.8787  0.8803  0.8803  0.8793
4     0.05  0.9893  0.9893  0.9893  0.9893  0.9893  0.9893  0.9893  0.9893  0.9893
5     0.05  1.0000  1.0000  1.0000  1.0000  1.0000  1.0000  1.0000  1.0000  1.0000
6     0.05  1.0000  1.0000  1.0000  1.0000  1.0000  1.0000  1.0000  1.0000  1.0000
7     0.05  1.0000  1.0000  1.0000  1.0000  1.0000  1.0000  1.0000  1.0000  1.0000
8     0.05  1.0000  1.0000  1.0000  1.0000  1.0000  1.0000  1.0000  1.0000  1.0000
9     0.05  1.0000  1.0000  1.0000  1.0000  1.0000  1.0000  1.0000  1.0000  1.0000
10    0.05  1.0000  1.0000  1.0000  1.0000  1.0000  1.0000  1.0000  1.0000  1.0000
}\SecondConNFiftyCubicKnotsZero

\pgfplotstableread{
delta alpha   P1      P2      P3      L1      L2      L3      C1      C2      C3
0     0.05  0.0710  0.0703  0.0703  0.0707  0.0703  0.0703  0.0727  0.0723  0.0707
1     0.05  0.1523  0.1517  0.1517  0.1523  0.1517  0.1517  0.1533  0.1527  0.1523
2     0.05  0.3130  0.3133  0.3127  0.3133  0.3137  0.3133  0.3150  0.3143  0.3133
3     0.05  0.5737  0.5733  0.5737  0.5743  0.5733  0.5733  0.5753  0.5747  0.5740
4     0.05  0.8273  0.8270  0.8270  0.8267  0.8270  0.8267  0.8270  0.8270  0.8273
5     0.05  0.9610  0.9613  0.9613  0.9613  0.9610  0.9613  0.9613  0.9617  0.9613
6     0.05  0.9957  0.9957  0.9957  0.9957  0.9957  0.9957  0.9957  0.9957  0.9957
7     0.05  1.0000  1.0000  1.0000  1.0000  1.0000  1.0000  1.0000  1.0000  1.0000
8     0.05  1.0000  1.0000  1.0000  1.0000  1.0000  1.0000  1.0000  1.0000  1.0000
9     0.05  1.0000  1.0000  1.0000  1.0000  1.0000  1.0000  1.0000  1.0000  1.0000
10    0.05  1.0000  1.0000  1.0000  1.0000  1.0000  1.0000  1.0000  1.0000  1.0000
}\SecondConNFiftyCubicKnotsOne

\pgfplotstableread{
delta alpha   P1      P2      P3      L1      L2      L3      C1      C2      C3
0     0.05  0.0680  0.0670  0.0670  0.0680  0.0670  0.0670  0.0690  0.0683  0.0677
1     0.05  0.1373  0.1367  0.1367  0.1373  0.1367  0.1367  0.1383  0.1380  0.1373
2     0.05  0.2623  0.2620  0.2620  0.2623  0.2620  0.2620  0.2627  0.2627  0.2623
3     0.05  0.4540  0.4537  0.4537  0.4540  0.4540  0.4537  0.4547  0.4543  0.4540
4     0.05  0.6843  0.6830  0.6830  0.6837  0.6837  0.6830  0.6860  0.6850  0.6837
5     0.05  0.8677  0.8677  0.8677  0.8677  0.8680  0.8677  0.8680  0.8680  0.8677
6     0.05  0.9720  0.9713  0.9710  0.9717  0.9710  0.9720  0.9717  0.9720  0.9720
7     0.05  0.9973  0.9973  0.9973  0.9973  0.9973  0.9973  0.9973  0.9973  0.9973
8     0.05  0.9997  0.9997  0.9997  0.9997  0.9997  0.9997  0.9997  0.9997  0.9997
9     0.05  1.0000  1.0000  1.0000  1.0000  1.0000  1.0000  1.0000  1.0000  1.0000
10    0.05  1.0000  1.0000  1.0000  1.0000  1.0000  1.0000  1.0000  1.0000  1.0000
}\SecondConNFiftyCubicKnotsTwo

\pgfplotstableread{
delta alpha   P1      P2      P3      L1      L2      L3      C1      C2      C3
0     0.05  0.0640  0.0623  0.0620  0.0630  0.0623  0.0620  0.0647  0.0640  0.0623
1     0.05  0.1080  0.1080  0.1070  0.1080  0.1080  0.1070  0.1090  0.1083  0.1080
2     0.05  0.1823  0.1807  0.1803  0.1817  0.1807  0.1803  0.1827  0.1823  0.1807
3     0.05  0.2850  0.2840  0.2837  0.2843  0.2840  0.2837  0.2860  0.2853  0.2840
4     0.05  0.4153  0.4143  0.4140  0.4147  0.4143  0.4140  0.4160  0.4153  0.4143
5     0.05  0.5577  0.5573  0.5563  0.5577  0.5573  0.5563  0.5577  0.5577  0.5573
6     0.05  0.7070  0.7070  0.7070  0.7070  0.7070  0.7070  0.7070  0.7070  0.7070
7     0.05  0.8223  0.8220  0.8220  0.8220  0.8220  0.8220  0.8223  0.8223  0.8220
8     0.05  0.9110  0.9110  0.9110  0.9110  0.9110  0.9110  0.9110  0.9110  0.9110
9     0.05  0.9643  0.9643  0.9643  0.9643  0.9643  0.9643  0.9643  0.9643  0.9643
10    0.05  0.9897  0.9897  0.9897  0.9897  0.9897  0.9897  0.9897  0.9897  0.9897
}\SecondMonConNFiveQuadraticKnotsZero

\pgfplotstableread{
delta alpha   P1      P2      P3      L1      L2      L3      C1      C2      C3
0     0.05  0.0677  0.0663  0.0663  0.0670  0.0663  0.0663  0.0693  0.0677  0.0663
1     0.05  0.0987  0.0977  0.0970  0.0980  0.0977  0.0970  0.0993  0.0987  0.0977
2     0.05  0.1363  0.1360  0.1353  0.1360  0.1360  0.1353  0.1363  0.1363  0.1360
3     0.05  0.1800  0.1797  0.1787  0.1797  0.1793  0.1787  0.1803  0.1800  0.1797
4     0.05  0.2550  0.2540  0.2540  0.2543  0.2540  0.2540  0.2557  0.2550  0.2543
5     0.05  0.3500  0.3500  0.3500  0.3497  0.3500  0.3500  0.3507  0.3500  0.3500
6     0.05  0.4667  0.4667  0.4667  0.4667  0.4667  0.4667  0.4667  0.4667  0.4667
7     0.05  0.5843  0.5843  0.5843  0.5843  0.5843  0.5843  0.5843  0.5843  0.5843
8     0.05  0.7013  0.7013  0.7013  0.7013  0.7013  0.7013  0.7013  0.7013  0.7013
9     0.05  0.7980  0.7980  0.7980  0.7980  0.7980  0.7980  0.7980  0.7980  0.7980
10    0.05  0.8817  0.8817  0.8817  0.8817  0.8817  0.8817  0.8817  0.8817  0.8817
}\SecondMonConNFiveQuadraticKnotsOne

\pgfplotstableread{
delta alpha   P1      P2      P3      L1      L2      L3      C1      C2      C3
0     0.05  0.0830  0.0823  0.0817  0.0830  0.0823  0.0817  0.0837  0.0830  0.0827
1     0.05  0.1003  0.1000  0.0997  0.1000  0.1000  0.0997  0.1010  0.1003  0.1000
2     0.05  0.1250  0.1243  0.1240  0.1243  0.1243  0.1240  0.1250  0.1250  0.1243
3     0.05  0.1553  0.1553  0.1550  0.1553  0.1553  0.1550  0.1553  0.1553  0.1553
4     0.05  0.2043  0.2043  0.2043  0.2043  0.2043  0.2043  0.2043  0.2043  0.2043
5     0.05  0.2700  0.2700  0.2700  0.2700  0.2700  0.2700  0.2700  0.2700  0.2700
6     0.05  0.3527  0.3520  0.3520  0.3520  0.3520  0.3520  0.3530  0.3527  0.3520
7     0.05  0.4383  0.4383  0.4383  0.4383  0.4383  0.4383  0.4383  0.4383  0.4383
8     0.05  0.5440  0.5440  0.5440  0.5440  0.5440  0.5440  0.5440  0.5440  0.5440
9     0.05  0.6500  0.6500  0.6500  0.6500  0.6500  0.6500  0.6500  0.6500  0.6500
10    0.05  0.7520  0.7520  0.7520  0.7520  0.7520  0.7520  0.7520  0.7520  0.7520
}\SecondMonConNFiveQuadraticKnotsTwo

\pgfplotstableread{
delta alpha   P1      P2      P3      L1      L2      L3      C1      C2      C3
0     0.05  0.0703  0.0697  0.0693  0.0700  0.0697  0.0693  0.0703  0.0703  0.0697
1     0.05  0.1000  0.1000  0.0997  0.1000  0.1000  0.0993  0.1003  0.1003  0.1000
2     0.05  0.1370  0.1363  0.1363  0.1367  0.1363  0.1363  0.1373  0.1370  0.1363
3     0.05  0.1947  0.1947  0.1930  0.1947  0.1943  0.1930  0.1950  0.1947  0.1947
4     0.05  0.2647  0.2647  0.2647  0.2647  0.2647  0.2647  0.2650  0.2647  0.2647
5     0.05  0.3737  0.3733  0.3733  0.3733  0.3733  0.3733  0.3737  0.3737  0.3733
6     0.05  0.4857  0.4853  0.4853  0.4853  0.4853  0.4853  0.4860  0.4857  0.4853
7     0.05  0.6007  0.6007  0.6007  0.6007  0.6007  0.6007  0.6007  0.6007  0.6007
8     0.05  0.7167  0.7163  0.7163  0.7167  0.7163  0.7163  0.7167  0.7167  0.7163
9     0.05  0.8153  0.8153  0.8153  0.8153  0.8153  0.8153  0.8153  0.8153  0.8153
10    0.05  0.8950  0.8950  0.8950  0.8950  0.8950  0.8950  0.8950  0.8950  0.8950
}\SecondMonConNFiveCubicKnotsZero

\pgfplotstableread{
delta alpha   P1      P2      P3      L1      L2      L3      C1      C2      C3
0     0.05  0.0860  0.0850  0.0850  0.0850  0.0850  0.0850  0.0860  0.0860  0.0850
1     0.05  0.0987  0.0987  0.0987  0.0987  0.0987  0.0987  0.0993  0.0990  0.0987
2     0.05  0.1280  0.1273  0.1267  0.1277  0.1273  0.1267  0.1283  0.1280  0.1273
3     0.05  0.1627  0.1627  0.1627  0.1627  0.1627  0.1627  0.1630  0.1627  0.1627
4     0.05  0.2130  0.2120  0.2117  0.2120  0.2120  0.2117  0.2130  0.2130  0.2120
5     0.05  0.2807  0.2800  0.2797  0.2803  0.2800  0.2797  0.2807  0.2807  0.2800
6     0.05  0.3600  0.3600  0.3597  0.3600  0.3600  0.3597  0.3600  0.3600  0.3600
7     0.05  0.4513  0.4513  0.4510  0.4513  0.4513  0.4510  0.4513  0.4513  0.4513
8     0.05  0.5677  0.5677  0.5677  0.5677  0.5677  0.5677  0.5677  0.5677  0.5677
9     0.05  0.6803  0.6800  0.6800  0.6800  0.6800  0.6800  0.6803  0.6803  0.6800
10    0.05  0.7730  0.7730  0.7727  0.7727  0.7727  0.7727  0.7730  0.7730  0.7730
}\SecondMonConNFiveCubicKnotsOne

\pgfplotstableread{
delta alpha   P1      P2      P3      L1      L2      L3      C1      C2      C3
0     0.05  0.0867  0.0863  0.0860  0.0863  0.0860  0.0860  0.0870  0.0867  0.0863
1     0.05  0.1040  0.1033  0.1033  0.1033  0.1033  0.1033  0.1040  0.1040  0.1033
2     0.05  0.1233  0.1227  0.1227  0.1227  0.1227  0.1227  0.1233  0.1233  0.1227
3     0.05  0.1550  0.1547  0.1543  0.1547  0.1547  0.1543  0.1553  0.1550  0.1547
4     0.05  0.1933  0.1933  0.1933  0.1933  0.1933  0.1933  0.1933  0.1933  0.1933
5     0.05  0.2453  0.2453  0.2450  0.2453  0.2453  0.2450  0.2453  0.2453  0.2453
6     0.05  0.3183  0.3180  0.3180  0.3180  0.3180  0.3180  0.3187  0.3183  0.3180
7     0.05  0.3973  0.3973  0.3970  0.3973  0.3970  0.3970  0.3973  0.3973  0.3973
8     0.05  0.4870  0.4867  0.4867  0.4867  0.4867  0.4867  0.4870  0.4870  0.4867
9     0.05  0.5847  0.5847  0.5847  0.5847  0.5847  0.5847  0.5847  0.5847  0.5847
10    0.05  0.6840  0.6840  0.6840  0.6840  0.6840  0.6840  0.6840  0.6840  0.6840
}\SecondMonConNFiveCubicKnotsTwo

\pgfplotstableread{
delta alpha   P1      P2      P3      L1      L2      L3      C1      C2      C3
0     0.05  0.0630  0.0617  0.0613  0.0620  0.0620  0.0613  0.0637  0.0630  0.0620
1     0.05  0.1147  0.1143  0.1143  0.1147  0.1143  0.1143  0.1153  0.1147  0.1143
2     0.05  0.2283  0.2277  0.2277  0.2283  0.2277  0.2277  0.2290  0.2283  0.2283
3     0.05  0.3740  0.3737  0.3737  0.3737  0.3737  0.3737  0.3743  0.3743  0.3737
4     0.05  0.5457  0.5453  0.5453  0.5453  0.5453  0.5453  0.5457  0.5457  0.5453
5     0.05  0.7290  0.7287  0.7283  0.7287  0.7287  0.7283  0.7290  0.7290  0.7287
6     0.05  0.8597  0.8597  0.8597  0.8597  0.8597  0.8597  0.8600  0.8597  0.8597
7     0.05  0.9450  0.9450  0.9450  0.9450  0.9450  0.9450  0.9450  0.9450  0.9450
8     0.05  0.9820  0.9820  0.9820  0.9820  0.9820  0.9820  0.9820  0.9820  0.9820
9     0.05  0.9957  0.9957  0.9957  0.9957  0.9957  0.9957  0.9957  0.9957  0.9957
10    0.05  0.9987  0.9987  0.9987  0.9987  0.9987  0.9987  0.9987  0.9987  0.9987
}\SecondMonConNSevenQuadraticKnotsZero

\pgfplotstableread{
delta alpha   P1      P2      P3      L1      L2      L3      C1      C2      C3
0     0.05  0.0740  0.0733  0.0733  0.0737  0.0733  0.0733  0.0740  0.0740  0.0733
1     0.05  0.1133  0.1130  0.1130  0.1133  0.1130  0.1130  0.1133  0.1133  0.1130
2     0.05  0.1633  0.1630  0.1627  0.1630  0.1630  0.1627  0.1633  0.1633  0.1630
3     0.05  0.2473  0.2467  0.2463  0.2473  0.2470  0.2463  0.2473  0.2473  0.2473
4     0.05  0.3537  0.3533  0.3533  0.3537  0.3537  0.3533  0.3543  0.3540  0.3537
5     0.05  0.4853  0.4850  0.4850  0.4853  0.4853  0.4850  0.4857  0.4857  0.4853
6     0.05  0.6243  0.6243  0.6243  0.6243  0.6247  0.6243  0.6243  0.6247  0.6243
7     0.05  0.7547  0.7547  0.7547  0.7547  0.7547  0.7547  0.7547  0.7547  0.7547
8     0.05  0.8587  0.8587  0.8587  0.8587  0.8587  0.8587  0.8587  0.8587  0.8587
9     0.05  0.9320  0.9320  0.9320  0.9320  0.9320  0.9320  0.9320  0.9320  0.9320
10    0.05  0.9743  0.9743  0.9743  0.9743  0.9743  0.9743  0.9743  0.9743  0.9743
}\SecondMonConNSevenQuadraticKnotsOne

\pgfplotstableread{
delta alpha   P1      P2      P3      L1      L2      L3      C1      C2      C3
0     0.05  0.0777  0.0770  0.0760  0.0770  0.0770  0.0760  0.0780  0.0777  0.0770
1     0.05  0.1013  0.1003  0.1003  0.1010  0.1007  0.1003  0.1020  0.1017  0.1007
2     0.05  0.1380  0.1377  0.1373  0.1377  0.1377  0.1373  0.1380  0.1380  0.1377
3     0.05  0.1903  0.1897  0.1897  0.1903  0.1897  0.1897  0.1903  0.1903  0.1903
4     0.05  0.2630  0.2623  0.2623  0.2627  0.2623  0.2623  0.2633  0.2630  0.2623
5     0.05  0.3587  0.3583  0.3583  0.3587  0.3583  0.3583  0.3587  0.3587  0.3583
6     0.05  0.4883  0.4883  0.4883  0.4883  0.4883  0.4883  0.4883  0.4883  0.4883
7     0.05  0.6177  0.6177  0.6177  0.6177  0.6177  0.6177  0.6177  0.6177  0.6177
8     0.05  0.7330  0.7330  0.7330  0.7330  0.7330  0.7330  0.7330  0.7330  0.7330
9     0.05  0.8307  0.8307  0.8307  0.8307  0.8307  0.8307  0.8307  0.8307  0.8307
10    0.05  0.9063  0.9063  0.9063  0.9063  0.9063  0.9063  0.9063  0.9063  0.9063
}\SecondMonConNSevenQuadraticKnotsTwo

\pgfplotstableread{
delta alpha   P1      P2      P3      L1      L2      L3      C1      C2      C3
0     0.05  0.0687  0.0667  0.0663  0.0680  0.0670  0.0663  0.0707  0.0690  0.0673
1     0.05  0.1083  0.1077  0.1067  0.1080  0.1077  0.1067  0.1100  0.1083  0.1077
2     0.05  0.1693  0.1687  0.1683  0.1687  0.1687  0.1683  0.1707  0.1693  0.1687
3     0.05  0.2593  0.2583  0.2583  0.2590  0.2583  0.2583  0.2603  0.2593  0.2583
4     0.05  0.3703  0.3700  0.3700  0.3700  0.3700  0.3700  0.3707  0.3703  0.3700
5     0.05  0.5070  0.5070  0.5070  0.5070  0.5070  0.5070  0.5070  0.5070  0.5070
6     0.05  0.6427  0.6427  0.6427  0.6427  0.6427  0.6427  0.6427  0.6427  0.6427
7     0.05  0.7620  0.7620  0.7620  0.7620  0.7620  0.7620  0.7623  0.7620  0.7620
8     0.05  0.8683  0.8683  0.8683  0.8683  0.8683  0.8683  0.8683  0.8683  0.8683
9     0.05  0.9387  0.9387  0.9387  0.9387  0.9387  0.9387  0.9387  0.9387  0.9387
10    0.05  0.9750  0.9750  0.9750  0.9753  0.9750  0.9750  0.9750  0.9750  0.9753
}\SecondMonConNSevenCubicKnotsZero

\pgfplotstableread{
delta alpha   P1      P2      P3      L1      L2      L3      C1      C2      C3
0     0.05  0.0783  0.0770  0.0767  0.0777  0.0770  0.0767  0.0797  0.0787  0.0773
1     0.05  0.1020  0.1013  0.1010  0.1013  0.1013  0.1010  0.1030  0.1027  0.1013
2     0.05  0.1390  0.1373  0.1373  0.1383  0.1377  0.1373  0.1393  0.1390  0.1377
3     0.05  0.1963  0.1960  0.1960  0.1963  0.1960  0.1960  0.1967  0.1963  0.1960
4     0.05  0.2803  0.2803  0.2803  0.2803  0.2803  0.2803  0.2807  0.2807  0.2803
5     0.05  0.3833  0.3833  0.3833  0.3833  0.3833  0.3833  0.3833  0.3833  0.3833
6     0.05  0.5083  0.5083  0.5083  0.5083  0.5083  0.5083  0.5083  0.5083  0.5083
7     0.05  0.6380  0.6387  0.6380  0.6387  0.6387  0.6380  0.6380  0.6383  0.6387
8     0.05  0.7557  0.7557  0.7557  0.7557  0.7557  0.7557  0.7557  0.7557  0.7557
9     0.05  0.8467  0.8467  0.8467  0.8467  0.8467  0.8467  0.8467  0.8467  0.8467
10    0.05  0.9207  0.9207  0.9207  0.9207  0.9207  0.9207  0.9207  0.9207  0.9207
}\SecondMonConNSevenCubicKnotsOne

\pgfplotstableread{
delta alpha   P1      P2      P3      L1      L2      L3      C1      C2      C3
0     0.05  0.0790  0.0787  0.0787  0.0790  0.0790  0.0787  0.0790  0.0790  0.0790
1     0.05  0.1013  0.1013  0.1013  0.1013  0.1013  0.1013  0.1030  0.1017  0.1013
2     0.05  0.1350  0.1333  0.1330  0.1333  0.1333  0.1330  0.1357  0.1350  0.1333
3     0.05  0.1770  0.1770  0.1770  0.1770  0.1770  0.1770  0.1773  0.1770  0.1770
4     0.05  0.2433  0.2430  0.2430  0.2430  0.2430  0.2430  0.2437  0.2433  0.2430
5     0.05  0.3303  0.3300  0.3300  0.3300  0.3300  0.3300  0.3307  0.3307  0.3300
6     0.05  0.4293  0.4293  0.4293  0.4293  0.4293  0.4293  0.4293  0.4293  0.4293
7     0.05  0.5430  0.5430  0.5427  0.5427  0.5430  0.5427  0.5430  0.5430  0.5430
8     0.05  0.6580  0.6577  0.6577  0.6577  0.6577  0.6577  0.6580  0.6580  0.6577
9     0.05  0.7633  0.7633  0.7633  0.7633  0.7633  0.7633  0.7633  0.7637  0.7633
10    0.05  0.8507  0.8507  0.8507  0.8507  0.8507  0.8507  0.8507  0.8507  0.8507
}\SecondMonConNSevenCubicKnotsTwo

\pgfplotstableread{
delta alpha   P1      P2      P3      L1      L2      L3      C1      C2      C3
0     0.05  0.0530  0.0523  0.0523  0.0530  0.0523  0.0523  0.0540  0.0530  0.0527
1     0.05  0.1220  0.1207  0.1207  0.1207  0.1207  0.1207  0.1223  0.1220  0.1207
2     0.05  0.2387  0.2380  0.2377  0.2387  0.2380  0.2377  0.2400  0.2390  0.2383
3     0.05  0.4260  0.4253  0.4253  0.4257  0.4257  0.4253  0.4267  0.4263  0.4260
4     0.05  0.6433  0.6433  0.6433  0.6433  0.6433  0.6433  0.6433  0.6433  0.6433
5     0.05  0.8260  0.8260  0.8260  0.8260  0.8260  0.8260  0.8260  0.8260  0.8260
6     0.05  0.9403  0.9403  0.9403  0.9403  0.9403  0.9403  0.9403  0.9403  0.9403
7     0.05  0.9800  0.9800  0.9800  0.9800  0.9800  0.9800  0.9800  0.9800  0.9800
8     0.05  0.9950  0.9950  0.9950  0.9950  0.9950  0.9950  0.9950  0.9950  0.9950
9     0.05  0.9990  0.9990  0.9990  0.9990  0.9990  0.9990  0.9990  0.9990  0.9990
10    0.05  1.0000  1.0000  1.0000  1.0000  1.0000  1.0000  1.0000  1.0000  1.0000
}\SecondMonConNTenQuadraticKnotsZero

\pgfplotstableread{
delta alpha   P1      P2      P3      L1      L2      L3      C1      C2      C3
0     0.05  0.0580  0.0577  0.0577  0.0577  0.0577  0.0577  0.0593  0.0583  0.0577
1     0.05  0.0980  0.0977  0.0973  0.0977  0.0977  0.0973  0.0983  0.0983  0.0977
2     0.05  0.1517  0.1513  0.1513  0.1517  0.1513  0.1513  0.1520  0.1517  0.1517
3     0.05  0.2613  0.2610  0.2607  0.2610  0.2610  0.2607  0.2617  0.2613  0.2610
4     0.05  0.3957  0.3960  0.3957  0.3960  0.3960  0.3953  0.3960  0.3960  0.3957
5     0.05  0.5570  0.5567  0.5563  0.5567  0.5563  0.5563  0.5570  0.5570  0.5567
6     0.05  0.7337  0.7337  0.7337  0.7337  0.7337  0.7337  0.7337  0.7337  0.7337
7     0.05  0.8697  0.8697  0.8697  0.8697  0.8697  0.8697  0.8697  0.8697  0.8697
8     0.05  0.9530  0.9530  0.9530  0.9530  0.9530  0.9530  0.9530  0.9530  0.9530
9     0.05  0.9833  0.9833  0.9833  0.9833  0.9833  0.9833  0.9833  0.9833  0.9833
10    0.05  0.9953  0.9953  0.9953  0.9953  0.9953  0.9953  0.9953  0.9953  0.9953
}\SecondMonConNTenQuadraticKnotsOne

\pgfplotstableread{
delta alpha   P1      P2      P3      L1      L2      L3      C1      C2      C3
0     0.05  0.0590  0.0583  0.0583  0.0590  0.0587  0.0583  0.0603  0.0590  0.0590
1     0.05  0.0867  0.0857  0.0853  0.0867  0.0860  0.0857  0.0870  0.0870  0.0863
2     0.05  0.1223  0.1220  0.1217  0.1223  0.1223  0.1217  0.1223  0.1223  0.1223
3     0.05  0.1870  0.1870  0.1867  0.1870  0.1867  0.1870  0.1873  0.1867  0.1867
4     0.05  0.2887  0.2887  0.2890  0.2890  0.2890  0.2887  0.2887  0.2887  0.2887
5     0.05  0.4073  0.4070  0.4070  0.4070  0.4070  0.4070  0.4073  0.4073  0.4070
6     0.05  0.5637  0.5637  0.5637  0.5637  0.5637  0.5637  0.5637  0.5637  0.5637
7     0.05  0.7160  0.7160  0.7160  0.7160  0.7160  0.7160  0.7160  0.7160  0.7160
8     0.05  0.8460  0.8460  0.8460  0.8460  0.8460  0.8460  0.8460  0.8460  0.8460
9     0.05  0.9317  0.9317  0.9317  0.9317  0.9317  0.9317  0.9317  0.9317  0.9317
10    0.05  0.9730  0.9730  0.9730  0.9730  0.9730  0.9730  0.9730  0.9730  0.9730
}\SecondMonConNTenQuadraticKnotsTwo

\pgfplotstableread{
delta alpha   P1      P2      P3      L1      L2      L3      C1      C2      C3
0     0.05  0.0567  0.0563  0.0563  0.0567  0.0563  0.0563  0.0570  0.0567  0.0567
1     0.05  0.0947  0.0943  0.0943  0.0940  0.0943  0.0943  0.0947  0.0947  0.0943
2     0.05  0.1640  0.1637  0.1633  0.1640  0.1637  0.1633  0.1643  0.1640  0.1637
3     0.05  0.2757  0.2750  0.2747  0.2753  0.2753  0.2747  0.2763  0.2757  0.2753
4     0.05  0.4137  0.4133  0.4133  0.4133  0.4133  0.4133  0.4137  0.4137  0.4133
5     0.05  0.5863  0.5860  0.5860  0.5863  0.5860  0.5860  0.5863  0.5863  0.5860
6     0.05  0.7513  0.7513  0.7513  0.7513  0.7513  0.7513  0.7513  0.7513  0.7513
7     0.05  0.8717  0.8717  0.8717  0.8717  0.8717  0.8717  0.8717  0.8717  0.8717
8     0.05  0.9533  0.9533  0.9533  0.9533  0.9533  0.9533  0.9533  0.9533  0.9533
9     0.05  0.9833  0.9833  0.9833  0.9833  0.9833  0.9833  0.9833  0.9833  0.9833
10    0.05  0.9953  0.9953  0.9953  0.9953  0.9953  0.9953  0.9953  0.9953  0.9953
}\SecondMonConNTenCubicKnotsZero

\pgfplotstableread{
delta alpha   P1      P2      P3      L1      L2      L3      C1      C2      C3
0     0.05  0.0593  0.0587  0.0587  0.0587  0.0587  0.0587  0.0603  0.0597  0.0587
1     0.05  0.0840  0.0833  0.0833  0.0833  0.0833  0.0833  0.0840  0.0840  0.0833
2     0.05  0.1303  0.1297  0.1297  0.1300  0.1300  0.1297  0.1303  0.1303  0.1300
3     0.05  0.1970  0.1967  0.1967  0.1967  0.1967  0.1967  0.1970  0.1970  0.1967
4     0.05  0.2997  0.2997  0.2993  0.2997  0.2997  0.2993  0.2997  0.2997  0.2997
5     0.05  0.4290  0.4287  0.4287  0.4290  0.4287  0.4287  0.4290  0.4290  0.4287
6     0.05  0.5853  0.5850  0.5850  0.5850  0.5850  0.5850  0.5853  0.5853  0.5850
7     0.05  0.7370  0.7370  0.7370  0.7370  0.7370  0.7370  0.7373  0.7370  0.7370
8     0.05  0.8623  0.8623  0.8623  0.8623  0.8623  0.8623  0.8623  0.8623  0.8623
9     0.05  0.9423  0.9420  0.9420  0.9420  0.9420  0.9420  0.9420  0.9420  0.9420
10    0.05  0.9783  0.9783  0.9783  0.9783  0.9783  0.9783  0.9783  0.9783  0.9783
}\SecondMonConNTenCubicKnotsOne

\pgfplotstableread{
delta alpha   P1      P2      P3      L1      L2      L3      C1      C2      C3
0     0.05  0.0643  0.0640  0.0640  0.0640  0.0640  0.0640  0.0647  0.0643  0.0640
1     0.05  0.0853  0.0847  0.0843  0.0850  0.0847  0.0843  0.0860  0.0853  0.0850
2     0.05  0.1230  0.1227  0.1227  0.1230  0.1227  0.1227  0.1233  0.1233  0.1227
3     0.05  0.1783  0.1783  0.1783  0.1783  0.1783  0.1783  0.1790  0.1787  0.1783
4     0.05  0.2600  0.2600  0.2600  0.2600  0.2600  0.2600  0.2600  0.2600  0.2600
5     0.05  0.3677  0.3680  0.3677  0.3680  0.3677  0.3680  0.3680  0.3683  0.3680
6     0.05  0.5043  0.5043  0.5043  0.5043  0.5043  0.5043  0.5043  0.5043  0.5043
7     0.05  0.6423  0.6423  0.6423  0.6423  0.6423  0.6423  0.6427  0.6427  0.6423
8     0.05  0.7697  0.7693  0.7693  0.7693  0.7693  0.7693  0.7697  0.7697  0.7693
9     0.05  0.8773  0.8773  0.8773  0.8773  0.8773  0.8773  0.8773  0.8773  0.8773
10    0.05  0.9400  0.9400  0.9400  0.9400  0.9400  0.9400  0.9400  0.9400  0.9400
}\SecondMonConNTenCubicKnotsTwo

\pgfplotstableread{
delta alpha   P1      P2      P3      L1      L2      L3      C1      C2      C3
0     0.05  0.0633  0.0633  0.0633  0.0633  0.0633  0.0633  0.0640  0.0637  0.0633
1     0.05  0.2763  0.2763  0.2767  0.2763  0.2767  0.2767  0.2777  0.2770  0.2763
2     0.05  0.7370  0.7367  0.7367  0.7363  0.7370  0.7363  0.7367  0.7367  0.7367
3     0.05  0.9763  0.9763  0.9763  0.9763  0.9763  0.9763  0.9763  0.9763  0.9763
4     0.05  1.0000  1.0000  1.0000  1.0000  1.0000  1.0000  1.0000  1.0000  1.0000
5     0.05  1.0000  1.0000  1.0000  1.0000  1.0000  1.0000  1.0000  1.0000  1.0000
6     0.05  1.0000  1.0000  1.0000  1.0000  1.0000  1.0000  1.0000  1.0000  1.0000
7     0.05  1.0000  1.0000  1.0000  1.0000  1.0000  1.0000  1.0000  1.0000  1.0000
8     0.05  1.0000  1.0000  1.0000  1.0000  1.0000  1.0000  1.0000  1.0000  1.0000
9     0.05  1.0000  1.0000  1.0000  1.0000  1.0000  1.0000  1.0000  1.0000  1.0000
10    0.05  1.0000  1.0000  1.0000  1.0000  1.0000  1.0000  1.0000  1.0000  1.0000
}\SecondMonConNFiftyQuadraticKnotsZero

\pgfplotstableread{
delta alpha   P1      P2      P3      L1      L2      L3      C1      C2      C3
0     0.05  0.0597  0.0590  0.0590  0.0597  0.0590  0.0590  0.0610  0.0610  0.0593
1     0.05  0.1717  0.1710  0.1710  0.1713  0.1710  0.1710  0.1717  0.1717  0.1710
2     0.05  0.4720  0.4720  0.4720  0.4720  0.4723  0.4720  0.4723  0.4723  0.4720
3     0.05  0.8290  0.8290  0.8290  0.8290  0.8290  0.8290  0.8290  0.8290  0.8290
4     0.05  0.9833  0.9833  0.9833  0.9833  0.9833  0.9833  0.9833  0.9833  0.9833
5     0.05  0.9997  0.9997  0.9997  0.9997  0.9997  0.9997  0.9997  0.9997  0.9997
6     0.05  1.0000  1.0000  1.0000  1.0000  1.0000  1.0000  1.0000  1.0000  1.0000
7     0.05  1.0000  1.0000  1.0000  1.0000  1.0000  1.0000  1.0000  1.0000  1.0000
8     0.05  1.0000  1.0000  1.0000  1.0000  1.0000  1.0000  1.0000  1.0000  1.0000
9     0.05  1.0000  1.0000  1.0000  1.0000  1.0000  1.0000  1.0000  1.0000  1.0000
10    0.05  1.0000  1.0000  1.0000  1.0000  1.0000  1.0000  1.0000  1.0000  1.0000
}\SecondMonConNFiftyQuadraticKnotsOne

\pgfplotstableread{
delta alpha   P1      P2      P3      L1      L2      L3      C1      C2      C3
0     0.05  0.0627  0.0620  0.0620  0.0627  0.0623  0.0620  0.0630  0.0627  0.0627
1     0.05  0.1403  0.1400  0.1400  0.1403  0.1400  0.1400  0.1410  0.1407  0.1403
2     0.05  0.3487  0.3487  0.3487  0.3487  0.3487  0.3487  0.3487  0.3487  0.3487
3     0.05  0.6753  0.6753  0.6753  0.6757  0.6750  0.6750  0.6757  0.6757  0.6757
4     0.05  0.9243  0.9243  0.9243  0.9243  0.9243  0.9243  0.9243  0.9243  0.9243
5     0.05  0.9963  0.9963  0.9963  0.9963  0.9963  0.9963  0.9963  0.9963  0.9963
6     0.05  1.0000  1.0000  1.0000  1.0000  1.0000  1.0000  1.0000  1.0000  1.0000
7     0.05  1.0000  1.0000  1.0000  1.0000  1.0000  1.0000  1.0000  1.0000  1.0000
8     0.05  1.0000  1.0000  1.0000  1.0000  1.0000  1.0000  1.0000  1.0000  1.0000
9     0.05  1.0000  1.0000  1.0000  1.0000  1.0000  1.0000  1.0000  1.0000  1.0000
10    0.05  1.0000  1.0000  1.0000  1.0000  1.0000  1.0000  1.0000  1.0000  1.0000
}\SecondMonConNFiftyQuadraticKnotsTwo

\pgfplotstableread{
delta alpha   P1      P2      P3      L1      L2      L3      C1      C2      C3
0     0.05  0.0543  0.0537  0.0537  0.0543  0.0537  0.0537  0.0547  0.0547  0.0543
1     0.05  0.1807  0.1803  0.1803  0.1807  0.1803  0.1803  0.1817  0.1810  0.1807
2     0.05  0.4913  0.4913  0.4910  0.4910  0.4913  0.4910  0.4910  0.4913  0.4910
3     0.05  0.8367  0.8367  0.8367  0.8367  0.8367  0.8367  0.8367  0.8367  0.8367
4     0.05  0.9853  0.9853  0.9853  0.9853  0.9853  0.9853  0.9853  0.9853  0.9853
5     0.05  0.9997  0.9997  0.9997  0.9997  0.9997  0.9997  0.9997  0.9997  0.9997
6     0.05  1.0000  1.0000  1.0000  1.0000  1.0000  1.0000  1.0000  1.0000  1.0000
7     0.05  1.0000  1.0000  1.0000  1.0000  1.0000  1.0000  1.0000  1.0000  1.0000
8     0.05  1.0000  1.0000  1.0000  1.0000  1.0000  1.0000  1.0000  1.0000  1.0000
9     0.05  1.0000  1.0000  1.0000  1.0000  1.0000  1.0000  1.0000  1.0000  1.0000
10    0.05  1.0000  1.0000  1.0000  1.0000  1.0000  1.0000  1.0000  1.0000  1.0000
}\SecondMonConNFiftyCubicKnotsZero

\pgfplotstableread{
delta alpha   P1      P2      P3      L1      L2      L3      C1      C2      C3
0     0.05  0.0660  0.0657  0.0657  0.0660  0.0657  0.0657  0.0667  0.0660  0.0660
1     0.05  0.1453  0.1450  0.1450  0.1450  0.1453  0.1453  0.1453  0.1457  0.1453
2     0.05  0.3670  0.3670  0.3670  0.3670  0.3670  0.3670  0.3673  0.3673  0.3670
3     0.05  0.7047  0.7047  0.7047  0.7047  0.7047  0.7047  0.7047  0.7047  0.7047
4     0.05  0.9353  0.9353  0.9353  0.9353  0.9353  0.9353  0.9353  0.9353  0.9353
5     0.05  0.9973  0.9973  0.9973  0.9973  0.9973  0.9973  0.9973  0.9973  0.9973
6     0.05  1.0000  1.0000  1.0000  1.0000  1.0000  1.0000  1.0000  1.0000  1.0000
7     0.05  1.0000  1.0000  1.0000  1.0000  1.0000  1.0000  1.0000  1.0000  1.0000
8     0.05  1.0000  1.0000  1.0000  1.0000  1.0000  1.0000  1.0000  1.0000  1.0000
9     0.05  1.0000  1.0000  1.0000  1.0000  1.0000  1.0000  1.0000  1.0000  1.0000
10    0.05  1.0000  1.0000  1.0000  1.0000  1.0000  1.0000  1.0000  1.0000  1.0000
}\SecondMonConNFiftyCubicKnotsOne

\pgfplotstableread{
delta alpha   P1      P2      P3      L1      L2      L3      C1      C2      C3
0     0.05  0.0570  0.0570  0.0570  0.0570  0.0570  0.0570  0.0580  0.0577  0.0570
1     0.05  0.1260  0.1257  0.1257  0.1260  0.1260  0.1257  0.1263  0.1260  0.1260
2     0.05  0.2957  0.2957  0.2953  0.2957  0.2960  0.2953  0.2957  0.2957  0.2960
3     0.05  0.6023  0.6023  0.6023  0.6023  0.6023  0.6023  0.6023  0.6023  0.6023
4     0.05  0.8740  0.8740  0.8740  0.8740  0.8740  0.8740  0.8740  0.8740  0.8740
5     0.05  0.9870  0.9870  0.9870  0.9870  0.9870  0.9870  0.9870  0.9870  0.9870
6     0.05  0.9997  0.9997  0.9997  0.9997  0.9997  0.9997  0.9997  0.9997  0.9997
7     0.05  1.0000  1.0000  1.0000  1.0000  1.0000  1.0000  1.0000  1.0000  1.0000
8     0.05  1.0000  1.0000  1.0000  1.0000  1.0000  1.0000  1.0000  1.0000  1.0000
9     0.05  1.0000  1.0000  1.0000  1.0000  1.0000  1.0000  1.0000  1.0000  1.0000
10    0.05  1.0000  1.0000  1.0000  1.0000  1.0000  1.0000  1.0000  1.0000  1.0000
}\SecondMonConNFiftyCubicKnotsTwo

\pgfplotstableread{
delta alpha   FiveOp  FiveUn  SevenOp  SevenUn  TenOp   TenUn
0     0.05    0.0610  0.0430  0.0660   0.0540   0.0540  0.0440
1     0.05    0.0920  0.0610  0.1000   0.0820   0.0960  0.0780
2     0.05    0.1300  0.0840  0.1670   0.1160   0.1520  0.1300
3     0.05    0.1960  0.1300  0.2430   0.1850   0.2380  0.1830
4     0.05    0.2600  0.1780  0.3450   0.2580   0.3800  0.2620
5     0.05    0.3460  0.2230  0.4750   0.3240   0.5210  0.3500
6     0.05    0.4560  0.2900  0.5810   0.4080   0.6530  0.4460
7     0.05    0.5590  0.3650  0.7200   0.4970   0.7980  0.5640
8     0.05    0.6780  0.4470  0.8230   0.5960   0.8970  0.6620
9     0.05    0.7780  0.5400  0.8900   0.7040   0.9720  0.7960
10    0.05    0.8410  0.6270  0.9410   0.7840   0.9930  0.8680
}\SecondLSWMon

\pgfplotstableread{
delta alpha  FiveOp  FiveUn  SevenOp  SevenUn   TenOp   TenUn
0     0.05   0.0590  0.0460  0.0710   0.0680    0.0620  0.0530
1     0.05   0.0660  0.0550  0.0760   0.0670    0.0610  0.0570
2     0.05   0.0780  0.0670  0.0880   0.0640    0.0700  0.0600
3     0.05   0.0840  0.0710  0.0960   0.0760    0.0800  0.0710
4     0.05   0.0990  0.0740  0.1110   0.0800    0.1030  0.0770
5     0.05   0.0950  0.0750  0.1180   0.0930    0.1160  0.0850
6     0.05   0.1050  0.0710  0.1300   0.0920    0.1320  0.0960
7     0.05   0.1150  0.0860  0.1490   0.1040    0.1540  0.1010
8     0.05   0.1320  0.0870  0.1720   0.1240    0.1740  0.1040
9     0.05   0.1400  0.0980  0.2070   0.1330    0.1920  0.1200
10    0.05   0.1600  0.1040  0.2290   0.1400    0.2370  0.1460
}\SecondLSWCon

\pgfplotstableread{
delta alpha  FiveOp  FiveUn  SevenOp   SevenUn  TenOp    TenUn
0     0.05   0.0150  0.0150  0.0120    0.0070   0.0190   0.0130
1     0.05   0.0150  0.0160  0.0120    0.0060   0.0210   0.0130
2     0.05   0.0170  0.0200  0.0130    0.0090   0.0210   0.0190
3     0.05   0.0200  0.0180  0.0140    0.0080   0.0230   0.0170
4     0.05   0.0180  0.0210  0.0150    0.0130   0.0200   0.0210
5     0.05   0.0210  0.0160  0.0160    0.0140   0.0220   0.0200
6     0.05   0.0210  0.0200  0.0170    0.0140   0.0260   0.0190
7     0.05   0.0210  0.0200  0.0160    0.0160   0.0240   0.0280
8     0.05   0.0230  0.0210  0.0190    0.0150   0.0290   0.0240
9     0.05   0.0270  0.0190  0.0220    0.0200   0.0330   0.0220
10    0.05   0.0310  0.0250  0.0300    0.0220   0.0370   0.0290
}\SecondLSWMonCon

\pgfplotstableread{
delta alpha   P1      P2      P3      L1      L2      L3      C1      C2      C3
0     0.05  0.0727  0.0723  0.0717  0.0727  0.0727  0.0717  0.0730  0.0727  0.0727
1     0.05  0.0760  0.0760  0.0760  0.0760  0.0760  0.0760  0.0767  0.0763  0.0760
2     0.05  0.0887  0.0883  0.0883  0.0883  0.0883  0.0883  0.0887  0.0887  0.0883
3     0.05  0.1150  0.1140  0.1140  0.1140  0.1140  0.1140  0.1153  0.1150  0.1140
4     0.05  0.1537  0.1533  0.1533  0.1537  0.1533  0.1533  0.1540  0.1537  0.1537
5     0.05  0.2253  0.2237  0.2237  0.2240  0.2237  0.2237  0.2260  0.2253  0.2237
6     0.05  0.3277  0.3263  0.3260  0.3267  0.3267  0.3260  0.3280  0.3277  0.3267
7     0.05  0.4537  0.4527  0.4527  0.4537  0.4530  0.4527  0.4543  0.4537  0.4533
8     0.05  0.6090  0.6080  0.6080  0.6087  0.6080  0.6080  0.6097  0.6090  0.6080
9     0.05  0.7420  0.7410  0.7407  0.7413  0.7410  0.7407  0.7433  0.7423  0.7410
10    0.05  0.8543  0.8533  0.8530  0.8540  0.8533  0.8530  0.8557  0.8547  0.8540
}\SlutskyNTenQuadraticKnotsZero

\pgfplotstableread{
delta alpha   P1      P2      P3      L1      L2      L3      C1      C2      C3
0     0.05  0.0920  0.0920  0.0920  0.0920  0.0920  0.0920  0.0920  0.0920  0.0920
1     0.05  0.1037  0.1037  0.1037  0.1037  0.1037  0.1037  0.1047  0.1040  0.1037
2     0.05  0.1087  0.1083  0.1083  0.1087  0.1083  0.1083  0.1090  0.1090  0.1087
3     0.05  0.1100  0.1097  0.1097  0.1097  0.1097  0.1097  0.1113  0.1107  0.1097
4     0.05  0.1273  0.1270  0.1270  0.1273  0.1270  0.1270  0.1280  0.1273  0.1273
5     0.05  0.1483  0.1477  0.1477  0.1480  0.1480  0.1477  0.1490  0.1483  0.1480
6     0.05  0.1703  0.1700  0.1700  0.1703  0.1703  0.1700  0.1713  0.1703  0.1703
7     0.05  0.2080  0.2070  0.2063  0.2077  0.2073  0.2063  0.2087  0.2080  0.2073
8     0.05  0.2567  0.2563  0.2560  0.2563  0.2563  0.2560  0.2570  0.2567  0.2563
9     0.05  0.3070  0.3063  0.3057  0.3067  0.3063  0.3057  0.3080  0.3073  0.3067
10    0.05  0.3833  0.3820  0.3817  0.3823  0.3820  0.3817  0.3840  0.3833  0.3820
}\SlutskyNTenQuadraticKnotsOne

\pgfplotstableread{
delta alpha   P1      P2      P3      L1      L2      L3      C1      C2      C3
0     0.05  0.0983  0.0983  0.0980  0.0983  0.0983  0.0980  0.0990  0.0983  0.0983
1     0.05  0.1093  0.1093  0.1090  0.1093  0.1093  0.1090  0.1093  0.1093  0.1093
2     0.05  0.1140  0.1137  0.1137  0.1140  0.1140  0.1137  0.1150  0.1147  0.1140
3     0.05  0.1230  0.1230  0.1230  0.1230  0.1230  0.1230  0.1233  0.1230  0.1230
4     0.05  0.1380  0.1377  0.1377  0.1380  0.1380  0.1377  0.1393  0.1387  0.1380
5     0.05  0.1587  0.1587  0.1580  0.1587  0.1587  0.1580  0.1590  0.1587  0.1587
6     0.05  0.1847  0.1843  0.1840  0.1847  0.1843  0.1840  0.1850  0.1850  0.1843
7     0.05  0.2260  0.2260  0.2253  0.2260  0.2260  0.2253  0.2263  0.2260  0.2260
8     0.05  0.2807  0.2803  0.2803  0.2807  0.2803  0.2803  0.2810  0.2807  0.2807
9     0.05  0.3400  0.3390  0.3387  0.3397  0.3390  0.3387  0.3413  0.3403  0.3390
10    0.05  0.4173  0.4153  0.4150  0.4163  0.4160  0.4150  0.4180  0.4180  0.4160
}\SlutskyNTenCubicKnotsZero

\pgfplotstableread{
delta alpha   P1      P2      P3      L1      L2      L3      C1      C2      C3
0     0.05  0.1553  0.1540  0.1533  0.1550  0.1540  0.1533  0.1563  0.1557  0.1550
1     0.05  0.1760  0.1753  0.1753  0.1760  0.1757  0.1753  0.1763  0.1763  0.1760
2     0.05  0.1747  0.1743  0.1743  0.1743  0.1743  0.1743  0.1753  0.1753  0.1743
3     0.05  0.1857  0.1847  0.1843  0.1847  0.1847  0.1843  0.1857  0.1857  0.1847
4     0.05  0.1963  0.1960  0.1960  0.1960  0.1960  0.1960  0.1963  0.1963  0.1960
5     0.05  0.2050  0.2050  0.2047  0.2050  0.2050  0.2047  0.2050  0.2050  0.2050
6     0.05  0.2150  0.2137  0.2137  0.2147  0.2140  0.2137  0.2153  0.2150  0.2147
7     0.05  0.2407  0.2400  0.2400  0.2403  0.2400  0.2400  0.2410  0.2407  0.2403
8     0.05  0.2627  0.2623  0.2620  0.2623  0.2623  0.2620  0.2633  0.2633  0.2623
9     0.05  0.2910  0.2903  0.2903  0.2910  0.2903  0.2903  0.2920  0.2910  0.2907
10    0.05  0.3260  0.3240  0.3240  0.3247  0.3240  0.3240  0.3267  0.3260  0.3243
}\SlutskyNTenCubicKnotsOne

\pgfplotstableread{
delta alpha   P1      P2      P3      L1      L2      L3      C1      C2      C3
0     0.05  0.0523  0.0523  0.0523  0.0523  0.0523  0.0523  0.0520  0.0520  0.0523
1     0.05  0.0633  0.0637  0.0637  0.0637  0.0637  0.0637  0.0633  0.0633  0.0637
2     0.05  0.0923  0.0923  0.0923  0.0923  0.0923  0.0923  0.0923  0.0923  0.0923
3     0.05  0.1507  0.1507  0.1507  0.1507  0.1507  0.1507  0.1500  0.1503  0.1507
4     0.05  0.2590  0.2593  0.2593  0.2590  0.2593  0.2593  0.2593  0.2593  0.2593
5     0.05  0.4340  0.4343  0.4343  0.4340  0.4343  0.4343  0.4333  0.4340  0.4343
6     0.05  0.6407  0.6410  0.6410  0.6407  0.6407  0.6410  0.6407  0.6407  0.6407
7     0.05  0.8343  0.8347  0.8347  0.8343  0.8347  0.8347  0.8350  0.8343  0.8347
8     0.05  0.9543  0.9543  0.9543  0.9543  0.9543  0.9543  0.9543  0.9543  0.9543
9     0.05  0.9927  0.9927  0.9927  0.9927  0.9927  0.9927  0.9927  0.9927  0.9927
10    0.05  1.0000  1.0000  1.0000  1.0000  1.0000  1.0000  1.0000  1.0000  1.0000
}\SlutskyNTwentyQuadraticKnotsZero

\pgfplotstableread{
delta alpha   P1      P2      P3      L1      L2      L3      C1      C2      C3
0     0.05  0.0693  0.0693  0.0693  0.0693  0.0693  0.0693  0.0693  0.0693  0.0693
1     0.05  0.0693  0.0693  0.0693  0.0693  0.0693  0.0693  0.0690  0.0690  0.0693
2     0.05  0.0810  0.0813  0.0813  0.0813  0.0813  0.0813  0.0810  0.0810  0.0813
3     0.05  0.0950  0.0950  0.0950  0.0950  0.0950  0.0950  0.0950  0.0950  0.0950
4     0.05  0.1250  0.1253  0.1253  0.1253  0.1253  0.1253  0.1250  0.1250  0.1253
5     0.05  0.1643  0.1653  0.1653  0.1650  0.1650  0.1653  0.1640  0.1640  0.1650
6     0.05  0.2197  0.2197  0.2197  0.2197  0.2197  0.2197  0.2197  0.2197  0.2197
7     0.05  0.3047  0.3047  0.3047  0.3047  0.3047  0.3047  0.3040  0.3043  0.3047
8     0.05  0.4140  0.4140  0.4140  0.4140  0.4140  0.4140  0.4140  0.4140  0.4140
9     0.05  0.5590  0.5593  0.5593  0.5590  0.5593  0.5593  0.5597  0.5597  0.5590
10    0.05  0.7137  0.7137  0.7137  0.7137  0.7137  0.7137  0.7137  0.7137  0.7137
}\SlutskyNTwentyQuadraticKnotsOne

\pgfplotstableread{
delta alpha   P1      P2      P3      L1      L2      L3      C1      C2      C3
0     0.05  0.0713  0.0713  0.0713  0.0713  0.0713  0.0713  0.0713  0.0713  0.0713
1     0.05  0.0670  0.0673  0.0673  0.0670  0.0670  0.0673  0.0673  0.0673  0.0670
2     0.05  0.0773  0.0777  0.0777  0.0777  0.0777  0.0777  0.0773  0.0777  0.0777
3     0.05  0.0980  0.0980  0.0980  0.0980  0.0980  0.0980  0.0983  0.0980  0.0980
4     0.05  0.1303  0.1303  0.1303  0.1303  0.1303  0.1303  0.1303  0.1303  0.1303
5     0.05  0.1820  0.1817  0.1817  0.1817  0.1817  0.1817  0.1813  0.1820  0.1817
6     0.05  0.2420  0.2420  0.2420  0.2420  0.2420  0.2420  0.2417  0.2420  0.2420
7     0.05  0.3413  0.3417  0.3417  0.3417  0.3417  0.3417  0.3413  0.3413  0.3417
8     0.05  0.4753  0.4757  0.4757  0.4757  0.4757  0.4757  0.4750  0.4750  0.4757
9     0.05  0.6287  0.6293  0.6293  0.6287  0.6293  0.6293  0.6293  0.6293  0.6287
10    0.05  0.7763  0.7760  0.7760  0.7763  0.7760  0.7760  0.7770  0.7767  0.7760
}\SlutskyNTwentyCubicKnotsZero

\pgfplotstableread{
delta alpha   P1      P2      P3      L1      L2      L3      C1      C2      C3
0     0.05  0.0880  0.0883  0.0883  0.0880  0.0880  0.0883  0.0880  0.0880  0.0880
1     0.05  0.0960  0.0960  0.0960  0.0960  0.0960  0.0960  0.0960  0.0960  0.0960
2     0.05  0.1017  0.1017  0.1017  0.1017  0.1017  0.1017  0.1017  0.1017  0.1017
3     0.05  0.1093  0.1093  0.1093  0.1093  0.1093  0.1093  0.1090  0.1093  0.1093
4     0.05  0.1240  0.1240  0.1240  0.1240  0.1240  0.1240  0.1240  0.1240  0.1240
5     0.05  0.1433  0.1433  0.1433  0.1433  0.1433  0.1433  0.1433  0.1433  0.1433
6     0.05  0.1817  0.1817  0.1817  0.1817  0.1817  0.1817  0.1817  0.1817  0.1817
7     0.05  0.2080  0.2083  0.2083  0.2080  0.2083  0.2083  0.2080  0.2080  0.2080
8     0.05  0.2507  0.2507  0.2507  0.2507  0.2507  0.2507  0.2503  0.2503  0.2507
9     0.05  0.3187  0.3190  0.3190  0.3187  0.3190  0.3190  0.3187  0.3187  0.3187
10    0.05  0.3857  0.3857  0.3857  0.3857  0.3857  0.3857  0.3853  0.3857  0.3857
}\SlutskyNTwentyCubicKnotsOne

\pgfplotstableread{
delta alpha   P1      P2      P3      L1      L2      L3      C1      C2      C3
0     0.05  0.0547  0.0540  0.0540  0.0547  0.0540  0.0540  0.0550  0.0547  0.0543
1     0.05  0.0713  0.0710  0.0710  0.0713  0.0710  0.0710  0.0717  0.0717  0.0710
2     0.05  0.1133  0.1127  0.1127  0.1133  0.1133  0.1127  0.1147  0.1133  0.1133
3     0.05  0.2203  0.2200  0.2200  0.2203  0.2200  0.2200  0.2220  0.2213  0.2203
4     0.05  0.4243  0.4237  0.4237  0.4240  0.4240  0.4237  0.4247  0.4247  0.4240
5     0.05  0.6893  0.6880  0.6880  0.6893  0.6887  0.6880  0.6897  0.6897  0.6893
6     0.05  0.8967  0.8960  0.8960  0.8960  0.8960  0.8960  0.8980  0.8967  0.8960
7     0.05  0.9803  0.9803  0.9803  0.9803  0.9803  0.9803  0.9807  0.9803  0.9803
8     0.05  0.9987  0.9987  0.9987  0.9987  0.9987  0.9987  0.9987  0.9987  0.9987
9     0.05  1.0000  1.0000  1.0000  1.0000  1.0000  1.0000  1.0000  1.0000  1.0000
10    0.05  1.0000  1.0000  1.0000  1.0000  1.0000  1.0000  1.0000  1.0000  1.0000
}\SlutskyNThirtyQuadraticKnotsZero

\pgfplotstableread{
delta alpha   P1      P2      P3      L1      L2      L3      C1      C2      C3
0     0.05  0.0650  0.0647  0.0647  0.0650  0.0647  0.0647  0.0650  0.0650  0.0650
1     0.05  0.0707  0.0700  0.0700  0.0703  0.0700  0.0700  0.0710  0.0707  0.0700
2     0.05  0.0847  0.0843  0.0843  0.0847  0.0843  0.0843  0.0847  0.0847  0.0847
3     0.05  0.1103  0.1097  0.1097  0.1103  0.1100  0.1097  0.1107  0.1107  0.1103
4     0.05  0.1630  0.1627  0.1627  0.1630  0.1627  0.1627  0.1630  0.1630  0.1630
5     0.05  0.2320  0.2313  0.2313  0.2320  0.2320  0.2313  0.2330  0.2323  0.2320
6     0.05  0.3440  0.3433  0.3433  0.3440  0.3437  0.3433  0.3450  0.3450  0.3440
7     0.05  0.5070  0.5047  0.5047  0.5067  0.5063  0.5047  0.5077  0.5073  0.5067
8     0.05  0.6883  0.6880  0.6880  0.6880  0.6880  0.6880  0.6890  0.6890  0.6880
9     0.05  0.8453  0.8447  0.8447  0.8450  0.8450  0.8447  0.8467  0.8457  0.8450
10    0.05  0.9483  0.9483  0.9483  0.9483  0.9483  0.9483  0.9493  0.9490  0.9483
}\SlutskyNThirtyQuadraticKnotsOne

\pgfplotstableread{
delta alpha   P1      P2      P3      L1      L2      L3      C1      C2      C3
0     0.05  0.0657  0.0650  0.0650  0.0653  0.0653  0.0650  0.0663  0.0660  0.0653
1     0.05  0.0673  0.0667  0.0667  0.0673  0.0670  0.0667  0.0673  0.0673  0.0673
2     0.05  0.0830  0.0830  0.0830  0.0830  0.0830  0.0830  0.0837  0.0837  0.0830
3     0.05  0.1123  0.1123  0.1123  0.1123  0.1123  0.1123  0.1123  0.1123  0.1123
4     0.05  0.1747  0.1740  0.1740  0.1747  0.1740  0.1740  0.1757  0.1757  0.1743
5     0.05  0.2603  0.2597  0.2597  0.2603  0.2600  0.2597  0.2613  0.2607  0.2600
6     0.05  0.3973  0.3963  0.3963  0.3970  0.3967  0.3963  0.3983  0.3983  0.3967
7     0.05  0.5740  0.5730  0.5730  0.5740  0.5730  0.5730  0.5757  0.5747  0.5740
8     0.05  0.7557  0.7553  0.7553  0.7557  0.7557  0.7553  0.7567  0.7560  0.7557
9     0.05  0.8960  0.8953  0.8953  0.8953  0.8953  0.8953  0.8973  0.8967  0.8953
10    0.05  0.9690  0.9680  0.9680  0.9690  0.9683  0.9680  0.9697  0.9693  0.9690
}\SlutskyNThirtyCubicKnotsZero

\pgfplotstableread{
delta alpha   P1      P2      P3      L1      L2      L3      C1      C2      C3
0     0.05  0.0840  0.0830  0.0830  0.0840  0.0833  0.0830  0.0843  0.0843  0.0840
1     0.05  0.0857  0.0857  0.0857  0.0857  0.0857  0.0857  0.0863  0.0860  0.0857
2     0.05  0.0917  0.0913  0.0913  0.0917  0.0913  0.0913  0.0917  0.0917  0.0917
3     0.05  0.1090  0.1087  0.1087  0.1090  0.1090  0.1087  0.1097  0.1093  0.1090
4     0.05  0.1303  0.1300  0.1300  0.1300  0.1300  0.1300  0.1303  0.1303  0.1300
5     0.05  0.1650  0.1647  0.1647  0.1650  0.1650  0.1647  0.1653  0.1653  0.1650
6     0.05  0.2090  0.2087  0.2087  0.2090  0.2087  0.2087  0.2090  0.2090  0.2087
7     0.05  0.2777  0.2770  0.2770  0.2773  0.2773  0.2770  0.2787  0.2780  0.2773
8     0.05  0.3620  0.3607  0.3607  0.3617  0.3610  0.3607  0.3623  0.3620  0.3617
9     0.05  0.4733  0.4723  0.4723  0.4730  0.4727  0.4723  0.4740  0.4733  0.4730
10    0.05  0.5987  0.5973  0.5973  0.5983  0.5977  0.5973  0.5997  0.5993  0.5980
}\SlutskyNThirtyCubicKnotsOne

\pgfplotstableread{
delta alpha   P1      P2      P3      L1      L2      L3      C1      C2      C3
0     0.05  0.0560  0.0560  0.0560  0.0560  0.0560  0.0560  0.0563  0.0563  0.0560
1     0.05  0.0757  0.0753  0.0753  0.0757  0.0757  0.0753  0.0757  0.0757  0.0757
2     0.05  0.1670  0.1667  0.1667  0.1670  0.1670  0.1667  0.1670  0.1670  0.1670
3     0.05  0.3743  0.3737  0.3737  0.3743  0.3743  0.3737  0.3750  0.3750  0.3743
4     0.05  0.7257  0.7247  0.7247  0.7253  0.7250  0.7247  0.7277  0.7273  0.7253
5     0.05  0.9457  0.9453  0.9453  0.9457  0.9453  0.9453  0.9463  0.9460  0.9457
6     0.05  0.9947  0.9947  0.9947  0.9947  0.9947  0.9947  0.9957  0.9950  0.9947
7     0.05  1.0000  1.0000  1.0000  1.0000  1.0000  1.0000  1.0000  1.0000  1.0000
8     0.05  1.0000  1.0000  1.0000  1.0000  1.0000  1.0000  1.0000  1.0000  1.0000
9     0.05  1.0000  1.0000  1.0000  1.0000  1.0000  1.0000  1.0000  1.0000  1.0000
10    0.05  1.0000  1.0000  1.0000  1.0000  1.0000  1.0000  1.0000  1.0000  1.0000
}\SlutskyNFiftyQuadraticKnotsZero

\pgfplotstableread{
delta alpha   P1      P2      P3      L1      L2      L3      C1      C2      C3
0     0.05  0.0673  0.0670  0.0670  0.0673  0.0670  0.0670  0.0680  0.0677  0.0673
1     0.05  0.0783  0.0783  0.0783  0.0783  0.0783  0.0783  0.0783  0.0783  0.0783
2     0.05  0.0967  0.0967  0.0967  0.0967  0.0967  0.0967  0.0970  0.0970  0.0967
3     0.05  0.1443  0.1443  0.1443  0.1443  0.1443  0.1443  0.1443  0.1443  0.1443
4     0.05  0.2423  0.2420  0.2420  0.2423  0.2423  0.2420  0.2427  0.2423  0.2423
5     0.05  0.3980  0.3970  0.3970  0.3977  0.3970  0.3970  0.3990  0.3983  0.3973
6     0.05  0.6373  0.6357  0.6357  0.6373  0.6367  0.6357  0.6390  0.6387  0.6373
7     0.05  0.8520  0.8513  0.8513  0.8520  0.8517  0.8513  0.8530  0.8527  0.8520
8     0.05  0.9640  0.9640  0.9640  0.9640  0.9640  0.9640  0.9640  0.9640  0.9640
9     0.05  0.9943  0.9943  0.9943  0.9943  0.9943  0.9943  0.9943  0.9943  0.9943
10    0.05  0.9993  0.9993  0.9993  0.9993  0.9993  0.9993  0.9993  0.9993  0.9993
}\SlutskyNFiftyQuadraticKnotsOne

\pgfplotstableread{
delta alpha   P1      P2      P3      L1      L2      L3      C1      C2      C3
0     0.05  0.0667  0.0660  0.0660  0.0663  0.0663  0.0660  0.0667  0.0667  0.0663
1     0.05  0.0750  0.0750  0.0750  0.0750  0.0750  0.0750  0.0750  0.0750  0.0750
2     0.05  0.1010  0.1010  0.1010  0.1010  0.1010  0.1010  0.1017  0.1013  0.1010
3     0.05  0.1583  0.1573  0.1573  0.1580  0.1573  0.1573  0.1587  0.1583  0.1580
4     0.05  0.2693  0.2677  0.2677  0.2690  0.2680  0.2677  0.2697  0.2697  0.2690
5     0.05  0.4600  0.4590  0.4590  0.4600  0.4593  0.4590  0.4620  0.4613  0.4593
6     0.05  0.7083  0.7080  0.7080  0.7080  0.7080  0.7080  0.7113  0.7097  0.7080
7     0.05  0.8993  0.8990  0.8990  0.8993  0.8990  0.8990  0.9007  0.8997  0.8993
8     0.05  0.9833  0.9833  0.9833  0.9833  0.9833  0.9833  0.9833  0.9833  0.9833
9     0.05  0.9987  0.9987  0.9987  0.9987  0.9987  0.9987  0.9987  0.9987  0.9987
10    0.05  0.9993  0.9993  0.9993  0.9993  0.9993  0.9993  0.9993  0.9993  0.9993
}\SlutskyNFiftyCubicKnotsZero

\pgfplotstableread{
delta alpha   P1      P2      P3      L1      L2      L3      C1      C2      C3
0     0.05  0.0670  0.0667  0.0667  0.0670  0.0670  0.0667  0.0677  0.0670  0.0670
1     0.05  0.0750  0.0747  0.0747  0.0750  0.0750  0.0747  0.0750  0.0750  0.0750
2     0.05  0.0933  0.0930  0.0930  0.0930  0.0930  0.0930  0.0933  0.0933  0.0930
3     0.05  0.1210  0.1210  0.1210  0.1210  0.1210  0.1210  0.1217  0.1213  0.1210
4     0.05  0.1567  0.1567  0.1567  0.1567  0.1567  0.1567  0.1573  0.1570  0.1567
5     0.05  0.2257  0.2257  0.2257  0.2257  0.2257  0.2257  0.2257  0.2257  0.2257
6     0.05  0.3267  0.3257  0.3257  0.3263  0.3263  0.3257  0.3287  0.3277  0.3263
7     0.05  0.4687  0.4677  0.4677  0.4687  0.4680  0.4677  0.4693  0.4690  0.4687
8     0.05  0.6310  0.6307  0.6307  0.6310  0.6310  0.6307  0.6323  0.6320  0.6310
9     0.05  0.8037  0.8037  0.8037  0.8037  0.8037  0.8037  0.8040  0.8037  0.8037
10    0.05  0.9177  0.9170  0.9170  0.9177  0.9173  0.9170  0.9180  0.9177  0.9177
}\SlutskyNFiftyCubicKnotsOne

\setcounter{table}{0}
\renewcommand{\thetable}{\thesection.\arabic{table}}

{
\scriptsize
\captionsetup[longtable]{name=Figure}

}

{
\scriptsize
\captionsetup[longtable]{name=Figure}
%
}

{
\scriptsize
\captionsetup[longtable]{name=Figure}
%
}

{
\def\showpgfcircle{\tikz[baseline=-0.5ex]\node[blue,mark size=2pt]{\pgfuseplotmark{halfcircle*}};}
\scriptsize
\captionsetup[longtable]{name=Figure}
%
}

{
\def\showpgfcircle{\tikz[baseline=-0.5ex]\node[blue,mark size=2pt]{\pgfuseplotmark{halfcircle*}};}
\scriptsize
\captionsetup[longtable]{name=Figure}
%
}

\end{appendices}

\clearpage 

\titleformat{\section}{\normalfont\Large\bfseries}{\thesection}{1em}{}

\addcontentsline{toc}{section}{References}
\putbib
\end{bibunit}


\end{document}